\RequirePackage{amsthm}
\documentclass[sn-mathphys,Numbered]{sn-jnl}% Math and Physical Sciences Reference Style

\usepackage{lineno}

\usepackage{geometry}
\usepackage{cancel}
\usepackage{tikz}
\usepackage{mathtools}
\definecolor{charcoal}{rgb}{0.21, 0.27, 0.31}
\usepackage{accents}
\usepackage{csquotes}
\newcommand{\ubar}[1]{\underaccent{\bar}{#1}}
\usepackage{graphicx}%
\usepackage{multirow}%
\usepackage{amsmath,amssymb,amsfonts,amsbsy,latexsym}%
\usepackage{amsthm}%
\usepackage{mathrsfs}%
\usepackage[title]{appendix}%
\usepackage{xcolor}%
\usepackage{textcomp}%
\usepackage{manyfoot}%
\usepackage{booktabs}%
\usepackage{algorithm}%
\usepackage{algorithmicx}%
\usepackage{algpseudocode}%
\usepackage{listings}%
\newcommand{\fatdot}{\raisebox{0.1ex}{\tikz\filldraw[black,x=1pt,y=1pt] (0,0) circle (0.7);}}

\newcommand{\R}{\mathbb{R}}

\newcommand{\gslash}{{\cancel{g}}}
\newcommand{\pslash}{\cancel{p}}

\newcommand{\pslashangle}{\cancel{\Phi}}

\renewcommand{\phi}{\varphi}

\DeclareMathOperator*{\sgn}{sgn}

\DeclareMathOperator*{\supp}{supp \,}

\DeclareMathOperator*{\vol}{vol}

\newcommand{\gERN}{\gRN}
\newcommand{\gRN}{g}
\newcommand{\OsqS}{\Osqrn}
\newcommand{\Osqern}{\Osqrn}
\newcommand{\Osqrn}{\Omega^2}
\newcommand{\OsqrnNEG}{\Omega^{-2}}
\newcommand{\Mschw}{\Mrn} % Schwarzschild exterior manifold
\newcommand{\Mern}{\Mrn} % ERN exterior manifold
\newcommand{\Mrn}{M}
\newcommand{\psuppconst}{C_p}
\newcommand{\rsuppconst}{C_r}
\newcommand{\bigc}{\mathfrak{C}}
\newcommand{\constbsl}{\mathcal{C}}
\newcommand{\decayrate}{\mathfrak{c}}
\newcommand{\decayrateERN}{\decayrate}
\newcommand{\indextrapped}{\mathrm{ph}}
\newcommand{\indexsmall}{{\mathcal{H}^+}}
\newcommand{\indexERNtrapped}{\indextrapped}
\newcommand{\indexERNsmall}{\indexsmall}
\newcommand{\trappedsupportset}{\mathfrak{Q}^\indextrapped}
\newcommand{\smallsupportset}{\mathfrak{Q}^\indexsmall}
\newcommand{\trappedsupportsetx}{\mathfrak{Q}^{\indextrapped}_{x}}
\newcommand{\smallsupportsetx}{\mathfrak{Q}^{\indexsmall}_{x}}
\newcommand{\slowsupportset}{\mathfrak{S}}
\newcommand{\slowsupportsettau}{\mathfrak{S}_\tauslow}
\newcommand{\slowsupportsetx}{\mathfrak{S}_{x}}
\newcommand{\slowsupportsettaux}{\mathfrak{S}_{\tauslow,x}}
\newcommand{\trapschw}{\mathfrak{t}}
\newcommand{\trapern}{\trapschw}
\newcommand{\ERNtrappedsupportset}{\mathfrak{Q}^\indexERNtrapped}
\newcommand{\ERNsmallsupportset}{\mathfrak{Q}^\indexERNsmall}
\newcommand{\ERNtrappedsupportsetx}{\mathfrak{Q}^{\indexERNtrapped}_{x}}
\newcommand{\badsetaplarge}{\mathfrak{B}}
\newcommand{\ERNsmallsupportsetx}{\mathfrak{Q}^{\indexERNsmall}_{x}}

\newcommand{\badset}{\pi_0(\slowsupportset)}
\newcommand{\badsettau}{\pi_0(\slowsupportset_\tauslow)}

\newcommand{\badsetallconst}{\pi_0(\slowsupportset_{C_1,C_2,\tauslow})}

\newcommand{\badsetapprox}{\mathfrak{b}}

\newcommand{\tauslow}{{\tau^{\fatdot}}}
\newcommand{\tauzero}{\tau_{\fatdot}}
\newcommand{\sphere}{\mathbb{S}^2}
\newcommand{\rnconstant}{\lambda}
\newcommand{\signpr}{\mathfrak{s}}
\newcommand{\dmux}{d\mu_{\mathcal{P}_x}}
\newcommand{\dmu}{d\mu_{\mathcal{P}}}

\theoremstyle{definition}
\newtheorem{defi}{Definition}[section]

\theoremstyle{thmstyleone}%
\newtheorem{thm}[defi]{Theorem}
\newtheorem{prop}[defi]{Proposition}
\newtheorem{assumption}[defi]{Assumption}
\newtheorem*{thm*}{Theorem}
\newtheorem{lem}[defi]{Lemma}
\newtheorem{cor}[defi]{Corollary}
\newtheorem{claim}[defi]{Claim}
\newtheorem*{claim*}{Claim}
\theoremstyle{remark}
\newtheorem{rem}[defi]{Remark}
\newenvironment{subproof}{\paragraph{Proof:}}{\hfill$\blacksquare$}

\begin{document}

\title[Decay and non-decay for the  massless Vlasov equation on subextremal and extremal black holes]{Decay and non-decay for the  massless Vlasov equation on subextremal and extremal Reissner--Nordstr\"om black holes}

%% \author*[1]{\pfx{Dr} \fnm{Joergen W.} \spfx{van der} \sur{Ploeg} \sfx{IV} \tanm{Poet Laureate} 

\author{\fnm{Max} \sur{Weissenbacher}}\email{mweissen@ic.ac.uk}

\affil{\orgdiv{Department of Mathematics}, \orgname{Imperial College London},
\orgaddress{\street{South Kensington Campus}, \city{London}, \postcode{SW7 2AZ}, \country{United Kingdom}}}


\abstract{We study the massless Vlasov equation on the exterior of the subextremal and extremal Reissner--Nordstr\"om spacetimes. We prove that moments decay at an exponential rate in the subextremal case and at a polynomial rate in the extremal case. This polynomial rate is shown to be sharp along the event horizon. In the extremal case we show that transversal derivatives of certain components of the energy momentum tensor do not decay along the event horizon if the solution and its first time derivative are initially supported on a neighbourhood of the event horizon. The non-decay of transversal derivatives in the extremal case is compared to the work of Aretakis on instability for the wave equation. Unlike Aretakis' results for the wave equation, which exploit a hierarchy of conservation laws, our proof is based entirely on a quantitative analysis of the geodesic flow and conservation laws do not feature in the present work.}

\maketitle

\tableofcontents

\section{Introduction}
Understanding the late time dynamics of the Einstein equations in the vicinity of black hole solutions is a highly active area of research. The Einstein equations are at the heart of the theory of general relativity and may be expressed as
\begin{equation}
    \text{Ric}(g)_{\mu \nu} - \frac{1}{2} g_{\mu \nu} R(g) = T_{\mu \nu},
\end{equation}
where $g_{\mu \nu}$ is the metric tensor, $T_{\mu \nu}$ is the energy-momentum tensor of an appropriate matter model and the equations close by specifying appropriate evolution equations for the matter model. See~\cite{general} for an introduction to general relativity. Despite tremendous progress on subextremal black holes, many open problems remain in the extremal limit. For instance, for the extremal Kerr spacetime, two key geometric phenomena complicate its study:
\begin{itemize}
    \item the fact that the red-shift effect degenerates and
    \item the coupling of trapping and superradiance.
\end{itemize}
The first effect is already present in the simpler extremal Reissner--Nordstr\"om spacetime and is known to lead to an instability for the scalar wave equation, see Section~\ref{section_aretakis}. By contrast, the effect of the lack of decoupling of trapping and superradiance is not understood even at the level of massless linear scalar fields.

%Alternative text here: The degeneracy of the red-shift effect is known to lead to an instability on the linear level. This so-called Aretakis instability is well understood for the scalar wave equation. The Aretakis instability is also present for the extremal Reissner--Nordstr\"om black hole, which does not exhibit the second geometric phenomenon outlined above.

\subsection{Main results}
In this work we consider the massless Vlasov equation on the exterior of the subextremal and extremal Reissner--Nordstr\"om solution $(\Mrn,\gRN)$. In its most familiar form the metric takes the form
\begin{equation} \label{RN_metric_intro}
    g = - \Osqrn dt^2 + \OsqrnNEG dr^2 + r^2 d \omega^2, \quad \Osqrn = 1 - \frac{2m}{r} + \frac{q^2}{r^2},
\end{equation}
where $d \omega^2$ is the usual round metric on $\sphere$ and we assume $m>0$ and $\left| q \right| \leq m$. For this range of parameters~\eqref{RN_metric_intro} is known to describe a black hole. The parameter $m$ represents the mass of the black hole and $q$ its electromagnetic charge. The Reissner--Nordstr\"om family is the unique (up to diffeomorphism) stationary spherically symmetric $2$-parameter family of solutions to the Einstein--Maxwell equations~\cite{uniqueness}. We say the solution is subextremal when $\left| q \right| < m$ and extremal when $\left| q \right| = m$. 

The massless Vlasov equation or collisionless Boltzmann equation is a kinetic particle model describing a distribution of collisionless particles moving at the speed of light in a given spacetime. Phenomena in general relativity are often understood by first studying massless linear fields without back-reaction, such as the massless Vlasov equation. Examples of nonlinear results that have their origins in this line of reasoning include the work of Poisson--Israel~\cite{poissonisreal1,poissonisreal2} on instability of black hole interiors and Moschidis'~\cite{moschidismasslessvlasov,moschidisnulldust} proof of the AdS instability conjecture~\cite{adsinstabilityconj}.

\begin{figure}
    \centering
    \includegraphics[scale=0.4]{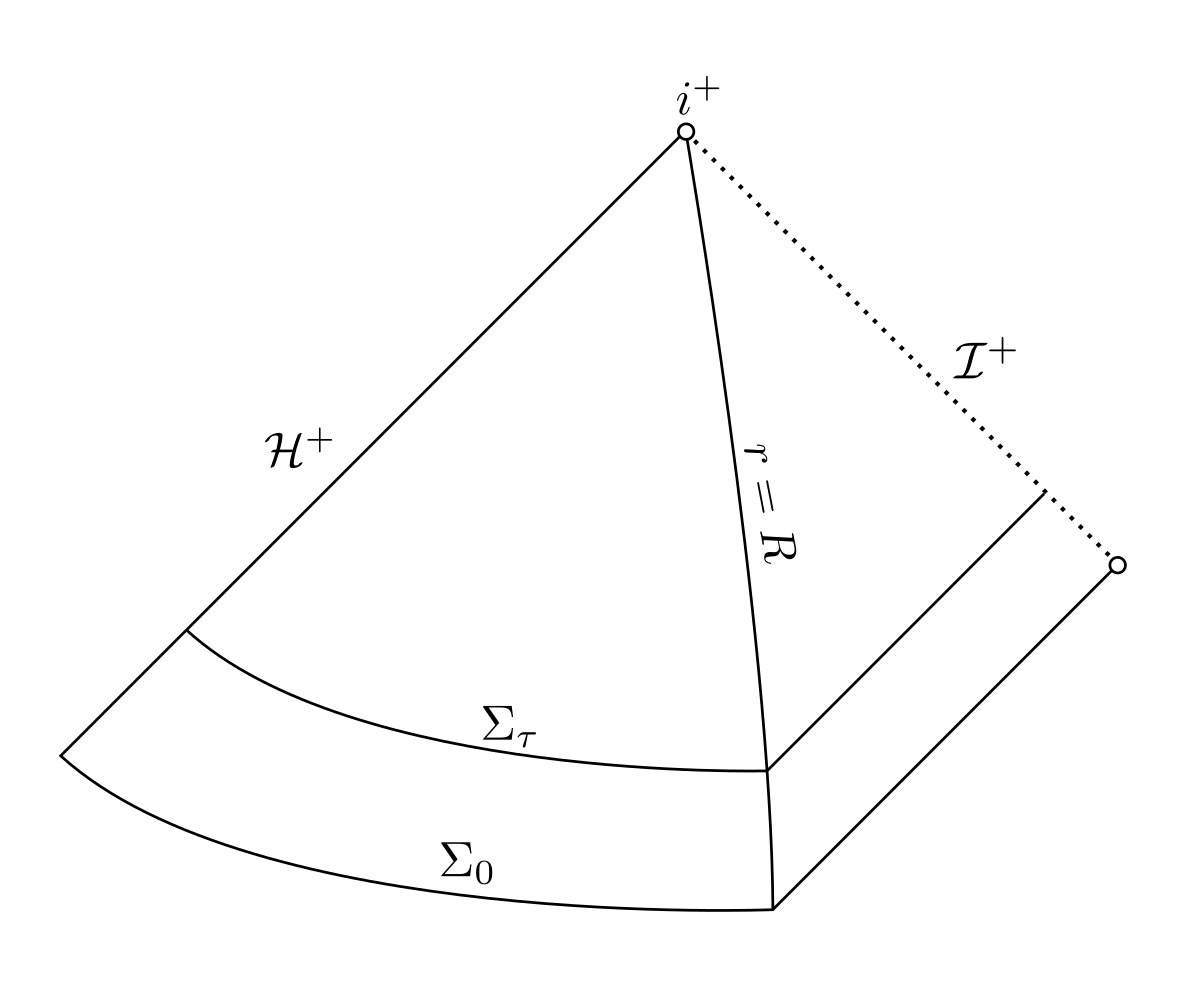}
    \caption{Penrose diagram of the Reissner--Nordstr\"{o}m black hole exterior. We denote the event horizon by $\mathcal{H}^+$ and future null infinity by $\mathcal{I}^+$. The hypersurfaces $\Sigma_0$ and $\Sigma_\tau$ for some $\tau > 0$ are being displayed. Refer to Section~\ref{timecoordinate} for the definition of the time function $\tau$ and the associated surfaces $\Sigma_\tau$.}
    \label{penrose_tau}
\end{figure}

Let $(\Mrn,\gRN)$ denote the Reissner--Nordstr\"om exterior. A coordinate system $(x^\mu)$ on $\Mrn$ induces the conjugate coordinate system $(x^\mu, p^\mu)$ on the tangent bundle $T \Mrn$ by representing each $p \in T_x \Mrn$ as $p = p^\mu \partial_\mu|_x$. We define the mass-shell $\mathcal{P} \subset T \Mrn$ as the set
\begin{equation}
\mathcal{P} = \Big\{ (x,p) \in T \Mrn : \, \gRN (x)(p,p)=0, \, p \text{ is future-directed} \Big\} .
\end{equation}
A function $f: \mathcal{P} \rightarrow \R_{\geq 0}$ solves the massless Vlasov equation if $f$ is conserved along the geodesic flow or equivalently if
\begin{equation}
    X(f) = \left( p^\mu \partial_\mu - \Gamma^\mu_{\alpha \beta} p^\alpha p^\beta \partial_{p^\mu} \right) f = 0,
\end{equation}
where $X$ denotes the geodesic spray, $\Gamma^\mu_{\alpha \beta}$ denote the Christoffel symbols in the coordinates $(x^\mu)$ and we have given the explicit expression of the geodesic spray in conjugate coordinates $(x^\mu, p^\mu)$. Let $\tau: \Mrn \rightarrow \R$ such that $\Sigma_0 = \{ \tau = 0 \}$ is a spherically symmetric Cauchy hypersurface which connects the event horizon and future null infinity and such that the hypersurfaces $\Sigma_\tau$ of constant $\tau > 0$ are obtained by propagating $\Sigma_0$ forward in time along the flow of the timelike Killing field. See Figure~\ref{penrose_tau} for an illustration and Section~\ref{timecoordinate} for a precise definition. If we denote the mass-shell over the initial Cauchy hypersurface $\Sigma_0$ by $\mathcal{P}_0 = \mathcal{P}|_{\Sigma_0}$ then we may treat the massless Vlasov equation as the initial value problem
\begin{equation}
	\begin{cases}
		X(f) = 0 \\
		f|_{\mathcal{P}_0} = f_0
	\end{cases} .
\end{equation}
For any sufficiently regular initial data $f_0: \mathcal{P}_0 \rightarrow \R_{\geq 0}$ there exists a unique solution $f$. This is readily seen by noting that for any $(x,p) \in \mathcal{P}$ there exists a unique null geodesic $\gamma$ intersecting $(x,p)$ and this geodesic intersects $\Sigma_0$ (since $\Sigma_0$ is a Cauchy hypersurface) in exactly one point, which we denote by $(x_0, p_0) \in \mathcal{P}_0$, see~\cite{oneill, waldbook}. The unique solution is then given by $f(x,p) = f_0(x_0, p_0)$. In Section~\ref{sec_vlasov} we define the natural measure $\dmux$ on the fibre $\mathcal{P}_x$ induced by the metric. Given a continuous weight $w: \mathcal{P} \rightarrow \R$, we define moments of a solution by $\int_{\mathcal{P}_x} w f \, \dmux$. We will be particularly interested in weights of the form $w = p^\alpha p^\beta$ for arbitrary spacetime indices $\alpha,\beta$. Their associated moments define the components of the energy momentum tensor $T^{\alpha \beta}[f] = \int_{\mathcal{P}_x} p^\alpha p^\beta f \, \dmux$ of the solution $f$. The energy momentum tensor is divergence free and is a natural object of interest since it describes the flux and density of energy in the spacetime and features in the coupled massless Einstein--Vlasov system.

%We define a time function with respect to which we measure decay.

%We refer the reader to Section~\ref{prelim} for a more detailed discussion of the geometry of the subextremal and extremal Reissner--Nordstr\"om solutions and the massless Vlasov equation.

\subsubsection{Exponential decay on subextremal Reissner--Nordstr\"om} \label{subsec_expdecay}

We may now state our main theorem on the subextremal Reissner--Nordstr\"om background. In Section~\ref{section_mainthms} below we will state a more precise version of Theorem~\ref{maintheorem} in the form of Theorem~\ref{maintheorem_precise}.

\begin{thm}[Exponential decay on subextremal Reissner--Nordstr\"om] \label{maintheorem}
Let $f_0: \mathcal{P}_0 \rightarrow [0,\infty)$ be smooth and compactly supported and let $f$ be the unique solution to the massless Vlasov equation on subextremal Reissner--Nordstr\"om with initial data $f_0$. Let $w: \mathcal{P} \rightarrow [0,\infty)$ be smooth and bounded in $x$ (see Definition~\ref{boundedness_in_x}). Then there exist constants $C = C(w, \supp(f_0), m,q)$ and $c = c(m,q)$ such that for all $x \in \Mrn$ with $\tau = \tau(x) \geq 0$, where $\tau$ is defined as above, the following decay estimate holds:
\begin{equation}
\int_{\mathcal{P}_x} w f \, \dmux \leq C \| f_0 \|_{L^{\infty}} \frac{1}{r^2} e^{-c \tau}.
\end{equation}
\end{thm}

\begin{rem}[Decay of the energy-momentum tensor] \label{rem_expdecay_tensor}
A direct consequence of Theorem~\ref{maintheorem} is that all components of the energy-momentum tensor decay at an exponential rate. We will compare this with the wave equation later in Section~\ref{section_aretakis}. By making use of the $(t^*,r)$-coordinate system defined in Section~\ref{rngeometry} and applying Theorem~\ref{maintheorem} with the weights $w = (p^{t^*})^2$ respectively $w = (p^r)^2$ we find that
\begin{equation}
    \sup_{\Sigma_\tau} \left| T^{\mu \nu}[f] \right| \leq C \| f_0 \|_{L^\infty} e^{-c \tau}, \quad \mu,\nu \in \{ t^*, r\}
\end{equation}
where $c = c(m,q)$ is as in Theorem~\ref{maintheorem} and $C = C(\supp(f_0),m,q)$ is the constant $C$ from above. By making use of the mass-shell relation, we find in addition for the angular component
\begin{equation}
    \sup_{\Sigma_\tau} \left| \gslash_{\alpha \beta} T^{\alpha \beta}[f] \right| \leq C \| f_0 \|_{L^\infty} e^{-c \tau},
\end{equation}
where $\gslash = r^2 d \omega $ denotes the angular component of the metric, see Section~\ref{rnmetricmanifold}.
\end{rem}

\begin{rem}
For certain weights $w$, the rate of decay can be improved. Furthermore as $r \rightarrow \infty$ one recovers the decay rates~in~$r$ from Minkowski space. We remind the reader at this point that the Reissner--Nordstr\"om solution is asymptotically flat. In Theorem~\ref{maintheorem_precise} we provide a complete characterisation of decay rates in $\tau$ and $r$ for polynomial weights. 
\end{rem}

\begin{rem}
The constant $C$ degenerates as $\left| q \right| \rightarrow m$. Its precise dependence on the support of the initial data $f_0$ and the weight $w$ is given in Theorem~\ref{maintheorem_precise} below. For reasons of clarity we have assumed $f_0$ and $w$ to be smooth, however it will become clear from the proof that far less regularity (in fact, little more than measurability) is sufficient.
\end{rem}

\begin{rem}
The assumption that $f_0$ be compactly supported may be relaxed to the assumption that $f_0(x,p)$ decays at a sufficient (for instance, exponential) rate in $r$ and the momentum $p$. This will follow immediately from Theorem~\ref{maintheorem_precise}, which is the more precise form of Theorem~\ref{maintheorem}, where we make explicit the dependence on the size of the support of $f_0$.
\end{rem}

We note the recent work by Bigorgne~\cite{leo} who proves faster than polynomial decay for moments of solutions to the massless Vlasov equation on Schwarzschild. The proof adapts the $r^p$-method of Dafermos--Rodnianski~\cite{rp} originally devised for the wave equation. We remark that the result generalises to the full subextremal Reissner--Nordstr\"om range. Velozo~\cite{renato} has shown nonlinear stability of Schwarzschild as a solution to the coupled spherically symmetric massless Einstein--Vlasov system. In~\cite{renato} exponential decay is independently shown for solutions to the massless Vlasov equation on spacetimes close to Schwarzschild by exploiting the hyperbolic nature of the geodesic flow around the photon sphere. Andersson--Blue--Joudioux~\cite{anderssonblue} have shown an integrated energy decay estimate for the massless Vlasov equation on very slowly rotating Kerr backgrounds.

\subsubsection{Polynomial decay on extremal Reissner--Nordstr\"om}

We now state our main results on the extremal Reissner--Nordstr\"om spacetime. In contrast to the subextremal case, moments only decay at an inverse polynomial rate in general. Responsible for this slower rate of decay is the existence of a family of null geodesics which approximate the generators of the extremal event horizon $\mathcal{H}^+ = \{ r = m \}$. Over $\Sigma_0$ these ``approximate generators'' lie in a one-parameter family $\badsetaplarge_\delta \subset \mathcal{P}_0$ defined explicitly in Definition~\ref{def_badsetapprox} below. We give a brief explanation of these sets here and refer the reader to Section~\ref{section_ERN} for more details. The parameter $\delta > 0$ measures ``closeness'' to generators of the event horizon. The sets $\badsetaplarge_\delta$ form a decreasing family as $\delta \searrow 0$ and $\bigcap_{\delta > 0} \badsetaplarge_\delta$ is supported over the event horizon $\mathcal{H}^+$ and only contains null generators of $\mathcal{H}^+$. Furthermore, each set $\badsetaplarge_\delta$ is compact with non-empty interior in $\mathcal{P}$. We will state a more precise version of the following result in the form of Theorems~\ref{maintheoremERNprecise} and~\ref{ERN_slowdecayprop} in Section~\ref{section_mainthms}.

\begin{thm}[Polynomial decay on extremal Reissner--Nordstr\"om] \label{maintheoremERN}
Let $f_0: \mathcal{P}_0 \rightarrow [0,\infty)$ be smooth and compactly supported and let $f$ be the unique solution to the massless Vlasov equation on extremal Reissner--Nordstr\"om with initial data $f_0$. Let $w: \mathcal{P} \rightarrow [0,\infty)$ be smooth and bounded in $x$ as in Definition~\ref{boundedness_in_x}. There exists a constant $C = C(w, \supp(f_0), m)$ such that for all $x \in \Mern$ with $\tau(x) \geq 0$ the following decay estimate holds
\begin{equation} \label{decay_eqn_mainthmERN}
\int_{\mathcal{P}_x} w f \, \dmux \leq C \| f_0 \|_{L^{\infty}} \frac{1}{r^2} \frac{1}{\tau^2}.
\end{equation}
Moreover the polynomial rate of decay is sharp along the event horizon, in the following sense: Let $\delta > 0$ and assume that $\badsetaplarge_\delta \subset \supp(f_0)$. Then for all $x \in \mathcal{H}^+$ with $\tau = \tau(x) \gtrsim \delta^{-1}$ we have
\begin{equation}
    \int_{\mathcal{P}_x} f \, \dmux \geq C \left( \inf_{(x,p)\in \badsetaplarge_\delta} {f_0(x,p)} \right) \frac{1}{\tau^{2}},
\end{equation}
for an appropriate constant $C = C(\supp(f_0),m)$.

Finally, if $f_0$ is supported away from the event horizon $\mathcal{H}^+ = \{ r = m \}$, then $\int_{\mathcal{P}_x} f \, \dmux$ decays at an exponential rate in $\tau = \tau(x)$ globally, similar to the subextremal case.
\end{thm}

\begin{rem}[Decay of the energy-momentum tensor] \label{rem_poldecay_tensor}
The rate of decay stated in equation~\eqref{decay_eqn_mainthmERN} holds for general weights and may be improved depending on the weight $w$ and the distance from the event horizon, for example. Making use of $(t^*,r)$-coordinates defined in Section~\ref{rngeometry}, consider the weight $w = (p^{t^*})^2$, whose associated moment defines a component of the energy-momentum tensor $T^{t^* t^*}[f]$. Then along the event horizon
\begin{equation} \label{remark_tstar_rate_horizon}
    C \left( \inf_{\badsetaplarge_\delta} {f_0} \right) \tau^{-2} \leq \sup_{\Sigma_\tau \cap \mathcal{H}^+} T^{t^* t^*}[f] \leq C \| f_0 \|_{L^\infty} \tau^{-2},
\end{equation}
where the constant $C = C(\supp(f_0), m)$ is as in Theorem~\ref{maintheoremERN} above. At a positive distance from the event horizon we have for $\delta > 0$
\begin{equation}
    \sup_{\Sigma_\tau \cap \{ r \geq (1+\delta)m \} } T^{t^* t^*}[f] \leq C \| f_0 \|_{L^\infty} \tau^{-4},
\end{equation}
where $C = C(\supp(f_0), m, \delta)$ is as in Theorem~\ref{maintheoremERN}. On the other hand, for the weight $w = p^{t^*} p^r$ and its associated moment $T^{t^* r}[f]$ we have
\begin{align}
    \sup_{\Sigma_\tau} T^{t^* r}[f] &\leq C \| f_0 \|_{L^\infty} \tau^{-4}, \\
    \sup_{\Sigma_\tau \cap \mathcal{H}^+} T^{t^* r}[f] &\geq C \left( \inf_{\badsetaplarge_\delta} {f_0} \right) \tau^{-4},  \label{eqn_tstar_r_ineq}
\end{align}
while for the weight $w = (p^r)^2$ and its associated moment $T^{rr}[f]$ we find
\begin{align}
    \sup_{\Sigma_\tau} T^{rr}[f] &\leq C \| f_0 \|_{L^\infty} \tau^{-6}, \\
    \sup_{\Sigma_\tau \cap \mathcal{H}^+} T^{rr}[f] &\geq C \left( \inf_{\badsetaplarge_\delta} {f_0} \right) \tau^{-6},  \label{eqn_rr_ineq}
\end{align}
for an appropriate constant $C = C(\supp(f_0), m)$. We point out that in contrast to $T^{t^* t^*}[f]$, the upper bound holds uniformly in the whole exterior up to and including the event horizon.
\end{rem}

\begin{rem}
As in the subextremal case, we recover the rates of $r$-decay from flat Minkowski spacetime as $r \rightarrow \infty$. We will provide a full characterisation of decay rates in both $\tau$ and $r$ for polynomial weights in Theorems~\ref{maintheoremERNprecise} and~\ref{ERN_slowdecayprop} below. Similar remarks about the dependence of the constant $C$ on the size of the support of $f_0$, as well as the smoothness of $f_0$ and the weight $w$ apply here as in the subextremal case.
\end{rem}

\begin{rem} \label{rem_intro_smoothness}
As shown in Theorem~\ref{maintheoremERN}, solutions with initial data supported at a positive distance from $\mathcal{H}^+$ decay at an exponential rate, whereas solutions supported on a neighbourhood $\badsetaplarge_\delta$ of the generators of $\mathcal{H}^+$ only decay at a polynomial rate. If the initial data $f_0$ are smooth and supported on the generators of the event horizon, then there exists $0 < \delta < \frac{1}{2}$ such that $\badsetaplarge_\delta \subset \supp(f_0)$.
\end{rem}

\begin{rem}
Based on the discussion in Remark~\ref{rem_intro_smoothness}, one expects the existence of a class of non-smooth data which interpolates between the class of data supported on the generators of $\mathcal{H}^+$ and the class of data supported away from $\mathcal{H}^+$ and for which any decay rate between a polynomial and exponential rate can be achieved. The author conjectures the existence of a two-parameter family of sets $\bar{\badsetaplarge}_{\alpha, \delta}, 0 \leq \alpha \leq \delta < \frac{1}{2}$ such that $\bar{\badsetaplarge}_{\alpha_1, \delta} \subset \bar{\badsetaplarge}_{\alpha_2, \delta}, \bar{\badsetaplarge}_{\delta, \delta} = \badsetaplarge_\delta, \vol \bar{\badsetaplarge}_{0, \delta} = 0$ for $\alpha_1 < \alpha_2$ and such that if $f_0$ satisfies $\inf_{\alpha < \delta} \inf_{\bar{\badsetaplarge}_{\alpha, \delta}} f_0 = 0, \sup_{\alpha < \delta} \inf_{\bar{\badsetaplarge}_{\alpha, \delta}} f_0 > 0$, then $f$ will decay at a sub-exponential but super-polynomial rate.
\end{rem}

\begin{rem}
Like in the case of Theorem \ref{maintheorem} above, we remark that the assumption of compact support of $f_0$ may be relaxed to the assumption that $f_0(x,p)$ decays at a sufficient (for instance, exponential) rate in $r$ and the momentum $p$. This will follow immediately from Theorem~\ref{maintheoremERNprecise}, which is the more precise form of Theorem~\ref{maintheoremERN}, where we provide an explicit bound in terms of the size of the support of $f_0$.
\end{rem}

%This leaves a gap of initial data which may satisfy $\supp(f_0) \cap \badsetaplarge_\delta = \emptyset$ for all $\delta > 0$ yet $\{0\} \times [m, (1+\eta)m] \times \sphere \subset \pi (\supp(f_0))$ for some $\eta > 0$, where $\pi: \mathcal{P} \rightarrow M$ denotes the standard projection from the tangent bundle and where we used $(t^*,r)$-coordinates. We conjecture that for such initial data, any rate interpolating between an inverse polynomial and exponential rate can be achieved. \textcolor{orange}{However such data will most likely not be smooth.}

\subsubsection{Non-decay for transversal derivatives on extremal Reissner--Nordstr\"om}

Finally, we prove the following result which holds along the event horizon of extremal Reissner--Nordstr\"om, see Theorem~\ref{ERN_nondecaytransversal} below for a more precise version of the result. It is most easily stated in the $(t^*,r)$-coordinates defined in Section~\ref{rngeometry} below. In these coordinates, the timelike Killing derivative may be expressed as $\partial_{t^*}$ and we note that $\partial_r$ is transversal to the event horizon.

\begin{thm}[Non-decay for transversal derivatives on extremal Reissner--Nordstr\"om] \label{maintheorem_ern_nondecay_rough}
Assume $f_0: \mathcal{P}_0 \rightarrow [0,\infty)$ is smooth and compactly supported and let $f$ be the unique solution to the massless Vlasov equation on extremal Reissner--Nordstr\"om with initial data $f_0$. If we assume in addition that there exists $0 < \delta$ such that $\badsetaplarge_\delta \subset \supp(f_0)$ and $\partial_{t^*} f(x,p) \neq 0$ for all $(x,p) \in \badsetaplarge_\delta$, then
\begin{equation} \label{eqn_lowerbound_der_ern}
    \left| \partial_r \int_{\sphere} T^{t^* t^*}[f] \, d \omega \right| \geq C  \inf_{(x,p) \in \badsetaplarge_\delta} \left| \partial_{t^*} f(x,p) \right|,
\end{equation}
for all $x \in \mathcal{H}^+$ with $\tau(x)$ sufficiently large and a suitable constant $C = C(\supp(f_0),m)$. Therefore transversal derivatives of the energy momentum tensor do not decay along the event horizon in general.
\end{thm}

\begin{rem}
We may interpret the timelike Killing derivative $\partial_{t^*} f |_{\mathcal{P}_0}$ as an operator acting on initial data by using the geodesic spray $X$ to express $\partial_{t^*}$ as a function of derivatives which are tangential to $\mathcal{P}_0$.
\end{rem}

\begin{rem}
The ${t^*t^*}$-component of the energy-momentum tensor is the only component whose transversal derivative along the event horizon is non-decaying. This will follow directly from the proof of Theorem~\ref{maintheorem_ern_nondecay_rough}, where we establish that taking a transversal derivative along the event horizon essentially amounts to a multiplication with $\tau^2$. This fact, combined with inequalities~\eqref{eqn_tstar_r_ineq} and~\eqref{eqn_rr_ineq} as well as the mass-shell relation yields the desired decay.
\end{rem}

\begin{rem}
One expects higher-order derivatives to grow along the event horizon:
\begin{equation} \label{expected_growth}
	\left| \partial_r^k \int_{\sphere} T^{t^* t^*}[f] \, d \omega \right| \geq \tau^{2k-2} C[f_0], \quad k \in \mathbb{Z}_{\geq 0},
\end{equation}
where $C[f_0]$ is a suitable higher-order term analogous to the term on the right hand side of equation~\eqref{eqn_lowerbound_der_ern} involving $k$ derivatives of $f$. This will follow straightforwardly given a bound which establishes the sharpness of inequality~\eqref{eqn_lowerbound_der_ern} (and its higher order analogues), which the author has not been able to attain using the present methods, see the detailed discussion of this in Remark~\ref{rem:difficulty}. Therefore, the case $k \geq 2$ is not considered in this work.
\end{rem}

\begin{rem} \label{rem_intro_nondecaythm}
In the same spirit as Remark~\ref{rem_intro_smoothness} above, we note that if we assume $f_0 \geq 0$ to be smooth, and that $f_0$ and $\partial_{t^*} f |_{\mathcal{P}_0}$ are nowhere vanishing on the generators of the event horizon, it follows that there exists a $\delta > 0$ such that $\badsetaplarge_\delta \subset \supp(f_0)$ and $\partial_{t^*} f(x,p) \neq 0$ for all $(x,p) \in \badsetaplarge_\delta$.
\end{rem}

\begin{rem}
The spherical average in~\eqref{eqn_lowerbound_der_ern} serves to streamline and simplify the proof of Theorem~\ref{maintheorem_ern_nondecay_rough}. A simple modification of the argument allows one to remove the spherical average and show corresponding pointwise estimates, see Remark~\ref{rem_remove_average} for a more detailed discussion.
\end{rem}

\subsection{Comparison with the wave equation and the Aretakis instability} \label{section_aretakis}
Classically, the most studied linear field model on black hole backgrounds is the wave equation
\begin{equation} \label{waveequn}
    \Box_g \psi = g^{\mu \nu} \nabla_\mu \nabla_\nu \psi = 0,
\end{equation}
going back to the result by Wald~\cite{wald} and Kay--Wald~\cite{kaywald} on boundedness of solutions to the wave equation on the Schwarzschild spacetime. Solutions to the scalar wave equation have since been shown to be bounded and decaying on the subextremal Kerr and subextremal Reissner--Nordstr\"om exterior~\cite{kerrsubextremalfull,dafermosrodnianskishlap,moschidis2016r} and precise late-time asymptotics (Price's law) have been established in~\cite{priceslawRN,priceslawKerr,priceslaw2018, hintz2022sharp}. Works on the wave equation have recently been upgraded to nonlinear results by Dafermos--Holzegel--Rodnianski--Taylor~\cite{schwarzschildnonlinearstable}, who show nonlinear stability of Schwarzschild as a solution to the Einstein vacuum equations, see also Giorgi--Klainerman--Szeftel~\cite{klainerman3}. In contrast to the subextremal case, the extremal Reissner--Nordstr\"om and extremal Kerr solutions admit a linear instability mechanism along their event horizons, leading to the so-called Aretakis instability~\cite{aretakis1,aretakis2, aretakis3,subextkerr}. 

Let us consider the wave equation in some more detail for sake of comparison with the results obtained in this work. Assume that $\psi$ is a smooth solution to the wave equation~\eqref{waveequn} on the subextremal or extremal Reissner--Nordstr\"om exterior with regular compactly supported initial data $\psi_0 = \psi |_{\Sigma_0},  \psi_1 = \partial_{t^*} \psi |_{\Sigma_0}$. The energy-momentum tensor for the wave equation is defined in coordinates by
\begin{equation}
T_W^{\mu \nu}[\psi] = \nabla^\mu \psi \nabla^\nu \psi - \frac{1}{2} g^{\mu \nu} (g^{\alpha \beta} \nabla_\alpha \psi \nabla_\beta \psi).
\end{equation}
We express the energy momentum tensor in $(t^*,r)$-coordinates here, see Section~\ref{rngeometry}. For the purpose of comparing the wave equation to the massless Vlasov equation, let us define $k$-th order transversal moments for the wave equation and the massless Vlasov equation as in Table~\ref{table_moments}. Note that along the event horizon, $T_W^{t^* t^*}[\psi] \big|_{r=r_+} \approx \sum_\mu \left| \partial_{\mu} \psi \right|^2$, so that $T_W^{t^* t^*}[\psi]$ controls all derivatives non-degenerately. We conclude $T_W^{t^* t^*}[\partial_r^{k-1} \psi] \big|_{r=r_+} = (\partial_r^{k} \psi)^2 + \text{l.o.t.}$ for all $k \geq 1$, where the lower order terms contain at most $k-1$ transversal derivatives. Finally we remark for the massless Vlasov equation that a simple computation shows $\partial_r^k T^{t^* t^*}[f] = T^{t^* t^*}[\partial_r^k f] + \text{l.o.t.}$, where again the lower order terms contain at most $k-1$ transversal derivatives. Therefore, for both the wave equation and the massless Vlasov equation, the order of the transversal moment corresponds to the number of transversal derivatives of the solution along the event horizon. 

For simplicity, we will only compare decay rates in a neighbourhood of the black hole, where $r \leq R$ for a suitably large radius $R$. On the subextremal Reissner--Nordstr\"om exterior, the work of Angelopoulos--Aretakis--Gajic~\cite{priceslawKerr,priceslaw2018} implies the following result:

\setlength{\arrayrulewidth}{0.2mm}
\setlength{\tabcolsep}{18pt}
\renewcommand{\arraystretch}{1.8}
\begin{table}
	\centering
	\begin{tabular}{ p{4.5cm}|c|c  }
		%\hline
		%\multicolumn{3}{|c|}{$k$-th order transversal moments} \\
		 & $k=0$ & $k \geq 1$ \\
		\hline
		Wave equation& $\left| \psi \right|^2$ & $T_W^{t^* t^*}[\partial_r^{k-1} \psi]$  \\
		\hline
		Massless Vlasov equation & $T^{t^* t^*}[f]$ & $\partial_r^k  T^{t^* t^*}[f]$ \\
		%\hline
	\end{tabular}
\caption{Definition of $k$-th order transversal moments for the wave equation and the massless Vlasov equation, for the sake of comparing the two models.}
\label{table_moments}
\end{table}
%We refer to $\left| \psi \right|^2$ as the zeroth-order moment for the wave equation, and to $T_W^{t^* t^*}[\partial_r^{k-1} \psi]$ for $k \geq 1$ as a $k$-th order transversal moment. Along the event horizon, the $k$-th order transversal moment reduces to $\big| \partial_r^k \psi \big|^2$ for all $k \geq 0$. For the massless Vlasov equation let us refer to $T^{t^* t^*}[\partial_r^k f] = \int_{\mathcal{P}_x} (p^{t^*})^2 \,  \partial_r^k f \, \dmux $ for $k \geq 0$ as a $k$-th order transversal moment.

\begin{thm}[\cite{priceslawKerr,priceslaw2018}] \label{aretakis_thm_1}
Let $\psi$ be a solution to the wave equation on the subextremal Reissner--Nordstr\"om exterior as above. Then
\begin{equation}
    \sup_{\Sigma_\tau \cap \{ r \leq R \} } \left| \psi \right|^2 \leq C \frac{E[\psi_0,\psi_1]}{\tau^6}, \quad \sup_{\Sigma_\tau \cap \{ r \leq R \} } T_W^{t^* t^*}[\psi] \leq C \frac{E[\psi_0,\psi_1]}{\tau^8},
\end{equation}
for an appropriate constant $C > 0$ and where $E[\psi_0,\psi_1]$ denotes a suitable weighted and higher order initial data norm. For generic initial data these rates are sharp along the event horizon and along constant area radius hypersurfaces on the exterior.
\end{thm}

Next, we discuss two results on the extremal Reissner--Nordstr\"om spacetime immediately implied by the work of Aretakis~\cite{aretakis1,aretakis2, aretakis3} and Angelopoulos--Aretakis--Gajic~\cite{angelopoulos2020late}. In the extremal case, we need to distinguish between the event horizon (located at radius $r=m$) and the region at a positive distance from the black hole, where $(1+\delta)m \leq r$ for some $\delta > 0$. In the following theorem, we consider the region away from the black hole.

\begin{thm}[\cite{aretakis1,aretakis2, aretakis3,angelopoulos2020late}] \label{aretakis_thm_2}
Let $\psi$ be a solution to the wave equation on the extremal Reissner--Nordstr\"om exterior as above. Then for $\delta > 0$
\begin{equation}
    \sup_{\Sigma_\tau \cap \{ (1+\delta)m \leq r \leq R \} } \left| \psi \right|^2 \leq C(\delta) \frac{E[\psi_0,\psi_1]}{\tau^4} ,\quad \sup_{\Sigma_\tau \cap \{ (1+\delta)m \leq r \leq R \} } T_W^{t^* t^*}[\psi] \leq C(\delta) \frac{E[\psi_0,\psi_1]}{\tau^{4}},
\end{equation}
for an appropriate constant $C = C(\delta) > 0$ and where $E[\psi_0,\psi_1]$ denotes a suitable weighted and higher order initial data energy. For generic initial data these rates are sharp along constant area radius hypersurfaces at a positive distance from the event horizon.
\end{thm}

We remark that there is an explicit characterisation for the class of data for which the rates in Theorems~\ref{aretakis_thm_1} and~\ref{aretakis_thm_2} are sharp in the cited works. Finally, along the extremal event horizon, for generic solutions higher-order transversal moments and higher do not decay:

\begin{thm}[\cite{aretakis1,aretakis2, aretakis3,angelopoulos2020late}] \label{aretakis_thm3}
Let $\psi$ be a solution to the wave equation on the extremal Reissner--Nordstr\"om exterior as above. Along the event horizon, the solution itself decays, while the first order transversal moment does not decay:
\begin{equation}
	\left| \psi \right|^2 \Big|_{r=m} \leq C_0 \frac{H_0[\psi]}{\tau^2}, \quad \int_{\sphere} T_W^{t^* t^*}[\psi] \, d \omega \Big|_{r=m} \geq C_1 (H_0[\psi])^2,
\end{equation}
for suitable constants $C_0,C_1 > 0$ and where the so-called horizon charge $H_0[\psi]$ is conserved along the event horizon and may be expressed in $(t^*,r)$-coordinates as
\begin{equation}
	H_0[\psi] = \frac{m^2}{4 \pi} \int_{\sphere} (\partial_{t^*} - \partial_r) (r \psi)|_{r=m} \, d \omega.
\end{equation}
In fact, higher order moments grow polynomially:
\begin{equation} \label{intro_compwave_growth}
\int_{\sphere} T_W^{t^* t^*}[\partial_r^{k-1} \psi] \, d \omega \Big|_{r=m} \geq C_k \tau^{2k-2} (H_0[\psi])^2, \quad k \geq 1,
\end{equation}
for suitable constants $C_k >0$. The growth~\eqref{intro_compwave_growth} is known as the Aretakis instability.
\end{thm}

Let us compare Theorems~\ref{aretakis_thm_1}-\ref{aretakis_thm3} with Remark~\ref{rem_expdecay_tensor}, Remark~\ref{rem_poldecay_tensor} and Theorem~\ref{maintheorem_ern_nondecay_rough}. We point out two key similarities between the wave equation and the massless Vlasov equation:
\begin{itemize}
	\item The rate of decay of transversal moments of a fixed order at a positive distance from the black hole is slower in the extremal case as compared to the subextremal case. Consider the case of zeroth-order moments as an example. For the massless Vlasov equation, $T^{t^* t^*}[f]$ decays at an exponential rate in the subextremal case, but only at a polynomial rate in the extremal case. Likewise, for the wave equation, $\left| \psi \right|^2$ decays at a slower polynomial rate in the extremal case as compared to the subextremal case.
	\item Along the event horizon of extremal Reissner--Nordstr\"om, zeroth-order moments decay, while first-order transversal moments (which involve one transversal derivative of the solution) are non-decaying. In fact, if we compare Theorem~\ref{aretakis_thm3} on the wave equation with the conjectured growth for higher order transversal moments for the massless Vlasov equation as stated in Remark~\ref{expected_growth}, we find that the rate of growth for higher-order transversal moments is the same for the wave equation and massless Vlasov equation.
\end{itemize}

We take note of a key difference between the proofs of Theorem~\ref{aretakis_thm3} and our Theorem~\ref{maintheorem_ern_nondecay_rough}. The proof of Theorem~\ref{aretakis_thm3} is based on an infinite hierarchy of conservation laws for solutions of the wave equation~\eqref{waveequn} on extremal Reissner--Nordstr\"om. Conservation laws do not feature in the proof of Theorem~\ref{maintheorem_ern_nondecay_rough}, however.

\subsection{Overview of the proofs}
In this subsection we explain the key ideas of the proofs of our main Theorems~\ref{maintheorem},~\ref{maintheoremERN} and~\ref{maintheorem_ern_nondecay_rough}. The argument rests on a precise understanding of the momentum support of a solution to the massless Vlasov equation. The core ideas are much the same for both the subextremal and extremal Reissner--Nordstr\"om spacetimes. We will therefore set out by explaining the main ideas for the subextremal case and point out the key differences in the extremal case later.

\subsubsection{Decay of momentum support implies decay of moments}
Consider a solution $f : \mathcal{P} \rightarrow \R$ to the massless Vlasov equation on the subextremal Reissner--Nordstr\"om exterior with compactly supported and bounded initial data $f_0$. Since $f$ satisfies a transport equation along future-directed null geodesics, the solution itself does not decay pointwise. Instead, as a first step, we estimate moments of the solution $f$ as follows
\begin{equation}
\left| \int_{\mathcal{P}_x} w f \, \dmux \right| \leq C(w) \| f_0 \|_{L^\infty} \int_{\supp{f(x, \cdot)}} 1 \, \dmux,
\end{equation}
where we made use of the fact that $\left| f \right| \leq \| f_0 \|_{L^\infty}$ since $f$ is transported along null geodesics and we have abbreviated $C(w) = \sup_{(x,p) \in \supp(f)} \left| w(x,p) \right|$. We will primarily be interested in the case of weights $w$ which are polynomial functions of the momentum. For such weights, if we assume that $f_0$ is compactly supported, we can show a quantitative upper bound on $C(w) < \infty$ in terms of the size of the initial support by making use of the fact that the momentum support of $f$ remains compact for all times. To prove decay of moments, it therefore suffices to show that the the volume of the momentum support of $f$ decays. We therefore aim to show that
\begin{equation}
\vol \, \supp{f(x, \cdot)} = \int_{\supp{f(x, \cdot)}} 1 \, \dmux \leq C e^{-c \tau(x)}, \quad \tau(x) \rightarrow \infty,
\end{equation}
for appropriate constants $C,c > 0$. The key to the proof of Theorem~\ref{maintheorem} is therefore to understand the decay of the phase-space volume of the momentum support $\supp{f(x, \cdot)} \subset \mathcal{P}_x$ of a solution $f$ as time tends to infinity.

\subsubsection{Geodesics in the momentum support are almost trapped}
By a careful study of the geodesic flow we will show that for points $x \in \Mrn$ with $\tau(x) \gg 1$, there are two types of momenta in the support of $f(x,\cdot)$. More formally, we prove that
\begin{equation} \label{supp_intro_ideaproof}
    \supp{f(x, \cdot)} \subset \trappedsupportsetx \cup \smallsupportsetx \subset \mathcal{P}_x,
\end{equation}
where the set $\trappedsupportsetx$ contains momenta of geodesics which are almost trapped at the photon sphere and $\smallsupportsetx$ contains momenta of geodesics which are almost trapped at the event horizon. We say that an affinely parametrised null geodesic is (exactly) trapped at a hypersurface if it is future complete and approaches the hypersurface as its affine parameter tends to infinity. We note that the set of geodesics which are exactly trapped at the photon sphere forms a co-dimension one submanifold of the mass-shell $\mathcal{P}$.\footnote{In other words, for each fixed point $x$ in the exterior, consider the set of null momenta $p \in \mathcal{P}_x$ with the property that the unique future-directed null geodesic with initial data $(x,p)$ is trapped at the photon sphere. Then this set of trapped momenta forms a two-dimensional subset of the three-dimensional cone $\mathcal{P}_x$.} A geodesic can only be trapped at the event horizon if it is a generator of the event horizon. The sets $\trappedsupportset$ and $\smallsupportset$ capturing the effect of trapping then turn out to be exponentially small neighbourhoods of the respective sets of exactly trapped geodesics. As a consequence, their phase-space volume is exponentially small in $\tau(x)$:
\begin{equation} \label{eqn_vol_trappedsets_intro}
\begin{gathered}
    \vol \trappedsupportsetx \sim e^{-c \tau(x)}, \\
    \vol \smallsupportsetx \sim e^{-c \tau(x)}.
\end{gathered}
\end{equation}
In order to prove~\eqref{supp_intro_ideaproof}, we require a quantitative estimate of how close to being trapped at the photon sphere or the event horizon a given geodesic which has not scattered to infinity or fallen into the black hole after a given time is.

\subsubsection{The almost-trapping estimate} \label{subsubsec_geodflow}
Consider a point $(x,p) \in \supp(f)$ with $\tau(x) \gg 1$ and let $\gamma$ denote the unique null geodesic determined by it. Since $f$ solves the massless Vlasov equation, $\gamma$ must intersect $\supp(f_0)$. Since the time function $\tau$ connects the event horizon and future null infinity, $\gamma$ has neither scattered to infinity nor fallen into the black hole before time $\tau(x)$, thus $\gamma$ can be thought of as \emph{almost trapped}. Understanding the momentum support $\supp{f(x,\cdot)}$ now amounts to estimating the components of $p = \dot{\gamma}$ in terms of initial data and $\tau(x)$.

The key to accomplishing this is the \emph{almost-trapping estimate}, which bounds the time that the geodesic $\gamma$ requires to cross a certain region of spacetime in terms of how close the geodesic is to being trapped. To make this precise, we introduce the trapping parameter $\trapschw : \mathcal{P} \rightarrow [-\infty,1]$ in Definition~\ref{def_eps_def}. The trapping parameter is conserved along the geodesic flow and measures how far a geodesic is from being future-trapped (or past-trapped) at the photon sphere. In particular a null geodesic $\gamma$ is future-trapped or past-trapped at the photon sphere if and only if $\trapschw(\gamma,\dot{\gamma}) = 0$. The \emph{almost-trapping estimate} may then be stated as
\begin{equation} \label{eqn_almosttrapping_intro}
	\tau(x) \leq C \left( 1 + ( \log \left| \trapschw \right| )_- + ( \log \, [(1+\left| \trapschw \right|) \OsqS(r_0) ] )_- \right),
\end{equation}
where $\trapschw = \trapschw(\gamma,\dot{\gamma})$ and $\gamma$ intersects $\Sigma_0$ at radius $r_0$.  We provide a sketch of the proof in Section~\ref{sec_radialgeod}.

The term $( \log \left| \trapschw \right| )_-$ arises from trapping at the photon sphere, while $( \log \, [(1+\left| \trapschw \right|) \OsqS(r_0) ] )_-$ arises from trapping at the event horizon. Indeed, assume $\tau(x) \geq 2C$. The inequality~\eqref{eqn_almosttrapping_intro} implies $( \log \left| \trapschw \right| )_- \geq c \tau(x)$ or $ ( \log \, [(1+\left| \trapschw \right|) \OsqS(r_0) ] )_- \geq c \tau(x)$, where $c =( 4C)^{-1}$. Let us consider the case where $( \log \left| \trapschw \right| )_- \geq c \tau(x)$ in some more detail. We find $ \left| \trapschw \right| \leq e^{- c\tau(x)}$ and thus $\gamma$ is exponentially close to being trapped at the photon sphere. By definition of the set $\trappedsupportsetx$, this implies $p = \dot{\gamma} \in \trappedsupportsetx$. We use the mass-shell relation in combination with the assumption that $\supp(f_0)$ is compact to bound the size of the components of $p$. This allows us to estimate the volume of the set $\trappedsupportsetx$. In the latter case, where $( \log \, [(1+\left| \trapschw \right|) \OsqS(r_0) ] )_- \geq c \tau(x)$, we readily conclude $(1+\left| \trapschw \right|) \OsqS(r(0)) \leq e^{-c\tau(x)}$, which may be shown to correspond to $\gamma$ being exponentially close to being trapped at the event horizon and $p = \dot{\gamma} \in \smallsupportsetx$. We argue analogously to the first case and thereby prove estimate~\eqref{eqn_vol_trappedsets_intro}.

%We next explain the interpretation of the almost trapping estimate~\eqref{eqn_almosttrapping_intro} and use it to prove~\eqref{eqn_vol_trappedsets_intro}.

%In a neighbourhood of the black hole, we use $(t^*,r)$-coordinates (see Section~\ref{rngeometry}) to define the time function as $\tau = t^*$. Assume that $\gamma$ is affinely parametrised in such a way that $(\gamma(0),\dot{\gamma}(0)) \in \supp(f_0)$ and $(\gamma(s),\dot{\gamma}(s)) = (x,p)$. If we use $(t^*,r)$-coordinates to express $\gamma = (t^*,r,\omega)$ and $\dot{\gamma} = (p^{t^*},p^r,\pslash)$ then
%\begin{equation}
%    t^*(s) - t^*(0) = \int_{0}^{s} \frac{dt^*}{d s} \, d s = \int_{0}^{s} p^{t^*} \, d s = \int_{r(0)}^{r(s)} \frac{p^{t^*}}{p^r} \, d r.
%\end{equation}
%Here we made use of the fact that the geodesic equations readily imply that there is at most one point where $p^r = 0$ and we split the integral along this point if necessary.

\subsubsection{Key differences between the extremal and subextremal case}
We next point out the similarities and differences between the null geodesic flow on the subextremal and extremal Reissner--Nordstr\"om exterior. While the structure of trapping at the photon sphere remains virtually identical, there are two key differences concerning trapping at the event horizon:
\begin{itemize}
	\item Outgoing null geodesics, which are almost trapped at the event horizon, leave the region close to the black hole at a slower rate in the extremal case as compared to the subextremal case, due to the subextremal geodesics being \emph{red-shifted}. This phenomenon is the reason that moments of solutions to the massless Vlasov equation decay at a slower rate in the extremal as compared to the subextremal case. We illustrate this for the special case of radial geodesics in Section~\ref{sec_radialgeod} below.
	\item There exist ingoing null geodesics in the extremal Reissner--Nordstr\"om spacetime which are almost trapped at the event horizon but fall into the black hole after a finite (long) time. These geodesics are unique to the extremal case and are key to proving non-decay of transversal derivatives of moments along the event horizon. Their subextremal analogues are untrapped, so that they fall into the black hole after an at most constant amount of time. Such geodesics are not radial, so the discussion is deferred until Section~\ref{psupporteventhorizon}, see specifically Remark~\ref{rem_bad_geodesics} for a detailed comparison of these geodesics and a comparison to their subextremal analogues.
\end{itemize}

\subsubsection{Proof of  the almost-trapping estimate} \label{sec_radialgeod}
\begin{figure}
	\centering
	\includegraphics[scale=0.4]{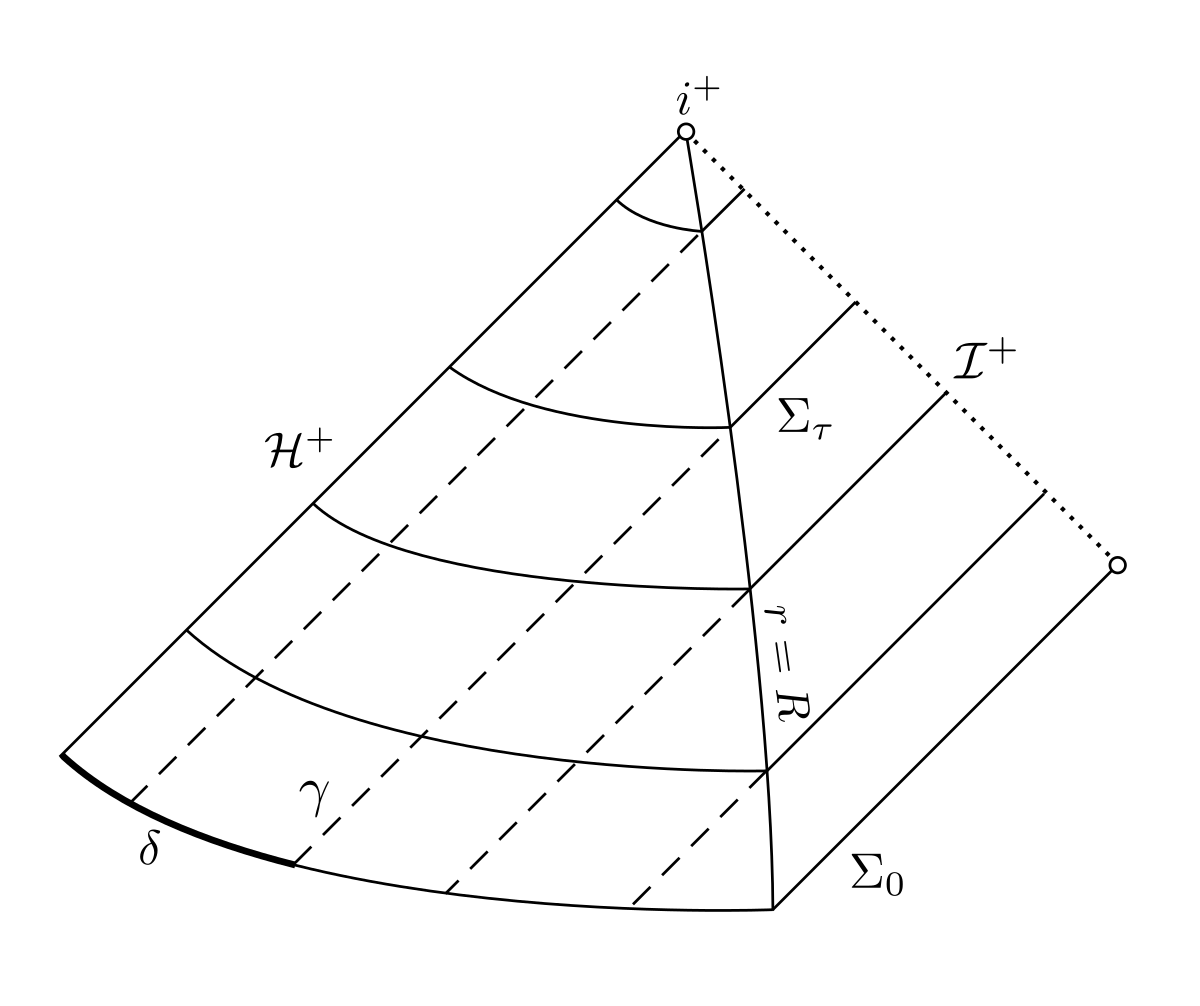}
	\caption{Propagation of radial null geodesics in the Penrose diagram of the Reissner--Nordstr\"{o}m exterior, see also Figure~\ref{penrose_tau} above. Several outgoing radial null geodesics $\gamma$ are depicted (dashed lines), together with the $\Sigma_\tau$-hypersurfaces which they coincide with in the region $r \geq R$. The initial distance from the event horizon is $\delta$. The time parameter $\tau$, for which $\gamma$ intersects $\Sigma_\tau$ at the area radius $r = R$ depends only on $\delta$ and satisfies $\tau \rightarrow \infty$ as $\delta \rightarrow 0$. The quantitative relation between $\delta$ and $\tau$ depends on whether the black hole is extremal or subextremal, see equation~\eqref{ineq_radial_time}.}
	\label{penrose_radial}
\end{figure}

In this subsection we obtain the almost-trapping estimate for the special case of radial geodesics. We discuss both the subextremal estimate~\eqref{eqn_almosttrapping_intro} and its extremal analogue~\eqref{eqn_almosttrapping_intro_ERN}, which is stated in Section~\ref{into_sec_extremal_upperbounds} below. This will serve to clarify the strategy of proof and to point out a key difference between the subextremal and extremal case. We invite the reader to compare the form of estimates~\eqref{eqn_almosttrapping_intro} and~\eqref{eqn_almosttrapping_intro_ERN}. Note that radial geodesics cannot be trapped at the photon sphere. Therefore, only the rightmost term on the right hand side of~\eqref{eqn_almosttrapping_intro} respectively~\eqref{eqn_almosttrapping_intro_ERN} is relevant for radial geodesics. However, radial geodesics exhibit the phenomenon of trapping at the event horizon in a simple way. See Figure~\ref{penrose_radial} for a visual guide to the behaviour of outgoing radial null geodesics. 

%We remind the reader of the mixed spacelike-null nature of the time function $\tau$, see Figure~\ref{penrose_tau}. We use the mass-shell relation in combination with considerations in the asymptotically flat region to establish the \emph{almost-trapping estimate} 

%We begin by recalling that the symmetries of the Reissner--Nordstr\"om metric are associated with conserved quantities along the geodesic flow: the energy $E$ (time invariance) and total angular momentum $L$ (spherical symmetry).  

Consider an affinely parametrised future-directed radial null geodesic $\gamma$ on the subextremal or extremal Reissner--Nordstr\"om exterior. We express $\gamma$ in $(t^*,r)$-coordinates as $\gamma(s) = (t^*(s),r(s),\omega(s))$ and its momentum as $\dot{\gamma}(s) = (p^{t^*}(s),p^r(s),\pslash(s))$. Since $\gamma$ is radial, we have $\left| \pslash(s) \right|_{\gslash} = 0$ for all $s$. We assume in addition that $\gamma$ is outgoing, so that $p^r(0) > 0$ and that it intersects $\Sigma_0$ at a distance $\delta$ from the event horizon, so $r(0) - r_+ = \delta$. If $\delta = 0$, then $\gamma$ is trapped at the event horizon, which means that $\gamma(s) \in \mathcal{H}^+$ for all $s \geq 0$. We may think of $\delta$ as measuring the distance to being trapped at the event horizon. Let us therefore assume that $\delta > 0$ is small. Using that $\gamma$ is radial, the geodesic equations reduce to
\begin{equation}
	\dot{p}^{t^*} = - \frac{1}{2} \frac{d \Osqrn}{d r} \left( p^{t^*} + p^r \right)^2, \quad \dot{p}^r = 0.
\end{equation}
Making use of the fact that $p^r(0) > 0$, we conclude that for all affine parameters $s \geq 0$,
\begin{equation} \label{eqn_radial_explicit_soln}
	p^r(s) = p^r(0), \quad p^{t^*}(s) = \frac{2-\Osqrn(r(s))}{\Osqrn(r(s))} p^r(s)	.
\end{equation}
By solving the geodesic equations in double null coordinates (introduced in Section~\ref{altcoordsystems}) or alternatively by visual inspection of Figure~\ref{penrose_radial}, we conclude that radial null geodesics propagate along the hypersurfaces $\Sigma_\tau \cap \{ r \geq R \}$ in the region where $r \geq R$. Therefore, we need only estimate the time $\gamma$ requires to travel outward from its initial radius $r(0)$ to the radius $R > r_+$. We follow the method outlined in Section~\ref{subsubsec_geodflow} above combined with equation~\eqref{eqn_radial_explicit_soln} to find
\begin{align} \label{eqn_overview_time}
	t^*(s) - t^*(0) &= \int_{r(0)}^{r(s)} \frac{p^{t^*}}{p^r} \, dr \sim  \int_{r(0)}^{R} \frac{1}{\Osqrn} \, dr \sim \int_{r(0)-r_+}^{R-r_+} \frac{1}{x(2\kappa+x)} \, dx \\
	&= \left[ \frac{\log x - \log(x+2\kappa)}{\kappa} \right]_{\delta}^{R-r_+}  \sim \frac{\log (\delta + 2\kappa) - \log(\delta)}{\kappa},
\end{align}
where we say $a \sim b$ if there exist constants $C_2 > C_1 > 0$ such that $C_1 a \leq b \leq C_2 a$ and where we have also introduced the \emph{surface gravity}
\begin{equation}
\kappa = \frac{1}{2} \frac{d}{dr} \Osqrn \bigg|_{r=r_+}.
\end{equation}
The surface gravity allows us to approximate the lapse $\Osqrn \approx (r-r_+)(r-r_+ + 2\kappa) = x (x+2 \kappa)$, where we introduced the change of coordinates $x = r-r_+$. Crucially, the surface gravity vanishes in the extremal case and is positive in the subextremal case, see Section~\ref{rnmetricmanifold} below. We conclude
\begin{equation} \label{ineq_radial_time}
	\tau(\gamma(s)) \sim \begin{cases}
		C(\kappa) \left|\log \delta \right| & \kappa > 0 \\
		C \delta^{-1} & \kappa = 0
	\end{cases}.
\end{equation}
This establishes the almost-trapping estimates~\eqref{eqn_almosttrapping_intro} respectively~\eqref{eqn_almosttrapping_intro_ERN} for radial geodesics. For non-radial geodesics, we follow a similar approach and consider the first equality in~\eqref{eqn_overview_time}. In lieu of solving the geodesic equations explicitly, we use the mass-shell relation to estimate the quotient $p^{t^*}/p^r$ as a function of the trapping parameter $\trapschw$ and radius $r$. Estimating the resulting integral leads to the non-radial analogue of~\eqref{eqn_overview_time}. In the asymptotically flat region, we show similarly that null geodesics propagate approximately along hypersurfaces $\Sigma_\tau \cap \{ r \geq R \}$. Finally we combine the non-radial analogue of~\eqref{eqn_overview_time} with the considerations in the asymptotically flat region to obtain the almost-trapping estimate. We treat the general case in Lemma~\ref{taubound} for the subextremal case and Lemma~\ref{tauestimate_ERN} for the extremal case. 

Finally we discuss how to bound the size of the momentum components, which is the key step in establishing~\eqref{eqn_vol_trappedsets_intro} respectively its extremal analogue~\eqref{eqn_vol_trappedsets_intro_ERN}. We make use of the fact that $\gamma$ is assumed to intersect the compact support of the initial distribution, so that $0 \leq p^r(0), p^{t^*}(0) \leq \psuppconst$ for some constant $\psuppconst > 0$. Combined with equation~\eqref{eqn_radial_explicit_soln} this implies
\begin{equation}
	\left| p^r(s) \right| = \left| p^r(0) \right| \leq \Osqrn(r(0)) p^{t^*}(0) \leq (r(0)-r_+)(r(0)-r_+ + 2 \kappa) p^{t^*}(0) \leq \delta (2 \kappa + \delta) \psuppconst.
\end{equation}
We now combine this bound with inequality~\eqref{ineq_radial_time}, which relates $\delta$ and $\tau(\gamma(s))$ to find
\begin{equation} \label{eqn_prbound_intro}
	\left| p^r(s) \right| \leq \begin{cases}
		C(\kappa) \psuppconst \delta \; \leq \; C(\kappa) \psuppconst e^{-c \tau(\gamma(s))} & \kappa > 0 \\
		C \psuppconst \delta^2 \, \quad \leq \; C \psuppconst \tau(\gamma(s))^{-2} & \kappa = 0
	\end{cases}.
\end{equation}
Inequality~\eqref{eqn_prbound_intro} is the key ingredient for proving estimate~\eqref{eqn_vol_trappedsets_intro} in the subextremal case, respectively its extremal analogue~\eqref{eqn_vol_trappedsets_intro_ERN}.

%\begin{equation}
%	p^v(s) = \frac{2}{\Osqrn} p^r(s), \quad p^u(s) = 0.
%\end{equation}

\subsubsection{The extremal case: Upper bounds} \label{into_sec_extremal_upperbounds}
The strategy of proof remains much the same in the subextremal case. At the heart of the argument lies again the almost-trapping estimate, which in the extremal case takes the form
\begin{equation} \label{eqn_almosttrapping_intro_ERN}
	\tau(x) \leq C \left( 1 + ( \log \left| \trapschw \right| )_- + \frac{1}{\sqrt{\OsqS(r_0)}} \right).
\end{equation}
We argue in a similar fashion to the subextremal case to obtain an analogue of inclusion~\eqref{supp_intro_ideaproof}, with two analogous sets $\ERNtrappedsupportsetx, \ERNsmallsupportsetx \subset \mathcal{P}_x$. The structure and volume of the set $\ERNtrappedsupportsetx$ of geodesics which are almost trapped at the photon sphere remains virtually identical to the subextremal case. Crucially however, the volume of the set $\ERNsmallsupportsetx$ of geodesics almost trapped at the event horizon only decays polynomially in $\tau(x)$:
\begin{equation} \label{eqn_vol_trappedsets_intro_ERN}
\begin{gathered}
    \vol \trappedsupportsetx \sim e^{-c \tau(x)}, \\
    \vol \smallsupportsetx \sim \frac{1}{\tau(x)^2}.
\end{gathered}
\end{equation}
This can already be appreciated at the level of radial geodesics, as shown in Section~\ref{sec_radialgeod}. The slower decay of the volume of $\smallsupportsetx$ directly translates into a slower rate of decay for moments of solutions. An additional key difference is that while in the subextremal case all momentum components decay, in the extremal case the $p^{t^*}$-component can only be shown to be bounded along the event horizon.

\subsubsection{The extremal case: Lower bounds}
In order to show lower bounds for moments of a solution $f$, we utilise the existence of slowly infalling geodesics to construct a family of geodesics crossing the event horizon at arbitrarily late times. The initial data of this family of geodesics is captured in the family of subsets $\badsetaplarge_\delta \subset \mathcal{P}|_{\Sigma_0}$. Consider a point along the event horizon $x \in \mathcal{H}^+$ with $\tau(x) \gg 1$. Then the set of all geodesics intersecting $\badsetaplarge_\delta$ and the point $x$ populates a subset $\slowsupportsetx \subset \smallsupportsetx$. This subset satisfies $\vol \slowsupportsetx \sim \vol \smallsupportsetx$ and any momentum $p \in \slowsupportsetx$ satisfies $p^{t^*} \sim 1$. This demonstrates that the upper bounds obtained before are in fact sharp, as long as $\badsetaplarge_\delta \subset \supp {f(x,\cdot)}$ for some $\delta > 0$.

\subsubsection{The extremal case: Growth for transversal derivatives} \label{subsec_growth_overview}

Finally, the proof of Theorem~\ref{maintheorem_ern_nondecay_rough} proceeds by first pulling the transversal $\partial_r$-derivative inside the energy momentum tensor in equation~\eqref{eqn_lowerbound_der_ern}. This incurs only lower order error terms which can be controlled. We then use the massless Vlasov equation to argue that, up to lower order error terms
\begin{equation}
\int_{\sphere} \int_{\mathcal{P}_x} \partial_r f \, \dmux d \omega = \int_{\sphere} \int_{\mathcal{P}_x} \frac{p^{t^*}}{p^r} \partial_{t^*} f \, \dmux d \omega + \dots,
\end{equation}
where $\partial_{t^*}$ is the stationary Killing derivative expressed in $(t^*,r)$-coordinates. Notice that $\partial_{t^*} f$ is again a solution to the massless Vlasov equation if $f$ is a solution. Now Theorem~\ref{maintheorem_ern_nondecay_rough} follows essentially by noticing that any $p = (p^{t^*},p^r,\pslash) \in \slowsupportsetx$ expressed in $(t^*,r)$-coordinates satisfies $p^{t^*} \sim 1$ and $\left| p^r \right| \leq C \tau(x)^{-2}$. Therefore the growth of the fraction $\frac{p^{t^*}}{\left| p^r \right|}$ exactly cancels the decay of the volume of the momentum support.

\subsubsection{Outlook: The Kerr spacetime}
We have already alluded to the Kerr spacetime in the introduction. The geodesic flow on the subextremal and extremal Kerr exterior is similarly determined in full by conserved quantities. In Kerr with $a \neq 0$ one encounters the difficulty that the energy associated to the stationarity is non-positive in the ergoregion, a phenomenon known as \emph{superradiance}. We nonetheless expect that the same principles of proof will apply to the subextremal Kerr solution. On the extremal Kerr spacetime one encounters further the fundamental phenomenon that trapping and superradiance do not decouple. This is the main challenge in understanding the behaviour of linear fields outside the axisymmetric class on extremal Kerr. We hope that the massless Vlasov equation, along with the techniques introduced in the present work, will provide insight into this phenomenon.

\subsection{Related work}

\subsubsection{Related results on the Vlasov equation}
Recall that we have already mentioned the work of Bigorgne~\cite{leo}, Velozo~\cite{renato} and Andersson--Blue--Joudioux~\cite{anderssonblue} on the massless Vlasov equation in Section~\ref{subsec_expdecay}. In the asymptotically flat case, outside the realm of black holes, the Minkowski spacetime has been shown to be stable as a solution to the coupled massless Einstein--Vlasov system. This was first accomplished by Taylor~\cite{martin} and subsequently by Joudioux--Thaller--Kroon~\cite{minkstab2} and Bigorgne--Fajman--Joudioux--Smulevici--Thaller~\cite{fajman}. These works were preceded by the result of Dafermos~\cite{mihalisSpherSymmVlasov}, who demonstrated stability of Minkowski as a solution to massless Einstein--Vlasov under the assumption of spherical symmetry. In the massive case, the first result is due to Rein--Rendall~\cite{reinrendall}, who showed stability of Minkowski space as a solution to the spherically symmetric massive Einstein--Vlasov system. Lindblad--Taylor~\cite{lindbladtaylor} and independently Fajman--Joudioux--Smulevici~\cite{fajman2} established stability of Minkowski space as a solution to the \emph{massive} Einstein--Vlasov system without assumptions of symmetry. Wang~\cite{wang2022global} provides an alternative proof of the stability of Minkowski space for the massive Einstein--Vlasov system which permits non-compactly supported.

For the connection to experimental results, we point out the work of Bieri--Garfinkle~\cite{bierigarfinkle}, who model neutrino radiation in general relativity as a coupled Einstein--null-fluid system and show that neutrino radiation enlarges the Christodoulou memory effect~\cite{christodoulougravwaves} of gravitational waves. The null-fluid matter model can be regarded as a limiting case of the massless Vlasov system.

\subsubsection{Related results on instability of extremal black holes}
We have discussed the work of Aretakis~\cite{aretakis1,aretakis2,aretakis3} on the horizon instability of extremal Reissner--Nordstr\"om for the wave equation in Section~\ref{section_aretakis} above. Since then, there has been a tremendous amount of progress expanding these results. Angelopoulos--Aretakis--Gajic~\cite{nonlinearwavesERN} have since shown that the Aretakis instability holds for a class of nonlinear wave equations satisfying the null condition on the extremal Reissner--Nordstr\"om spacetime. Recently, Apetroaie~\cite{apetroaie} has shown that instability results along the event horizon of extremal Reissner--Nordstr\"om persist for the linearised Einstein equations. Aretakis~\cite{aretakisgeneralhorizons} has generalised the instability result for the wave equation to a more general class of axisymmetric extremal horizons, which includes as special cases the event horizons of the extremal Reissner--Nordstr\"om and extremal Kerr spacetimes. Teixeira~da~Costa~\cite{rita} has shown that despite the presence of a linear instability, there are no exponentially growing modes for the Teukolsky equation on extremal Kerr black holes. Lucietti--Reall~\cite{reall} have extended the conservation law that lies at the heart of the horizon instability to any extremal black hole and a larger class of theories in several dimensions. 

For the extremal Kerr spacetime, Aretakis~\cite{subextkerr} has obtained boundedness and decay of solutions to the wave equation under the assumption of axisymmetry. A new instability result for higher azimuthal mode solutions to the wave equation on extremal Kerr has recently been obtained by Gajic~\cite{dejanforthcoming}, who proved the existence of stronger asymptotic instabilities for non-axisymmetric solutions to the wave equation. Understanding linear fields in generality, in particular including the regime in which trapping meets superradiance, is one of the biggest open problems in extremal black hole dynamics. 

Despite the fact that massless linear fields on extremal Reissner--Nordstr\"om are well understood, an understanding of the behaviour of the solution on the nonlinear level is still lacking. Dafermos--Holzegel--Rodnianski--Taylor~\cite{schwarzschildnonlinearstable} conjecture a weak form of nonlinear asymptotic stability of the extremal Reissner--Nordstr\"om spacetime as a solution to the Einstein--Maxwell equations. More precisely, they conjecture that on a suitable finite codimension `submanifold' of the moduli space of initial data the maximal Cauchy development asymptotes an extremal Reissner--Nordstr\"om solution, while satisfying only weaker decay along the event horizon and suitable higher order quantities blow up polynomially along the event horizon (growing `horizon hair').

\subsection{Outline}
In Section~\ref{prelim} we discuss some preliminaries on the massless Vlasov equation and the geometry of the subextremal and extremal Reissner--Nordstr\"om solution. In Section~\ref{section_mainthms} we state our main Theorems~\ref{maintheorem},~\ref{maintheoremERN},~\ref{maintheorem_ern_nondecay_rough} in a more precise form. Section~\ref{section_proof} is devoted to the subextremal case and the proof of Theorem~\ref{maintheorem}. In Section~\ref{section_ERN}, we turn towards the extremal case and provide the proofs of Theorems~\ref{maintheoremERN} and~\ref{maintheorem_ern_nondecay_rough}.

\subsection{Acknowledgments}
The author would like to thank Martin Taylor and Gustav Holzegel for their expert advice and many insightful discussions. The author would also like to thank the Department of Mathematics at the University of M\"unster for their hospitality during several research visits.

\subsection{Declarations}

\subsubsection{Funding and competing interests}
This research has been funded through ERC Consolidator Grant 772249 and Germany’s Excellence Strategy EXC 2044~390685587, Mathematics M\"unster: Dynamics--Geometry--Structure. The author has no competing interests to declare that are relevant to the content of this article.

\subsubsection{Data availability statement}
Data sharing not applicable to this article as no datasets were generated or analysed during the current study.

\section{Geometry of the Reissner--Nordstr\"om solution} \label{prelim}
In this section we discuss the necessary preliminaries and definitions to state and prove our main theorems in a more precise way. We introduce the Reissner--Nordstr\"om family of solutions and express the metric in several coordinate systems which we shall use throughout the work in Section~\ref{rngeometry}. We then define the time function used to measure decay in Section~\ref{timecoordinate} and introduce the massless Vlasov equation on the Reissner--Nordstr\"om exterior in Section~\ref{sec_vlasov}. In the final subsection~\ref{sec_subsets} we introduce a family of subsets of the mass-shell which will play a crucial role in characterising the momentum support of a solution to the massless Vlasov equation in the proofs.

\subsection{The Reissner--Nordstr\"om metric} \label{rnmetricmanifold} \label{rngeometry}
The Reissner--Nordstr\"om metric is a stationary spherically symmetric solution to the Einstein--Maxwell equations and models the exterior of a non-rotating charged black hole. See~\cite{oneill, waldbook, dafermosETHnotes, aretakisbook} for a thorough discussion of the geometric concepts used here. 

Let $\sphere$ denote the standard $2$-sphere equipped with the standard round metric $d \omega^2$. For the most part we will not require an explicit choice of coordinates on the sphere $\sphere$, although we will occasionally make use of the standard spherical coordinates $(\theta,\phi)$. If we do not require explicit coordinates on the $2$-sphere we will denote points on the sphere by $\omega \in \sphere$. 

We introduce the exterior of the Reissner--Nordstr\"om black hole in $(t^*,r)$-coordinates, which shall be the coordinate system we will mainly make use of in this discussion. In these coordinates the metric takes the form
\begin{equation}
\gRN = - \Osqrn (dt^*)^2 + 2 (1-\Osqrn) dt^* dr + (2-\Osqrn) dr^2 + r^2 d \omega^2, \quad \Osqrn = 1 - \frac{2m}{r} + \frac{q^2}{r^2},
\end{equation}
where the parameter $m > 0$ represents the black hole's mass and $\left| q \right| \leq m$ represents the black hole's electromagnetic charge. We denote the spherical part of the metric by $\gslash = r^2 d \omega^2$. Note that $\Osqrn$ has two roots which we denote by $0 < r_- \leq r_+$. In this study we are concerned with the exterior of the black hole, which is the manifold with boundary
\begin{equation}
(t^*,r,\omega) \in \Mrn = \R \times [r_+,\infty) \times \sphere.
\end{equation}
The stationarity of the metric gives rise to the timelike Killing vector field $T =  \partial_{t^*}$, while the spherical symmetry induces the usual rotational Killing vector fields $\{ \Omega_i \}_{i=1,2,3}$ generating a representation of the Lie algebra $\text{so}(3)$. We call the null hypersurface
\begin{equation}
\mathcal{H}^+ = \Big\{ r = r_+, t^* > 0 \Big\} \subset \Mrn
\end{equation}
the future event horizon of the black hole. The normal to $\mathcal{H}^+$ is easily computed to be the stationary timelike Killing vector field $T$. A computation reveals that along $\mathcal{H}^+$
\begin{equation}
    \nabla_T T = \kappa T, \quad \kappa = \frac{1}{2} \frac{d \Osqrn}{dr}\bigg|_{r=r_+} .
\end{equation}
We say that the solution is subextremal when $\kappa > 0$, which is equivalent to $r_- < r_+$ and to $ \left| q \right| < m$. We say the solution is extremal when $\kappa = 0$ or equivalently $r_- = r_+$, which is further equivalent to $\left| q \right| = m$. See also~\cite[Section 1.5]{aretakisbook} for an in-depth discussion of the notions of extremal and subextremal black holes. The timelike hypersurface $\left\{ r = r_{ph} , t^* > 0 \right\} \subset \Mrn$ with 
\begin{equation}
    r_{ph} = \frac{3m}{2}\left( 1 + \sqrt{1- \frac{8 q^2}{9 m^2}} \right)
\end{equation}
is called the photon sphere of the black hole. Both the event horizon and the photon sphere permit trapped null geodesics, see Section~\ref{prelim_massshell_estimates} below. This trapping effect is key to understanding the decay of moments of solutions to the massless Vlasov equation. 

From Section~\ref{section_mainthms} onward we will be mostly concerned with two special cases of the Reissner--Nordstr\"om family: the Schwarzschild solution, obtained by setting $q=0$ and the extremal Reissner--Nordstr\"om (ERN) solution, obtained by setting $\left| q \right| = m$. For the reader's convenience, we summarise the relevant metric and geometric quantities here:
\begin{align}
    \text{Schwarzschild: }\quad &\Osqrn = 1 - \frac{2m}{r}, & &r_+ = 2m, & &r_{ph} = 3m, & & \kappa = \frac{1}{4m}, \label{table_geom_values_schw}\\
    \text{ERN: }\quad &\Osqrn = \left( 1 - \frac{m}{r} \right)^2, & &r_+ = m, & &r_{ph} = 2m, & & \kappa = 0 . \label{table_geom_values_ERN}
\end{align}

\begin{rem}
By a slight abuse of notation, we will use the same symbols to denote any geometric quantities related to either the Schwarzschild or the extremal Reissner--Nordstr\"om solution. It will be apparent from the context which member of the Reissner--Nordstr\"om family we are referring to. Thus, if we are discussing the extremal Reissner--Nordstr\"om solution, then $\Osqrn, r_+, r_{ph}$ etc. should all be set to their corresponding values given in~\eqref{table_geom_values_ERN} above. Likewise for the Schwarzschild solution with~\eqref{table_geom_values_schw}.
\end{rem}

\subsubsection{Alternative coordinate systems} \label{altcoordsystems}
Let us restrict to the region $\{r > r_+ \} \subset \Mrn$ in the Reissner--Nordstr\"om exterior and apply the coordinate transformation
\begin{equation}
t^* = t + F(r), \quad \frac{dF}{dr} = \frac{1-\Osqrn}{\Osqrn}, \quad \lim_{r \rightarrow \infty} F(r) = 0.
\end{equation}
In this way we obtain the local Boyer--Lindquist coordinates $(t,r,\omega)$, which are defined only in the range $t \in \R,r>r_+, \omega \in \sphere$. In these coordinates the metric may be expressed as
\begin{equation}
\gRN = - \Osqrn dt^2 + \OsqrnNEG dr^2 + r^2 d \omega^2.
\end{equation}
We may solve for $F$ explicitly. We provide the closed form of $F$ for the Schwarzschild exterior and the extremal Reissner--Nordstr\"om exterior:
\begin{equation}
    F(r) = \begin{cases}
        2m \log(r-2m) & q = 0 \\
        2m \log(r-m) - \frac{m^2}{r-m} & \left| q \right| = m
    \end{cases}.
\end{equation}
We note that the metric may be smoothly extended to the region where $0 < r < \infty$, so that it is the Boyer--Lindquist coordinate system that becomes singular as $r \rightarrow r_+$, not the spacetime itself. 

Let us introduce two further useful coordinate systems, Regge--Wheeler coordinates and double null coordinates. If $(t,r,\omega)$ denote Boyer--Lindquist coordinates as above, we define
\begin{equation} \label{eqn_rstar_defn}
    \frac{d r^*}{dr} = \frac{1}{\Osqrn}, \quad r^*(r_{ph}) = 0.
\end{equation}
to obtain Regge--Wheeler coordinates $(t,r^*,\omega)$. The exterior excluding the event horizon then corresponds to the region $(t,r^*,\omega) \in \R \times \R \times \sphere$ and the metric takes the form
\begin{equation}
\gRN = \Osqrn \left( - dt^2 + (d r^*)^2 \right) + r^2 d \omega^2.
\end{equation}
We may solve equation~\eqref{eqn_rstar_defn} explicitly and find that in the special cases of the Schwarzschild and extremal solutions
\begin{equation}
r^* = \begin{cases}
        r + 2m \log(r-2m) -3m - 2m \log m & q = 0 \\
        r + 2m \log(r-m) - \frac{m^2}{r-m} - 2m \log(m) - m & \left| q \right| = m
    \end{cases}.
\end{equation}
For the double null coordinate system, we use Regge--Wheeler coordinates and define
\begin{equation} \label{regwheelertodoublenull}
u = t-r^*, \quad v = t+r^*,
\end{equation}
to obtain the double null coordinate system $(u,v,\omega)$. The coordinates are defined over the same region and the metric takes the form
\begin{equation}
\gRN = -\Osqrn du dv + r^2 d \omega^2.
\end{equation}
We refer to the idealised null hypersurface $\mathcal{I}^+ = \{ v = \infty \}$ as future null infinity. Let us emphasise again at this point that all the coordinate systems discussed in this section break down at the event horizon.

\subsection{The time function $\tau$} \label{timecoordinate}
Let us define the specific time function with respect to which we will state the decay result. For our proofs to work, we need only choose a function $\tau: \Mrn \rightarrow \R$ such that $\nabla \tau$ is future-directed causal and such that the hypersurfaces $\Sigma_{\tau'} = \{ x \in \Mrn \, | \, \tau(x) = \tau' \}$ are Cauchy hypersurfaces invariant under the action of the rotation group and connect the event horizon $\mathcal{H}^+$ and future null infinity $\mathcal{I}^+$. For concreteness, consider the following
\begin{defi} \label{defn_timecoord_Schw}
Let $R > r_{ph}$ and define the time function $\tau: \Mrn \rightarrow \R$ by
\begin{equation}
\tau(x) = \begin{cases}
	t^*(x) & r_+ \leq r(x) \leq R \\
	u(x) + c(R) & R \leq r(x)
\end{cases},
\end{equation}
where we use $(t^*,r)$-coordinates in the region $r_+ \leq r \leq R$ and double null coordinates $(u,v)$ in the region $r \geq R$. The constant $c(R)$ is chosen to ensure continuity of $\tau$. Explicitly in the Schwarzschild respectively extremal Reissner--Nordstr\"om spacetime
\begin{equation}
    c(R) = \begin{cases}
        R + 4m \log(R-2m) - 3m - 2m \log(m) & q = 0 \\
        R + 4m \log(R-m) - \frac{2m^2}{R-m} - 2m \log(m) - m & \left| q \right| = m
    \end{cases} .
\end{equation}
We denote hypersurfaces of constant $\tau$ by $\Sigma_\tau$, see Figure~\ref{penrose_tau}.
\end{defi}

Note carefully that we have only chosen $\tau$ to be continuous. The derivative $\nabla \tau(x)$ is not defined at points $x \in \Mrn$ where $r(x) = R$. However both a left-sided and a right-sided derivative of $\tau$ may be defined and they are both causal. It is with respect to $\tau$ that we will state the decay of the momentum support of $f$ below. We will subsequently refer to the region $\mathcal{A} := \{ r \geq R \}$ as the asymptotically flat region, or in other words the region which is in some sense far away from the black hole. We will frequently suppress the argument of $\tau$ when it is clear from the context that we are referring to $\tau(x)$ for a given $x \in \Mrn$.

\subsection{The mass-shell and the massless Vlasov equation} \label{sec_vlasov}
In this subsection we introduce the massless Vlasov equation on the Reissner--Nordstr\"om exterior $\Mrn$ and discuss some important associated definitions.

\subsubsection{The geodesic flow on Reissner--Nordstr\"om}
Let us denote by $\{ S_s \}_{s \in \R}$ the geodesic flow on $\Mrn$, so that for any point $(x,p) \in T \Mrn$
\begin{equation} \label{geodesicflow}
S_s(x,p) = (\gamma_{(x,p)}(s),\dot{\gamma}_{(x,p)}(s)),
\end{equation}
where $\gamma_{(x,p)}$ is the unique affinely parametrised geodesic with $\gamma_{(x,p)}(0) = x, \dot{\gamma}_{(x,p)}(0) = p$ and we only define $S_s(x,p)$ if $\gamma_{(x,p)}(s) \in \Mrn$. The generator of the geodesic flow is a vector field on the tangent bundle $TM$ called the geodesic spray $X$. Let $(x^\mu)$ denote one of the coordinate systems on the exterior $\Mrn$ defined above. Then we can obtain an associated coordinate system $(x^\mu, p^\mu)$ on the tangent bundle $T \Mrn$ by representing any tangent vector $p \in T_x \Mrn$ as $p = p^\mu \partial_\mu$, so that $(x,p) \in T \Mrn$ has the coordinates $( x^\mu, p^\mu )$. In the coordinates on $T \Mrn$ associated to $(x^\mu)$ the geodesic spray takes the form
\begin{equation}
	X = p^\mu \partial_\mu - \Gamma^\mu_{\alpha \beta} p^\alpha p^\beta \partial_{p^\mu},
\end{equation}
where $\Gamma^\mu_{\alpha \beta}$ denote the Christoffel symbols of $g$ in $(x^\mu)$-coordinates.

\subsubsection{The mass-shell and massless Vlasov equation}
The bundle of future light-cones or mass-shell on $(\Mrn,\gRN)$ is defined as
\begin{equation}
\mathcal{P} = \Big\{ (x,p) \in T \Mrn : \, \gRN(x)(p,p)=0, \, p \text{ is future-directed} \Big\} .
\end{equation}
The condition $\gRN(x)(p,p)=0$ is referred to as the mass-shell relation. If we use $(t^*,r)$-coordinates $(x^\mu) = (t^*,r,\theta,\phi)$ with standard spherical coordinates $(\theta,\phi)$ on $\sphere$, we denote the conjugate coordinates on the tangent bundle by $(x^\mu,p^\mu) = (t^*,r,\theta,\phi,p^{t^*},p^r,p^{\theta},p^{\phi})$. Often times, we do not require an explicit choice of coordinates on $\sphere$. In this case, we simply denote points on $\sphere$ by $\omega = (\theta,\phi)$ and write $\pslash = (p^{\theta},p^{\phi})$ for the angular momentum components, so that the conjugate coordinates on the tangent bundle become $(x^\mu,p^\mu) = (t^*,r,\omega,p^{t^*},p^r,\pslash)$. We also introduce the notation $\left| \pslash \right|_{\gslash}^2 = \gslash_{AB} p^A p^B$, where $A,B \in \{ \theta, \phi \}$. We may then express the mass-shell relation in $(t^*,r)$-coordinates as
\begin{equation}
\gRN(x)(p,p)= - \Osqrn (p^{t^*} )^2 + 2(1-\Osqrn) p^{t^*} p^r + (2-\Osqrn) (p^r)^2 + \left| \pslash \right|_{\gslash}^2 = 0.
\end{equation}
Note that the geodesic flow preserves lengths, $X(g(p,p)) = 0$, so that the geodesic spray restricts to a vector field on the mass-shell $\mathcal{P}$. We can now state the following

\begin{defi}[Massless Vlasov equation]
Let $f: \mathcal{P} \rightarrow \R_{\geq 0}$ be smooth. We say that $f$ solves the massless Vlasov equation if $X(f) = 0$ or equivalently if $f$ is conserved  along the null geodesic flow.
\end{defi}

\begin{rem}
If $\Sigma_0 \subset M$ denotes a Cauchy hypersurface we define $\mathcal{P}_0 = \mathcal{P}|_{\Sigma_0}$. We may then specify initial data $f_0: \mathcal{P}_0 \rightarrow \R_{\geq 0}$ and obtain a unique solution $f: \mathcal{P} \rightarrow \R_{\geq 0}$ attaining these initial data by transporting along future-directed null geodesics. It may be directly verified that if $f_0 \in C^k(\mathcal{P}_0; \R)$, the space of $k$-times continuously differentiable functions, then also $f \in C^k(\mathcal{P}_0; \R)$. Therefore the massless Vlasov equation is well-posed.
\end{rem}

\begin{assumption}[Compactness of support of initial distribution] \label{assumption_support}
	Let $f_0: \mathcal{P}_0 \rightarrow \R$ be a continuous bounded initial distribution on the (subextremal or extremal) Reissner--Nordstr\"om background. We assume the existence of constants $\rsuppconst > 0$ and $\psuppconst > 0$ such that
	\begin{equation} \label{eqn_assumption_support}
		\supp(f_0) \subset \Big\{ (x,p) \in \mathcal{P} : x \in \Sigma_0, \; r \leq \rsuppconst, \; p^{t^*}, \left| p^r \right|, \left| \pslash \right|_{\gslash} \leq \psuppconst \Big\},
	\end{equation}
	where we have used $(t^*,r)$-coordinates and their associated coordinates on the tangent bundle.
\end{assumption}

\begin{rem}
The constant $\rsuppconst$ bounds the spatial support of $f_0$, while $\psuppconst$ provides a bound on the size of the momentum support of $f_0$. We further note that Assumption~\ref{assumption_support} makes use of the $(t^*,r)$-coordinates and is therefore dependent on this specific choice of coordinates. However, it follows immediately that if $(x^\mu)$ denotes another system of coordinates which remains regular up to and including the event horizon and $(x^\mu, p^\mu)$ denotes the associated conjugate coordinate system on the tangent bundle, then $f_0$ satisfies Assumption~\ref{assumption_support} with some $\rsuppconst, \psuppconst$ if and only if
\begin{equation}
		\supp(f_0) \subset \Big\{ (x,p) \in \mathcal{P} : x \in \Sigma_0, \; \left| x^\mu \right| \leq \tilde{\rsuppconst}, \;  \left| p^\mu \right| \leq \tilde{\psuppconst} \Big\},
	\end{equation}
 for some constants $\tilde{\rsuppconst}, \tilde{\psuppconst}$. Therefore, the assumption of compact support is not dependent on a choice of coordinates, although the precise constants $\rsuppconst, \psuppconst$ are.
\end{rem}

\begin{rem}
We will often say that a null geodesic segment $\gamma: [s_1,s_2] \rightarrow \Mrn$ satisfies Assumption~\ref{assumption_support}, by which we mean that there exists $s \in [s_1,s_2]$ for which $\gamma(s) \in \Sigma_0$ and $(\gamma(s),\dot{\gamma}(s))$ is contained in the set on the right hand side of~\eqref{eqn_assumption_support}. The constants $\rsuppconst$ and $\psuppconst$ defined above will appear throughout the proofs.
\end{rem}

\subsubsection{Conserved quantities of the geodesic flow}
The symmetries of the Reissner--Nordstr\"om metric imply the existence of quantities which are conserved along the geodesic flow. The stationarity respectively spherical symmetry of the Reissner--Nordstr\"om metric imply that the energy $E: \mathcal{P} \rightarrow \R_{\geq 0}$ respectively the total angular momentum $L^2: \mathcal{P} \rightarrow \R_{\geq 0}$ are conserved, or in other words $X(E) = X(L) = 0$. Let us use $(t^*,r)$-coordinates and denote the conjugate coordinates on the tangent bundle by $(t^*,r,\omega,p^{t^*},p^r,\pslash)$. Then we may express the energy and angular momentum explicitly as
\begin{equation} \label{def_energy}
	E = \Osqrn p^{t^*} - (1-\Osqrn) p^r, \quad L^2 = r^2 \left| \pslash \right|^2_{\gslash}.
\end{equation}
Combining these explicit expressions with the mass-shell relation shows that for all points on the mass-shell $(x,p) = (t^*,r,\omega,p^{t^*},p^r,\pslash) \in \mathcal{P}$ expressed in $(t^*,r)$-coordinates the conservation of energy identity holds,
\begin{equation} \label{Schw::energy_conservation}
	E^2 = (p^r)^2 + \frac{\Osqrn}{r^2} L^2.
\end{equation}

\begin{defi}[Trapping parameter] \label{def_eps_def}
We define the trapping parameter $\trapschw: \mathcal{P} \rightarrow [-\infty,1]$ to be the conserved quantity satisfying the following equality
\begin{equation} \label{def_eps}
	E^2 - \rnconstant \frac{L^2}{m^2} = \trapschw E^2,
\end{equation}
where $\rnconstant$ depends only on the member of the Reissner--Nordstr\"om family under consideration and is explicitly defined as
\begin{equation}
	\rnconstant = \rnconstant \left( \frac{\left| q \right|}{m} \right) = \frac{m^2}{r^2} \Osqrn \Bigg|_{r = r_{ph}} = \begin{cases}
		\frac{1}{27} & q = 0 \\
		\frac{1}{16} & \left| q \right| = m
	\end{cases}.
\end{equation}
\end{defi}

\begin{rem}
The trapping parameter measures the failure of a null geodesic $\gamma$ to be trapped at the photon sphere. Consider a future-directed null geodesic $\gamma(s) = (t^*(s),r(s),\omega(s))$ with affine parameter $s$ expressed in $(t^*,r)$-coordinates. Then $\gamma$ is trapped at the photon sphere if $\gamma$ is future complete and $r(s) \rightarrow r_{ph}$ as $s \rightarrow \infty$. It is a classic fact that a future-directed null geodesic $\gamma$ is trapped at the photon sphere if and only if $\trapschw(\gamma,\dot{\gamma}) = 0$. It follows that at any point $x \in \Mrn \setminus \mathcal{H}^+$, the subset of all $p \in \mathcal{P}_x$ with the property that the geodesic with initial data $(x,p)$ is trapped at the photon sphere, is identical with the codimension one subset $\{ p \in \mathcal{P}_x : \trapern(x,p) = 0 \} \subset \mathcal{P}_x$.
\end{rem}

\begin{rem}
Conservation of energy~\eqref{Schw::energy_conservation} immediately implies $\frac{L^2}{E^2} \leq \frac{r^2}{\Omega^2}$ so that
\begin{equation}  \label{epsbound}
- \infty \leq - \frac{1}{\Osqrn} \frac{r^2}{m^2} \left( \rnconstant - \frac{m^2}{r^2} \Osqrn \right) \leq \trapschw \leq 1,
\end{equation}
where we note that by definition
\begin{equation}
\rnconstant - \frac{m^2}{r^2} \Osqrn \sim \left( 1 - \frac{r_{ph}}{r} \right)^2.
\end{equation}
This bound may be interpreted as saying that for a given radius $r$, the trapping parameter $\trapschw$ is restricted to a finite range depending on the value of $r$. Alternatively, we may interpret this inequality as saying that for fixed $\trapschw < 0$, the radius $r$ is restricted. 
\end{rem}

\begin{defi}
Assume that $-\infty < \trapschw < 0$, then the equation
\begin{equation}
	\frac{1}{\Osqrn} \frac{r^2}{m^2} \left( \rnconstant - \frac{m^2}{r^2} \Osqrn \right) = \left| \trapschw \right|
\end{equation}
has exactly two real solutions $r_{\text{min}}^-(\trapschw)$ and $r_{\text{min}}^+(\trapschw)$, labelled such that
\begin{equation}
	r_+ < r_{\text{min}}^-(\trapschw) < r_{ph} < r_{\text{min}}^+(\trapschw) < \infty.
\end{equation}
We call $r_{\text{min}}^-(\trapschw), r_{\text{min}}^+(\trapschw)$ the radii of closest approach to the photon sphere for a given negative trapping parameter.
\end{defi}

\subsubsection{Parametrising the mass-shell} \label{sec_parametrising_massshell}
To obtain a parametrisation of the mass-shell $\mathcal{P}$ we consider $(t^*,r)$-coordinates $(x^\mu) = (t^*,r,\omega)$ and their associated coordinates $(x^\mu, p^\mu) = (t^*,r,\omega,p^{t^*},p^r,\pslash)$ on the tangent bundle. We then eliminate the variable $p^{t^*}$ by using the mass-shell relation and express it as a function of $r,p^r$ and $\pslash$. We make this explicit in the following
\begin{lem} \label{express_ptstar}
For $(x,p) = (t^*,r,\omega, p^{t^*},p^r,\pslash) \in \mathcal{P}$ with $r > r_+$ we have
\begin{equation} \label{pt_expressed_via_pr}
	p^{t^*} = \frac{(1-\Osqrn) p^r + \sqrt{(p^r)^2 + \Osqrn \left| \pslash \right|_{\gslash}^2 }}{\Osqrn} = \frac{E + (1-\Osqrn) p^r}{\Osqrn}.
\end{equation}
Furthermore when we assume that $p^r \leq 0$ and $r \geq r_+$ we find the alternative expression
\begin{equation} \label{express_ptstar_neg}
	p^{t^*} = \left| p^r \right| + \frac{\left| \pslash \right|_{\gslash}^2}{E + \left| p^r \right|}.
\end{equation}
Along the event horizon $r = r_+$ and we necessarily have $p^r \leq 0$ and
\begin{equation} \label{pt_expressed_via_pr_horizon}
	p^{t^*} = \left| p^r \right| +\frac{\left| \pslash \right|_{\gslash}^2}{2 E}.
\end{equation}
In particular $p^{t^*} \geq 0$ for all $r \geq r_+$ and $p^{t^*} = 0$ if and only if $p = 0$.
\end{lem}
\begin{proof}
Let us begin by noting that the definition of energy $E$ implies
\begin{equation}
	E = \OsqS p^{t^*} - (1-\Osqrn) p^r,
\end{equation}
which we may rearrange in order to immediately conclude the relation
\begin{equation}
	p^{t^*} = \frac{E + (1-\Osqrn) p^r}{\OsqS},
\end{equation}
whenever $r > r_+$. Conservation of energy~\eqref{Schw::energy_conservation} now implies
\begin{equation}
	E = \sqrt{ (p^r)^2 + \frac{\OsqS}{r^2} L^2 } =\sqrt{ (p^r)^2 + \OsqS \left| \pslash \right|_{\gslash}^2 } ,
\end{equation}
which inserted into the relation above immediately allows us to conclude~\eqref{pt_expressed_via_pr}. If we now assume that $p^r \leq 0$ and $r > r_+$, note that again using conservation of energy we may rewrite
\begin{equation}
	p^{t^*} = \frac{E - (1-\Osqrn) \left| p^r \right|}{\OsqS} = \left| p^r \right| + \frac{E - \left| p^r \right|}{\OsqS} =\left| p^r \right| + \frac{1}{\OsqS} \frac{E^2 - (p^r)^2}{E + \left| p^r \right|} = \left| p^r \right| + \frac{\left| \pslash \right|_{\gslash}^2}{E + \left| p^r \right|}.
\end{equation}
In particular we note that at the event horizon, future-orientedness of $p$ implies that $- p^r = E \geq 0$. Therefore this relation holds for all $r \geq r_+$ and evaluating it at $r = r_+$ recovers equation~\eqref{pt_expressed_via_pr_horizon}. Alternatively this may be seen by an application of L'Hôpital's rule. Note that as a consequence of the mass-shell relation $p^{t^*} = 0$ implies $L=p^r=0$ and therefore $p^\mu = 0$ for all $\mu$. 
\end{proof}

\begin{rem}
We denote the resulting parametrisation of $\mathcal{P}$ by $(t^*,r,\omega,p^r,\pslash)$. Note carefully that if $r= r_+$ and $p^r = 0$ the mass-shell relation implies $\pslash = 0$ and the coordinate system $(p^r,\pslash)$ on $\mathcal{P}_{(t^*,r_+,\omega)}$ over a point on the event horizon degenerates. This will however not cause any issues in practice, as the reader may readily verify that the set $\{ (p^{t^*},p^r,\pslash) \in \mathcal{P}_{(t^*,r_+,\omega)} : p^r = 0 \}$ has vanishing measure, see equation~\eqref{dmu_first_comp_schw} for the definition of the volume form.
\end{rem}

\subsubsection{Preliminary estimates for points in the mass-shell} \label{prelim_massshell_estimates} \label{prelimestimates_ERN}
In this section we prove estimates for momenta in the mass-shell over the subextremal or extremal Reissner--Nordstr\"om exterior. We express our results in $(t^*,r)$-coordinates and consider the conjugate coordinates $(x,p) = (t^*,r,\omega,p^{t^*},p^r,\pslash)$ as functions on the mass-shell $\mathcal{P}$. Lemma~\ref{ptstar_bound} establishes a bound for the $p^{t^*}$-component and Lemma~\ref{lem_quotientbound} provides a bound for the quotient $\frac{p^{t^*}}{p^r}$.

\begin{lem} \label{ptstar_bound} \label{ptstar_bounds_ern}
	Let  $(x,p) = (t^*,r,\omega, p^{t^*},p^r,\pslash) \in \mathcal{P}$, where $\mathcal{P}$ denotes the mass-shell over the subextremal or extremal Reissner--Nordstr\"om exterior. Then if $p^r > 0$ and $r > r_+$ we have the bound
	\begin{equation} \label{ptstar_bound_eq1}
		\frac{m^2}{r^2} E \left( 1 + \left| \trapschw \right| \right) \lesssim p^{t^*} \leq \frac{2 E}{\OsqS},
	\end{equation}
	where $\trapschw$ denotes the trapping parameter introduced in Definition~\ref{def_eps_def}. If $p^r \leq 0$ and $r \geq r_+$ then
	\begin{equation} \label{ptstar_bound_eq3}
		\frac{m^2}{r^2} E \left( 1 - \trapschw \right) \lesssim p^{t^*} \lesssim \left( 1 + \frac{m^2}{r^2} \left| \trapschw \right| \right) E \lesssim \frac{E}{\OsqS}.
	\end{equation}
	In particular, we may conclude that if $\delta > 0$ and $r_+ + \delta m \leq r$ we have
	\begin{equation}
		p^{t^*} \lesssim_\delta E.
	\end{equation}
\end{lem}
\begin{proof}
	We use Lemma~\ref{express_ptstar} to express $p^{t^*}$ as
	\begin{equation}
		p^{t^*} = \frac{E + (1-\Osqrn) p^r}{\OsqS}.
	\end{equation}
	Let us first assume that $p^r > 0$ and $r > r_+$. Using the fact that conservation of energy~\eqref{Schw::energy_conservation} implies $\left| p^r \right| \leq E$ we readily conclude
	\begin{equation}
		\frac{E}{\OsqS} \leq p^{t^*} \leq 2 \frac{E}{\OsqS}.
	\end{equation}
	Now recall inequality~\eqref{epsbound} which immediately implies $\OsqS \frac{m^2}{r^2} \left( 1 + \left| \trapschw \right| \right) \lesssim 1$. Therefore we can conclude the bound
	\begin{equation}
		\frac{m^2}{r^2} \left( 1 + \left| \trapschw \right| \right) E \lesssim \frac{E}{\OsqS} \leq p^{t^*}.
	\end{equation}
	Let us turn to the case that $p^r \leq 0$ and $r \geq r_+$. In this case, we again apply Lemma~\ref{express_ptstar} to represent $p^{t^*}$ as in equality~\eqref{express_ptstar_neg} in order to deduce
	\begin{equation}
		p^{t^*} = \left| p^r \right| + \frac{\left| \pslash \right|_{\gslash}^2}{E + \left| p^r \right|} \leq E +  \frac{\left| \pslash \right|_{\gslash}^2}{E} = E \left( 1 + \frac{1}{r^2} \frac{L^2}{E^2} \right) \lesssim E \left( 1 + \frac{m^2}{r^2} \left| \trapschw \right| \right),
	\end{equation}
	where we have used the inequality $\frac{1}{m^2} \frac{L^2}{E^2} \lesssim 1 + \left| \trapschw \right|$ by definition of the trapping parameter. In the other direction we find
	\begin{equation}
		p^{t^*} \geq \frac{\left| \pslash \right|_{\gslash}^2}{2 E} = \frac{1}{2 r^2} \frac{L^2}{E^2} E \gtrsim \frac{m^2}{r^2} \left( 1 - \trapschw \right) E.
	\end{equation}
	This concludes the proof.
\end{proof}

\begin{lem} \label{lem_quotientbound} \label{prptstar_ern}
	Let $\mathcal{P}$ denote the mass-shell over the subextremal or extremal Reissner--Nordstr\"om exterior. For $(x,p) = (t^*,r,\omega, p^{t^*},p^r,\pslash) \in \mathcal{P}$ with $p^r \neq 0$ and $\signpr = \sgn(p^r)$, we have
	\begin{equation} \label{lem_quotientbound_eqn}
		\frac{p^r}{p^{t^*}} \sim \signpr \sqrt{(r-r_{ph})^2+ \trapschw \mathfrak{a}} \, \frac{1}{r} \mathfrak{A}_{\signpr},
	\end{equation}
	where $\trapschw$ denotes the trapping parameter from Definition~\ref{def_eps_def}, $\mathfrak{a}: [r_+, \infty) \rightarrow \R, r \mapsto \mathfrak{a}(r)$ is an explicit function which satisfies $\mathfrak{a} \sim m^2 \OsqS$ and $\mathfrak{A}_{\signpr}: [r_+, \infty) \times \R \rightarrow \R, (r, \trapschw) \mapsto \mathfrak{A}_{\signpr}(r, \trapschw)$ is an explicit function satisfying for any $0 < \delta < 1$
	\begin{equation}
		\mathfrak{A}_{\signpr} \sim_\delta \begin{cases}
			1 & r \geq r_+ + \delta m \\
			\begin{cases}
				\frac{1}{1 + \left| \trapschw \right|} & \signpr=-1 \\
				\OsqS & \signpr=1
			\end{cases} & r_+ \leq r \leq r_+ + \delta m
		\end{cases}.
	\end{equation}
	In particular, away from the horizon the quotient takes an identical form for $\signpr \in \{-1,+1\}$ apart from the obvious difference in sign. As an immediate consequence, we conclude that $\left| p^r \right| \lesssim p^{t^*}$.
\end{lem}

\begin{rem} \label{rem_prelimestimates_continuity}
	The quotient $\frac{p^r}{p^{t^*}}$ is continuous as a function on the mass-shell $\mathcal{P}$ up to and including the event horizon. It may however be easily verified that the function on the right hand side of equation~\eqref{lem_quotientbound_eqn} is discontinuous at all points $(x,p) = (t^*,r,\omega,p^{t^*},p^r,\pslash) \in \mathcal{P}$ with $r= r_+$ and $p^r = 0$. We will however only be interested in evaluating equation~\eqref{lem_quotientbound_eqn} along null geodesics, and the reader may readily verify that the right hand side is continuous along every future-directed null geodesic.
\end{rem}

\begin{proof}
	We express both sides of the expression~\eqref{def_eps} above in $(t^*,r)$-coordinates and collect terms to find the following quadratic equation:
	\begin{equation} \label{eqn_quadratic}
		\left( b - \trapschw (1-\Osqern)^2 \right) (p^r)^2 +2(1-\Osqern) \left( a + \trapschw \Osqern  \right) p^{t^*} p^r - \Osqern  \left( a + \trapschw \Osqern  \right) (p^{t^*})^2 = 0,
	\end{equation}
	where we have abbreviated
	\begin{equation} \label{def_ab_ern}
		a = \rnconstant \frac{r^2}{m^2} - \Osqern \sim \left(1-\frac{r_{ph}}{r}\right)^2 \frac{r^2}{m^2}, \quad b = (1-\Osqern)^2 + \rnconstant \frac{r^2}{m^2} (2-\Osqern),
	\end{equation}
    where $\rnconstant = \rnconstant \left( \frac{\left| q \right|}{m} \right)$ is as in Definition~\ref{def_eps_def}. We may solve~\eqref{eqn_quadratic} for $p^r$ as a function of $p^{t^*},r,\trapschw$.  Noting that Lemma~\ref{express_ptstar} implies that $p^{t^*} > 0$ unless $p^\mu = 0$ for all $\mu \in \{t^*,r,\omega\}$,  we find
    \begin{equation} \label{eqn_ern_prs}
		p^r = \frac{-(1-\Osqern)(a + \trapschw \Osqern) + \mathfrak{s} \sqrt{(a+\trapschw \Osqern) \left( a (1-\Osqern)^2 + b \Osqern \right)}}{ \left( b - \trapschw (1-\Osqern)^2 \right)} p^{t^*},
	\end{equation}
    where $\mathfrak{s} \in \{-1,+1 \}$ is the usual sign in the solution formula for quadratic equations and we made use of the fact that inequality~\eqref{epsbound} implies $a + \trapern \Osqern \geq 0$. From the definition of $b$ it follows in particular that $b-\trapschw (1-\Osqern)^2 \geq 0$, from which we readily deduce that $\sgn(p^r) = \mathfrak{s}$.

    In order to bound the quotient $\frac{p^r}{p^{t^*}}$, we bound the numerator and denominator of~\eqref{eqn_ern_prs} separately. Let $0 < \delta < 1$, then the definition of $b$ implies the following bound for the denominator of~\eqref{eqn_ern_prs}:
	\begin{equation} \label{eqn_denombound}
		b- \trapschw (1-\Osqern )^2 = (1- \trapschw) (1-\Osqern )^2 + \rnconstant \frac{r^2}{m^2} \left( 2- \Osqern \right) \sim_\delta \begin{cases}
			1  + \left| \trapschw \right| & r_+ \leq r \leq r_+ + \delta m \\
			\frac{r^2}{m^2} & r_+ + \delta m \leq r
		\end{cases}.
	\end{equation}
	For the numerator of~\eqref{eqn_ern_prs}, we first note the identity $a ( 1 - \Osqern )^2 + b \Osqern = \lambda \frac{r^2}{m^2}$. Next, we observe
	\begin{equation} \label{eqn_aplust}
		a + \trapschw \Osqern \sim \frac{1}{m^2} \left( (r-r_{ph})^2 + \mathfrak{a} \trapschw \right), 
	\end{equation}
	for a certain function $\mathfrak{a} \sim m^2 \Osqrn$. Let us now bound the numerator in the case that $\mathfrak{s}=-1$:
	\begin{equation}
		-(1-\Osqern)(a + \trapschw \Osqern) - \sqrt{(a+\trapschw \Osqern) \left( a (1-\Osqern)^2 + b \Osqern \right)} \sim - \sqrt{a+\trapschw \Osqern} \frac{r}{m},
	\end{equation}
	where we have made use of the fact that in the whole exterior $1-\Osqern \sim \frac{m}{r}$. Combining this with~\eqref{eqn_aplust} and~\eqref{eqn_denombound} this implies the claimed bound for the quotient in the case $\mathfrak{s}=-1$. Let us now estimate the numerator in the case $\mathfrak{s}=1$:
	\begin{equation} \label{eqn_bigbound}
    \begin{aligned}
		&-(1-\Osqern)(a + \trapschw \Osqern) + \sqrt{(a+\trapschw \Osqern) \left( a (1-\Osqern)^2 + b \Osqern \right)} \\
		%=& \sqrt{a+\trapschw \Osqern} \left[ \sqrt{a (1-\Osqern)^2 + b \Osqern} - \sqrt{\left( 1-\Osqern \right)^2 \left( a+ \trapschw \Osqern \right)} \right] \\
		\sim& \sqrt{a+\trapschw \Osqern} \frac{r}{m} \left[ 1 - \sqrt{\frac{(1-\Osqern)^2 (a + \trapern \Osqern)}{a (1-\Osqern)^2 + b \Osqern}} \right] \\
		\sim& \sqrt{a+\trapschw \Osqern} \frac{r}{m} \left[ 1-\frac{(1-\Osqern)^2 (a + \trapern \Osqern)}{a (1-\Osqern)^2 + b \Osqern} \right] \\
		\sim & \sqrt{a+\trapschw \Osqern} \frac{m}{r} \Osqern \left( b - \trapschw \left(1 - \Osqern \right) \right),
	\end{aligned}
    \end{equation}
    where we used in the third line that the relation $1- z \sim 1 - \sqrt{z}$ holds for $0 \leq z \leq 1$, combined with the bound
    \begin{equation}
		0 \leq \frac{(1-\Osqern)^2 (a + \trapern \Osqern)}{a (1-\Osqern)^2 + b \Osqern} \leq (1 - \Osqern)^2 \leq  1.
	\end{equation}
	Therefore there is a cancellation with the denominator in this case and by combining~\eqref{eqn_bigbound} with~\eqref{eqn_aplust}, we again conclude the desired bound.
\end{proof}

\subsubsection{Moments of solutions to the massless Vlasov equation} \label{sec_moments_solutions}
The metric induces a volume form on each fibre $\mathcal{P}_x$ for $x \in M$. In order to define the volume form, we choose explicit coordinates on the sphere. Let $(\theta,\phi)$ denote the usual spherical coordinates\footnote{Note that since spherical coordinates do not cover the whole $2$-sphere, we need to repeat the argument with rotated versions of the coordinate system twice. The argument remains however identical so that we will only give it once.} on $\sphere$ so that the $(t^*,r)$-coordinates now become $(t^*,r,\theta,\phi)$. The conjugate parametrisation of the mass-shell is now denoted by $(t^*,r,\theta,\phi,p^r,p^\theta,p^\phi)$. The integration measure $\dmux$ on $\mathcal{P}_x$ is then readily computed to be
\begin{equation} \label{dmu_first_comp_schw}
	\dmux = \frac{r^2 \sin \theta}{\Osqrn p^{t^*} - \left( 1 - \Osqrn \right) p^r} \, d p^r d p^\theta d p^\phi,
\end{equation}
where $p^{t^*}$ is understood to be expressed as a function of the remaining variables as in Lemma~\ref{express_ptstar}. We note that future directedness of $p$ implies $\Osqrn p^{t^*} - \left( 1 - \Osqrn \right) p^r = - \gRN(\partial_{t^*},p) \geq 0$. Using this measure on $\mathcal{P}_x$ we may now define moments of the distribution $f$.

\begin{defi}[Moments of $f$]
Let $w,f: \mathcal{P} \rightarrow \R$ be smooth. Then the moment of $f$ associated to the weight $w$ is defined as
\begin{equation}
\int_{\mathcal{P}_x} w f \, \dmux \in [-\infty,+\infty].
\end{equation}
\end{defi}

\begin{rem}
Note that a priori the moment associated to a weight $w$ might not be finite. However, it will follow from Lemma~\ref{lem_boundedness_moments_schw} that $\int_{\mathcal{P}_x} w f \, \dmux \in \R$ if $f$ is a solution to the massless Vlasov equation with compactly supported initial data.
\end{rem}

\begin{defi}[Boundedness in $x$] \label{boundedness_in_x}
Let $C > 0$ be a constant and consider the set
\begin{equation}
B_C = \Big\{ (x,p) \in \mathcal{P} : p^{t^*}, \left| p^r \right|, \left| \pslash \right|_{\gslash} \leq C \Big\} \subset \mathcal{P}.
\end{equation}
Let $w: \mathcal{P} \rightarrow \R$ be a smooth weight. Then we say that $w$ is bounded in $x$ if for all $C>0$
\begin{equation}
    \sup_{(x,p) \in B_C} \left| w(x,p) \right| < \infty.
\end{equation}
\end{defi}

\begin{defi}[Energy momentum tensor]
Let $(x^\mu)$ denote a coordinate system on $\Mrn$ and $(x^\mu,p^\mu)$ the associated coordinate system on $T \Mrn$. Then
\begin{equation} \label{energymomentum_defn}
T^{\alpha \beta}[f] = \int_{\mathcal{P}_x} p^\alpha p^\beta f(x,p) \, \dmux
\end{equation}
defines the \emph{energy momentum tensor} associated to the solution $f$ of the massless Vlasov equation in the coordinates $( x^\mu)$.
\end{defi}

\begin{rem}
For smooth $f$ we have the divergence identity
\begin{equation}
\nabla_\mu T^{\mu \nu}[f] =  \int_{\mathcal{P}_x} X(f) p^\nu \, \dmux,
\end{equation}
so that if $X(f) = 0$, the energy momentum tensor is conserved. See~\cite{martin,riosecosarbach,sarbachzannias} for a general discussion on how to relate derivatives of the energy momentum tensor to derivatives of $f$. In addition, the energy momentum tensor has the following positivity property: If $X,Y$ are both future-directed and causal, then $T_{\mu \nu} X^{\mu} Y^{\nu} \geq 0$ (\emph{weak energy condition}). In addition we have $T_{\mu \nu} X^{\mu} X^{\nu} \geq 0$ for any vector field $X$ (\emph{non-negative pressure condition}). Furthermore the mass-shell relation implies that $T$ is \emph{trace-free}, in other words $T^{\mu}_{\mu} = 0$.
\end{rem}

\subsubsection{Angular coordinates on the mass-shell} \label{sec_angular_coords_massshell}
For our later convenience we introduce a change of variables on the mass-shell here. Recall the parametrisation $(t^*,r,\omega,p^r,\pslash)$ introduced in Section~\ref{sec_parametrising_massshell} above. We now introduce the change of variables $\pslash \mapsto (p^1,p^2)$ such that $L^2 = r^2 \left| \pslash \right|_{\gslash}^2 = (p^1)^2 + (p^2)^2$. This change of variables may be explicitly realised in terms of spherical coordinates on $\mathbb{S}^2$ and the associated conjugate coordinates as $p^1 = p^\theta, p^2 = \sin(\theta) p^\phi$ (see Section~\ref{sec_moments_solutions}). Next, we recall equation~\eqref{dmu_first_comp_schw}, which allows us to express the volume form $\dmux$ explicitly as
	\begin{equation}
		\dmux =  \frac{r^2 \sin \theta}{\OsqS p^{t^*} - \frac{2m}{r} p^r} \, d p^r d p^\theta d p^\phi =  \frac{1}{r^2} \frac{1}{E} \, d p^r d p^1 d p^2 .
	\end{equation}
Next we introduce coordinates $(L,\pslashangle)$ such that $(p^1,p^2) = (L \cos \pslashangle, L \sin \pslashangle)$, or in other words we use radial coordinates in the angular variables $(p^1,p^2)$. Note that by definition $\pslashangle \in [0,2 \pi)$. We may then express the volume form $\dmux$ explicitly as
\begin{equation} \label{eqn_volume_form_pslashangle}
    \dmux =  \frac{1}{r^2} \frac{L}{E} \, d p^r d L d \pslashangle .
\end{equation}
We shall make use of this parametrisation of the mass-shell repeatedly in later sections.

\subsection{The set of almost-trapped geodesics} \label{sec_subsets}
In this section we define various subsets of the mass-shell $\mathcal{P}$ that will play a central role later in the proofs of the main theorems.

\begin{defi}[Almost trapping at the photon sphere] \label{defn_trapping_ph}
Let $\mathcal{P}$ denote the mass-shell on the subextremal or extremal Reissner--Nordstr\"om exterior. Fix constants $\bigc,\decayrate > 0, \tauzero > m$ and recall the constants $\rsuppconst, \psuppconst$ introduced in Assumption~\ref{assumption_support}. We define
\begin{align} \label{defn_trappedsupportset}
	\trappedsupportset_{\bigc,\decayrate,\tauzero} = \left\{ (x,p) \in \mathcal{P} \; \Big| \; \begin{matrix}
		\left| \trapschw \right| \leq \bigc e^{-\frac{\decayrate}{2m}(\tau(x)-\tauzero)}, \quad
		\left| \pslash \right|_{\gslash} \leq \frac{\rsuppconst \psuppconst}{r} \\
		\left| m \frac{\left| p^r \right|}{L} - \sqrt{\rnconstant - \frac{m^2}{r^2} \Osqrn} \right| \leq \bigc e^{-\frac{\decayrate}{2m}(\tau(x)-\tauzero)} \\
		\left| m \frac{ p^{t^*}}{L} - \left( \frac{\sgn(p^r) (1-\Osqrn) \sqrt{\rnconstant - \frac{m^2}{r^2} \Osqrn} + \sqrt{\rnconstant}}{\Osqrn} \right) \right| \leq \bigc e^{-\frac{\decayrate}{2m}(\tau(x)-\tauzero)}
	\end{matrix} \right\},
\end{align}
where we used $(t^*,r)$-coordinates and their induced coordinates on the tangent bundle to represent each point as $(x,p) = (t^*,r,\omega,p^{t^*},p^r,\pslash)$ and the constant $\rnconstant$ is defined as in Definition~\ref{def_eps_def}. For a point $x \in \Mrn$ we denote the fibre by $\trappedsupportsetx = \trappedsupportset \cap \mathcal{P}_x$.
\end{defi}

\begin{rem}
We remark that the expression
\begin{equation} \label{exprn_defn_trappedset}
	\frac{1}{\Osqrn} \left( \sgn(p^r) (1-\Osqrn) \sqrt{\rnconstant- \frac{m^2}{r^2} \Osqrn} + \sqrt{\rnconstant} \right),
\end{equation}
used in the definition of the set $\trappedsupportset$, has a finite limit as $r \rightarrow r_+$ if and only if $p^r \leq 0$. Note that  if $(x,p) \in \mathcal{P}$ and $x \in \mathcal{H}^+$, then by future-directedness of $p$ we must have $p^r \leq 0$. Therefore, the set $\trappedsupportset$ is well-defined on the whole exterior, but it is not continuously defined up to and including the event horizon $\mathcal{H}^+$. However, we will only need to evaluate the expression~\eqref{exprn_defn_trappedset} along null geodesics, and it follows from the properties of the geodesic flow that the limit $r \rightarrow r_+$ can only occur if $p^r \leq 0$.
\end{rem}

\begin{defi}[Almost trapping at the event horizon] \label{defn_trapping_hor}
Let $\mathcal{P}$ denote the mass-shell on the subextremal or extremal Reissner--Nordstr\"om exterior. Fix constants $\bigc,\decayrate > 0, \tauzero > m$ and recall the constants $\rsuppconst, \psuppconst$ introduced in Assumption~\ref{assumption_support}. In the subextremal case we define
\begin{align} \label{defn_smallsupportset}
	\smallsupportset_{\bigc,\decayrate,\tauzero} = \left\{ (x,p) \in \mathcal{P} \; \Bigg| \; \begin{matrix}
		\sqrt{1+ \left| \trapschw \right|} \left| \pslash \right|_{\gslash} \leq \bigc \rsuppconst \psuppconst \frac{1}{r} e^{- \frac{\decayrate}{2m}(\tau(x)-\tauzero)} \\
		\left( 1+ \left| \trapschw \right| \right) \left| p^r \right| \leq \bigc \psuppconst e^{-\frac{\decayrate}{m}(\tau(x)-\tauzero)} \\
		p^{t^*} \leq \bigc \psuppconst e^{-\frac{\decayrate}{m}(\tau(x)-\tauzero)}
	\end{matrix} \right\},
\end{align}
whereas in the extremal case we define
\begin{align} \label{defn_smallsupportsetERN}
	\ERNsmallsupportset_{\bigc,\decayrate,\tauzero} = \left\{ (x,p) \in \mathcal{P} \; \Bigg| \; \begin{matrix}
		\left| \pslash \right|_{\gslash} \leq \bigc \psuppconst \frac{\rsuppconst}{r} \frac{m}{\left| \tau(x)-\tauzero \right|} \\
		\left| p^r \right| \leq \bigc \psuppconst \frac{m^2}{(\tau(x)-\tauzero)^2} \\
		p^{t^*} \leq \bigc \psuppconst \min \left( \frac{1}{\Osqern} \frac{m^2}{(\tau(x)-\tauzero)^2}, 1 \right)
	\end{matrix} \right\},
\end{align}
where in both cases we used $(t^*,r)$-coordinates as in the preceding definition. For a point $x \in \Mschw$ we denote the fibre by $\smallsupportsetx = \smallsupportset \cap \mathcal{P}_x$.
\end{defi}

\begin{rem} 
Let us elaborate on the meaning of the constant $\tauzero$ involved in the definition of the sets $\trappedsupportset_{\bigc,\decayrate,\tauzero}$ and $\ERNsmallsupportset_{\bigc,\decayrate,\tauzero}$. By definition, the momenta contained in these sets satisfy estimates of the form $p^{t^*} \leq C e^{-\frac{c}{m} (\tau(x) - \tauzero)}$, or similar. Firstly, note that an estimate of the form $p^{t^*} \leq C e^{-\frac{c}{m} (\tau(x) - \tauzero)}$ is equivalent to an estimate of the form $p^{t^*} \leq C' e^{-\frac{c}{m} \tau(x) }$ if we set $C' = C e^{\frac{c}{m} \tauzero }$. Later in the proofs, the cutoff time $\tauzero$ will serve the purpose of splitting the evolution of a solution to the massless Vlasov equation into two parts: Initially, when $0 \leq \tau \leq \tauzero$, most geodesics either fall into the black hole or scatter to infinity. After the time $\tauzero$, the remaining geodesics must then be either nearly trapped at the photon sphere or at the event horizon, allowing us to draw conclusions about the size of their momentum components. It will follow from the proof that $\tauzero$ will be proportional to the size of initial support of the solution.
\end{rem}

\begin{rem}
Note that the inequalities in Definition~\ref{defn_trapping_hor} involve various weights which are a function of the trapping parameter $\trapschw$. These estimates are simply the ones which arise naturally from a detailed analysis of the geodesic flow.
\end{rem}

\begin{rem}
We will from now on omit the dependence of the sets $\trappedsupportset_{\bigc,\decayrate,\tauzero}$ and $\ERNsmallsupportset_{\bigc,\decayrate,\tauzero}$ on the constants $\bigc,\decayrate,\tauzero$ and simply write $\trappedsupportset_{\bigc,\decayrate,\tauzero} = \trappedsupportset$ and $\ERNsmallsupportset_{\bigc,\decayrate,\tauzero} = \ERNsmallsupportset$. The proofs of the main theorems will reveal how the constants $\bigc,\decayrate,\tauzero$ must be chosen. In particular, $\bigc,\decayrate$ are universal constants depending only on the geometry of Reissner--Nordstr\"om through the quotient $\frac{\left| q \right|}{m}$, while $\tauzero$ will depend only on the size of the initial support of the solution to the massless Vlasov equation under consideration.
\end{rem}

\begin{figure} \label{figure_fibres}
\includegraphics[width = \textwidth]{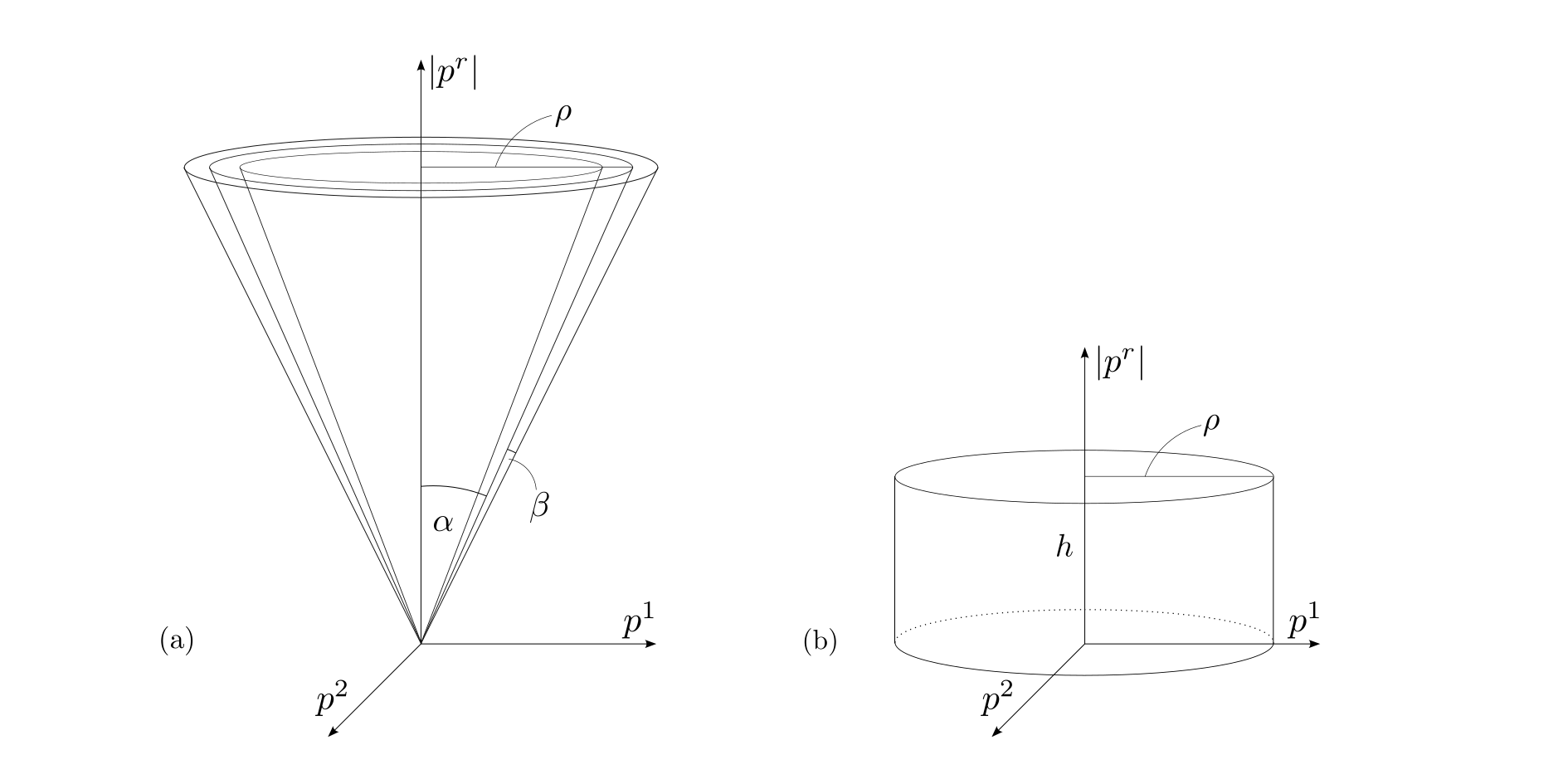}
\centering
\caption{A plot of the fibres $\trappedsupportsetx$ in Figure~(a) and $\smallsupportsetx$ in Figure~(b) for a point $x \in \Mschw$. We parametrise the fibre $\mathcal{P}_x$ by $(p^r,p^1,p^2) \in \R^3$. This coordinate system is obtained from the coordinates on the tangent bundle associated to $(t^*,r)$-coordinates by using the mass-shell relation to eliminate the $p^{t^*}$-variable. Then we choose $(p^1,p^2)$ such that $L^2 = r^2 \left| \pslash \right|_{\gslash}^2 = (p^1)^2 + (p^2)^2$. The set $\trappedsupportsetx$ is the region between the inner truncated cone (with apex angle $\alpha - \beta$) and the outer truncated cone (with apex angle $\alpha + \beta$). The centre truncated cone (with apex angle $\alpha$) is the subset of momenta which generate exactly trapped geodesics. We explain how $\alpha,\beta$ and $\rho$ relate to the quantities in Definition~\ref{defn_trapping_ph}. The apex angle $\alpha$ of the centre truncated cone may be seen to scale like $\tan \alpha \sim \sqrt{\rnconstant - \frac{m^2}{r^2} \Osqrn} \sim \left( 1 - \frac{r_{ph}}{r} \right)^{-1}$. Moreover, the centre cone has a maximal radius of $\rho \leq \rsuppconst \psuppconst$. In particular over the photon sphere, where $r = r_{ph}$, the apex angle is $\alpha = \frac{\pi}{2}$ and the centre cone flattens to the disk $\{ p^r = 0, L \leq \rsuppconst \psuppconst \}$. The angle $\beta$ is exponentially small in $\tau(x)$, so that $\trappedsupportsetx$ approximates the centre cone as $\tau(x) \rightarrow \infty$. The set $\smallsupportsetx$ is a small cylinder around the origin with radius $\rho$ and height $h$. Height and radius relate to each other like $h \sim \rho^2$. The height $h$ is proportional to the right hand side of the upper bound for the $p^r$-component in Definition~\ref{defn_trapping_hor}. Therefore, $h$ is exponentially small in $\tau(x)$ in the subextremal case and inverse quadratic in $\tau(x)$ in the extremal case.}
\end{figure}

\begin{rem}
The set $\trappedsupportset \subset \mathcal{P}$ contains geodesics which are almost trapped at the photon sphere, while $\smallsupportset \subset \mathcal{P}$ contains those geodesics which are almost trapped at the event horizon. Geometrically, each fibre $\trappedsupportsetx$ is an approximate $2$-cone in $\mathcal{P}_x$, while $\smallsupportsetx$ is a small cylinder around the origin, see Figure~\ref{figure_fibres}. We will later show that if a geodesic has not scattered to infinity or fallen into the black hole after a finite (fixed) time $\tauzero$, then it must be contained in either $\trappedsupportset$ or $\smallsupportset$ after this time. This is accomplished in Proposition~\ref{psupport_prop} for the subextremal case and in Proposition~\ref{psupport_propERN} for the extremal case and forms the core of the proof of decay of moments of solutions to the massless Vlasov equation. See Sections~\ref{section_proof} and~\ref{section_ERN}.
\end{rem}

\begin{rem}
Both sets involve a choice of certain constants $\bigc, \decayrateERN > 0, \tauzero > m$, which will be chosen appropriately in the proofs of Proposition~\ref{psupport_prop} for the subextremal case respectively Proposition~\ref{psupport_propERN} for the extremal case.
\end{rem}

The following definitions only apply to the \emph{extremal} Reissner--Nordstr\"om spacetime.

\begin{defi} \label{def_badsetapprox}
Let $\mathcal{P}$ denote the mass-shell on the extremal Reissner--Nordstr\"om exterior. Let $0 < \delta < \frac{1}{2}$ and $c_2 > c_1 > 0$ and let $\psuppconst$ be as in Assumption~\ref{assumption_support}. Define the following subset of the mass-shell over the initial hypersurface $\Sigma_0$
\begin{equation}
		\badsetapprox_{c_1,c_2,\delta} = \left\{ (x,p) \in \mathcal{P}_0 \; \Bigg| \; m \leq r \leq (1 + \delta) m, \; \begin{matrix}
		\left| p^r \right| \leq c_2 \psuppconst \Osqern(r) \\
		c_1 \psuppconst \sqrt{\Osqern(r)} \leq \left| \pslash \right|_{\gslash} \leq c_2 \psuppconst \sqrt{\Osqern(r)}
	\end{matrix} \right\} \subset \mathcal{P}_0.
 \end{equation}
Let furthermore $\constbsl > 0$ and define the following subset of $\mathcal{P}_0$
	\begin{equation}
		\badsetaplarge_{\constbsl,\delta} = \left\{ (x,p) \in \mathcal{P}_0 \; \Bigg| \; m \leq r \leq (1 + \delta) m, \; \begin{matrix}
			- \constbsl \psuppconst \sqrt{\Osqern(r)} \leq p^r \leq \psuppconst \Osqern(r) \\
			\left| \pslash \right|_{\gslash} \leq \constbsl \psuppconst (\Osqern(r))^{\frac{1}{4}} \\
            p^{t^*} \leq \psuppconst
		\end{matrix} \right\} \subset \mathcal{P}_0.
	\end{equation}
\end{defi}

\begin{rem}
We will at times omit the dependence on the constants $c_1,c_2$ and $\constbsl$ and prefer to think of the sets $\badsetapprox_{c_1,c_2,\delta}, \badsetaplarge_{\constbsl,\delta}$ as one-parameter families parametrised by $0 < \delta < \frac{1}{2}$ with an appropriate choice of constants $c_1, c_2$ and $\constbsl$, respectively. For $\badsetapprox_{c_1,c_2,\delta}$, this choice will be specified in the proof of Lemma~\ref{sharpness_lemma}, while for $\badsetaplarge_{\constbsl,\delta}$ the choice of $\constbsl$ will be made in the proof of Proposition~\ref{psupport_propERN}. We will then simply write $\badsetapprox_{\delta} = \badsetapprox_{c_1,c_2,\delta}$ and $ \badsetaplarge_{\delta} =  \badsetaplarge_{\constbsl,\delta}$.
\end{rem}

\begin{rem}
The sets $\badsetapprox_{\delta}$ and $\badsetaplarge_{\delta}$ represent two neighbourhoods of the set of generators of the extremal event horizon. In particular, if $f_0$ is a smooth initial distribution that is supported on the generators of the event horizon, it follows that there must exist $\delta > 0$ such that $\badsetapprox_{\delta} \subset \badsetaplarge_{\delta} \subset \supp(f_0)$ and in fact $\inf_{\badsetaplarge_{\delta}} f_0 > 0$. See also Remark~\ref{rem_intro_smoothness} and Remark~\ref{rem_intro_nondecaythm}.
\end{rem}

\begin{rem}
We will later choose the constants $c_1,c_2,\constbsl$ from Definition~\ref{def_badsetapprox} such that $\constbsl > c_2$, so that $\badsetapprox_{c_1,c_2,\delta} \subset \badsetaplarge_{\constbsl,\delta}$ for all $0 < \delta < \frac{1}{2}$. Moreover, the fibres of $\badsetaplarge_\delta$ are uniformly bounded. More precisely, every point $(x,p) \in \badsetaplarge_\delta$ satisfies Assumption~\ref{assumption_support}. Therefore, the assumption that $f_0$ is compactly supported, as defined in Assumption~\ref{assumption_support}, is compatible with the assumption that $\badsetaplarge_\delta \subset \supp(f_0)$.
\end{rem}

\begin{defi} \label{defi_slowsupportset}
Let $\mathcal{P}$ denote the mass-shell on the extremal Reissner--Nordstr\"om exterior. Let $C_2 > C_1 > 0$ and $\tauslow > m$ be constants and recall the constant $\psuppconst$ from Assumption~\ref{assumption_support}. We define the following subset of the mass-shell over the event horizon
	\begin{equation}
		\slowsupportset_{C_1,C_2,\tauslow} = \left\{ (x,p) \in \mathcal{P} \; \Bigg| \; \tau(x) > \tauslow, r=m, \; \begin{matrix}
			C_1 \psuppconst \frac{m}{\tau(x)} \leq \left| \pslash \right|_{\gslash} \leq C_2 \psuppconst \frac{m}{\tau(x)} \\
			C_1 \psuppconst \frac{m^2}{\tau(x)^2} \leq \left| p^r \right| \leq C_2 \psuppconst \frac{m^2}{\tau(x)^2}
		\end{matrix} \right\} \subset \mathcal{P} |_{\mathcal{H}^+}.
	\end{equation}
\end{defi}

\begin{rem}
	We will often omit the dependence on $C_1,C_2$ and consider $\slowsupportset_\tauslow = \slowsupportset_{C_1,C_2,\tauslow}$ as a one-parameter family parametrised by $\tauslow$ for an appropriate choice of constants $C_1,C_2$. At times we will omit all constants and merely write $\slowsupportset = \slowsupportset_{C_1,C_2,\tauslow}$.
\end{rem}

\begin{defi} \label{defn_pi0}
Consider the mass-shell $\mathcal{P}$ on the extremal Reissner--Nordstr\"om exterior and the mass-shell $\mathcal{P}_0 = \mathcal{P} |_{\Sigma_0}$ over the initial hypersurface $\Sigma_0$. We define the following map
	\begin{equation}
		\pi_0 : \mathcal{P} \rightarrow \mathcal{P}_0, \quad (x,p) \mapsto S_{s_0}(x,p),
	\end{equation}
	where $S_s$ denotes the geodesic flow on the mass-shell $\mathcal{P}$ as defined in~\eqref{geodesicflow} above and $s_0$ is the unique affine parameter time such that $S_{s_0}(x,p) \in \mathcal{P}_0$.
\end{defi}

\begin{rem}
The set $\badsettau$ is the set of initial data of the geodesics which generate the set $\slowsupportset_{\tauslow}$ over the event horizon. We will prove in Lemma~\ref{sharpness_lemma} that the sets $\badsettau$ and $\badsetapprox_{\delta}$ are comparable if we assume $\tauslow \sim \delta^{-1}$. To make this a bit more precise, Lemma~\ref{sharpness_lemma} establishes the existence of a suitable choice of constants $c_1, c_2, C_1,C_2$ such that for any of $0 < \delta < \frac{1}{2}$, there exists $\tauslow \sim \delta^{-1}$ such that $\badsetallconst \approx \badsetapprox_{c_1,c_2,\delta}$. Likewise, in Proposition~\ref{psupport_propERN} we show that (roughly speaking) $\pi_0(\smallsupportset_{\tauzero} \cap \left\{ \tau \geq \bar{\tau} \right\}) \subset \badsetaplarge_\delta$ for any $ \bar{\tau} > \tauzero$ if we assume $\delta \sim \bar{\tau}^{-1}$ and choose the remaining constants appropriately.
\end{rem}

\subsection{A note on constants and notation}
Unless otherwise stated, the letter $C>0$ will denote a constant which depends only on the member of the Reissner--Nordstr\"om family under consideration through the ratio $\left| q \right| / m$. The symbol $\lesssim$ is used to express inequalities which hold up to multiplicative constants. We say $a \sim b$ for $a,b > 0$ if and only if there exist constants $0 < c < C < \infty$ such that $c \leq \frac{a}{b} \leq C$. If the constants $c,C$ in this inequality depend on some parameter $\lambda$, we reflect this in the notation by $a \sim_\lambda b$. We denote the negative part of a real number by $\left( x \right)_{-} := - \min(x, 0)$.

In Assumption~\ref{assumption_support} we introduced the constants $\rsuppconst,\psuppconst$, which bound the size of the spatial support respectively the size of the momentum support of an initial distribution. These constants appear throughout the proof and we always make dependence on them explicit in our estimates. Several definitions in Section~\ref{sec_subsets} require a choice of constants $\bigc, \decayrate, C_1, C_2, c_1, c_2, \constbsl, \tauslow, \tauzero$. These will be chosen as appropriate functions of the Reissner--Nordstr\"om parameters $(m,q)$ and $\rsuppconst, \psuppconst$ in the course of the argument. We will make the dependence on these constants explicit in estimates where necessary and suppress it when doing so would not yield any more insight.

%\label{defi_trappedsets}
%\label{def_ERN_subsets}

\section{The main theorems} \label{section_mainthms}
In this section, we state precise versions of Theorems~\ref{maintheorem},~\ref{maintheoremERN} and~\ref{maintheorem_ern_nondecay_rough}. We first state the simple Lemma~\ref{prop_compactness_support_general}, which we appeal to several times in the following theorems. Lemma~\ref{prop_compactness_support_general} states that the momentum support of a solution with compactly supported initial data remains compact for all times and it follows immediately from Assumption~\ref{assumption_support}. In particular, it provides a quantitative and uniform bound for the supremum norm $W$ of a polynomial weight $w$ on the support of a solution to the massless Vlasov equation with compactly supported initial data. It also immediately follows that, under the assumptions of Lemma~\ref{prop_compactness_support_general}, moments of solutions to the massless Vlasov equation with polynomial weights are well-defined. We prove Lemma~\ref{prop_compactness_support_general} in Lemma~\ref{psupport_bounded} for the subextremal case and remark that the proof in the extremal case is nearly identical.

\begin{lem}[Compactness of momentum support] \label{prop_compactness_support_general}
	Let $f_0 \in L^\infty(\mathcal{P}_0)$ satisfy Assumption~\ref{assumption_support} and assume that $f$ is the unique solution to the massless Vlasov equation on subextremal or extremal Reissner--Nordstr\"om such that $f|_{\mathcal{P}_0} = f_0$. Then the momentum support of $f$ remains compact for all times: there exists a dimensionless constant $C > 0$ such that
	\begin{equation}
		\supp(f) \subset \Bigg\{ (x,p) \in \mathcal{P} : p^{t^*} \leq C \psuppconst, \; \left| p^r \right| \leq \psuppconst, \; \left| \pslash \right|_{\gslash} \leq \frac{\rsuppconst}{r} \psuppconst \Bigg\} ,
	\end{equation}
    where $\psuppconst,\rsuppconst$ are as in Assumption~\ref{assumption_support}. In particular, if $w: \mathcal{P} \rightarrow [0,\infty)$ is bounded polynomially, i.e.~$\left| w(x,p) \right| \lesssim (p^{t^*})^a \left| p^r \right|^b \left| \pslash \right|^c_{\gslash}$ for $a,b,c \geq 0$, then we have the following quantitative bound
	\begin{equation}
		W := \max_{(x,p) \in \supp(f)} \left| w(x,p) \right| \lesssim_{a+b+c} \psuppconst^{a+b+c} \rsuppconst^c .
	\end{equation}
	More generally, if $w$ is bounded in $x$ according to Definition~\ref{boundedness_in_x}, then
	\begin{equation}
		W = \max_{(x,p) \in \supp(f)} \left| w(x,p) \right| < \infty.
	\end{equation}
\end{lem}

We give a precise version of Theorem~\ref{maintheorem} on the subextremal Reissner--Nordstr\"om exterior.

\begin{thm}[Exponential decay on subextremal Reissner--Nordstr\"om] \label{maintheorem_precise}
	Let $f_0 \in L^\infty(\mathcal{P}_0)$ satisfy Assumption~\ref{assumption_support} with constants $\rsuppconst$ and $\psuppconst$ and let $f$ be the unique solution to the massless Vlasov equation on subextremal Reissner--Nordstr\"om such that $f|_{\mathcal{P}_0} = f_0$. Let $w: \mathcal{P} \rightarrow [0,\infty)$ be smooth and bounded in $x$ according to Definition~\ref{boundedness_in_x}, so that Lemma~\ref{prop_compactness_support_general} applies and in particular $W < \infty$.
	
	There exist dimensionless constants $C,C_0,\decayrate>0$ such that if $\tauzero = C_0 \left( m + \frac{\rsuppconst^2}{m} \right)$, then for times $\tau(x) \geq \tauzero$ the moment with weight $w$ decays at an exponential rate:
	\begin{equation}
		\int_{\mathcal{P}_x} w f \, \dmux \leq C W \| f_0 \|_{L^\infty} \psuppconst^2 \rsuppconst^2 \frac{1}{r^2} e^{-\frac{\decayrate}{m} (\tau(x) - \tauzero)}.
	\end{equation}
	
	Moreover, we have the following refinements: If $0 < \delta < 1$ and $\tau(x) \geq \tauzero - \ln \delta$ then in the region where $r \notin [r_{ph}-\delta m, r_{ph}+\delta m]$, the rate of decay may be improved to
	\begin{equation}
		\int_{\mathcal{P}_x} w f \, \dmux \leq C W \| f_0 \|_{L^\infty} \psuppconst^2 \rsuppconst^2 \frac{1}{r^2} e^{-\frac{2 \decayrate}{m} (\tau(x) - \tauzero)}.
	\end{equation}
	In addition, for certain weights, an improved decay estimate holds. Let $a,b \geq 0$ and consider
	\begin{equation}
		w_{a,b} = \left| \left| p^r \right| - \frac{L}{m}  \sqrt{\frac{m^2}{r^2} \OsqS \Bigg|_{r=r_{ph}} - \frac{m^2}{r^2} \OsqS} \, \right|^a \left| p^r \right|^b,
	\end{equation}
	and define the constant $\decayrate_{a,b} = \decayrate \min \left\{ 1+a, 3 + 2 b \right\}$ with $\decayrate$ as above. Then there exists a constant $C = C(a+b) > 0$ such that the following improved decay estimate holds uniformly in $r \geq r_+$,
	\begin{equation}
		\int_{\mathcal{P}_x} w_{a,b} f \, \dmux \leq C \| f_0 \|_{L^\infty} \frac{\rsuppconst^{2+a} \psuppconst^{2 + a + b}}{m^a} \frac{1}{r^2} e^{- \frac{\decayrate_{a,b}}{m} (\tau(x)-\tauzero)}.
	\end{equation}
	Along the photon sphere, this bound may be further improved: For every $\kappa \geq 0$, there exists a constant $C = C(\kappa) > 0$ such that if $x \in \Mrn$ is chosen along the photon sphere as $\tau \rightarrow \infty$, i.e.~$r = r_{ph}$ remains fixed, then this estimate may be further improved to
	\begin{equation}
		\int_{\mathcal{P}_x} \left| p^r \right|^\kappa f \, \dmux \Big|_{r=r_{ph}} \leq C \| f_0 \|_{L^\infty} \frac{\rsuppconst^{2+\kappa} \psuppconst^{2+\kappa}}{m^{2+\kappa}} e^{- \frac{\decayrate}{m} (1+\kappa)(\tau(x)-\tauzero)},
	\end{equation}
	In the asymptotically flat region, we can show an improved rate of decay for certain weights expressed in double null coordinates: For all $a,b,c \geq 0$, there exists a constant $C = C(a+b+c) > 0$ such that if $x \in \Mrn$ satisfies $\tau(x) \geq \tauzero$ and $r \geq R$, where $R$ is as in Definition~\ref{defn_timecoord_Schw}, then
	\begin{equation}
		\int_{\mathcal{P}_x} \left| p^v \right|^a \left| p^u \right|^b \left| \pslash \right|_{\gslash}^{c} f \, \dmux \leq C \| f_0 \|_{L^\infty} \psuppconst^{2 + a + b + c}
		\frac{\rsuppconst^{2(1+b) + c}}{r^{2(1+ b) + c}} e^{-\frac{\decayrate}{m} (\tau(x) - \tauzero)}.
	\end{equation}
\end{thm}

\begin{rem}
We note that the dimensionless constants $C, C_0, \decayrate$ can be quantified from the proof. Importantly, we note here that these constants are a function of the (dimensionless) ratio $\frac{q}{m} < 1$ and \emph{degenerate} as $\frac{q}{m} \rightarrow 1$, so that the corresponding estimates do not generalise to the extremal case.
%After a finite time, the `untrapped' parts of the solution have either escaped to future null infinity or fallen into the black hole, so that we are left to deal with the `almost trapped' parts of the solution, which present main difficulty in showing decay.
\end{rem}

%\begin{rem}
%Along the photon sphere weights of the form $w \sim \left|p^r \right|^\alpha, \alpha > 0$ decay at a faster rate, as stated above in the Theorem. The fact that weights that decay better along the photon sphere should exist can already be deduced from a Morawetz estimate for the massless Vlasov equation.
%\end{rem}

Next we turn to the extremal Reissner--Nordstr\"om spacetime. We provide a precise version of Theorem~\ref{maintheoremERN} in the form of Theorems~\ref{maintheoremERNprecise} and~\ref{ERN_slowdecayprop}. The following Theorem~\ref{maintheoremERNprecise} shows that moments of solutions to the massless Vlasov equation decay at least at a polynomial rate, while Theorem~\ref{ERN_slowdecayprop} shows that this decay rate is sharp along the extremal event horizon.

\begin{thm}[Polynomial decay on extremal Reissner--Nordstr\"om] \label{maintheoremERNprecise}
	Assume $f_0 \in L^\infty(\mathcal{P}_0)$ satisfies Assumption~\ref{assumption_support} with constants $\rsuppconst$ and $\psuppconst$ and let $f$ be the unique solution to the massless Vlasov equation on the extremal Reissner--Nordstr\"om background such that $f|_{\mathcal{P}_0} = f_0$. Let $w: \mathcal{P} \rightarrow [0,\infty)$ be smooth and bounded in $x$ as in Definition~\ref{boundedness_in_x}. Then Lemma~\ref{prop_compactness_support_general} applies and in particular $W < \infty$.
	
	There exist dimensionless constants $C,C_0>0$ such that if $\tauzero = C_0 \left( m + \frac{\rsuppconst^2}{m} \right)$, then for times $\tau(x) \geq \tauzero$ the moment associated to $w$ decays at least at a quadratic rate:
	\begin{equation}
		\int_{\mathcal{P}_x} w f \, \dmux \leq C W \| f_0 \|_{L^\infty} \frac{\psuppconst^2 \rsuppconst^2}{r^2} \frac{m^2}{(\tau(x)-\tauzero)^2}.
	\end{equation}
	
	Moreover, certain weights decay at a faster rate, depending on the distance from the event horizon: For all $a,b,c \geq 0$ there exists a constant $C = C(a+b+c) > 0$ such that if we assume $r \geq (1+ \delta) m$ for some $0< \delta < 1$, then
	\begin{equation}
		\int_{\mathcal{P}_x} \left| p^r \right|^a (p^{t^*})^b \left| \pslash \right|_{\gslash}^c f \, \dmux \leq \frac{C}{\delta^2} \| f_0 \|_{L^\infty} \frac{\psuppconst^{2+a+b+c} \rsuppconst^{2+c}}{r^{2+c}} \left( \frac{m}{\tau(x) - \tauzero} \right)^{2(1+a+b)+c},
	\end{equation}
	whereas if we assume that $m \leq r \leq (1+\delta) m$, then we obtain the bound
	\begin{equation} \label{ERN_momentsthm_inequ}
		\int_{\mathcal{P}_x} \left| p^r \right|^a (p^{t^*})^b \left| \pslash \right|_{\gslash}^c f \, \dmux \leq C \| f_0 \|_{L^\infty} \frac{\psuppconst^{2+a+b+c} \rsuppconst^{2+c}}{r^{2+c}} \left( \frac{m}{\tau(x) - \tauzero} \right)^{2(1+a)+c}.
	\end{equation}
	Furthermore, in the asymptotically flat region certain weights expressed in double null coordinates decay at a faster rate: For all $a,b,c \geq 0$, there exists a constant $C = C(a+b+c) > 0$ such that if $x \in \Mern$ satisfies $\tau(x) \geq \tauzero$ and $r \geq R$, then
	\begin{equation}
		\int_{\mathcal{P}_x} \left| p^v \right|^a \left| p^u \right|^b \left| \pslash \right|_{\gslash}^c f \, \dmux \leq C \| f_0 \|_{L^\infty} \frac{\psuppconst^{2+a+b+c} \rsuppconst^{2(1+b)+c}}{r^{2(1+b)+c}} \left( \frac{m}{\tau(x) - \tauzero} \right)^{2(1+a+b)+c}.
	\end{equation}
	
	If the initial distribution is supported away from the event horizon, we recover an exponential rate of decay. More precisely, let $0 < \delta < 1$ and assume $\supp(f_0) \subset \{ (x,p) \in \mathcal{P} : (1+\delta)m \leq r \}$. Then there exists a dimensionless constant $\decayrate>0$, such that we have the improved decay rate
	\begin{equation}
		\int_{\mathcal{P}_x} w f \, \dmux \leq C W \| f_0 \|_{L^\infty} \frac{\rsuppconst^2 \psuppconst^2}{r^2} e^{-\frac{\decayrate}{m} \left( \tau(x) - \tauzero \right)}.
	\end{equation}
\end{thm}

\begin{thm}[Sharpness of polynomial decay on extremal Reissner--Nordstr\"om] \label{ERN_slowdecayprop}
	Under the same assumptions as in Theorem~\ref{maintheoremERNprecise}, for all $0 < \delta < \frac{1}{2}$ and $a,b,c \geq 0$, there exist a time $\tauslow = \tauslow(\delta) \sim m \delta^{-1}$ and dimensionless constants $C = C(a+b+c), \constbsl > 0$, such that for all points along the extremal event horizon $x \in \mathcal{H}^+$ with $\tau(x) > \tauslow$, the following lower bounds holds:
	\begin{equation} \label{ERN_lowerbound}
		\int_{\mathcal{P}_x} \left| p^r \right|^a \left| \pslash \right|_{\gslash}^b (p^{t^*})^c f \, \dmux \geq C \psuppconst^{2+a+b+c} \frac{m^{2(a+1) + b}}{\tau(x)^{2(a+1)+b}} \left( \inf_{(x,p) \in \badsetaplarge_\delta} \left| f_0(x,p) \right| \right),
	\end{equation}
	where we have abbreviated $\badsetaplarge_\delta = \badsetaplarge_{\constbsl,\delta}$.
\end{thm}

We recall here that if $f_0$ is smooth and supported on the generators of the event horizon, then $\inf_{\badsetaplarge_\delta} f_0 > 0$ for some $\delta > 0$, see Remark~\ref{rem_intro_smoothness}. By combining Theorems~\ref{maintheoremERNprecise} and~\ref{ERN_slowdecayprop} we immediately arrive at the conclusion that different components of the energy momentum tensor decay at different rates along the event horizon.

%Theorems~\ref{maintheoremERNprecise} and~\ref{ERN_slowdecayprop} immediately imply

%We recall here that if $f_0$ is smooth and supported on the generators of the event horizon, then $\badsetaplarge_\delta \subset \supp(f_0)$ for some $\delta > 0$. See Remark~\ref{rem_intro_smoothness}. By combining Theorems~\ref{maintheoremERNprecise} and~\ref{ERN_slowdecayprop} we immediately arrive at the conclusion that different components of the energy momentum tensor decay at different rates along the event horizon.

\begin{cor}
	Assume $f_0 \in L^\infty(\mathcal{P}_0)$ satisfies Assumption~\ref{assumption_support} with constants $\rsuppconst$ and $\psuppconst$ and let $f$ be the unique solution to the massless Vlasov equation on the extremal Reissner--Nordstr\"om background such that $f|_{\mathcal{P}_0} = f_0$. Then for all $x \in \mathcal{H}^+$ with $\tau(x)$ sufficiently large, we have
    \begin{equation}
        T^{t^* t^*} \sim_{f_0} \frac{1}{\tau(x)^2}, \quad \left| T^{t^* r} \right| \sim_{f_0} \frac{1}{\tau(x)^4}, \quad T^{rr} \sim_{f_0} \frac{1}{\tau(x)^6},
    \end{equation}
    where $a \sim_{f_0} b$ is taken to mean $\psuppconst^4 \left( \inf\nolimits_{\badsetaplarge_\delta} \left| f_0 \right| \right) \lesssim \frac{a}{b} \lesssim \frac{\psuppconst^4 \rsuppconst^2}{m^2} \| f_0 \|_{L^\infty}$ for some $0 < \delta < \frac{1}{2}$ here.
\end{cor}

The following Theorem~\ref{ERN_nondecaytransversal} is the precise version of Theorem~\ref{maintheorem_ern_nondecay_rough} and shows that \emph{transversal derivatives of the energy momentum tensor do not decay along the extremal event horizon}.

\begin{thm}[Non-decay of transversal derivatives on extremal Reissner--Nordstr\"om] \label{ERN_nondecaytransversal}
	Under the same assumptions and using the same notation as in Theorem~\ref{maintheoremERNprecise}, let $0 < \delta < \frac{1}{2}$ and let us abbreviate $\badsetaplarge_\delta = \badsetaplarge_{\constbsl,\delta}$. Assume in addition
	\begin{itemize}
		\item $f_0 \in C^{1}(\mathcal{P}_0)$ and $f_0 \geq 0$,
		\item $\inf\nolimits_{\badsetaplarge_\delta} f_0 > 0$ and $\partial_{t^*} f |_{\mathcal{P}_0} (x,p) \neq 0$ for all $(x,p) \in \badsetaplarge_\delta$.
	\end{itemize}
	Then there exist dimensionless constants $C,C_0 > 0$ independent of $\delta$ such that for all $x \in \mathcal{H}^+$ with $\tau(x) \geq C_0 \left( \tauzero + m \delta^{-1} + m \left( \inf\nolimits_{\badsetaplarge_\delta} \left| \partial_{t^*} f |_{\mathcal{P}_0} \right| \right)^{-\frac{1}{2}} \right)$,
	\begin{equation} \label{eqn_ERN_nondecay_eqn}
		\left| \partial_r \int_{\sphere} T^{t^* t^*}[f] \, d \omega \right| \geq C \psuppconst^4 \left( \inf\nolimits_{\badsetaplarge_\delta} \left| \partial_{t^*} f |_{\mathcal{P}_0} \right| \right).
	\end{equation}
\end{thm}

\begin{rem}
We recall that if $f_0 \geq 0$ is $C^1$ and $f_0$ and $\partial_{t^*} f |_{\mathcal{P}_0}$ are nowhere vanishing on the generators of the event horizon, then the assumptions of the theorem are satisfied, see Remark~\ref{rem_intro_nondecaythm}. We remark that since $\badsetaplarge_\delta$ is a compact set, the assumptions of the theorem readily imply $\inf\nolimits_{\badsetaplarge_\delta} \left| \partial_{t^*} f |_{\mathcal{P}_0} \right| > 0$. Note also that by a slight abuse of notation, we have denoted the complete lift of the stationary Killing vector field $\partial_{t^*}$ to the mass-shell by the same symbol.
\end{rem}

\begin{rem}
Instead of assuming a lower bound for $\tau$ like stated in Theorem~\ref{ERN_nondecaytransversal}, the proof reveals that we may instead assume $\tau(x) \geq C_0 \left( \tauzero + m \delta^{-1} \right)$ to obtain the weaker lower bound
\begin{equation}
	\left| \partial_r \int_{\sphere} T^{t^* t^*}[f] \, d \omega \right| \geq C \psuppconst^4 \left[ \left( \inf\nolimits_{\badsetaplarge_\delta} \left| \partial_{t^*} f |_{\mathcal{P}_0} \right| \right) - \frac{\rsuppconst^2 \left( \| f_0 \|_{L^{\infty}} + \| \partial_{t^*} f_0 \|_{L^{\infty}} \right)}{(\tau(x) - \tauzero)^2} \right].
\end{equation}
Therefore, if $\tau$ is assumed to be large enough (in relation to certain initial data quantities), we may conclude the lower bound~\eqref{eqn_ERN_nondecay_eqn}.
\end{rem}

\begin{rem} \label{rem_remove_average}
The spherical average in~\eqref{eqn_ERN_nondecay_eqn} has been included to streamline and simplify the proof of Theorem~\ref{ERN_nondecaytransversal}. By providing suitable decay estimates of spherical derivatives of $T^{t^* t^*}[f]$, one may remove the spherical estimate and show the pointwise bound
\begin{equation} \label{eqn_ERN_nondecay_eqn}
		\left| \partial_r T^{t^* t^*}[f] \right| \geq C \psuppconst^4 \left( \inf\nolimits_{\badsetaplarge_\delta} \left| \partial_{t^*} f |_{\mathcal{P}_0} \right| \right),
\end{equation}
for $\tau \gg 1$ sufficiently large. Bounding angular derivatives of $T^{t^* t^*}[f]$ is possible due to the spherical symmetry of the Reissner--Nordstr\"om family which implies that angular derivates are Killing. Since this is a simple modification of the proof of Theorem~\ref{ERN_nondecaytransversal}, it shall suffice to make this remark here.
\end{rem}

\section{Decay for massless Vlasov on Schwarzschild} \label{section_proof}
In this section we provide the proof of Theorem~\ref{maintheorem_precise} on the subextremal Reissner--Nordstr\"om spacetime. We will only show the result for the Schwarzschild spacetime, which is the special case $q=0$. The proof generalises almost verbatim to the full subextremal range $\left| q \right| < m$, the main difference being that all dimensionless constants then depend on the quotient $\left| q \right| / m$.

The main step in the proof consists in obtaining control over the momentum support of a solution $f$ to the massless Vlasov equation with compactly supported initial data. This is the content of the main Proposition~\ref{psupport_prop} of this section.

%More precisely, we aim to understand the volume of the set $\supp(f(x,\cdot)) \subset \mathcal{P}_x$ at any given point $x \in \Mschw$ with $\tau(x) \gg 1$.

Before stating Proposition~\ref{psupport_prop}, we  briefly recall the definition of the sets $\trappedsupportset$ and $\smallsupportset$ in Section~\ref{sec_subsets}. The set $\trappedsupportset \subset \mathcal{P}$ contains those geodesics which are almost trapped at the photon sphere. Such geodesics spend a lot of time in the neighbourhood of the photon sphere. Let $(x,p) \in \trappedsupportset$ such that $\tau(x)$ is large. Then $p$ approximately lies on a truncated $2$-cone in $\mathcal{P}_x$. The distance of $p$ to this $2$-cone is exponentially small in the time $\tau(x)$. The set $\smallsupportset \subset \mathcal{P}$ contains those geodesics which are almost trapped at the event horizon. The geodesics populating this set originate close to the event horizon, are initially outgoing and spend a lot of time to in the vicinity of the black hole. If $(x,p) \in \smallsupportset$, then $p$ lies on the inside of a cylinder around the origin with exponentially small height and radius. See Figure~\ref{figure_fibres}. 

We remind the reader that $\trappedsupportset$ and $\smallsupportset$ depend on a choice of constants $\bigc, \decayrate,\tauzero$. These constants will be chosen appropriately in the proof of Proposition~\ref{psupport_prop}. Recall also the constant $\rsuppconst$ from Assumption~\ref{assumption_support}.

\begin{prop} \label{psupport_prop}
	Let $f_0 \in L^\infty(\mathcal{P}_0)$ satisfy Assumption~\ref{assumption_support} and let $f$ be the unique solution to the massless Vlasov equation on the Schwarzschild exterior with initial distribution $f_0$. Then there exist dimensionless constants $\bigc, \decayrate, C_0 > 0$ such that if we choose $\tauzero \geq C_0  \frac{\rsuppconst^2}{m}$, then
	\begin{equation}
		\supp(f) \cap \left\{ (x,p) \in \mathcal{P} \; | \; \tau(x) \geq \tauzero \right\} \subset \trappedsupportset \cup \smallsupportset,
	\end{equation}
	where we have abbreviated $\trappedsupportset = \trappedsupportset_{\bigc,\decayrate,\tauzero}$ and $\smallsupportset = \smallsupportset_{\bigc,\decayrate,\tauzero}$.
\end{prop}

%\begin{rem} \label{remark_explain_psupport}
%In the case that $(x,p) \in \trappedsupportset$ the rate stated above at which $p^{t^*}$ and $p^r$ decay to their limit values as $\tau(x) \rightarrow \infty$ may be improved in certain cases. Specifically, let $0 < \delta' < 1$ and assume that  $r \notin [(3-\delta')m,(3+\delta')m]$. If we in addition assume that $\tau(x) \gtrsim \tauzero - \ln \delta'$ then we may in fact improve the decay rate from $\frac{c}{2}$ to $c$. However it will be demonstrated in the proof below that the decay rate as stated may not be improved with our methods at the photon sphere.
%\end{rem}

Having stated the main proposition, let us outline the strategy of proof. In Section~\ref{section_tstar} we then consider a fixed null geodesic $\gamma: [s_0,s] \rightarrow \Mschw$ such that $\gamma(s_0) \in \Sigma_0$ and use the estimates from Section~\ref{prelim_massshell_estimates} to obtain a bound on the time required by the geodesic to cross a certain region of spacetime. Specifically, we bound $\tau(\gamma(s))$ in terms of the radii $r(s_0),r(s)$ and the trapping parameter of the geodesic. This is the \emph{almost-trapping estimate}, which we introduced in Section~\ref{subsubsec_geodflow} and which is central to proving Proposition~\ref{psupport_prop}. See Section~\ref{sec_radialgeod} for an outline of the proof of the almost-trapping estimate.

In Section~\ref{section_psupport} we then use the almost-trapping estimate to prove Proposition~\ref{psupport_prop}. The argument may be outlined as follows: Given a point $x \in \Mschw$ with $\tau(x)$ large, suppose a geodesic populates the momentum support at $x \in \Mschw$. The almost-trapping estimate then allows us to relate the time $\tau(x)$ with the trapping parameter and initial data of the geodesic. Combined with our assumption of compactly supported initial data, this allows us to derive bounds on the momentum of the geodesic at the point $x \in \Mschw$.

Finally, we prove Theorem~\ref{maintheorem_precise} in Section~\ref{section_integralestimate}. Given Proposition~\ref{psupport_prop}, it only remains to show that the volume of the sets $\trappedsupportset$ and $\smallsupportset$ is exponentially small in $\tau$, which amounts to a simple computation involving the estimation of some integrals.

\subsection{The almost-trapping estimate}  \label{section_tstar}
In this section we estimate the time a null geodesic requires to cross a certain region of spacetime in terms of the radius travelled by the geodesic and the value of its trapping parameter. The main result of this section is

\begin{lem}[The almost-trapping estimate] \label{taubound}
	Let $\gamma: [s_0,s_1] \rightarrow \Mrn$ be  an affinely parameterised future-oriented null geodesic. Let us express $\gamma$ in $(t^*,r)$-coordinates as $\gamma(s) = (t^*(s),r(s),\omega(s))$ with momentum $\dot{\gamma}(s) = (p^{t^*}(s),p^r(s),\pslash(s))$. Assume $\gamma$ intersects $\Sigma_0$ at radius $r_0 = r(s_0) \leq \rsuppconst$. Let $0 < \delta < 1$ and denote $\mathfrak{s} = \chi_{(0,\infty)}\left( p^r(s_0) \right)$, where $\chi_{(0,\infty)}$ denotes the characteristic function of the set $(0,\infty) \subset \R$. There exists a constant $C > 0$ such that if we assume $\tauzero \geq C(m + \frac{\rsuppconst^2}{m})$ then for all $s \in [s_0,s_1]$ with $2m \leq r(s) \leq (2+\delta)m$ we have the bound
	\begin{equation}
		\frac{1}{m} \left( \tau(\gamma(s)) - \tauzero \right) \lesssim_{\delta}
		\begin{cases}
			\left| \log \left( \frac{\OsqS(r_0)}{\OsqS(r(s))} \right) \right| & \text{if } p^r(s) > 0 \\
			\mathfrak{s} \left( \log \, (1+\left| \trapschw \right|) \OsqS(r_0) \right)_- + \left( \log \left| \trapschw \right| \right)_-  & \text{if } p^r(s) \leq 0
		\end{cases},
	\end{equation}
	and if $r(s) \geq (2+\delta)m$ we have
	\begin{equation}
		\frac{1}{m} \left( \tau(\gamma(s)) - \tauzero \right) \lesssim_{\delta} \left( \log \left| \trapschw \right| \right)_- + \mathfrak{s} \left( \log \, (1+\left| \trapschw \right|) \OsqS(r_0) \right)_-,
	\end{equation}
	independent of the sign of $p^r(s)$, where we have abbreviated $\trapschw = \trapschw(\gamma,\dot{\gamma})$.
\end{lem}

The proof will proceed by first establishing a bound for the time measured in the $t^*$-coordinate introduced in Section~\ref{rnmetricmanifold}. Specifically, let $\gamma: [s_1,s_2] \rightarrow \Mschw$ be an affinely parameterised future-directed null geodesic segment with affine parameter $s$ and let us express $\gamma(s) = (t^*(s),r(s),\omega(s))$ in $(t^*,r)$-coordinates. We derive an estimate for the time $t^*(s_2) - t^*(s_1)$ in terms of the radii $r(s_1),r(s_2)$ and the trapping parameter $\trapschw(\gamma)$ in Lemma~\ref{tstar_lem_schw}. Because of the mixed spacelike-null nature of the time function $\tau$, we consider the asymptotically flat region separately in Lemma~\ref{tauestimate_far}.
We prove that in this region, the behaviour of geodesics is comparable to the flat (Minkowski) case. Finally we combine the $t^*$-estimates with the estimates in the asymptotically flat region to obtain a bound on the time $\tau(\gamma(s_2)) - \tau(\gamma(s_1))$ in terms of $r(s_1),r(s_2)$ and $\trapschw(\gamma)$ and prove Lemma~\ref{taubound}. See Section~\ref{sec_radialgeod} for an informal discussion of the proof in the special case of radial geodesics.

\subsubsection{The $t^*$-time estimate} \label{sec_tstarestimate}
In this section, we derive a bound for the time a null geodesic requires to cross a certain region of spacetime, where we measure time using the $t^*$-coordinate. Let us split the black hole exterior in several regions to simplify our statement of the time estimate. Fix three constants $0< \delta_1,\delta_2 < 1$ and $\delta_3 > 0$ such that $0 < \delta_1 + \delta_2 < 1$. Consider the following intervals of Schwarzschild radius
\begin{equation}
	\begin{aligned}
		\mathfrak{I}_{\mathcal{H}^+} &= [2m, (2+\delta_1)m], \\
		\mathfrak{I}_{\text{int}} &= [(2+\delta_1)m, (3-\delta_2)m], \\
		\mathfrak{I}_{\text{ps}} &= [(3-\delta_2)m, (3+\delta_3)m], \\
		\mathfrak{I}_{\text{flat}} &= [(3+\delta_3)m, \infty),
	\end{aligned}
\end{equation}
where we have suppressed the dependence on $\delta_1,\delta_2,\delta_3$ in the notation. Therefore $r \in \mathfrak{I}_{\mathcal{H}^+}$ signifies closeness to the event horizon and $r \in \mathfrak{I}_{\text{ps}}$ closeness to the photon sphere, whereas the interval $\mathfrak{I}_{\text{int}}$ denotes the region between the horizon and photon sphere (at positive distance from both) and $\mathfrak{I}_{\text{flat}}$ denotes the asymptotically flat region.

\begin{lem} \label{tstar_lem_schw}
	Denote by $\gamma$ an affinely parameterised future-oriented null geodesic expressed in $(t^*,r)$-coordinates as $\gamma(s) = (t^*(s),r(s),\omega(s))$ with momentum $\dot{\gamma}(s) = (p^{t^*}(s),p^r(s),\pslash(s))$. Assume that $p^r \neq 0$ in the interval of affine parameter time $[s_1,s_2]$, so that the radius $r$ is a strictly monotone function of $s$. Let us assume that $r(s_i) = r_i$ for $i=1,2$ and let $0 < \delta_1,\delta_2,\delta_3$ be as above. Then
	\begin{equation} \label{t*bound}
		\frac{t^*(s_2) - t^*(s_1)}{m} \lesssim_{\delta_1,\delta_2,\delta_3 } \begin{cases}
			\begin{cases}
				1 & p^r \leq 0 \text{ on } [r_2,r_1] \\
				1 + \left| \log \left( \frac{\OsqS(r_1)}{\OsqS(r_2)} \right) \right| & p^r > 0 \text{ on } [r_1,r_2]
			\end{cases} & r_1,r_2 \in \mathfrak{I}_{\mathcal{H}^+} \\
			1 & r_1,r_2 \in \mathfrak{I}_{\text{int}} \\
			1 + \left( \log \left| \trapschw \right| \right)_- & r_1,r_2 \in \mathfrak{I}_{\text{ps}} \\
			1 + \frac{\max(r_1,r_2)}{m} & r_1,r_2 \in \mathfrak{I}_{\text{flat}}
		\end{cases}
	\end{equation}
	where we have used the shorthand $\trapschw = \trapschw(\gamma(s),\dot{\gamma}(s))$.
\end{lem}
\begin{proof}
	For the entirety of the proof, all momenta and radii are considered along the geodesic $\gamma$. We will for the most part omit the dependence on the affine parameter $s$ and write for instance $p^r = p^r(s)$ to simplify notation. Since we have $p^r \neq 0$ in the interval of affine parameter time $[s_1,s_2]$, the radius $r$ is a strictly monotone function of $s$ and we have
	\begin{equation}
		t^*(s_2) - t^*(s_1) = \int_{s_1}^{s_2} \frac{dt^*}{d s} \, d s = \int_{s_1}^{s_2} p^{t^*} \, d s = \int_{r_1}^{r_2} \frac{p^{t^*}}{p^r} \, d r,
	\end{equation}
	where we carefully note that if $p^r < 0$, the integration boundaries will be such that $r_2 < r_1$ so that the orientation of the integral ensures the correct (positive) sign. We will now use Lemma~\ref{lem_quotientbound} to bound the integrand in different regions of spacetime.
	
	\paragraph{Region close to horizon.}
	Let us begin by assuming that $r_1,r_2 \in \mathcal{I}_{\mathcal{H}^+}$ and simply write $\delta = \delta_1$. We will distinguish between the two cases that $p^r \leq 0$ and $p^r \geq 0$.

    \subparagraph{Case 1: $p^r \leq 0$}
    We note that in this case, the geodesic will eventually fall into the black hole. We may therefore estimate
	\begin{equation}
		\int_{r_1}^{r_2} \frac{p^{t^*}}{p^r} \, d r \leq \int_{2m}^{r_1} \frac{p^{t^*}}{\left| p^r \right|} \, d r.
	\end{equation}
    We now estimate the integral by distinguishing based on the sign of the trapping parameter $\trapschw$. Before going further, we remind the reader of Lemma~\ref{lem_quotientbound} and the associated definitions of the function $\mathfrak{a}$.

    \subparagraph{Case 1.1: $\trapschw \geq 0$}
 
	Here $(r-3m)^2 + \trapschw \mathfrak{a} \sim_\delta m^2$ and $\frac{r}{r^2 - \trapschw m^2} \sim_\delta \frac{1}{m(2- \trapschw)} \sim \frac{1}{m}$, so that $\frac{p^{t^*}}{|p^r|} \sim_\delta 1$. Hence
	\begin{equation}
		\int_{2m}^{r_1} \frac{p^{t^*}}{|p^r|} \, dr \sim_\delta r_1-2m \lesssim_\delta m,
	\end{equation}
	so that nothing more needs to be shown in this case.

    \subparagraph{Case 1.2: $\trapschw < 0$}
    In this case we rewrite 
	\begin{equation}
		(r-3m)^2 + \trapschw \mathfrak{a} = (r-3m)^2 \left( 1 + \trapschw (r-2m) \mathfrak{b} \right) \sim_\delta m^2 \left( 1 + \trapschw (r-2m) \mathfrak{b} \right) = m^2 \left( 1 + \trapschw x \mathfrak{b} \right),
	\end{equation}
	where we have introduced the change of coordinates $x=r-2m$ and defined the coefficient
    \begin{equation}
        \mathfrak{b}(x) = 27m^2 (x+8m)^{-1}(x-m)^{-2}.
    \end{equation}
    Note that $m \mathfrak{b} \sim_\delta 1$ in the region of integration. We find
	\begin{equation}
		\int_{2m}^{r_1} \frac{p^{t^*}}{|p^r|} \, dr \sim_\delta \int_{2m}^{r_1} \frac{2-\trapschw}{\sqrt{1+\trapschw (r-2m) \mathfrak{b}}} \, dr = \int_{0}^{r_1-2m} \frac{2-\trapschw}{\sqrt{1+\trapschw x \mathfrak{b}}} \, dx \sim_\delta \int_0^{y_1} \frac{2-\trapschw}{\sqrt{1+\trapschw y}} \, dy,
	\end{equation}
	where we have made the change of coordinates $y = x \mathfrak{b}(x)$ and $y_1 = (r_1-2m) \mathfrak{b}(r_1-2m)$. In the last step, we made use of the fact that the Jacobian satisfies
	\begin{equation}
		\left|\mathfrak{b} + x \mathfrak{b}' \right|^{-1} = \frac{(x-m)^3 (x+8m)^2}{54 m^2 (x+2m)^2} \sim_{\delta} m
	\end{equation}
	in the region of concern. Finally we compute
	\begin{align} \label{corrboundSS}
		\int_0^{y_1} \frac{2-\trapschw}{\sqrt{1+\trapschw y}} \, dy &= \frac{2 + |\trapschw|}{|\trapschw|} \int_0^{|\trapschw| y_1} \frac{1}{\sqrt{1-z}} \, dz = 2 \frac{2 + |\trapschw|}{|\trapschw|} \left( 1-\sqrt{1-|\trapschw| y_1} \right) \\
		&\sim \frac{2 + |\trapschw|}{|\trapschw|} |\trapschw| y_1 = (2 + |\trapschw|) y_1 \lesssim_\delta 1.
	\end{align}
	In the last step we used that Lemma \ref{lem_quotientbound} and the definition of $y$ imply $0 \leq 1 + \trapschw y = 1 - | \trapschw | y$ so that $0 \leq y_1 \leq \frac{1}{|\trapschw|}$. In total it follows that regardless of the sign and size of $\trapschw$ if $p^r \leq 0$ we have
	\begin{equation} \label{time::infalling_horizon}
		\int_{2m}^{r_1} \frac{p^{t^*}}{|p^r|} \, dr \lesssim_\delta m.
	\end{equation}

    \subparagraph{Case 2: $p^r \geq 0$}
	We again distinguish two further sub-cases by the sign of $\trapschw$.
 
    \subparagraph{Case 2.1: $\trapschw \geq 0$} We note that the estimate $(r-3m)^2 + \trapschw \mathfrak{a} \sim_\delta m^2$ holds here, like in the case where $p^r \leq 0$. In the present case however it follows that $\frac{p^{t^*}}{|p^r|} \sim_\delta \frac{1}{\OsqS}$. Therefore
	\begin{equation}
		\int_{r_1}^{r_2} \frac{p^{t^*}}{|p^r|} \, dr \sim_\delta \int_{r_1}^{r_2} \frac{1}{\OsqS} \, dr \sim_\delta m \int_{\OsqS(r_1)}^{\OsqS(r_2)} \frac{1}{z} \, dz = m \left| \log \left( \frac{\OsqS(r_1)}{\OsqS(r_2)} \right) \right|,
	\end{equation}
	where we introduced the change of coordinates $z = \OsqS(r)$ and made use of the fact that the Jacobian satisfies $\frac{r^2}{2m} \sim_\delta m$ in the region of integration. This concludes Case 2.1. 

    \subparagraph{Case 2.2: $\trapschw < 0$}
    Here, we rewrite as above
	\begin{equation}
		(r-3m)^2 + \trapschw \mathfrak{a} = (r-3m)^2 \left( 1 + \trapschw (r-2m) \mathfrak{b} \right) \sim_\delta m^2 \left( 1 + \trapschw (r-2m) \mathfrak{b} \right) = m^2 \left( 1 + \trapschw x \mathfrak{b} \right),
	\end{equation}
	where we use the change of coordinates $x=r-2m$ and define $\mathfrak{b}(x) = 27m^2 (x+8m)^{-1}(x-m)^{-2}$. Note that $m \mathfrak{b} \sim_\delta 1$ in the region of integration. We compute
	\begin{equation}
		\int_{r_1}^{r_2} \frac{p^{t^*}}{|p^r|} \, dr \sim_\delta \int_{r_1-2m}^{r_2-2m} \frac{1}{x \sqrt{1+\trapschw x \mathfrak{b}(x)}} \, dx \sim_\delta m \int_{y_1}^{y_2} \frac{1}{y \sqrt{1+\trapschw y}} \, dy,
	\end{equation}
	where we have made the change of variables $y = x \mathfrak{b}(x)$, set $y_i = (r_i-2m) \mathfrak{b}(r_i-2m)$ for $i=1,2$ and made use of the fact that the Jacobian $\left|\mathfrak{b} + x \mathfrak{b}' \right|^{-1} \sim_{\delta} m$ in the region of integration. Observing that $0 \leq 1 + \trapschw y \leq 1$, we then find
	\begin{equation}
		\int_{y_1}^{y_2} \frac{1}{y \sqrt{1+\trapschw y}} \, dy = -2 \left[ \tanh ^{-1}\left(\sqrt{1+\trapschw y}\right) \right]_{y_1}^{y_2} = \left[ \log \left( \frac{1-\sqrt{1+\trapschw y}}{1+\sqrt{1+\trapschw y}} \right) \right]_{y_1}^{y_2}.
	\end{equation}
	Note carefully that since $y_1 < y_2$ we find that $1 \leq (1+\sqrt{1+\trapschw y_1})(1+\sqrt{1+\trapschw y_2})^{-1} \leq 2$ and
	\begin{equation}
		\left[ \log \left( \frac{1-\sqrt{1+\trapschw y}}{1+\sqrt{1+\trapschw y}} \right) \right]_{y_1}^{y_2} \leq \log 2 + \log \left( \frac{1-\sqrt{1-\left|\trapschw \right| y_2}}{1-\sqrt{1- \left| \trapschw \right| y_1}} \right) \leq 2 \log 2 + \log \left( \frac{y_2}{y_1} \right),
	\end{equation}
	where we used that $\left| \trapschw \right| y_i \leq 1$. Plugging in the definition of $y$ we finally find
	\begin{equation}
		\log \left( \frac{y_2}{y_1} \right) \lesssim 1 + \log \left( \frac{\OsqS(r_2)}{\OsqS(r_1)} \right),
	\end{equation}
	which allows us to conclude the desired bound.
	
	\paragraph{Region far from horizon}
	Let us now assume that $r_1,r_2 \in \mathfrak{I}_{\text{int}} \cup \mathfrak{I}_{\text{ps}} \cup \mathfrak{I}_{\text{flat}}$ and let us again for simplicity write $\delta = \delta_1$. Since the quotient $\frac{p^{t^*}}{|p^r|}$ is comparable in case $p^r \leq 0$ and $p^r \geq 0$, we do not need to distinguish different cases based on the sign of $p^r$. Notice that $\mathfrak{a} \sim_\delta m^2$ in this region. We introduce the change of variables $x=r-3m$ so that $\sqrt{(r-3m)^2 + \trapschw \mathfrak{a}} = \sqrt{x^2 + \trapschw \mathfrak{a}}$. We then introduce a further change of variable $y = \mathfrak{a}(x)^{-\frac{1}{2}} x$ and compute
	\begin{equation} \label{prooftstar_schw_eq}
		\int_{r_1}^{r_2} \frac{p^{t^*}}{p^r} \, dr \sim \left| \int_{r_1-3m}^{r_2-3m} \frac{x+3m}{\sqrt{x^2 + \trapschw \mathfrak{a}}} \, dx \right| = \left| \int_{y_1}^{y_2} \frac{y + \frac{3m}{\sqrt{\mathfrak{a}}}}{\sqrt{y^2 + \trapschw}} \, \mathfrak{J}(y) \, dy \right| \sim_\delta m \left| \int_{y_1}^{y_2} \frac{|y| + 1}{\sqrt{y^2 + \trapschw}} \, dy \right| ,
	\end{equation}
	where we made use of the fact that the Jacobian satisfies
	\begin{equation}
		\mathfrak{J} = \left| \mathfrak{a}^{-\frac{1}{2}}  \left( 1- \frac{1}{2} \frac{\mathfrak{a}' }{\mathfrak{a}} x \right) \right|^{-1} = \mathfrak{a}^{\frac{1}{2}} \frac{(x+m)(x+9m)}{(x+3m)^2} \sim_{\delta} m.    
	\end{equation}
	Note that $\sgn(y) = \sgn(r-3m)$, so that $r_i < 3m $ if and only if $y_i < 0$. It follows from equation~\eqref{prooftstar_schw_eq} that we may assume $y_1 < y_2$ or equivalently $p^r > 0$ without loss of generality. We estimate the integral in~\eqref{prooftstar_schw_eq} by considering three separate cases, based on the size of $| \trapschw | $ and $y_1, y_2$ relative to $\eta$.

    \subparagraph{Case 1: $4 |\trapschw| \, \geq \eta^2 >0$}

	%\begin{equation}
	%	\int_{r_1}^{r_2} \frac{p^{t^*}}{|p^r|} \, dr \lesssim_{\eta,\delta} 1 + r_2 .
	%\end{equation}
	%In other words, if either $2 \sqrt{|\trapschw|} \geq \eta > 0$ or if $|\trapschw|$ is small but the integration boundaries satisfy either $y_1 \leq y_2 \leq - \eta < 0$ or $y_2 \geq y_1 \geq \eta >0$, then the above bound holds.
    We introduce the rescaled variable $y = \sqrt{|\trapschw|} z$, so that
	\begin{equation}
		\int_{y_1}^{y_2} \frac{|y| + 1}{\sqrt{y^2 + \trapschw}} \, dy \leq  \sqrt{|\trapschw|} \int_{z_1}^{z_2} \frac{|z| + \frac{2}{\eta}}{\sqrt{z^2 + \sgn(\trapschw)}} \, dz,
	\end{equation}
	where $ \sqrt{|\trapschw|} z_i = y_i$ for $i=1,2$. We notice that in case $\trapschw > 0$, the integrand is bounded by a constant for the argument $z \in \R$. Therefore we find the bound
	\begin{equation}
		\int_{r_1}^{r_2} \frac{p^{t^*}}{p^r} \, dr \lesssim_{\eta,\delta} \sqrt{|\trapschw|} (z_2-z_1) = y_2 - y_1 \sim_\delta \left| r_2 - r_1 \right|.
	\end{equation}
	In case $\trapschw < 0$, let us first consider the case that $z_1 \leq z_2 \leq -1$ and note immediately that since we assume $r_1 \geq (2+\delta)m$, it follows that $\sqrt{|\trapschw|} \leq \left|y_1 \right| = - y_1 \lesssim_\delta 1$. Therefore in this region we have $1 \lesssim_\eta \left|\trapschw \right| \lesssim_\delta 1$ and therefore also $\left|z_1 \right| \sim_{\delta,\eta} \left| y_1 \right|$ so that
	\begin{equation}
		\int_{z_1}^{z_2} \frac{|z| + \frac{2}{\eta}}{\sqrt{z^2 -1}} \, dz \leq \int_{z_1}^{-1} \frac{|z| + \frac{2}{\eta}}{\sqrt{z^2 -1}} \, dz \lesssim_{\delta,\eta} 1,
	\end{equation}
	where we merely exploited the local integrability of the integrand. Let us now turn to the case that $\trapschw < 0$ and $1 \leq z_1 \leq z_2$. We notice that
	\begin{equation}
		\mathfrak{I} := \frac{|z| + \frac{2}{\eta}}{\sqrt{z^2 -1}} \sim_\eta \begin{cases}
			\frac{1}{\sqrt{z-1}} & z \in [1,2] \\
			1 & z \in [2, \infty]
		\end{cases}.
	\end{equation}
	Hence upon integration, we find that if $z_2 \leq 2$, then
	\begin{equation}
		\int_{z_1}^{z_2} \mathfrak{I} \, dz \sim_\eta \int_{z_1}^{z_2} \frac{1}{\sqrt{z-1}} \, dz \leq \sqrt{z_2-z_1} ,   
	\end{equation}
	where we made use of the inequality $\sqrt{x}-\sqrt{y} \leq \sqrt{x-y}$ for $x \geq y \geq 0$. If $z_1 \geq 2$, we find
	\begin{equation}
		\int_{z_1}^{z_2} \mathfrak{I} \, dz \sim_\eta \int_{z_1}^{z_2} 1 \, dz = z_2-z_1.
	\end{equation}
	The case that $z_1 \leq 2$ and $z_2 \geq 2$ is then covered by summing the above two bounds. In summary we obtain
	\begin{equation}
		\int_{y_1}^{y_2} \frac{|y| + 1}{\sqrt{y^2 + \trapschw}} \, dy \leq \begin{cases}
			|\trapschw|^{\frac{1}{4}} \sqrt{y_2-y_1} & \text{if } y_2,y_1 \sim \sqrt{|\trapschw|} \\
			y_2-y_1 & \text{if } y_2,y_1 \geq 2 \sqrt{|\trapschw|}
		\end{cases} \; \leq 1 + y_2 \lesssim_\delta 1 + \frac{r_2}{m}.
	\end{equation}
	We conclude the bound as claimed above. The above computation also reveals the reason why we only show the upper bound $1+y_2$ instead of $y_2-y_1$ for the integral: Suppose $d > 0$, then the time it takes to move from $y_2 = (1+d) \sqrt{|\trapschw|}$ to $y_1 = \sqrt{|\trapschw|}$ is like $\sqrt{|\trapschw|} \sqrt{d} \gg \sqrt{|\trapschw|} d = y_2-y_1$ when $d$ is small.

    \subparagraph{Case 2: $4 |\trapschw| \, \leq \eta^2$ and $\min (|y_1|, |y_2|) \geq \eta > 0$}
 
    In this case, the assumptions on $\trapschw, y_1, y_2$ readily imply the bound $y^2 + \trapschw \geq \frac{y^2}{2} \geq \frac{\eta^2}{2}$. It follows that
	\begin{equation}
		\frac{|y| + 1}{\sqrt{y^2 + \trapschw}} \lesssim_{\eta} 1,
	\end{equation}
	for $y \notin [-\eta,\eta]$. Therefore we again conclude the better bound 
	\begin{equation} \label{timebound_away_from_photonsphere}
		\int_{r_1}^{r_2} \frac{p^{t^*}}{p^r} \, dr \lesssim_{\eta,\delta} \left| r_2 - r_1 \right|.
	\end{equation}

    \subparagraph{Case 3: $4 |\trapschw| \, \leq \eta^2$ and $\min (|y_1|, |y_2|) \leq \eta$} 
    Consider the case that $\trapschw \leq 0$ first. It follows readily from the geodesic equations that the geodesic cannot cross the photon sphere, so that we must either have $- \eta \leq y_1 \leq y_2 \leq -\sqrt{|\trapschw|}$ or $\sqrt{|\trapschw|} \leq y_1 \leq y_2 \leq \eta$. By symmetry of the integral we may assume the latter without loss of generality. We bound
	\begin{equation}
		\int_{y_1}^{y_2} \frac{y + 1}{\sqrt{y^2 - |\trapschw|}} \, dy \leq \int_{y_1}^{\eta} \frac{y + 1}{\sqrt{y^2 - |\trapschw|}} \, dy = \left[ \sqrt{y^2 - |\trapschw|} + \log \left( y + \sqrt{y^2 - |\trapschw|} \right) \right]_{y_1}^{\eta}.
	\end{equation}
	We may choose $\eta$ sufficiently small and assume $\left| \trapschw \right| \leq \frac{\eta^2}{4}$ so that the relation
	\begin{gather}
		\left( 1 + \sqrt{\frac{3}{4}} \right) \eta \leq \eta + \sqrt{\eta^2 - |\trapschw|} \leq 2 \eta < \frac{1}{2},
	\end{gather}
	holds. This immediately implies that $\left| \log \left( \eta + \sqrt{\eta^2 - |\trapschw|} \right) \right| \geq \left| \log ( 2 \eta ) \right| > \eta \geq \sqrt{\eta^2 - |\trapschw|}$ and therefore
	\begin{equation}
		\sqrt{\eta^2 - |\trapschw|} + \log \left( \eta + \sqrt{\eta^2 - |\trapschw|} \right) < 0.
	\end{equation}
	Using this bound, we find
	\begin{align}
		\left[ \sqrt{y^2 - |\trapschw|} + \log \left( y + \sqrt{y^2 - |\trapschw|} \right) \right]_{y_1}^{\eta} \leq \left| \log \left( y_1 + \sqrt{y_1^2 - |\trapschw|} \right) \right| \leq \left| \log(y_1) \right| \leq \frac{1}{2} \left| \log( \trapschw ) \right|,
	\end{align}
	where we used that $y_1 \geq \sqrt{|\trapschw|}$ in the last step. Let us now turn to the case where $\trapschw \geq 0$. In this case the geodesic may cross the photon sphere. By symmetry of the integrand around the origin we may therefore assume without loss of generality that either $0 \leq y_1 \leq y_2 \leq \eta$ or $-\eta \leq y_1 < 0 \leq y_2 \leq \eta$. In the former case, we find similar to above
	\begin{equation}
		\int_{y_1}^{y_2} \frac{y + 1}{\sqrt{y^2 + \trapschw}} \, dy \leq \int_{y_1}^{\eta} \frac{y + 1}{\sqrt{y^2 + \trapschw}} \, dy = \left[ \sqrt{y^2 + \trapschw} + \log \left( y + \sqrt{y^2 + \trapschw} \right) \right]_{y_1}^{\eta}.
	\end{equation}
	Again we assume  $\left| \trapschw \right| \leq \frac{\eta^2}{4}$ and choose $\eta$ sufficiently small so that similar to above
	\begin{equation} \label{yintegralbound}
		\left[ \sqrt{y^2 + \trapschw} + \log \left( y + \sqrt{y^2 + \trapschw} \right) \right]_{y_1}^{\eta} \leq \left| \log \left( y_1 + \sqrt{y_1^2 + \trapschw} \right) \right| \leq \frac{1}{2} \left| \log ( \trapschw ) \right|.
	\end{equation}
	Assume now that $y_1<0$, this means that the geodesic crosses the photon sphere in the affine parameter time interval under consideration. We split the range of integration in two parts $[y_1,y_2] = [y_1,0) \cup [0,y_2]$. For the first part $[y_1,0)$ we apply the transformation $y \mapsto -y$ and therefore find for both $i=1$ and $i=2$ that
	\begin{equation}
		\int_{0}^{|y_i|} \frac{y + 1}{\sqrt{y^2 + \trapschw}} \, dy \leq \int_{0}^{\eta} \frac{y + 1}{\sqrt{y^2 + \trapschw}} \, dy \leq \frac{1}{2} \left| \log ( \trapschw ) \right|,
	\end{equation}
	where this bound is simply obtained from~\eqref{yintegralbound} by setting $y_i=0$. This concludes the proof.
\end{proof}

\subsubsection{The asymptotically flat region}

We now turn our attention to the asymptotically flat region $\mathcal{A} = \{ r \geq R \} \cap \Mschw$. Let $\tauzero, R_0 \geq m$ be constants to be chosen later and define
\begin{equation}
	\mathcal{A}_{0} = \left( \left\{ \tau \geq \tauzero \right\} \cup \left\{ r \geq R_0 \right\} \right) \cap \mathcal{A} \subset \mathcal{A}.
\end{equation}
We first establish that for a suitable choice of constants $\tauzero$ and $R_0$, every geodesic within the region $\mathcal{A}_{0}$ is outgoing and its energy and radial momentum are comparable.

\begin{lem} \label{pr_like_E}
	Let us denote by $\gamma: [s_0,s_1) \rightarrow \Mschw$ an affinely parametrised future-oriented null geodesic segment expressed in $(t^*,r)$-coordinates as in Lemma~\ref{tstar_lem_schw}.  Assume that $\gamma(s_0) \in \Sigma_0$ and $r_0 = r(s_0) \leq \rsuppconst$. Let us assume in addition that $\gamma$ is maximally defined, so that either $s_1 < \infty$ and $\lim_{s \rightarrow s_1} r(s) = 2m$ or $s_1 = \infty$. Let $\tauzero \geq C \rsuppconst$ with a suitable constant $C > 1$ and $R_0 = \max(2 \rsuppconst, R)$. Then for all $s \in [s_0,s_1)$ such that $\gamma(s) \in \mathcal{A}_0$, we have
	\begin{equation}
		E \geq p^r(s) \geq \frac{1}{2} E > 0,
	\end{equation}
	where we abbreviated $E = E(\gamma,\dot{\gamma})$. In particular, all geodesics which intersect $\Sigma_0$ at radius $r_0 \leq \rsuppconst$ are outgoing in the region $\mathcal{A}_0$.
\end{lem}
\begin{proof}
	We will begin by considering the time $\gamma$ requires to cross a certain region measured in $t^*,u$ and $\tau$-time and establishing a relationship between these times. Let $[s_3,s_4] \subset [s_0,s_1)$ and for any function $F: \Mschw \rightarrow \R$ let us introduce the shorthand $\Delta F(s_3,s_4) = \left| F(\gamma(s_4)) - F(\gamma(s_3)) \right|$. Assume that $[s_3,s_4] \subset [s_0,s_1)$ is such that $p^r(s) \neq 0$ for $s \in [s_3,s_4]$. Recalling the definition of double null coordinates on $\Mschw$ we then have
	\begin{equation}
		\begin{aligned} \label{delta_u_estimate}
			\Delta u(s_3,s_4) &= \int_{s_3}^{s_4} p^u(s) \, ds = \int_{r(s_3)}^{r(s_4)} \frac{p^u}{p^r} \, dr \sim \int_{r(s_3)}^{r(s_4)} \frac{p^{t^*}}{p^r} - \frac{r+2m}{r-2m} \, dr \\
			&\sim \Delta t^*(s_3,s_4) - \int_{r(s_3)}^{r(s_4)}  \frac{r+2m}{r-2m} \, dr,
		\end{aligned}
	\end{equation}
	where we carefully note that if $p^r < 0$, the integration boundaries will be such that $r(s_3) < r(s_4)$ so that the orientation of the integral ensures the correct (positive) sign. If $r(s_3),r(s_4) \geq R > 3m$, then by definition of the time function $\tau$ we have $\Delta \tau(s_3,s_4) = \Delta u(s_3,s_4)$. We conclude that
	\begin{align}
		\Delta \tau(s_3,s_4) + \Delta r(s_3,s_4) \sim \Delta t^*(s_3,s_4) \quad &\text{if } p^r(s) > 0 \; \forall s \in [s_3,s_4], \label{timerelations_outgoing} \\
		\Delta \tau(s_3,s_4) \sim \Delta t^*(s_3,s_4) + \Delta r(s_3,s_4) \quad &\text{if } p^r(s) < 0 \; \forall s \in [s_3,s_4]. \label{timerelations_incoming}
	\end{align}
	Next we show that if we choose $\tauzero$ large enough, then for all $s \in [s_0,s_1)$ such that $\gamma(s) \in \mathcal{A}$ and $\tau(\gamma(s)) \geq \tauzero$ we have $p^r(s) \geq 0$. Carefully note that the geodesic equations imply
	\begin{equation} \label{geodequnsign}
		\sgn \left( \frac{d}{ds} p^r(s) \right) = 1 \quad \forall \, s \in [s_0,s_1) \text{ with } r(s) > 3m.
	\end{equation}
	Therefore if $s^* \in [s_0,s_1)$ is such that $r(s^*) \geq R$ and $p^r(s^*) \geq 0$ then $p^r(s) \geq 0$ for all $s \geq s^*$. Suppose there exists $s^* \in [s_0,s_1)$ such that $p^r(s^*) < 0$. Then it must be true that $p^r(s) < 0$ for all $s \in [s_0,s^*]$. Furthermore it follows that there must exist a time $s' \geq s^*$ such that either $p^r(s') = 0$ or $r(s') = R$. Equation~\eqref{geodequnsign} then implies that for all $s > s'$ we either have $r(s) < R$ or $r(s) \geq R$ and $p^r(s) \geq 0$, so that we need only estimate $\Delta \tau(s_0,s')$. Using equation~\eqref{timerelations_incoming} and Lemma~\ref{tstar_lem_schw} we find that
	\begin{equation}
		\Delta \tau(s_0,s') \lesssim \Delta t^*(s_0,s') + \Delta r(s_0,s') \lesssim 1 + 2 r(s_0) \lesssim \rsuppconst,
	\end{equation}
	where we used that by assumption $\gamma$ intersects $\Sigma_0$ at radius $r(s_0) = r_0 \leq \rsuppconst$. Therefore we have shown the existence of a constant $C > 0$ such that for any parameter time $s$ with $r(s) \geq R$ and $\tau(\gamma(s)) \geq C \rsuppconst$, we must have $p^r(s) \geq 0$. Let us denote the smallest parameter time $s \in [s_0,s_2)$ such that $r(s) \geq R$ and $p^r(s) \geq 0$ by $s'$ (assuming it exists). Then we have just shown that $\tau(\gamma(s')) \leq C \rsuppconst$ and further that $R \leq r(s') \leq \rsuppconst$. 
	
	Next we show that we only need to wait a finite further time until $p^r \geq \frac{1}{2} E$. Recall that energy conservation implies the bound $\frac{r^2}{\OsqS} \geq \left( \frac{L}{E} \right)^2$ for every point on the mass-shell. For any point $(x,p) \in \mathcal{P}$ the inequality $p^r \geq \frac{1}{2} E$ is then equivalent to the improved bound
	\begin{equation}\label{improvedbound}
		\frac{3}{4} \frac{r^2}{\OsqS} \geq \left( \frac{L}{E} \right)^2.
	\end{equation}
	Suppose this bound holds when evaluated at the point $(\gamma(\tilde{s}),\dot{\gamma}(\tilde{s}))$ for some $\tilde{s} \geq s'$, then it will hold for all future times $s \geq \tilde{s}$ since $p^r \geq 0$ and the function $\frac{r^2}{\OsqS}$ is increasing in $r$ when $r \geq 3m$. Let therefore $\tilde{s}$ be the parameter time such that the radius $\tilde{r} = r(\tilde{s})$ satisfies the equality
	\begin{equation}
		\frac{3}{4} \frac{r^2}{\OsqS}\Bigg|_{r=\tilde{r}} = \frac{r^2}{\OsqS} \Bigg|_{r=r(\tau')}.
	\end{equation}
	Then the inequality $\frac{r^2}{\OsqS} \geq \left( \frac{L}{E} \right)^2$ which holds for every point on the mass-shell implies that the improved bound~\eqref{improvedbound} holds for every time $s \geq \tilde{s}$. Our choice of $\tilde{r}$ implies that $\tilde{r} \sim r(\tau') \lesssim \rsuppconst$. From~\eqref{timerelations_outgoing} it then follows that $\Delta \tau(s',\tilde{s}) \lesssim \Delta t^*(s',\tilde{s}) \lesssim 1 + \tilde{r} \lesssim \rsuppconst$ as claimed. Therefore there exists a constant $C > 0$ such that for any parameter time $s$ with $r(s) \geq R$ and $\tau(\gamma(s)) \geq C \rsuppconst$, we must have $p^r(s) \geq \frac{E}{2}$. Let us therefore define $\tauzero = C \rsuppconst$. 
	
	Finally let us show that if we assume $r$ is sufficiently large but do not place a restriction on the time $\tau$ we may still conclude $p^r \geq \frac{1}{2}E$. To see this, recall that this inequality is equivalent to the improved bound
	\begin{equation}
		\frac{3}{4} \frac{r^2}{\OsqS} \geq \left( \frac{L}{E} \right)^2.
	\end{equation}
	Therefore it is enough to choose $R_0$ so that
	\begin{equation} \label{eqn_schw_asflat_rbound}
		\frac{3}{4} \frac{r^2}{\OsqS}\Bigg|_{r=R_0} \geq \frac{r^2}{\OsqS} \Bigg|_{r=
			\rsuppconst},
	\end{equation}
	since this implies for all $r \geq R_0$
	\begin{equation}
		\frac{3}{4} \frac{r^2}{\OsqS}\Bigg|_{r} \geq \frac{3}{4} \frac{r^2}{\OsqS}\Bigg|_{r=R_0} \geq \frac{r^2}{\OsqS} \Bigg|_{r=
			\rsuppconst} \geq \frac{r^2}{\OsqS} \Bigg|_{r=r(0)} \geq \left( \frac{L}{E} \right)^2.
	\end{equation}
	Therefore we may choose $R_0 = 2\rsuppconst$, which implies that inequality~\eqref{eqn_schw_asflat_rbound} is satisfied.
\end{proof}

We will now show that geodesics in the region $\mathcal{A}_0$ propagate approximately along $\Sigma_\tau \cap \{ r \geq R \}$. Note that the complement of $\mathcal{A}_0$ in the asymptotically flat region $\mathcal{A}$ is a compact subset of the exterior $\Mschw$. We show in Lemma~\ref{psupport_bounded} below that solutions with compactly supported initial data have compact momentum support for all times. This immediately implies that all moments of solutions to the massless Vlasov equation remain bounded in any compact region of spacetime.

\begin{lem} \label{tauestimate_far}
	Denote by $\gamma: [s_0,s_1] \rightarrow \Mschw$ an affinely parametrised future-oriented null geodesic segment. Assume that $\gamma(s_0) \in \Sigma_0$ and $r_0 = r(s_0) \leq \rsuppconst$. Let $s^* \in [s_0,s_1]$ be such that $\gamma([s^*,s_1]) \subset \mathcal{A}_0$. If we express $\gamma$ in double null coordinates, then the $p^v$-component remains approximately constant along $\gamma$ in the region $\mathcal{A}_0$. In fact for all $s \in [s^*,s_1]$:
	\begin{equation}
		p^v(s) \sim E(\gamma,\dot{\gamma}).
	\end{equation}
	Furthermore for all $s \in [s^*,s_1]$ we have the bound
	\begin{equation}
		r(s)^2 p^u(s) \lesssim \rsuppconst^2 p^v(s).
	\end{equation}
	In addition for all $[s_3,s_4] \subset [s^*,s_1]$ we have the bound
	\begin{equation} \label{taubound_far}
		\tau(\gamma(s_4)) - \tau(\gamma(s_3)) \lesssim m \left( 1 + \frac{\rsuppconst^2}{m^2} \right),
	\end{equation}
	or in other words outgoing null geodesics in this region approximately move along $\Sigma_\tau \cap \{ r \geq R \}$.
\end{lem}

\begin{rem}
	We conclude that, measured in double null coordinates, outgoing null geodesics in the asymptotically flat region behave essentially like in flat Minkowski space, see also the discussion in~\cite[Section 7]{martin}. Recall however that there is a logarithmic divergence between double null coordinates in Schwarzschild and Minkowski space.
\end{rem}

\begin{proof}
	Recall that by Lemma~\ref{pr_like_E} we may in particular assume all geodesics in the region $\mathcal{A}_0$ to be outgoing. We begin by showing that along $\gamma$ the $p^v$ component remains approximately constant while the $p^u$ component decays in $r$ as $r \rightarrow \infty$. By Lemma~\ref{pr_like_E} we have the bound $\frac{1}{2} \leq \frac{p^r(s)}{E} \leq 1$ for $s \in [s^*,s_1]$ or equivalently when $\gamma(s) \in \mathcal{A}_0$. Let us express the fraction $\frac{p^r}{E}$ first in $(t,r^*)$-coordinates and then in double null coordinates:
	\begin{equation}
		\frac{p^r}{E} = \frac{\OsqS p^{r^*}}{\OsqS p^t} = \frac{p^v-p^u}{p^v+p^u} = \frac{1 - \frac{p^u}{p^v}}{1 + \frac{p^u}{p^v}},
	\end{equation}
	which upon noticing that the function $x \mapsto \frac{1-x}{1+x}$ is self-inverse gives for $s \in [s^*,s_1]$
	\begin{equation} \label{p3p4}
		\frac{p^u(s)}{p^v(s)} = \frac{1 - \frac{p^r(s)}{E}}{1 + \frac{p^r(s)}{E}} \leq \frac{1}{2}.
	\end{equation}
	Let us furthermore express the energy in double null coordinates as $E = \OsqS \left( p^u + p^v \right)$. Recall that future-directedness of the geodesic and the mass-shell relation imply $E \geq 0$ and $p^u p^v \geq 0$, which combined allows us to conclude that $p^u(s) \geq 0$ and $p^v(s) \geq 0$. Therefore we immediately have the bound $\OsqS(r(s)) p^v(s) \leq E$. Now using that we have shown above $2 p^u(s) \leq p^v(s)$ we find
	\begin{equation}
		\frac{2}{3} E \leq \OsqS(r(s)) p^v(s) \leq E.
	\end{equation}
	Since we assumed $R > 3m$ we have that $\frac{1}{3} \leq \OsqS \leq 1$ in the asymptotically flat region. Therefore
	\begin{equation}
		\frac{2}{3} E \leq p^v(s) \leq 3 E.
	\end{equation}
	To show that $r^2 p^u \lesssim p^v$ in the region $\mathcal{A}_0$, the mass-shell relation implies
	\begin{equation}
		0 \leq 1 - \left| \frac{p^r(s)}{E} \right| \leq \frac{\OsqS}{r^2} \left( \frac{L}{E} \right)^2 \lesssim \frac{ \rsuppconst^2}{r(s)^2},   
	\end{equation}
    where we used that $\left( \frac{L}{E} \right)^2 \lesssim r(s)^2$ holds for all $s \in [s_0, s_1]$, so that in particular the bound holds for the parameter time $s$ where $r(s) = R \leq \rsuppconst$. Using equation~\eqref{p3p4} we find
	\begin{equation} \label{p3p42}
		\frac{p^u(s)}{p^v(s)} = \frac{1 - \frac{p^r(s)}{E}}{1 + \frac{p^r(s)}{E}} \lesssim \frac{\rsuppconst^2}{r(s)^2}.
	\end{equation}
	Finally let us show the bound for the $\tau$-time. Let $[s_3,s_4] \subset [s^*,s_1]$ and notice that by definition $\tau(\gamma(s_4)) - \tau(\gamma(s_3)) = u(s_4) - u(s_3)$, so that
	\begin{equation}
		u(s_4) - u(s_3) = \int_{s_3}^{s_4} p^u(s) \, ds = \int_{r(s_3)}^{r(s_4)} \frac{p^u}{p^r } \, dr \sim \int_{r(s_3)}^{r(s_4)} \frac{p^u}{p^v} \, dr \lesssim \rsuppconst^2 \int_{r(s_3)}^{r(s_4)} \frac{1}{r^2} \, dr \lesssim \frac{\rsuppconst^2}{R},
	\end{equation}
	where we have used that in the region $\mathcal{A}_0$ the relation $p^r \sim E \sim p^v$ holds. Finally noting that we chose $R > 3m$, the claimed bound readily follows.
\end{proof}

\subsubsection{Proof of Lemma~\ref{taubound}}

We now combine the estimates established in the previous sections in order to prove Lemma~\ref{taubound}. The proof will also make reference to certain basic facts about the geodesic flow on the Schwarzschild exterior, see~\cite[Chapter 6.3]{waldbook} for a comprehensive discussion.

\begin{proof}[Proof of Lemma~\ref{taubound}]
	To simplify notation we will adopt the convention that all of the bounds given in this proof will depend on $\delta > 0$, although it will be obvious which bounds exactly will depend on $\delta$. Like in the proof of Lemma~\ref{tstar_lem_schw} we will consider the regions close to and far from the event horizon separately.
	
	\paragraph{Region close to horizon.}
	Let us first treat the case that $2m \leq r(s) \leq (2+\delta)m$. Then if $p^r(s) > 0$, it follows readily from the form of the geodesic equations that $p^r(s') > 0$ and $r_0 \leq r(s') \leq (2+\delta)m$ for all $s' \in [s_0,s]$. Therefore by definition
	\begin{equation}
		\tau(\gamma(s)) - \tauzero \leq t^*(s) = t^*(s) - t^*(s_0),
	\end{equation}
	which we may directly bound using Lemma~\ref{tstar_lem_schw} if we set $\delta_1 = \delta$ there, so that
	\begin{equation}
		\frac{1}{m} \left( \tau(\gamma(s)) - \tauzero \right) \lesssim 1 + \left| \log \left( \frac{\OsqS(r_0)}{\OsqS(r(s))} \right) \right|.
	\end{equation}
	In the other case that $p^r(s) \leq 0$ we need to distinguish by the sign of $\trapschw$. Let us recall that if $\trapschw \geq 0$, the sign of $p^r$ will not change along $\gamma$, whereas in the case that $\trapschw < 0$, $p^r$ must in fact change sign along $\gamma$, see~\cite[Chapter 6.3]{waldbook}. Thus if $\trapschw \geq 0$, we must have $p^r(s') \leq 0$ for all $s' \in [s_0,s]$ and $r_0 \geq r(s)$. If we choose $\tauzero$ as in Lemma~\ref{pr_like_E}, the lemma implies that for $s' \in [s_0,s_1]$ with $\gamma(s') \in \mathcal{A}_0$ necessarily $p^r(s') > 0$. In other words, for times $\tau \geq \tauzero$ geodesics may cross from the region $\{ 2m \leq r \leq R_0 \}$ into the asymptotically flat region but not the other way around. Therefore for all $s' \leq s$ such that $\tau(\gamma(s')) \geq \tauzero$ we have $2m \leq r(s') \leq R_0$ for $s' \in [\tilde{s},s]$. Therefore we may again apply Lemma~\ref{tstar_lem_schw} directly and find
	\begin{equation}
		\frac{1}{m} \left( \tau(\gamma(s)) - \tauzero \right) \lesssim 1 + \left( \log \left| \trapschw \right| \right)_-.
	\end{equation}
	If $\trapschw < 0$ then it must be the case that $2m \leq r(s') \leq 3m$ for all $s' \in [s_0,s]$. Let us consider the simpler case that $p^r(s_0) \leq 0$ first. Since $\trapschw < 0$ it follows that $r_0 \leq r_{\text{min}}^-(\trapschw)$ and $p^r(s') \leq 0$ for all $s' \in [s_0, s]$. We may then apply Lemma~\ref{tstar_lem_schw} to find
    \begin{equation}
		\frac{1}{m} \left( \tau(\gamma(s)) - \tau(\gamma(s_0)) \right) \lesssim 1 + \left( \log \left| \trapschw \right| \right)_-.
	\end{equation}
    If we instead assume that $p^r(s_0) > 0$, then there must exist $\tilde{s} \in (s_0,s)$ with $p^r(\tilde{s}) = 0$, separating the evolution into an initial period where $p^r > 0$ and the geodesic travels outward from $r_0$ to the radius of closest approach to the photon sphere $r_{\text{min}}^-(\trapschw)$, followed by a period where $p^r \leq 0$ during which the geodesic travels inward from $r_{\text{min}}^-(\trapschw)$ to $r$. Noting that the geodesic remains inside the photon sphere, we need to apply Lemma~\ref{tstar_lem_schw} to both those periods and add the resulting bounds. For the period $[\tilde{s},s]$, where $p^r \leq 0$, Lemma~\ref{tstar_lem_schw} immediately implies that
	\begin{equation}
		\frac{1}{m} \left( \tau(\gamma(s)) - \tau(\gamma(\tilde{s})) \right) \lesssim 1 + \left( \log \left| \trapschw \right| \right)_-.
	\end{equation}
	For the initial period $[s_0,\tilde{s})$ where $p^r > 0$ let us distinguish between the two cases that $r_0 \leq (2+\delta)m$ and $r_0 > (2+\delta)m$. In the latter case, we may apply Lemma~\ref{tstar_lem_schw} with $\delta_1=\delta$ to find
	\begin{equation}
		\frac{1}{m} \left( \tau(\gamma(\tilde{s})) - \tau(\gamma(s_0)) \right) = \frac{\tau(\gamma(\tilde{s}))}{m} \lesssim 1 + \left( \log \left| \trapschw \right| \right)_-.
	\end{equation}
	Assume therefore that $r_0 \leq (2+\delta)m$. In this case let us further distinguish the two cases that $r_{\text{min}}^-(\trapschw) > (2+\delta)m$ and $r_{\text{min}}^-(\trapschw) \leq (2+\delta)m$. In the first case, we apply Lemma~\ref{tstar_lem_schw} with $\delta_1=\delta$ to find
	\begin{equation}
		\begin{aligned}
			\frac{\tau(\gamma(\tilde{s}))}{m} &\lesssim 1 + \left| \log \left( \frac{\OsqS(r_0)}{\OsqS((2+\delta)m)} \right) \right| + \left( \log \left| \trapschw \right| \right)_- \\
			&\lesssim 1 + \left| \log \left( \frac{\OsqS(r_0)}{\OsqS(r_{\text{min}}^-(\trapschw))} \right) \right| + \left( \log \left| \trapschw \right| \right)_-,
		\end{aligned}
	\end{equation}
	where we used that $\OsqS(r_{\text{min}}^-(\trapschw)) > \OsqS((2+\delta)m)$ by assumption. If $r_{\text{min}}^-(\trapschw) \leq (2+\delta)m$, Lemma~\ref{tstar_lem_schw} with $\delta_1=\delta$ gives 
	\begin{equation}
		\frac{\tau(\gamma(\tilde{s}))}{m} \lesssim 1 + \left| \log \left( \frac{\OsqS(r_0)}{\OsqS(r_{\text{min}}^-(\trapschw))} \right) \right|.
	\end{equation}
	In total we find, whenever $r_0 \leq (2+\delta)m$, that we have the bound
	\begin{equation}
		\frac{1}{m} \left( \tau(\gamma(s)) - \tauzero \right) \lesssim 1 + \left| \log \left( \frac{\OsqS(r_0)}{\OsqS(r_{\text{min}}^-(\trapschw))} \right) \right| + \left( \log \left| \trapschw \right| \right)_-.
	\end{equation}
	Recall that energy conservation~\eqref{Schw::energy_conservation} implies that the bound $\frac{L^2}{E^2} \leq \frac{r^2}{\OsqS}$ holds for any point on the mass-shell. The definition of $\trapschw$ in equation~\eqref{def_eps} directly implies  $\frac{1}{27m^2} \frac{L^2}{E^2} = 1 - \trapschw$. Therefore for any $2m \leq r \leq 3m$ and $\trapschw < 0$ we find
	\begin{equation}
		\frac{1}{\OsqS} \geq \frac{1}{r^2} \frac{L^2}{E^2} \geq \frac{1}{9m^2} \frac{L^2}{E^2} \geq 3 \left( 1 + \left| \trapschw \right| \right) \geq 1 + \left| \trapschw \right|.
	\end{equation}
	A direct consequence of this inequality is the bound
	\begin{equation}
		\frac{1}{\OsqS(r_{\text{min}}^-(\trapschw))} \geq 1 + \left| \trapschw \right|,
	\end{equation}
	where we note that in fact $\OsqS(r_{\text{min}}^-(\trapschw)) \rightarrow \frac{1}{3}$ as $\trapschw \rightarrow 0$ and $\OsqS(r_{\text{min}}^-(\trapschw)) \rightarrow 0$ as $\trapschw \rightarrow - \infty$ so that this bound is in fact asymptotically sharp. Note also that from~\eqref{epsbound} it readily follows that for $2m \leq r \leq 3m$ we have $(1 + \left| \trapschw \right|) \OsqS(r) \leq \frac{2}{3}$. We find
	\begin{equation}
		\frac{1}{m} \left( \tau(\gamma(s)) - \tauzero \right) \lesssim 1 + \left| \log \left( (1 + \left| \trapschw \right|) \OsqS(r_0) \right) \right| + \left( \log \left| \trapschw \right| \right)_-.
	\end{equation}
	Note finally that we may multiply the second summand with $\mathfrak{s} = \chi_{(0,\infty)}\left( p^r(0) \right)$ since we have shown that we must have $p^r(0) > 0$ in this case.
	
	\paragraph{Region far from the horizon.}
	Let us now treat the case that $(2+\delta)m \leq r(s) < \infty$. The discussion is broken into two further parts: the case where $(2+\delta)m \leq r(s) \leq 3m$ and the case where $r(s) > 3m$.
	
	When $(2+\delta)m \leq r(s) \leq 3m$ the structure of the argument is almost identical to the case where $2m 
	\leq r(s) \leq (2+\delta)m$. Indeed, the case that $p^r(s) \leq 0$ is treated identically to above. If $p^r(s) > 0$ the argument needs to be modified slightly. It follows as above that $p^r(s') > 0$ for all $s' \in [s_0,s]$. Therefore the geodesic is steadily outgoing from its initial radius $r_0$ to $r(s)$. Let us distinguish the two cases that $r_0 > (2+\delta)m$ and $r_0 \leq (2+\delta)m$. In the first case Lemma~\ref{tstar_lem_schw} implies
	\begin{equation}
		\frac{1}{m} \left( \tau(\gamma(s)) - \tauzero \right) \lesssim 1 + \left( \log \left| \trapschw \right| \right)_-.
	\end{equation}
	In the second case the radius of the geodesic first increases from $r_0$ to $(2+\delta)m$ and then from $(2+\delta)m$ to $r(s)$, where we now note that $r(s)$ may be close to $3m$. Therefore Lemma~\ref{tstar_lem_schw} implies 
	\begin{equation}
		\begin{aligned}
			\frac{1}{m} \left( \tau(\gamma(s)) - \tauzero \right) &\lesssim 1 + \left| \log \left( \frac{\OsqS(r_0)}{\OsqS((2+\delta)m)} \right) \right| + \left( \log \left| \trapschw \right| \right)_- \\
			&\leq 1 + \left| \log \left( \frac{\OsqS(r_0)}{\OsqS(r(s))} \right) \right| + \left( \log \left| \trapschw \right| \right)_-,
		\end{aligned}
	\end{equation}
	where we used that $\OsqS(r(s)) \geq \OsqS((2+\delta)m)$. Let us now see how we can simplify the second summand further. We begin by noting that since $r(s) \geq (2+\delta)m$ certainly $\frac{\delta}{3} \leq \OsqS(r(s)) \leq 1$. Inequality~\eqref{epsbound} allows us to conclude that $-\frac{4}{\delta} \leq \trapschw \leq 1$. Therefore
	\begin{equation}
		\frac{\OsqS(r_0)}{\OsqS(r(s))} \sim_\delta \OsqS(r_0) \sim_\delta \left( 1 + \left| \trapschw \right| \right) \OsqS(r_0).
	\end{equation}
	As above we note that~\eqref{epsbound} readily implies that for $2m \leq r \leq 3m$ we have $(1 + \left| \trapschw \right|) \OsqS(r) \leq \frac{2}{3}$. We have therefore simplified our estimate to
	\begin{equation}
		\frac{1}{m} \left( \tau(\gamma(s)) - \tauzero \right) \lesssim 1 + \left| \log \left( (1 + \left| \trapschw \right|) \OsqS(r_0) \right) \right| + \left( \log \left| \trapschw \right| \right)_-.
	\end{equation}
	As above we note that since we assume $p^r > 0$ in this case we may multiply the second summand with the factor $\mathfrak{s}$. 
	
	Let us therefore assume that $3m < r(s) < \infty$. Depending on the values of $\trapschw, r_0$ and $p^r(s_0)$, the geodesic may cross any region of the exterior before arriving at the radius $r(s)$. Let us distinguish the two cases that $2m \leq r_0 \leq 3m$ and $r_0 > 3m$. In the first case, it must be the case that $p^r(s_0) > 0$ and $\trapschw > 0$, since the geodesic must cross the photon sphere eventually. It follows that $p^r(s') > 0$ for all $s' \in [s_0,s_1]$. Let us denote by $s_0 \leq \tilde{s} < s$ the parameter time such that $r(\tilde{s}) = 3m$. We have shown above that
	\begin{equation}
		\frac{1}{m} \left( \tau(\gamma(\tilde{s})) - \tau(\gamma(s_0)) \right) = \frac{\tau(\gamma(\tilde{s}))}{m} \lesssim 1 + \left| \log \left( (1 + \left| \trapschw \right|) \OsqS(r_0) \right) \right| + \left( \log \left| \trapschw \right| \right)_-.
	\end{equation}
	It only remains to consider the period of parameter time $[\tilde{s},s]$. We again distinguish between the two further cases that $r(s) \leq R$ and $r(s) > R$. If $r(s) \leq R$ we invoke Lemma~\ref{tstar_lem_schw} with $\delta_3 = \frac{R}{m} -3$ so that $(3+\delta_3)m=R$. We obtain the bound
	\begin{equation}
		\frac{1}{m} \left( \tau(\gamma(s)) - \tau(\gamma(\tilde{s})) \right) \lesssim 1 + \left( \log \left| \trapschw \right| \right)_-.
	\end{equation}
	In case that $r(s) > R$ the geodesic will require additional time to cross the region between $R$ and $r(s)$, which may be bounded as in Lemma~\ref{tauestimate_far}. We obtain
	\begin{equation}
		\frac{1}{m} \left( \tau(\gamma(s)) - \tau(\gamma(\tilde{s})) \right) \lesssim 1 + \left( \log \left| \trapschw \right| \right)_- + \frac{\rsuppconst^2}{m^2}.
	\end{equation}
	Adding up the bounds for the two periods we considered, we find
	\begin{equation}
		\frac{\tau(\gamma(s))}{m} \lesssim 1 + \frac{\rsuppconst^2}{m^2} + \left| \log \left( (1 + \left| \trapschw \right|) \OsqS(r_0) \right) \right| + \left( \log \left| \trapschw \right| \right)_-.
	\end{equation}
	Finally we note that if we assume $\tauzero \gtrsim \frac{1}{m} \rsuppconst^2$ with a sufficiently large constant, we may absorb the term $\frac{1}{m^2} \rsuppconst^2$ to the left hand side and obtain
	\begin{equation}
		\frac{1}{m} \left( \tau(\gamma(s)) - \tauzero \right) \lesssim 1 + \left| \log \left( (1 + \left| \trapschw \right|) \OsqS(r_0) \right) \right| + \left( \log \left| \trapschw \right| \right)_-.
	\end{equation}
	Let us now consider the remaining case that $r_0 > 3m$. We can now distinguish the two cases that $p^r(s_0) \geq 0$ or $p^r(s_0) < 0$. In the first case it follows that $p^r(s') \geq 0$ for all $s' \in [s_0, s_1]$ and therefore we may easily conclude
	\begin{equation}
		\frac{1}{m} \left( \tau(\gamma(s)) - \tauzero \right) \lesssim 1 + \left( \log \left| \trapschw \right| \right)_-,
	\end{equation}
	from Lemma~\ref{tstar_lem_schw} with $\delta_3$ as above, Lemma~\ref{tauestimate_far} and where we assume $\tauzero$ as above to absorb the term proportional to $\frac{1}{m^2} \rsuppconst^2$. In the case that $p^r(s_0) < 0$ let us distinguish between the two cases that $\trapschw \geq 0$ and $\trapschw < 0$. If $\trapschw \geq 0$, we must have $p^r(s') < 0$ for all $s' \in [s_0,s]$ and therefore $r(s') \leq r_0$. If $R \leq r(s) \leq r_0$ we note that we have chosen $\tauzero$ large enough so that
	\begin{equation}
		\frac{1}{m} \left( \tau(\gamma(s)) - \tauzero \right) \lesssim 1.
	\end{equation}
	If $3m < r(s) \leq R$, the geodesic requires time at most $\tauzero$ to reach the radius $R$, and then we may apply Lemma~\ref{tstar_lem_schw} with $\delta_3$ as above to find
	\begin{equation}
		\frac{1}{m} \left( \tau(\gamma(s)) - \tauzero \right) \lesssim 1 + \left( \log \left| \trapschw \right| \right)_-.
	\end{equation}
	If $\trapschw < 0$, the geodesic will initially be ingoing until it reaches the radius $r^+_{\text{min}}(\trapschw)$ at parameter time $\tilde{s}$, after which time it will scatter off the photon sphere to infinity and cross the region between $r^+_{\text{min}}(\trapschw)$ and $r(s)$. We argue in a similar fashion to above to conclude that
	\begin{align}
		\frac{1}{m} \left( \tau(\gamma(\tilde{s})) - \tau(\gamma(s_0)) \right) &\lesssim 1 + \tauzero + \left( \log \left| \trapschw \right| \right)_- ,\\
		\frac{1}{m} \left( \tau(\gamma(s)) - \tau(\gamma(\tilde{s})) \right) &\lesssim 1 + \left( \log \left| \trapschw \right| \right)_- + \frac{\rsuppconst^2}{m^2}.
	\end{align}
	Again by assuming $\tauzero$ to be large enough as above and adding the two estimates, we conclude
	\begin{equation}
		\frac{1}{m} \left( \tau(\gamma(s)) - \tauzero \right) \lesssim 1 + \left( \log \left| \trapschw \right| \right)_-.
	\end{equation}
	Finally we note that we may absorb any constant terms on the right hand side of the estimates to the left hand side by choosing $\tauzero$ suitably large. This concludes the proof of Lemma~\ref{taubound}.
\end{proof}

\subsection{Estimating the momentum support} \label{section_psupport}
In this section, we prove Proposition~\ref{psupport_prop}. The proof makes fundamental use of the almost-trapping estimate established in Section~\ref{section_tstar}. Before we give the proof of Proposition~\ref{psupport_prop}, we show that the momentum support of a solution with compactly supported initial data remains compact for all times and we also establish a useful bound concerning the energy and trapping parameter for points in the support of a solution with compactly supported initial data.

\subsubsection{Compactness of momentum support} \label{sec_compmomsupport}
In this subsection, we first establish a useful bound concerning the energy and trapping parameter on the mass-shell $\mathcal{P}$ in Lemma~\ref{E_eps_bound_schw}. Then we show that the momentum support of a solution to the massless Vlasov equation with compactly supported initial data remains compact for all times in Lemma~\ref{psupport_bounded}.

\begin{lem} \label{E_eps_bound_schw}
	Let $f$ be the solution to the massless Vlasov equation with initial data $f_0 \in L^\infty(\mathcal{P}_0)$ that satisfy Assumption~\ref{assumption_support}. Then there exists a dimensionless constant $C>0$ such that for all $(x,p) = (t^*,r,\omega, p^{t^*},p^r,\pslash) \in \supp(f) \subset \mathcal{P}$ with $\tau(x) \geq 0$ we have $E \left( 1 + \left| \trapschw \right| \right) \leq C \frac{\rsuppconst^2}{m^2} \psuppconst$.
\end{lem}
\begin{proof}
	We begin by noting that since both $E$ and $\trapern$ are conserved quantities, it suffices to consider a point $(x,p) = (t^*,r,\omega,p^{t^*},p^r,\pslash) \in \supp(f_0)$. Inequality~\eqref{epsbound} implies that
	\begin{equation}
		\left( 1 + \left| \trapern \right| \right) \lesssim \frac{r^2}{m^2} \frac{1}{\Osqern(r)} \leq \frac{\rsuppconst^2}{m^2} \frac{1}{\Osqern(r)}.
	\end{equation}
	We distinguish the two cases that $p^r > 0$ and $p^r \leq 0$. In case $p^r > 0$ we find from the explicit coordinate expression of $E$ that $E \leq \Osqern(r) \psuppconst$ so that we conclude the claimed bound. If $p^r \leq 0$ we use Lemma~\ref{express_ptstar} to find
	\begin{equation}
		\frac{L^2}{E} \leq 2 r^2 p^{t^*}, \quad E \leq \psuppconst,
	\end{equation}
	which allows us to conclude
	\begin{equation}
		E \left( 1 + \left| \trapern \right| \right) \lesssim E + \frac{1}{m^2} \frac{L^2}{E} \lesssim E + \frac{r^2}{m^2} p^{t^*}.
	\end{equation}
	By making use of Assumption~\ref{assumption_support} again, we conclude the bound as claimed.
\end{proof}

\begin{lem} \label{psupport_bounded}
	Let $f$ be the solution to the massless Vlasov equation with initial data $f_0 \in L^\infty(\mathcal{P}_0)$ that satisfy Assumption~\ref{assumption_support}. Then there exists a dimensionless constant $C>0$ such that for all $(x,p) = (t^*,r,\omega, p^{t^*},p^r,\pslash) \in \supp(f) \subset \mathcal{P}$ with $\tau(x) \geq 0$, we have
	\begin{equation}
		p^{t^*} \leq C \psuppconst, \quad \left| p^r \right| \leq \psuppconst, \quad \left| \pslash \right|_{\gslash} \leq \frac{\rsuppconst}{r} \psuppconst.
	\end{equation}
\end{lem}
\begin{proof}
	Consider a future-directed maximally defined null geodesic $\gamma: [0,s_1) \rightarrow \Mschw$ with affine parameter $s$. Let us express $\gamma(s) = x(s) = (t^*(s),r(s),\omega(s))$ and $\dot{\gamma}(s) = p(s) = (p^{t^*}(s),p^r(s),\pslash(s))$ in $(t^*,r)$-coordinates and assume that $(x(0),p(0)) \in \supp(f_0)$. We begin by noting that Assumption~\ref{assumption_support} on the initial support allows us to conclude that
	\begin{equation}
		E = \OsqS(r(0)) p^{t^*}(0) - \frac{2m}{r(0)} p^r(0) \leq \left( \OsqS(r(0)) + \frac{2m}{r(0)} \right) \psuppconst = \psuppconst
	\end{equation}
	and likewise $L^2 = r(0)^2 \left| \pslash(0) \right|_{\gslash}^2 \leq \rsuppconst^2 \psuppconst^2$. From conservation of energy~\eqref{Schw::energy_conservation} and the definition of $L^2$ in equation~\eqref{def_energy} it immediately follows that for all $s \geq 0$ the following bounds are satisfied
	\begin{equation}
		\left| p^r(s) \right| \leq E \leq \psuppconst, \quad \left| \pslash(s) \right|_{\gslash} = \frac{L}{r(s)} \leq \frac{\rsuppconst}{r(s)} \psuppconst.
	\end{equation}
	In order to bound the $p^{t^*}$-component let us choose $0 < \delta < 1$ and distinguish between the regions $2m \leq r \leq (2+\delta)m$ and $r \geq (2+\delta)m$. For any affine parameter time $s$ such that $r(s) \geq (2+\delta)m$ then Lemma~\ref{express_ptstar} implies
	\begin{equation}
		p^{t^*}(s) \lesssim_\delta E \leq \psuppconst.
	\end{equation}
	Let us therefore consider the case that $2m \leq r(s) \leq (2+ \delta)m$. We distinguish the two cases that $p^r(s) > 0$ and $p^r(s) \leq 0$. If we assume that $p^r(s) > 0$, then necessarily $p^r(s') > 0$ for all $0 \leq s' \leq s$. In this case we use Lemma~\ref{express_ptstar} to represent
	\begin{equation}
		p^{t^*}(s) = \frac{E + \frac{2m}{r(s)} p^r(s)}{\OsqS(r(s))}.
	\end{equation}
	The geodesic equations readily imply that $\partial_s p^r(s') < 0$ for $s' \in [0, s)$ and since $p^r(s') > 0$ we also conclude $\partial_s \OsqS \geq 0$ for $s' \in [0, s)$. Therefore we conclude that $\partial_s p^{t^*}(s') \leq 0$. We deduce that
	\begin{equation}
		p^{t^*}(s) \leq p^{t^*}(0) \leq \psuppconst.
	\end{equation}
	Let us now consider the case that $p^r(s) \leq 0$. In this case we first apply Lemma~\ref{ptstar_bound} to find
	\begin{equation}
		p^{t^*}(s) \lesssim \left( 1 + \left| \trapschw \right| \right) E.
	\end{equation}
	Let us define
	\begin{equation}
		\trapschw_\delta = -\frac{1}{\OsqS} \left( 1-\frac{3m}{r} \right)^2 \left( 1+\frac{6m}{r} \right) \frac{r^2}{27m^2} \Bigg|_{r = (2+\delta)m}.
	\end{equation}
	If we assume that $\left| \trapschw \right| \leq 1 + \left| \trapschw_\delta \right|$ then we immediately conclude that
	\begin{equation}
		p^{t^*}(s) \lesssim \left( 1 + \left| \trapschw \right| \right) E \lesssim_\delta \psuppconst.
	\end{equation}
	Let us therefore consider the remaining case that $p^r(s) \leq 0$ and $\trapschw < \trapschw_\delta < 0$. In this case inequality~\eqref{epsbound} implies that it must be true that $2m \leq r(s') \leq (2+\delta)m$ for $s' \geq 0$. We now distinguish further between the two cases that $p^r(0) > 0$ and $p^r(0) \leq 0$. If we first assume that $p^r(0) \leq 0$, it follows that $p^r(s') \leq 0$ for $s' \in [0,s]$. We may therefore use equality~\eqref{express_ptstar_neg} from Lemma~\ref{express_ptstar} to express $p^{t^*}$ at parameter time $s$ and at parameter time $0$ to find
	\begin{align}
		p^{t^*}(0) &= \left|  p^r(0) \right| + \frac{\left| \pslash(0) \right|_{\gslash}^2}{E + \left| p^r(0) \right|} \geq \frac{1}{2 r(0)^2} \frac{L^2}{E}, \\
		p^{t^*}(s) &= \left| p^r(s) \right| + \frac{\left| \pslash(s) \right|_{\gslash}^2}{E + \left| p^r(s) \right|} \leq E + \frac{1}{r(s)^2} \frac{L^2}{E},
	\end{align}
	so that we may conclude
	\begin{equation}
		p^{t^*}(s) \leq E + 2 \frac{r(0)^2}{r(s)^2} p^{t^*}(0) \lesssim \psuppconst.
	\end{equation}
	Finally let us turn to the remaining case that $p^r(0) > 0$. Since $p^r(s) \leq 0$ we conclude the existence of $s^* \in (0,s]$ such that $p^r(s^*) = 0$, $p^r(s') > 0$ for $s' \in [0,s^*)$ and $p^r(s') \leq 0$ for $s' \in [s^*,s]$. At affine parameter time $s^*$ we find furthermore that $r(s^*) = r_{\text{min}}^-(\trapschw)$, which allows us to conclude $\OsqS(r(s^*))  \left( 1 + \left| \trapschw \right| \right) \sim 1$. We may therefore conclude
	\begin{equation}
		p^{t^*}(s) \lesssim E \left( 1 + \left| \trapschw \right| \right) \sim \frac{E}{\OsqS(r(s^*))} = p^{t^*}(s^*).
	\end{equation}
	Now we note that for $s' \in [0,s^*)$ we may argue as above to conclude that
	\begin{equation}
		p^{t^*}(s^*) \leq p^{t^*}(0) \leq \psuppconst,
	\end{equation}
	which finally allows us to conclude
	\begin{equation}
		p^{t^*}(s) \lesssim \psuppconst,
	\end{equation}
	as claimed.
\end{proof}

\subsubsection{Proof of Proposition~\ref{psupport_prop}}

We can now finally provide the proof of Proposition~\ref{psupport_prop}. We remind the reader that the definitions of the sets $\trappedsupportset, \smallsupportset \subset \mathcal{P}$ involve a choice of constants $\bigc, \decayrate$ and $\tauzero$, see Section~\ref{sec_subsets}. The proof makes use of the almost-trapping estimate established in Lemma~\ref{taubound} together with the assumption of initially compact support in order to derive bounds on the momenta of a given null geodesic. 

%\begin{prop}
%Assume that $f_0 \in L^\infty(\mathcal{P}_0)$ satisfies Assumption~\ref{assumption_support} and let $f$ be the unique solution to the massless Vlasov equation with $f|_{\Sigma_0} = f_0$. Then there exist dimensionless constants $\bigc > 0, \decayrate > 0, C_0 > 0$ so that if $\tauzero = C_0 \frac{\rsuppconst^2}{m}$, then all $(x,p) = (t^*,r,\omega,p^{t^*},p^r,\pslash) \in \supp(f)$ such that $\tau(x) \geq \tauzero$ satisfy
%\begin{equation} \label{eqn_prop41_refined}
%	(x,p) \in \trappedsupportset \cup \smallsupportset.
%\end{equation}
%Furthermore we can refine this estimate as follows: Let $0 < \delta < 1$. Then we can choose $\decayrate = \decayrate(\delta) > 0$ such that inclusion~\eqref{eqn_prop41_refined} still holds for all $(x,p) \in \supp(f)$ and in addition if $2m \leq r \leq (2+\delta)m$ and $p^r > 0$, then we in fact know that $(x,p) \in \smallsupportset$.
%\end{prop}

\begin{proof}[Proof of Proposition~\ref{psupport_prop}]
	Consider an affinely parametrised  future-directed null geodesic segment $\gamma: [0,s] \rightarrow \Mschw$ expressed in $(t^*,r)$-coordinates. Assume that $(x(0),p(0)) \in \supp(f_0)$. Let us write $(x(s),p(s)) = (x,p) = (t^*,r,\omega,p^{t^*},p^r,\pslash)$ and $\tau(x(s)) = \tau$, where we suppress the dependence on the affine parameter $s$ for the sake of simplicity. Therefore the geodesic $\gamma$ meets the hypersurface $\Sigma_\tau$ at the point $x \in \Mschw$ and populates the $p$-support of $f$ at this point. We will repeatedly apply Lemma~\ref{taubound} in the course of the proof. For simplicity of notation we will omit the dependence of the bounds on $\delta > 0$ in this proof, or alternatively we may make an explicit choice for $\delta$, i.e. $\delta= \frac{1}{2}$. We will distinguish between the two cases that $p^r > 0$ and $p^r \leq 0$ at time $\tau$. Let us consider first
	
	\paragraph{Case 1: $p^r \leq 0$ at time $\tau$.}
	In this case we may apply Lemma~\ref{taubound} to conclude the bound
	\begin{equation} \label{tauestimate_in_proof}
		\frac{1}{m} \left( \tau - \tauzero \right) \lesssim \left( \log \left| \trapschw \right| \right)_- + \mathfrak{s} \left( \log \, (1+\left| \trapschw \right|) \OsqS(r(0)) \right)_-,
	\end{equation}
	which holds regardless of the value of $2m \leq r < \infty$ and where we define $\mathfrak{s} = \chi_{(0,\infty)}\left( p^r(0) \right)$ as in the lemma to ensure that the second term only appears when $p^r(0) > 0$ and $2m < r(0) < 3m$. Therefore at least one of the two terms must be larger than a multiple of $\frac{1}{m} \left( \tau - \tauzero \right)$ and we will further distinguish two cases based on which term satisfies the lower bound.
	
	\paragraph{Case 1.1: $\gamma$ is almost trapped.}
	In this case we may assume that the first term is large so that $\left| \trapschw \right| < 1$ and
	\begin{equation}
		\frac{1}{m} \left( \tau - \tauzero \right) \lesssim \left| \log \left| \trapschw \right| \right|.
	\end{equation}
    This directly implies
    \begin{equation} \label{ineq_trap_first}
        \left| \trapschw \right| \leq e^{- \frac{\decayrate}{m} (\tau-\tauzero)},
    \end{equation}
	for an appropriate constant $\decayrate$ arising from the constant implicit in inequality~\eqref{tauestimate_in_proof}. We briefly remark on the interpretation of~\eqref{ineq_trap_first} here. Assume that inequality~\eqref{ineq_trap_first} holds. Then, since we also assume that $\gamma$ passes through a point $x \in \Mschw$ with $\tau = \tau(x) \gg 1$, $\gamma$ must be very nearly trapped at the photon sphere (otherwise, it would have scattered off the photon sphere at an earlier time). Therefore we are concluding a bound on a conserved quantity along the geodesic (the trapping parameter) from observing that the geodesic intersects a certain point $x$ with $\tau(x) \gg 1$. Continuing with our argument, from the definition~\eqref{def_eps} of $\trapschw$ we immediately find the upper bound
	\begin{equation} \label{ELquotient_bound}
		\left| 1 - \frac{1}{\sqrt{27}m} \frac{L}{E} \right| \leq \left| \trapschw \right| \leq e^{- \frac{\decayrate}{m} (\tau-\tauzero)}, 
	\end{equation}
	as well as the following two upper bounds
	\begin{align}
		\left| E - \frac{L}{\sqrt{27}m} \right| &\leq E \left| \trapschw \right| \leq \psuppconst  \left| \trapschw \right| \leq  \psuppconst e^{- \frac{\decayrate}{m} (\tau-\tauzero)} \label{ELquotient_bound2}, \\
		\left| E^2 - \frac{L^2}{27 m^2} \right| &= E^2 \left| \trapschw \right| \leq \psuppconst^2 \left| \trapschw \right| \leq \psuppconst^2 e^{- \frac{\decayrate}{m} (\tau-\tauzero)} \label{ELquotient_bound3},
	\end{align}
	where we used the bound $E \leq \psuppconst$, which we claim follows from our assumptions on the initial support of $f$ as given in Assumption~\ref{assumption_support}. To see this, recall the definition of $E$ in $(t^*,r)$-coordinates as $E = \OsqS p^{t^*} - \frac{2m}{r} p^r$. We conclude $E \leq \psuppconst$ regardless of the sign of $p^r$. Similarly it follows that $L \leq 2 \rsuppconst^2 \psuppconst$. Since we have in fact assumed $\tau \geq m + \tauzero$ so that $e^{- \frac{\decayrate}{m} (\tau-\tauzero)} \leq e^{- \decayrate} < 1$ and we may conclude from~\eqref{ELquotient_bound} that $\frac{E}{L} \lesssim \frac{1}{m}$. 
	
	We next express the $p^r$-component in terms of the energy $E$ and angular momentum $L$. Energy conservation~\eqref{Schw::energy_conservation} expressed in $(t^*,r)$-coordinates immediately implies that
	\begin{equation}
		(p^r)^2 = E^2 \trapschw + \left( \frac{1}{27m^2} - \frac{\OsqS}{r^2} \right) L^2.
	\end{equation}
	By dividing this relation by $\frac{L^2}{m^2}$ and noting that $\frac{E}{L} \lesssim \frac{1}{m}$, it follows that
	\begin{equation}
		\left| m^2 \frac{\left| p^r \right|^2}{L^2} - \left(\frac{1}{27} - \frac{m^2}{r^2} \OsqS \right) \right| \lesssim m^2 \frac{E^2}{L^2} e^{-\frac{\decayrate}{m}(\tau-\tauzero)} \lesssim e^{-\frac{\decayrate}{m}(\tau-\tauzero)}.
	\end{equation}
	Using the fact that $\left| a^2-b^2 \right| \leq \eta$ implies $\left| a-b \right| \leq \sqrt{\eta}$ for any $a,b \in \R_{>0}$, we find
	\begin{equation}
		\left| m \frac{\left| p^r \right|}{L} - \sqrt{\frac{1}{27} - \frac{m^2}{r^2} \OsqS} \right| \lesssim m \frac{E}{L} e^{-\frac{\decayrate}{2m}(\tau-\tauzero)} \lesssim e^{-\frac{\decayrate}{2m}(\tau-\tauzero)}.
	\end{equation}
	It is easy to see that (given a decay rate for $\trapschw$) this is the optimal decay rate at the photon sphere since there the relation $(p^r)^2 = \trapschw E^2$ or equivalently $\left| p^r \right| = \sqrt{\left| \trapschw \right|} E$ holds. Therefore at the photon sphere, $p^r$ must decay at half the rate at which $\trapschw$ decays. However in any region where $E \sim \left| p^r \right|$ holds we may conclude the better bound
	\begin{equation}
		\left| m \frac{\left| p^r \right|}{L} - \sqrt{\frac{1}{27} - \frac{m^2}{r^2} \OsqS} \right| \lesssim e^{-\frac{\decayrate}{m}(\tau-\tauzero)},
	\end{equation}
	by making use of the fact that for any $a,b \in \R_{>0}$ such that $\left| a^2 - b^2 \right| \leq a^2 \eta$ we have $\left| a - b \right| \leq a \eta$. To this end let us note that energy conservation may be reformulated as 
	\begin{equation} \label{Elikepr}
		\left[ \left( 1 - \frac{3m}{r} \right)^2 \left( 1 + \frac{6m}{r} \right) + 27 m^2 \frac{\OsqS} {r^2} \trapschw \right] E^2 = (p^r)^2.
	\end{equation}
	Therefore we immediately see that when $r=2m$ and in the limit as $r \rightarrow \infty$ we have $E \sim \left| p^r \right|$ and therefore the better decay rate. In fact, let $0 < \delta' < 1$ be fixed and consider the region where $r \notin [(3-\delta')m,(3+\delta')m]$. Then clearly for $\tau \gtrsim \tauzero - \ln \delta'$ we have $E^2 \sim (p^r)^2$ from~\eqref{Elikepr} and therefore again the better decay rate. Recall also that we have shown above that $\frac{1}{2}E \leq p^r \leq E$ in the region $\mathcal{A}_0$. 
	
	To obtain a bound on the $p^{t^*}$-component, we apply Lemma~\ref{express_ptstar} and express $p^{t^*}$ in terms of the remaining momentum components using the mass-shell relation. Since $p^r \leq 0$, we have
	\begin{equation}
		p^{t^*} = \frac{-\frac{2m}{r} \frac{\left| p^r \right|}{L} + \sqrt{\frac{\left| p^r \right|^2}{L^2} + \frac{\OsqS}{r^2}}}{\OsqS} L.
	\end{equation}
	We note that by an application of L'Hôpital's rule the quotient in the expression has a regular limit as $r \rightarrow 2m$ for any values of $\left| p^r \right|$ and $L$. Therefore this expression is regular for all $2m \leq r < \infty$. However, in order to show the required bounds, we find it convenient to consider the region close to and far from the event horizon separately. 
	
	Let $0 < \delta < 1$ as in the statement of the proposition and assume first that $r \geq (2+ \delta)m$. Here $\OsqS \sim_\delta 1$ and using the fact that $\left| \sqrt{a^2+x} - \sqrt{b^2+x} \right| \leq \left| a-b \right|$ for all $x \geq 0$ and $a,b \in \R$ we immediately deduce the bound
	\begin{equation}
		\left| m \frac{ p^{t^*}}{L} - \left( \frac{- \frac{2m}{r} \sqrt{\frac{1}{27} - \frac{m^2}{r^2} \OsqS} + \sqrt{\frac{1}{27}}}{\OsqS} \right) \right| \lesssim \left| m \frac{\left| p^r \right|}{L} - \sqrt{\frac{1}{27} - \frac{m^2}{r^2} \OsqS} \right|.
	\end{equation}
	We conclude that in the region $r \geq (2+ \delta)m$, $p^{t^*}$ decays to its limit at the same rate as $p^r$.
	
	Let us now assume that $2m \leq r \leq (2+ \delta)m$. We will prove the following claim in slightly greater generality than necessary here, so that it also applies to the extremal case in the proof of the corresponding Proposition~\ref{psupport_propERN}:
	
	\begin{claim} \label{lemma::F}
		Let $x_0 \geq \frac{1}{\sqrt{2}}$ and assume that $\Psi: [x_0,\infty) \rightarrow \R$ is a smooth function satisfying the conditions $\Psi(x_0) = 0$, $\Psi(x) > 0$ for all $x > x_0$ and $\Psi'(x) > 0$ for all $x \geq x_0$. Let $0 < c < 1$ and consider the function $F: [c,\infty) \times [x_0,\infty) \rightarrow \R$ defined by
		\begin{equation}
			F(a,x) = \frac{-\left( 1- \Psi \right) a + \sqrt{a^2 + \frac{1}{x^2} \Psi}}{\Psi}.
		\end{equation}
		Then $F$ is continuous and Lipschitz in the first parameter, more precisely for $a \geq b \geq c$ we have
		\begin{equation}
			- \frac{1}{c^2} (a-b) \leq F(a,r) - F(b,r) \leq a-b.
		\end{equation}
	\end{claim}
	\begin{subproof}
		We begin by noting it suffices to show continuity of $F$ when $x = x_0$, since $\Psi(x) > 0$ for $x > x_0$. An application of L'Hôpital's rule together with the assumption $\Psi'(x_0) > 0$ yields that $F$ is continuous up to and including $x=x_0$. Let $a \geq b \geq c > 0$, then the inequality $F(a,r) - F(b,r) \leq a-b$ is equivalent to
		\begin{equation}
			\sqrt{a^2 + \frac{1}{x^2} \Psi} - \sqrt{b^2 + \frac{1}{x^2} \Psi} \leq a-b.
		\end{equation}
		This follows immediately from the inequality $\sqrt{a^2 + y} - \sqrt{b^2 + y} \leq (a-b)$ which holds for all $a \geq b \geq 0$ and $y \geq 0$. The lower bound $F(a,r) - F(b,r) \geq - \frac{1}{c^2}(a-b)$ is equivalent to
		\begin{equation}
			\sqrt{a^2 + \frac{1}{x^2} \Psi} - \sqrt{b^2 + \frac{1}{x^2} \Psi} \geq \left[ 1 - \Psi \left( 1 + \frac{1}{c^2} \right) \right] (a-b).
		\end{equation}
		Evaluated at $x=x_0$ both sides agree. To see why the inequality holds for all $x \geq x_0$, note that both sides are decreasing in $x$. It suffices to show that the left hand side decreases at a slower rate than the right side. Taking derivatives on both sides we see it suffices to show
		\begin{equation}
			\left[ \frac{\sqrt{a^2 + \frac{1}{x^2} \Psi} - \sqrt{b^2 + \frac{1}{x^2} \Psi}}{\sqrt{ \left( a^2 + \frac{1}{x^2} \Psi \right) \left( b^2 + \frac{1}{x^2} \Psi \right)} } \right] \frac{1}{2} \frac{d}{dx} \left( \frac{1}{x^2} \Psi \right) \leq \left( 1+ \frac{1}{c^2} \right) (a-b) \frac{d}{dx} \Psi.
		\end{equation}
		Again using the inequality $\sqrt{a^2 + y} - \sqrt{b^2 + y} \leq (a-b)$ for all $a \geq b \geq 0$ and $y \geq 0$ we see that
		\begin{equation}
			\frac{\sqrt{a^2 + \frac{1}{x^2} \Psi} - \sqrt{b^2 + \frac{1}{x^2} \Psi}}{\sqrt{ \left( a^2 + \frac{1}{x^2} \Psi \right) \left( b^2 + \frac{1}{x^2} \Psi \right)} } \leq \frac{a-b}{b^2},
		\end{equation}
		so that it suffices to show
		\begin{equation}
			\frac{1}{2 b^2} \frac{d}{dx} \left( \frac{1}{x^2} \Psi \right) \leq \left( 1+ \frac{1}{c^2} \right) \frac{d}{dx} \Psi.
		\end{equation}
		Using that $b \geq c$ it is in fact enough to show
		\begin{equation}
			\frac{d}{dx} \left( \frac{1}{x^2} \Psi \right) \leq 2 \frac{d}{dx} \Psi.
		\end{equation}
		By simply expanding the left hand side we find
		\begin{equation}
			\frac{d}{dx} \left( \frac{1}{x^2} \Psi \right) = \frac{1}{x^2} \frac{d}{dx} \Psi - \frac{2}{x^3} \Psi \leq \frac{1}{x_0^2} \frac{d}{dx} \Psi \leq 2 \frac{d}{dx} \Psi,
		\end{equation}
		since we assumed that $x \geq x_0 \geq \frac{1}{\sqrt{2}}$. This concludes the proof of the claim.
	\end{subproof} 
	
	We now apply this claim with the substitution $x = \frac{r}{m}$ and choose $x_0 = 2$ and $\Psi = \OsqS$, noting that $\OsqS$ can indeed readily be expressed as a function of $x$. With these substitutions and choices, note that we then have
	\begin{equation}
		F \left( m \frac{\left| p^r \right|}{L}, r \right) \frac{L}{m} = p^{t^*}.
	\end{equation}
	Therefore we need only insert the limit value of $m \frac{\left| p^r \right|}{L}$ as $\tau \rightarrow \infty$ to find the limit of $p^{t^*}$. Let us remark that
	\begin{equation}
		F \left( \sqrt{\frac{1}{27} - \frac{m^2}{r^2} \OsqS} , r \right) = \frac{- \frac{2m}{r} \sqrt{\frac{1}{27} - \frac{m^2}{r^2} \OsqS} + \sqrt{\frac{1}{27}}}{\OsqS} \rightarrow \frac{35}{24 \sqrt{3}} ,
	\end{equation}
	as $r \rightarrow 2m$. Therefore we find
	\begin{equation}
		\left| m \frac{p^{t^*}}{L} - \left( \frac{- \frac{2m}{r} \sqrt{\frac{1}{27} - \frac{m^2}{r^2} \OsqS} + \sqrt{\frac{1}{27}}}{\OsqS} \right) \right| \leq \left| F \left( m \frac{\left| p^r \right|}{L}, r \right) - F \left( \sqrt{\frac{1}{27} - \frac{m^2}{r^2} \OsqS} , r \right) \right|.
	\end{equation}
	Note that since we assume $2m \leq r \leq (2+\delta)m$ we have that $\sqrt{\frac{1}{27} - \frac{m^2}{r^2} \OsqS} \geq \frac{1-\delta}{10} > 0$ and by assuming that $\tauzero$ is sufficiently large we may assume that also $m \frac{\left| p^r \right|}{L} \geq \frac{1-\delta}{20}$ for $\tau \geq \tauzero$ and $2m \leq r \leq (2+\delta)m$. Therefore we may apply Claim~\ref{lemma::F} above to see
	\begin{equation}
		\left| F \left( m \frac{\left| p^r \right|}{L}, r \right) - F \left( \sqrt{\frac{1}{27} - \frac{m^2}{r^2} \OsqS} , r \right) \right| \lesssim_\delta \left| m \frac{\left| p^r \right|}{L} - \sqrt{\frac{1}{27} - \frac{m^2}{r^2} \OsqS} \right|,
	\end{equation}
	so that we again see that the $p^{t^*}$ component decays to its limit value at the same rate as the $p^r$ component. In particular, the same remarks on the decay rate apply as above. Finally note that to bound the angular momentum component we need merely note that by Assumption~\ref{assumption_support} we have $L \leq \rsuppconst \psuppconst$ and therefore
	\begin{equation}
		\left| \pslash \right|_{\gslash} = \frac{L}{r} \leq \rsuppconst \psuppconst \frac{1}{r}.
	\end{equation}
	
	\paragraph{Case 1.2: $\gamma$ is initially outgoing and starts close to $\mathcal{H}^+$.}
	In this case we may assume that the second summand is large, so that in particular we may assume that $p^r(0) > 0$ and we have the bound
	\begin{equation} \label{case2_bound_prneg}
		\frac{1}{m} \left(\tau - \tauzero \right) \lesssim \left( \log \, (1+\left| \trapschw \right|) \OsqS(r(0)) \right)_-,
	\end{equation}
    which immediately implies the bound
    \begin{equation}
        \left( 1 + \left|\trapschw\right| \right) \OsqS(r(0)) \lesssim e^{- \frac{\decayrate}{m} (\tau-\tauzero)}
    \end{equation}
	for an appropriate constant $\decayrate$ arising from the constant implicit in inequality~\eqref{tauestimate_in_proof}. Note that we do not obtain a bound on the size of $\left| \trapschw \right|$ directly in this case. 
	
	Let us recall the definition of $E$ in $(t^*,r)$-coordinates as $E = \OsqS p^{t^*} - \frac{2m}{r} p^r$. Above we obtained bounds for all signs of $p^r(0)$. Combined with future-directedness which implies $E \geq 0$ we in fact find the better bounds $E \leq \OsqS(r(0)) \psuppconst$ and $\frac{2m}{r(0)} p^r(0) \leq \OsqS(r(0)) \psuppconst$. Using energy conservation and the bound on $E$ we also find $L \leq \rsuppconst \psuppconst \sqrt{\OsqS(0)}$ regardless of the size and sign of the trapping parameter $\trapschw$. 
	
	To bound the $p^r$ component, note that energy conservation immediately implies $\left| p^r \right| \leq E$. Therefore an application of inequality~\eqref{case2_bound_prneg} gives
	\begin{equation}
		\left( 1+ \left| \trapschw \right| \right) \left| p^r \right| \leq \left( 1+ \left| \trapschw \right| \right) E \leq \psuppconst \left( 1+ \left| \trapschw \right| \right) \OsqS(r(0)) \lesssim \psuppconst e^{-\frac{\decayrate}{m} (\tau-\tauzero)}.
	\end{equation}
	For the $p^{t^*}$ component we again need to distinguish between the region close to and far from the event horizon. Let $0 < \delta < 1$ be as in the statement of the proposition and let us consider the case $r \geq (2+\delta)m$ first. By Lemma~\ref{ptstar_bound} we know that $p^{t^*} \lesssim_\delta E$ so that by the same argument we find
	\begin{equation}
		p^{t^*} \leq \left( 1+ \left| \trapschw \right| \right) p^{t^*} \lesssim_\delta  \psuppconst e^{-\frac{\decayrate}{m} (\tau-\tauzero)}.
	\end{equation}
	In particular the decay rate in $\tau$ is the same as for the $p^r$ component. Let us now assume that $2m \leq r \leq (2+\delta)m$. We again apply Lemma~\ref{ptstar_bound} to deduce the bound
	\begin{equation}
		p^{t^*} \lesssim (1 + \left| \trapschw \right|) E \leq \psuppconst \left( 1 + \left| \trapschw \right| \right) \OsqS(r(0)) \lesssim \psuppconst e^{-\frac{\decayrate}{m} (\tau-\tauzero)},
	\end{equation}
	so that we obtain again the same decay rate as for the $p^r$ component. Finally we find for the angular momentum
	\begin{equation}
		\sqrt{1 + \left| \trapschw \right|} L \leq \rsuppconst \psuppconst \sqrt{\left( 1 + \left| \trapschw \right| \right) \OsqS(r(0))} \lesssim \rsuppconst \psuppconst e^{-\frac{\decayrate}{2m}(\tau - \tauzero)},
	\end{equation}
	which immediately gives the desired bound for $\pslash$ when applying the identity $\left| \pslash \right|_{\gslash} = \frac{L}{r}$. Therefore the angular component of the momentum decays exponentially, although its exponential decay rate is half the exponential decay rate of the $p^r$ and $p^{t^*}$ components.
	
	\paragraph{Case 2: $p^r > 0$ at time $\tau$.}
	In this case we will find it necessary to consider the region close to the event horizon and far from it separately. Let $0 < \delta < 1$ as in the statement of the proposition and let us assume first that $r \geq (2+\delta)m$. Here the situation is very similar to the case already considered above where $p^r \leq 0$. By Lemma~\ref{taubound} we have
	\begin{equation}
		\frac{1}{m} \left(\tau - \tauzero \right) \lesssim \left( \log \left| \trapschw \right| \right)_- + \mathfrak{s} \left( \log \, (1+\left| \trapschw \right|) \OsqS(r(0)) \right)_-.
	\end{equation}
	Again we consider two cases as above, depending on which of the two summands is large.
	
	\paragraph{Case 2.1: $\gamma$ is almost trapped and $r \geq (2+\delta)m$.} In this case we may assume that
	\begin{equation}
		\frac{1}{m} \left(\tau - \tauzero \right) \lesssim \left| \log \left| \trapschw \right| \right| ,
	\end{equation}
    which directly implies the bound
    \begin{equation}
        \left| \trapschw \right| \leq e^{- \frac{\decayrate}{m} (\tau-\tauzero)}.
    \end{equation}
	The argument in this case will proceed along parallel lines to that of Case~1.1. As above, we may conclude the bounds~\eqref{ELquotient_bound},~\eqref{ELquotient_bound2},~\eqref{ELquotient_bound3}. To obtain bounds for the $p^r$ component, we note that the analogous argument used in Case~1.1 did not depend on $\sgn(p^r)$. Therefore we readily conclude as above the bound
	\begin{equation}
		\left| m \frac{\left| p^r \right|}{L} - \sqrt{\frac{1}{27} - \frac{m^2}{r^2} \OsqS} \right|  \lesssim e^{-\frac{\decayrate}{2m}(\tau-\tauzero)},
	\end{equation}
	and note that the same remarks on the rate of decay apply as above. In order to obtain a bound for the $p^{t^*}$ component, let us express $p^{t^*}$ using the mass-shell relation in equation~\eqref{pt_expressed_via_pr}. Since we are in the case that $p^r > 0$ we find
	\begin{equation} \label{ptstar_massshell_case21}
		p^{t^*} = \frac{\frac{2m}{r} \frac{p^r}{L} + \sqrt{\frac{\left( p^r \right)^2}{L^2} + \frac{\OsqS}{r^2}}}{\OsqS} L.
	\end{equation}
	As above using the fact that $\left| \sqrt{a^2+x} - \sqrt{b^2+x} \right| \leq \left| a-b \right|$ for all $x \geq 0$ we immediately deduce the bound
	\begin{equation}
		\left| m \frac{ p^{t^*}}{L} - \left( \frac{ \frac{2m}{r} \sqrt{\frac{1}{27} - \frac{m^2}{r^2} \OsqS} + \sqrt{\frac{1}{27}}}{\OsqS} \right) \right| \lesssim \left| m \frac{\left| p^r \right|}{L} - \sqrt{\frac{1}{27} - \frac{m^2}{r^2} \OsqS} \right|.
	\end{equation}
	We again conclude that in the region $r \geq (2+ \delta)m$, $p^{t^*}$ decays to its limit at the same rate as the $p^r$ component. Carefully note that in the case where $p^r > 0$, the expression~\eqref{ptstar_massshell_case21} does not have a finite limit as $r \rightarrow 2m$. The bound on the angular component of the momentum is obtained in an identical manner to Case~1.1 above.
	
	\paragraph{Case 2.2: $\gamma$ is initially outgoing, starts close to $\mathcal{H}^+$ and $r \geq (2+\delta)m$.}
	In this case we may assume that $p^r(0) > 0$ and we have the bound
	\begin{equation}
		\frac{1}{m} \left(\tau - \tauzero \right) \lesssim \left( \log \, (1+\left| \trapschw \right|) \OsqS(r(0)) \right)_-,
	\end{equation}
    which allows us to conclude the inequality
    \begin{equation}
         \left( 1 + \left|\trapschw\right| \right) \OsqS(0) \lesssim e^{-\frac{\decayrate}{m} (\tau-\tauzero)}
    \end{equation}
	for an appropriate constant $\decayrate$ again arising from the constant implicit in inequality~\eqref{tauestimate_in_proof}. Identically to Case~1.2 above, from the fact that $p^r(0) > 0$ we obtain the bounds $E \leq \OsqS(r(0)) \psuppconst$ and $L \leq \rsuppconst \psuppconst \sqrt{\OsqS(r(0))}$. The bounds for the $p^r$, $p^{t^*}$ and angular momentum components are shown virtually identically to Case~1.2 above. 
	
	To bound the $p^r$ component, note that energy conservation implies $\left| p^r \right| \leq E$. Therefore
	\begin{equation}
		\left( 1+ \left| \trapschw \right| \right) \left| p^r \right| \leq \left( 1+ \left| \trapschw \right| \right) E \leq \psuppconst \left( 1+ \left| \trapschw \right| \right) \OsqS(r(0)) \lesssim \psuppconst e^{-\frac{\decayrate}{m} (\tau-\tauzero)}.
	\end{equation}
	For the $p^{t^*}$ component we apply Lemma~\ref{ptstar_bound} as above to deduce
	\begin{equation}
		p^{t^*} \lesssim_\delta E \lesssim \psuppconst e^{-\frac{\decayrate}{m} (\tau-\tauzero)}.
	\end{equation}
	For the angular momentum $L$ we argue identically to Case~1.2 above. 
	
	\paragraph{Case 2.3: $\gamma$ is initially outgoing, starts close to $\mathcal{H}^+$ and $2m \leq r \leq (2+\delta)m$.}
	In this case we may assume that $p^r(0) > 0$ and $2m \leq r(0) \leq r \leq (2+\delta)m$ and therefore in fact $p^r > 0$ for all times up to and including $\tau$. Lemma~\ref{taubound} implies the bound
	\begin{equation}
		\frac{1}{m} \left( \tau - \tauzero \right) \lesssim \left| \log \left( \frac{\OsqS(r(0))}{\OsqS(r)} \right) \right|,
	\end{equation}
    which in turn implies the bound
    \begin{equation}
        \frac{\OsqS(r(0))}{\OsqS(r)} \lesssim e^{-\frac{\decayrate}{m} (\tau-\tauzero)},
    \end{equation}
	where $\decayrate$ is an appropriately chosen constant. As in the preceding section we may conclude that the bounds $E \leq \OsqS(r(0)) \psuppconst$ and $L \leq \rsuppconst \psuppconst \sqrt{\OsqS(r(0))}$ hold from $p^r(0) > 0$ and future-directedness of the geodesic. 
	
	To bound the $p^r$ component, we again make use of conservation of energy~\eqref{Schw::energy_conservation} together with inequality~\eqref{epsbound} to find
	\begin{equation}
		\left( 1+ \left| \trapschw \right| \right) p^r \lesssim \frac{p^r}{\OsqS(r)} \leq \frac{E}{\OsqS(r)} \leq \psuppconst \frac{\OsqS(r(0))}{\OsqS(r)} \lesssim \psuppconst e^{-\frac{\decayrate}{m} (\tau-\tauzero)}.
	\end{equation}
	In order to obtain a bound for the $p^{t^*}$ component we make use of Lemma~\ref{ptstar_bound} again to find
	\begin{equation}
		p^{t^*} \lesssim \frac{E}{\OsqS(r)} \leq \psuppconst \frac{\OsqS(r(0))}{\OsqS(r)} \lesssim \psuppconst e^{-\frac{\decayrate}{m} (\tau-\tauzero)}.
	\end{equation}
	To bound the angular momentum we note as before
	\begin{equation}
		\sqrt{1 + \left| \trapschw \right|} L \lesssim \frac{L}{ \sqrt{\OsqS(r)}} \leq \rsuppconst \psuppconst \sqrt{\frac{\OsqS(0)}{\OsqS(r)}} \lesssim \rsuppconst \psuppconst e^{- \frac{\decayrate}{2m} (\tau-\tauzero)}.
	\end{equation}
	We obtain the desired bound by finally again making use of the relation $\left| \pslash \right|_{\gslash} = \frac{L}{r}$. This concludes the proof.
\end{proof}

\subsection{Exponential decay of moments} \label{section_integralestimate}
In this section we provide the proof of our main Theorem~\ref{maintheorem_precise}. The proof fundamentally makes use of Proposition~\ref{psupport_prop} to estimate moments of a solution $f$ to the massless Vlasov equation. Before establishing the proof of Theorem~\ref{maintheorem_precise}, we first show the simpler result that the volume of $\supp(f(x,\cdot)) \subset \mathcal{P}_x$ remains finite for all times for a solution $f$ with compactly supported initial data.

\subsubsection{Finiteness of volume of the momentum support}

\begin{lem} \label{lem_boundedness_moments_schw}
	Assume that $f$ solves the massless Vlasov equation on Schwarzschild and its initial distribution $f_0: \mathcal{P}_0 \rightarrow \R$ satisfies Assumption~\ref{assumption_support}. Then for all $x \in \Mschw$ with $\tau(x) \geq 0$:
	\begin{equation}
		\vol (\supp(f(x,\cdot))) = \int_{\mathcal{P}_x} \chi_{\supp(f(x,\cdot))} \, \dmux \lesssim \frac{\rsuppconst^2}{r^2} \psuppconst^2.
	\end{equation}
\end{lem}
\begin{proof}
    We start by making the change of variables $\pslash \mapsto (L, \pslashangle)$ introduced in Section~\ref{sec_angular_coords_massshell}. Recall that by virtue of Lemma~\ref{psupport_bounded} any $(x,p) \in \supp(f)$ must satisfy the bounds
	\begin{equation}
		p^{t^*} \lesssim \psuppconst, \quad \left| p^r \right| \leq \psuppconst, 
		\quad L \leq \rsuppconst \psuppconst.
	\end{equation}
	Using the explicit expression for the volume form derived in Section~\ref{sec_angular_coords_massshell} we find 
	\begin{equation}
		\int_{\supp(f(x,\cdot))} 1 \, \dmux \leq \int_{-\psuppconst}^{\psuppconst} \int_{0}^{\rsuppconst \psuppconst} \int_0^{2 \pi} \frac{1}{r^2} \frac{L}{E} \, d \pslashangle d L d p^r \lesssim \int_{-\psuppconst}^{\psuppconst} \int_{0}^{\rsuppconst \psuppconst} \frac{1}{r^2} \frac{L}{E} \, d L d p^r.
	\end{equation}
	Recall further from Lemma~\ref{E_eps_bound_schw} that the bound $E (1 + \left| \trapschw \right|) \lesssim \frac{\rsuppconst^2}{m^2} \psuppconst$ holds for all $(x,p) \in \supp(f)$. Therefore we find that
	\begin{equation}
		\frac{1}{r^2} \frac{L}{E} \lesssim \frac{m}{r^2} \sqrt{1 + \left| \trapschw \right|} \lesssim \frac{\rsuppconst}{r^2} \psuppconst^{\frac{1}{2}} \frac{1}{\sqrt{\left| p^r \right|}}.
	\end{equation}
	We conclude the bound
	\begin{equation}
		\int_{\supp(f(x,\cdot))} 1 \, \dmux \lesssim \frac{\rsuppconst}{r^2} \psuppconst^{\frac{1}{2}} \int_{-\psuppconst}^{\psuppconst} \int_{0}^{\rsuppconst \psuppconst} \frac{1}{\sqrt{\left| p^r \right|}} \, d L d p^r \lesssim \frac{\rsuppconst^2}{r^2} \psuppconst^2,
	\end{equation}
	as claimed.
\end{proof}

\subsubsection{Proof of Theorem~\ref{maintheorem_precise}}

\begin{proof}[Proof of Theorem~\ref{maintheorem_precise}]
	We again parametrise the mass-shell $\mathcal{P}$ in $(t^*,r)$-coordinates as discussed in Section~\ref{sec_parametrising_massshell}. If we introduce spherical coordinates on the sphere $\sphere$ the mass-shell is then explicitly parametrised by $(t^*,r,\theta,\phi,p^r,p^\theta,p^\phi)$. Recall that we may then express the induced volume form on each fibre $\mathcal{P}_x$ in these coordinates as
	\begin{equation}
		\dmu =  \frac{r^2 \sin \theta}{\OsqS p^{t^*} - \frac{2m}{r} p^r} \, d p^r d p^\theta d p^\phi =  \frac{r^2 \sin \theta}{E} \, d p^r d p^\theta d p^\phi,
	\end{equation}
	according to equation~\eqref{dmu_first_comp_schw}. To improve upon the bounds we obtained in the above Lemma~\ref{lem_boundedness_moments_schw} we need to make use of our knowledge of the momentum support of the solution $f$. Recall that Proposition~\ref{psupport_prop} implies
	\begin{equation}
		\supp(f) \cap \left\{ (x,p) \in \mathcal{P} \; | \; \tau(x) \geq \tauzero \right\} \subset \trappedsupportset \cup \smallsupportset.
	\end{equation}
	Let us assume in this proof that $w \geq 0$, since otherwise we may replace $w$ by $\left| w \right|$. We will slightly abuse notation and write $\tau = \tau(x)$ for simplicity. Since certainly $\left| f(x,p) \right| \leq \| f_0 \|_{L^\infty}$ for all $(x,p) \in \supp(f)$, we immediately conclude the bound
	\begin{equation}
		\int_{\mathcal{P}_x} w f \, \dmux \leq \| f_0 \|_{L^\infty} \left( \int_{\trappedsupportsetx} w \, \dmux + \int_{\smallsupportsetx} w \, \dmux \right) .
	\end{equation}
	Recall that heuristically, for large times $\tau$ the set $\trappedsupportsetx \subset \mathcal{P}_x$ asymptotically approaches a cone, whereas the set $\smallsupportsetx \subset \mathcal{P}_x$ is contained in an exponentially small neighbourhood of the origin $p=0$. Let us begin as in the proof of Lemma~\ref{lem_boundedness_moments_schw} by introducing the change of coordinates we will use to estimate both integrals. Assume that $r \geq 2m$ and $p \in \mathcal{P}_x$ and consider the transformation
	\begin{equation}
		(p^r, l, \pslashangle) \mapsto (p^r,p^\theta,p^\phi) = (p^r, l \cos \pslashangle, l \sin \pslashangle),
	\end{equation}
	where we note that $\pslashangle \in [0,2 \pi)$. If we assume $p \in \trappedsupportsetx$ then the length $l$ satisfies $l = \frac{1}{r^2} L \leq \frac{\rsuppconst \psuppconst}{r^2}$ and $p^r \in [p^r_-, p^r_+]$ where we have abbreviated
	\begin{equation}
		p^r_{\pm} = \frac{L}{m} \left( \sqrt{\frac{1}{27} - \frac{m^2}{r^2} \OsqS} \pm \bigc e^{-\frac{\decayrate}{2m}(\tau - \tauzero)} \right).
	\end{equation}
	If we assume that $p \in \smallsupportsetx$ the length $l$ satisfies the bound $l = \frac{1}{r^2} L \leq \bigc \frac{\rsuppconst \psuppconst}{r^2} e^{- \frac{\decayrate}{2m} (\tau - \tauzero)}$ and we know $\left( 1 + \left| \trapschw \right| \right) \left| p^r \right| \leq \bigc \psuppconst e^{-\frac{\decayrate}{m} (\tau - \tauzero)}$. Let us first consider the task of estimating the integral
	\begin{equation}
		\int_{\trappedsupportsetx} w \, \dmux.
	\end{equation}
	Let us recall our assumption on $w$ and apply Lemma~\ref{psupport_bounded} to conclude that
	\begin{equation}
		W := \max_{(x,p) \in \supp(f)} \left| w(x,p) \right| < \infty.
	\end{equation}
	We therefore conclude the estimate
	\begin{equation}
		\begin{aligned}
			\int_{\trappedsupportsetx} w \, \dmux &\lesssim W \int_0^{2 \pi} \int_0^{\frac{\rsuppconst \psuppconst}{r^2}} \int_{p^r_{-}}^{p^r_{+}} \frac{r^2 l}{E} \, d p^r d l d \pslashangle \\
			&= W \int_0^{2 \pi} \int_0^{\rsuppconst \psuppconst} \int_{p^r_{-}}^{p^r_{+}} \frac{1}{r^2} \frac{L}{E} \, d p^r d L d \pslashangle,
		\end{aligned}
	\end{equation}
	Now note that for any $(x,p) \in \trappedsupportset$ with $2m \leq r < \infty$, we have $\left| \trapschw \right| \leq \bigc e^{-\frac{\decayrate}{m}(\tau - \tauzero)} \leq \bigc$ so that
	\begin{equation}
		\frac{1}{r^2} \frac{L}{E} \lesssim \frac{m}{r^2} \sqrt{1 + \left| \trapschw \right|} \leq \frac{m}{r^2} \sqrt{1+ \bigc}.
	\end{equation}
	We therefore use the change of coordinates
	\begin{equation} \label{prtilde}
		(p^r, L) \mapsto \left( m \frac{p^r}{L} - \sqrt{\frac{1}{27} - \frac{m^2}{r^2} \OsqS}, L \right) = (\tilde{p}^r, \tilde{L})
	\end{equation}
	in order to compute the following integral
	\begin{equation}
		\int_0^{\rsuppconst \psuppconst} \int_{p^r_{-}}^{p^r_{+}} 1 \, d p^r d L = \left( \int_0^{\rsuppconst \psuppconst} \frac{\tilde{L}}{m} \, d \tilde{L} \right) \left( \int_{-\bigc e^{-\frac{\decayrate}{2m}(\tau - \tauzero)}}^{\bigc e^{-\frac{\decayrate}{2m}(\tau - \tauzero)}} 1 \, d \tilde{p}^r \right) = \frac{\bigc \rsuppconst^2 \psuppconst^2}{m} e^{-\frac{\decayrate}{2m}(\tau - \tauzero)}.
	\end{equation}
	In summary we have therefore obtained the bound
	\begin{equation}
		\int_{\trappedsupportsetx} w \, \dmux \lesssim W \rsuppconst^{2} \psuppconst^{2} \frac{1}{r^2} e^{-\frac{\decayrate}{2m}(\tau - \tauzero)}.
	\end{equation}
	Let us now turn to the remaining task of estimating the integral
	\begin{equation}
		\int_{\smallsupportsetx} w \, \dmux.
	\end{equation}
	As above we simply use the fact that we may bound $w$ in the support of $f$ by a constant in order to find the bound
	\begin{equation}
		\int_{\smallsupportsetx} w \, \dmux \lesssim W \int_0^{2 \pi} \int_{0}^{\bigc \rsuppconst \psuppconst e^{-\frac{\decayrate}{2m}(\tau - \tauzero)}} \int_0^{\bigc \psuppconst e^{-\frac{\decayrate}{m}(\tau - \tauzero)}} \frac{1}{r^2} \frac{L}{E} \, d p^r d L d \pslashangle.
	\end{equation}
    We carefully note that the angular momentum $L$ decays at half the rate of the $p^r$-component, as captured in the definition of the set $\smallsupportsetx$ in Definition~\ref{defn_trapping_hor}, or can be seen directly in the proof of Proposition~\ref{psupport_prop}. It follows from the definition of the set $\smallsupportset$ that $\left( 1 + \left| \trapschw \right| \right) \left| p^r \right| \leq \bigc \psuppconst e^{-\frac{\decayrate}{m}(\tau - \tauzero)}$ for every $p \in \smallsupportsetx$, from which we deduce
	\begin{equation}
		\frac{1}{r^2} \frac{L}{E} \lesssim \frac{m}{r^2} \sqrt{1 + \left| \trapschw \right|} \leq \sqrt{\bigc \psuppconst} e^{-\frac{\decayrate}{2m}(\tau - \tauzero)} \frac{m}{r^2} \frac{1}{\sqrt{\left| p^r \right|}}.
	\end{equation}
	Therefore we conclude the bound
	\begin{align}
		\int_{\smallsupportsetx} w \, \dmux &\lesssim W \sqrt{\psuppconst} \frac{m}{r^2} e^{-\frac{\decayrate}{2m}(\tau - \tauzero)} \left( \int_{0}^{\bigc \rsuppconst \psuppconst e^{-\frac{\decayrate}{2m}(\tau - \tauzero)}} 1 \, d L \right) \left( \int_0^{\bigc \psuppconst e^{-\frac{\decayrate}{m}(\tau - \tauzero)}} \frac{1}{\sqrt{\left| p^r \right|}} \, d p^r \right) \\
		&\lesssim W \psuppconst^2 \rsuppconst \frac{m}{r^2} e^{-\frac{3}{2} \frac{\decayrate}{m} (\tau - \tauzero)}.
	\end{align}
	Note carefully that we have shown that the decay rate for the integral $\int_{\smallsupportsetx} w \, \dmux$ is always strictly better than the rate for $\int_{\trappedsupportsetx} w \, \dmux$. Making use of the fact that certainly $\rsuppconst \geq 2m$ we therefore obtain the estimate
	\begin{equation}
		\int_{\supp(f(x,\cdot))} w \, \dmux \lesssim W \frac{\psuppconst^2 \rsuppconst^2}{r^2} e^{-\frac{\decayrate}{2m}(\tau - \tauzero)}.
	\end{equation}
	The decay rate in $\tau$ here is the decay rate we can show for the $p^r$ component in the set $\trappedsupportset$. Therefore if we let $0 < \delta' < 1$ and assume that $r \notin [(3-\delta')m,(3+\delta')m]$ and $\tau \gtrsim \tauzero - \ln \delta'$ then the rate of decay may be improved to
	\begin{equation}
		\int_{\supp(f(x,\cdot))} w \, \dmux \lesssim W \frac{\psuppconst^2 \rsuppconst^2}{r^2} e^{-\frac{\decayrate}{m}(\tau - \tauzero)}.
	\end{equation}
	Let us now turn to the question of which weights $w$ improve the decay rate further. Recall the definition of the weights $w_{a,b}$ as well as the constant $c_{a,b}$ from the statement of the theorem. Note carefully that for all $r \geq 2m$,
	\begin{equation}
		\sqrt{\frac{1}{27} - \frac{m^2}{r^2} \OsqS} \sim \left| 1 - \frac{3m}{r} \right|.
	\end{equation}
	In particular, when evaluated at $r = 3m$, the weight $w_{a,b}$ reduces to $w_{a,b} = \left| p^r \right|^{a + b}$. We will now show that the moment associated to the weight $w_{a,b}$ decays at an improved exponential rate. Proceeding as above, we bound the two integrals $\int_{\trappedsupportsetx} w \, \dmux$ and $\int_{\smallsupportsetx} w \, \dmux$ separately. Let us begin by noting that for $(x,p) \in \trappedsupportset$,
	\begin{equation}
		w_{a,b} \leq \psuppconst^b \left| \left| p^r \right| - \sqrt{\frac{1}{27} - \frac{m^2}{r^2} \OsqS} \frac{L}{m} \right|^a \leq \left( \frac{\rsuppconst}{m} \right)^a \psuppconst^{a + b} \left| \tilde{p}^r \right|^a,
	\end{equation}
	where we have used the variable $\tilde{p}^r$ defined above in equation~\eqref{prtilde} as well as Lemma~\ref{psupport_bounded}. Proceeding identically to the way we estimated $\int_{\trappedsupportsetx} w \, \dmux$ above, we therefore find
	\begin{equation} \label{eqn_schw_betterbd1}
		\begin{aligned}
			\int_{\trappedsupportsetx} w_{a,b} \, \dmux &\lesssim \frac{m}{r^2} \left( \frac{\rsuppconst}{m} \right)^a \psuppconst^{a + b} \left( \int_0^{\rsuppconst \psuppconst} \frac{\tilde{L}}{m} \, d \tilde{L} \right) \left( \int_{-\bigc e^{-\frac{\decayrate}{2m}(\tau - \tauzero)}}^{\bigc e^{-\frac{\decayrate}{2m}(\tau - \tauzero)}} \left| \tilde{p}^r \right|^a \, d \tilde{p}^r \right) \\
			&\lesssim_{a} \frac{1}{r^2} \frac{\rsuppconst^{2+a} \psuppconst^{2+a+b}}{m^a} e^{- \frac{\decayrate}{2m}(1+a)(\tau-\tauzero)}.
		\end{aligned}
	\end{equation}
	Similarly we see that for any $(x,p) \in \smallsupportset$ the following bound holds
	\begin{equation}
		w_{a,b} \lesssim_{a+b} \left( \frac{\rsuppconst}{m} \right)^a \psuppconst^a \left| p^r \right|^b.
	\end{equation}
	Again following the same steps of computation as above we find
	\begin{equation} \label{eqn_schw_betterbd2}
		\begin{aligned}
			\int_{\smallsupportsetx} w_{a,b} \, \dmux &\lesssim_{a+b} \frac{m}{r^2} \frac{\rsuppconst^{1+a} \psuppconst^{\frac{3}{2} + a}}{m^a} e^{-\frac{\decayrate}{m} (\tau - \tauzero)} \left( \int_0^{\bigc \psuppconst e^{-\frac{\decayrate}{m}(\tau - \tauzero)}} \left| p^r \right|^{b-\frac{1}{2}} \, d p^r \right) \\
			&\lesssim_{a+b} \frac{m}{r^2} \frac{\rsuppconst^{1+a} \psuppconst^{2 + a + b}}{m^a} e^{-\frac{\decayrate}{m} (\frac{3}{2} + b)(\tau - \tauzero)}.
		\end{aligned}
	\end{equation}
	Carefully note that these bounds hold uniformly in $r \geq 2m$. We may summarise the two bounds~\eqref{eqn_schw_betterbd1} and~\eqref{eqn_schw_betterbd2} in the following way
	\begin{equation}
		\int_{\supp(f(x,\cdot))} w_{a,b} \, \dmux \lesssim_{a+b} \frac{1}{r^2} \frac{\rsuppconst^{2+a} \psuppconst^{2 + a + b}}{m^a} e^{-\frac{\decayrate_{a,b}}{2m} (\tau-\tauzero)}.
	\end{equation}
	Next let us note that in the special case where we consider the decay of the moment along the photon sphere, i.e. we fix $r= 3m$ as $\tau \rightarrow \infty$ we note that the weight simplifies to
	\begin{equation}
		w_{a,b} = \left| p^r \right|^{a + b} = \left| p^r \right|^{\kappa}.
	\end{equation}
	Noting that at $r=3m$ we have $\tilde{p}^r \frac{L}{m} = p^r$, we find the improved bounds
	\begin{align}
		\int_{\trappedsupportsetx} \left| p^r \right|^{\kappa} \, \dmux \Big|_{r=3m} &\lesssim_\kappa \frac{1}{r^2} \frac{\rsuppconst^{2+\kappa} \psuppconst^{2+\kappa}}{m^{\kappa}} e^{-\frac{\decayrate}{2m}(1+\kappa)(\tau-\tauzero)}, \\
		\int_{\smallsupportsetx} \left| p^r \right|^{\kappa} \, \dmux \Big|_{r=3m} &\lesssim_\kappa \frac{1}{r^2} \rsuppconst^2 \psuppconst^{2+\kappa} e^{-\frac{\decayrate}{m}(\frac{3}{2} + \kappa) (\tau-\tauzero)},
	\end{align}
	by a computation completely analogous to above. Therefore we conclude for $\kappa \geq 0$ the bound
	\begin{equation}
		\int_{\supp(f(x,\cdot))} \left| p^r \right|^\kappa \, \dmux \Big|_{r=3m} \lesssim_\kappa \frac{1}{r^2} \frac{\rsuppconst^{2+\kappa} \psuppconst^{2+\kappa}}{m^{\kappa}} e^{-\frac{\decayrate}{2m}(1+\kappa)(\tau-\tauzero)}.
	\end{equation}
	Finally let us turn to weights of the form $w = \left| p^v \right|^a \left| p^u \right|^b \left| \pslash \right|_{\gslash}^c$ for $a,b,c \geq 0$, where we made use of double null coordinates and the induced coordinates on $\mathcal{P}$ to express a point $(x,p) \in \mathcal{P}$ as $(x,p) = (u,v,\omega,p^u,p^v,\pslash)$. We compute the change of coordinates and find that $p^v = p^{t^*} + p^r$. Therefore Lemma~\ref{tauestimate_far}, Lemma~\ref{psupport_bounded} and our choice of $\tauzero$ together imply the bounds
	\begin{equation}
		p^u \lesssim \frac{\rsuppconst^2}{r^2} p^v \lesssim \frac{\rsuppconst^2}{r^2} \psuppconst, \quad p^v \lesssim \psuppconst, \quad \left| \pslash \right|_{\gslash} \leq \frac{\rsuppconst \psuppconst}{r},
	\end{equation}
	for all $(x,p) \in \supp(f)$ such that $\tau(x) \geq \tauzero$ and $r \geq R$, where $r$ denotes the Schwarzschild radius of the point $x \in \Mschw$ and $C> 0$ is a suitable constant. Inserting this bound in the integral and then bounding the remaining integral as above allows us to conclude the better bound
	\begin{equation}
		\int_{\supp(f(x,\cdot))} \left| p^v \right|^a \left| p^u \right|^b \left| \pslash \right|_{\gslash}^c \, \dmux \lesssim_{a+b+c} \psuppconst^{2 + a + b + c}
		\frac{\rsuppconst^{2(1+b) + c}}{r^{2(1+ b) + c}} e^{-\frac{\decayrate}{m} (\tau(x) - \tauzero)}.
	\end{equation}
	The theorem as stated now follows after rescaling the constant $\decayrate$ appropriately.
\end{proof}

\section{Decay for massless Vlasov on extremal Reissner--Nordstr\"om} \label{section_ERN}
This section is dedicated to the proofs of Theorems~\ref{maintheoremERNprecise},~\ref{ERN_slowdecayprop} and~\ref{ERN_nondecaytransversal}. Recall that Theorem~\ref{maintheoremERNprecise} asserts that moments of solutions to the massless Vlasov equation decay at least at a polynomial rate, Theorem~\ref{ERN_slowdecayprop} establishes the sharpness of that rate of decay along the event horizon under certain conditions, while Theorem~\ref{ERN_nondecaytransversal} addresses the non-decay of transversal derivatives along the event horizon. Similarly to the case of the Schwarzschild solution, the main step in the proof is obtaining control over the momentum support of a solution. This will be the content of the two main Propositions~\ref{psupport_propERN} and~\ref{sharpness_prop_rough} of this section.

Proposition~\ref{psupport_propERN} is the extremal analogue of Proposition~\ref{psupport_prop} in the subextremal case and its statement analogously involves two subsets $\ERNtrappedsupportset$ and $\ERNsmallsupportset$ of the mass-shell, which were defined in Section~\ref{sec_subsets}. The set $\ERNtrappedsupportset \subset \mathcal{P}$ contains the geodesics which are almost trapped at the photon sphere and its definition is virtually identical to the Schwarzschild case. Geodesics which are trapped at the photon sphere spend a lot of time in the vicinity of the photon sphere. The set $\ERNtrappedsupportsetx$ approximates a $2$-cone in $\mathcal{P}_x$ at an exponential rate as $\tau(x) \rightarrow \infty$. The set $\ERNsmallsupportset \subset \mathcal{P}$ contains those geodesics which are almost trapped at the event horizon. As in the subextremal case, it is made up of geodesics which originate close to the event horizon and are slowly outgoing. In addition, $\ERNsmallsupportset$ contains geodesics which are slowly in-falling towards the event horizon. This latter type of geodesic is unique to the extremal case and does not exist in the subextremal case. The presence of these geodesics is crucial for the proof of non-decay of transversal derivatives and the slower polynomial decay of moments in the extremal case. Geometrically, the set $\ERNsmallsupportsetx$ may again be described as a small cylinder. In contrast to the subextremal case however, the height and radius of this cylinder are no longer exponentially small in $\tau(x)$ but inverse polynomial. See Figure~\ref{figure_fibres}. 

Recall from Section~\ref{sec_subsets} that the definitions of the sets $\ERNtrappedsupportset$ and $\ERNsmallsupportset$ involve a choice of certain constants $\bigc, \decayrateERN > 0, \tauzero > m$, and the family $\badsetaplarge_\delta$ requires a choice of constant $\constbsl$, all of which will be chosen appropriately in the proof. We also recall the constant $\rsuppconst$ from Assumption~\ref{assumption_support}.

%\newpage

\begin{prop} \label{psupport_propERN}
	Let $f_0 \in L^\infty(\mathcal{P}_0)$ satisfy Assumption~\ref{assumption_support} and let $f$ be the unique solution to the massless Vlasov equation on extremal Reissner--Nordstr\"om with initial distribution $f_0$. Then there exist dimensionless constants $\bigc, \decayrate, C_0 > 0$ such that if we choose $\tauzero \geq C_0 \frac{\rsuppconst^2}{m}$,
	\begin{equation}
		\supp(f) \cap \left\{ (x,p) \in \mathcal{P} \; | \; \tau(x) \geq \tauzero \right\} \subset \trappedsupportset_{\bigc, \decayrate, \tauzero} \cup \smallsupportset_{\bigc, \decayrate, \tauzero},
	\end{equation}
	In addition, there exists a dimensionless constant $\constbsl > 0$ such that for every $\bar{\tau} > \tauzero$ there exists $\delta > 0$ satisfying $\delta \sim m \bar{\tau}^{-1}$ so that the following inclusion holds
	\begin{equation} \label{pi0inclusion}
		\pi_0 \Big( (\smallsupportset \setminus \trappedsupportset) \cap \supp(f) \cap \left\{ (x,p) \in \mathcal{P} \; | \; \tau(x) \geq \bar{\tau} \right\} \Big) \subset \badsetaplarge_{\constbsl,\delta}.
	\end{equation}
\end{prop}

%\begin{rem}
%Like in the Schwarzschild case discussed above, in case that $(x,p) \in \ERNtrappedsupportset$ the rate stated above at which $p^{t^*}$ and $p^r$ decay to their limit values as $\tau \rightarrow \infty$ may be improved in certain cases. Since the components of the momentum for a point $(x,p) \in \ERNsmallsupportset$ only decay at an inverse polynomial rate however, this fact will not be of use here.
%\end{rem}

While Proposition~\ref{psupport_propERN} is the central component for proving upper bounds for moments of a solution to the massless Vlasov equation, Proposition~\ref{sharpness_prop_rough} is the key ingredient in proving lower bounds, as well as non-decay of transversal derivatives. As explained above, one way in which the set $\ERNsmallsupportset$ differs from its subextremal counterpart is the rate at which the height and radius of the cylinder $\ERNsmallsupportsetx$ shrink as $\tau(x) \rightarrow \infty$. Another crucial difference is that even though both the radial and angular momentum components are small, the $p^{t^*}$-component is not. In fact we will show the existence of a family of geodesics which cross the event horizon at arbitrarily late times while satisfying $p^{t^*} \sim \psuppconst$ on the event horizon. The initial data of these geodesics are (essentially) contained in the family of subsets $\badsetapprox_\delta$ defined in Definition~\ref{def_badsetapprox} and populate the set $\slowsupportset \subset \mathcal{P}_{\mathcal{H}^+}$ defined in  Definition~\ref{defi_slowsupportset}. Our second main Proposition~\ref{sharpness_prop_rough} now makes this precise.

%More precisely, if $(x,p) \in \ERNsmallsupportset$ is expressed in $(t^*,r)$-coordinates as $(x,p) = (t^*,r,\omega,p^{t^*},p^r,\pslash) \in \ERNsmallsupportset$ then
%\begin{equation} \label{ineq_ptstar_intro_secextremal}
%	p^{t^*} \lesssim \min \left( \frac{1}{\Osqern(r)} \frac{m^2}{(\tau(x)-\tauzero)^2}, 1 \right) \psuppconst.
%\end{equation}
%Inequality~\eqref{ineq_ptstar_intro_secextremal} is indicative of the fact that the $p^{t^*}$-component does not decay along the event horizon. Let us elaborate on this. At a positive distance from the event horizon, inequality~\eqref{ineq_ptstar_intro_secextremal} implies decay of the $p^{t^*}$-component: if $0 < \delta < 1$ and $r \geq (1+\delta)m$ then
%\begin{equation}
%	p^{t^*} \lesssim \frac{1}{\delta^2} \frac{m^2}{(\tau(x)-\tauzero)^2} \psuppconst.
%\end{equation}
%Close to or on the event horizon, inequality~\eqref{ineq_ptstar_intro_secextremal} only implies boundedness of the $p^{t^*}$-component.

%is as in Definition~\ref{defi_slowsupportset}

\begin{prop} \label{sharpness_prop_rough}
    There exist dimensionless constants $C_1,C_2,c_1,c_2,\constbsl > 0$ such that for all $0 < \delta < \frac{1}{2}$ there exists $\tauslow \sim \frac{m}{\delta}$ such that the following chain of inclusion holds:
    \begin{equation}
		\badsetallconst \subset \badsetapprox_{c_1, c_2, \delta} \subset \badsetaplarge_{\constbsl,\delta}.
	\end{equation}
	Furthermore, each point $(x,p) \in \badsetaplarge_{\constbsl,\delta}$ satisfies Assumption~\ref{assumption_support} and the choice of constant $\constbsl$ is compatible with the choice made in Proposition~\ref{psupport_propERN}. If we abbreviate $\slowsupportset = \slowsupportset_{C_1,C_2,\tauslow}$, then for every point on the event horizon $x \in \mathcal{H}^+$ with $\tau(x) \geq \tauslow$ the fibre $\slowsupportsetx$ satisfies the following:
	\begin{itemize}
		\item Every $p \in \slowsupportsetx$ expressed in $(t^*,r)$-coordinates as $p = (p^{t^*}, p^r, \pslash)$ satisfies $p^{t^*} \sim \psuppconst$.
		\item The subset $\slowsupportsetx \subset \mathcal{P}_x$ has volume $\vol \slowsupportsetx \sim \tau(x)^{-2}$.
	\end{itemize}
	In particular, the geodesics with initial data in $\badsetapprox_\delta$ which cross the point $x$ populate the set $\slowsupportsetx$ and every geodesic with initial data in $\badsetapprox_\delta$ will eventually cross the event horizon $\mathcal{H}^+$.
\end{prop}

%Older way to state the same Proposition
%\begin{prop} \label{sharpness_prop_rough}
%For all $0 < \delta < \frac{1}{2}$ there exists a time $\tauslow \sim \frac{m}{\delta}$ such that for any point $x \in \mathcal{H}^+$ with $\tau(x) \in (\tauslow, \infty)$ the following hold:
%\begin{itemize}
%    \item There exists a family of geodesics with initial data in $\badsetapprox_\delta$ which eventually cross the point $x$.
%   \item The momenta of these geodesics populate an open subset $\slowsupportsetx \subset \mathcal{P}_x$ such that every $p \in \slowsupportsetx$ expressed in $(t^*,r)$-coordinates as $p = (p^{t^*}, p^r, \pslash)$ satisfies $p^{t^*} \sim \psuppconst$.
%   \item The subset $\slowsupportsetx \subset \mathcal{P}_x$ has volume $\vol \slowsupportsetx \sim \tau(x)^{-2}$.
%\end{itemize}
%Furthermore, every point in $\badsetapprox_\delta$ satisfies Assumption~\ref{assumption_support}.
%\end{prop}

Throughout the discussion we will be careful to point out similarities with and differences from the corresponding estimates and results on the Schwarzschild exterior. Often times we will refrain from giving proofs when it is clear that they are analogous to the Schwarzschild case.

We first bound the time required by a given geodesic to cross a certain region of spacetime in Section~\ref{timeestimate_section_ERN}, entirely analogous to the corresponding estimate on the Schwarzschild exterior. This is the extremal analogue of the \emph{almost-trapping estimate} and the key to proving both Proposition~\ref{psupport_propERN} and Proposition~\ref{sharpness_prop_rough}, see Section~\ref{sec_radialgeod}.

Using the almost-trapping estimate, we are then able to prove Proposition~\ref{psupport_propERN} in Section~\ref{psupport_section_ERN}. In Section~\ref{psupporteventhorizon} we then prove a precise version of Proposition~\ref{sharpness_prop_rough}, the main ingredient necessary to allow us to show a lower bound for the rate of decay.

Using the knowledge gained about the momentum support of a solution, we then apply Proposition~\ref{psupport_propERN} and estimate the volume of the sets $\ERNtrappedsupportset$ and $\ERNsmallsupportset$ to prove Theorem~\ref{maintheoremERNprecise} on upper bounds in Section~\ref{estimating_moments_section_ERN}. In Section~\ref{sec_proof_lowerboundERN} we apply Proposition~\ref{sharpness_prop_rough} to prove Theorem~\ref{ERN_slowdecayprop} on lower bounds. Finally we study transversal derivatives of the energy momentum tensor in Section~\ref{section_derivativesERN} and provide the proof of Theorem~\ref{ERN_nondecaytransversal}, which will again make substantial use of Proposition~\ref{sharpness_prop_rough}.

\subsection{The almost-trapping estimate} \label{timeestimate_section_ERN}
In this section we will prove a bound for the time a given geodesic requires to cross a certain region of spacetime in terms of the radius the geodesic occupies at the beginning and at the end of the segment and the value of its trapping parameter. The main result of this section is Lemma~\ref{tauestimate_ERN}.

\begin{lem}[The almost-trapping estimate] \label{tauestimate_ERN}
	Let $\gamma: [s_0,s_1] \rightarrow \Mrn$ be  an affinely parameterised future-oriented null geodesic. Let us express $\gamma$ in $(t^*,r)$-coordinates as $\gamma(s) = (t^*(s),r(s),\omega(s))$ with momentum $\dot{\gamma}(s) = (p^{t^*}(s),p^r(s),\pslash(s))$. Assume $\gamma$ intersects $\Sigma_0$ at radius $r_0 = r(s_0) \leq \rsuppconst$. Let $0 < \delta < 1$ and denote $\mathfrak{s} = \chi_{(0,\infty)}\left( p^r(s_0) \right)$, where $\chi_{(0,\infty)}$ denotes the characteristic function of the set $(0,\infty) \subset \R$. There exists a constant $C_0 > 0$ such that if we assume $\tauzero \geq C_0 \frac{\rsuppconst^2}{m}$ then for all $s \in [s_0,s_1]$ with $m \leq r(s) \leq (1+\delta)m$ we have the bound
	\begin{equation}
		\frac{1}{m} \left( \tau(\gamma(s)) - \tauzero \right) \lesssim_{\delta} \begin{cases}
			\mathfrak{s} \frac{1}{\sqrt{\Osqern(r_0)}} & \text{if } p^r(s) > 0 \\
			\left( 1 + \left| \trapschw \right| \right) \sqrt{\Osqern(r_0)} + \mathfrak{s} \frac{1}{\sqrt{\Osqern(r_0)}} + \left( \log \left| \trapern \right| \right)_-  & \text{if } p^r(s) \leq 0
		\end{cases}.
	\end{equation}
	If $r(s) \geq (1+\delta)m$ we have
	\begin{equation}
		\frac{1}{m} \left( \tau(\gamma(s)) - \tauzero \right) \lesssim_{\delta} \left( \log \left| \trapern \right| \right)_- + \mathfrak{s} \frac{1}{\sqrt{\Osqern(r_0)}},
	\end{equation}
	for both signs of $p^r$. We have abbreviated $\trapern = \trapern(\gamma,\dot{\gamma})$.
\end{lem}

The strategy of proof is the same as in Section~\ref{section_tstar}. We first obtain a bound for the time a geodesic requires to cross a certain region of spacetime as measured in the $t^*$-coordinate in Lemma~\ref{tstar_estimate_ern}. This is the extremal analogue of Lemma~\ref{tstar_lem_schw}. Since the geodesic flow at a positive distance from the event horizon is entirely comparable to the geodesic flow on Schwarzschild at a positive distance from the event horizon, we do not prove bounds for the region away from the event horizon. In fact, it is straightforward to verify that both Lemma~\ref{pr_like_E} and Lemma~\ref{tauestimate_far} hold verbatim in the extremal Reissner--Nordstr\"om case. Having established these results, the argument to prove Lemma~\ref{tauestimate_ERN} proceeds along identical lines as for the corresponding Lemma~\ref{taubound} in the Schwarzschild case and we do not reproduce it here. It therefore only remains to show a bound for the $t^*$-time in Lemma~\ref{tstar_estimate_ern}. See Section~\ref{sec_radialgeod} for an informal discussion of the proof in the special case of radial geodesics.

\subsubsection{The $t^*$-time estimate}
As in Section~\ref{sec_tstarestimate} we split the black hole exterior in several regions to simplify our statement of the $t^*$-time estimate. Fix three constants $0< \delta_1,\delta_2 < 1$ and $\delta_3 > 0$ such that $0 < \delta_1 + \delta_2 < 1$. Consider the following intervals
\begin{equation}
	\begin{aligned}
		\mathfrak{I}_{\mathcal{H}^+} &= [m, (1+\delta_1)m], \\
		\mathfrak{I}_{\text{int}} &= [(1+\delta_1)m, (2-\delta_2)m], \\
		\mathfrak{I}_{\text{ps}} &= [(2-\delta_2)m, (2+\delta_3)m], \\
		\mathfrak{I}_{\text{flat}} &= [(2+\delta_3)m, \infty),
	\end{aligned}
\end{equation}
where we have suppressed the dependence on $\delta_1,\delta_2,\delta_3$ in the notation.

\begin{lem} \label{tstar_estimate_ern}
	Denote by $\gamma$ an affinely parameterised future-oriented null geodesic expressed in $(t^*,r)$-coordinates as $\gamma(s) = (t^*(s),r(s),\omega(s))$ with momentum $\dot{\gamma}(s) = (p^{t^*}(s),p^r(s),\pslash(s))$. Assume that $p^r \neq 0$ in the interval of affine parameter time $[s_1,s_2]$, so that the radius $r$ is a strictly monotonical function of $s$. Let us assume that $r(s_i) = r_i$ for $i=1,2$ and let $\delta_1,\delta_2,\delta_3 > 0$ be as above. Then
	\begin{equation}
		\frac{t^*(s_2) - t^*(s_1)}{m} \lesssim_{\delta_1,\delta_2,\delta_3 } \begin{cases}
			\begin{cases}
				\left( 1 + \left| \trapschw \right| \right) \sqrt{\Osqern(r_1)} & p^r \leq 0 \text{ on } [r_2,r_1] \\
				\frac{1}{\sqrt{\Osqern(r_1)}} & p^r > 0 \text{ on } [r_1,r_2]
			\end{cases} & r_1,r_2 \in \mathfrak{I}_{\mathcal{H}^+} \\
			1 & r_1,r_2 \in \mathfrak{I}_{\text{int}} \\
			1 + \left( \log \left| \trapschw \right| \right)_- & r_1,r_2 \in \mathfrak{I}_{\text{ps}} \\
			1 + \frac{\max(r_1, r_2)}{m} & r_1,r_2 \in \mathfrak{I}_{\text{flat}}
		\end{cases}
	\end{equation}
	where we have used the shorthand $\trapschw = \trapschw(\gamma(s),\dot{\gamma}(s))$. In addition if we assume that $p^r \leq 0$ for $s \in [s_1,s_2]$ and that $r_2 = m, r_1 \in \mathfrak{I}_{\mathcal{H}^+}$ we in fact also have the lower bound
	\begin{equation} \label{tstar_lowerbound_ern}
		\frac{t^*(s_2) - t^*(s_1)}{m} \gtrsim_{\delta_1} \left( 1 + \left| \trapschw \right| \right) \sqrt{\Osqern(r_1)}.
	\end{equation}
	If we assume instead that $r_2 > m, r_1 \in \mathfrak{I}_{\mathcal{H}^+}$ then the above bound readily implies the upper bound
	\begin{equation} \label{tstar_upperbound_triv}
		\frac{t^*(s_2) - t^*(s_1)}{m} \lesssim_{\delta_1} \frac{1}{\sqrt{\Osqern(r_2)}}.
	\end{equation}
	Furthermore if we assume that $p^r > 0$ for $s \in [s_1,s_2]$ and $\trapschw \ll -1$ then we may assume that $r_{\text{min}}^-(\trapern) \in \mathcal{I}_{\mathcal{H}^+}$. If we assume $r_2 =  r_{\text{min}}^-(\trapern)$ and $r_1 \leq \frac{1}{2} (r_2 + m)$ then we in fact also have the lower bound
	\begin{equation}
		\frac{t^*(s_2) - t^*(s_1)}{m} \gtrsim_{\delta_1} \frac{1}{\sqrt{\Osqern(r_1)}}.
	\end{equation}
\end{lem}

\begin{rem} \label{rem_bad_geodesics}
	Let us compare Lemma~\ref{tstar_estimate_ern} to the corresponding Lemma~\ref{tstar_lem_schw} in the subextremal case. If $\gamma$ is an in-falling geodesic close to the event horizon with $\trapern(\gamma) \ll -1$, a qualitatively different behaviour is exhibited in the extremal as compared to the subextremal case. To be more precise, if we assume $\gamma: [s_1,s_2] \rightarrow \Mern$ is an in-falling null geodesic on extremal Reissner--Nordstr\"om and is such that $r(s_1) = r_{\text{min}}^-(\trapern), r(s_2) = m$, then Lemma~\ref{tstar_estimate_ern} above shows that $t^*(s_2) - t^*(s_1) \sim m \sqrt{ 1 + \left| \trapern(\gamma) \right| }$ if $\trapern(\gamma) \ll -1$. The corresponding bound in the subextremal case for the time a geodesic requires to fall into the black hole however remains constant as $\trapschw(\gamma) \rightarrow - \infty$. More precisely, if we analogously let $\gamma: [s_1,s_2] \rightarrow \Mschw$ be an in-falling geodesic on Schwarzschild with $r(s_1) = r_{\text{min}}^-(\trapern), r(s_2) = 2m$, then Lemma~\ref{tstar_lem_schw} implies $t^*(s_2) - t^*(s_1) \lesssim m$. The heuristic reason for this is the degeneracy of the red-shift effect. Note carefully that for a geodesic to be slowly in-falling on the extremal Reissner--Nordstr\"om exterior, we must have $\trapern(\gamma) \ll -1$, which by definition of the trapping parameter implies that the geodesic cannot be radial (since $L=0$ implies $\trapern = 0$). Note that apart from the amount of time these geodesics require to reach the event horizon, there are no other differences between the subextremal and extremal case. Finally, if instead we let $\gamma$ be an outgoing geodesic originating close to the event horizon, the behaviour is qualitatively entirely comparable to the subextremal case. However, the time required to leave the region close to the event horizon is inverse linear in the initial distance from $\mathcal{H}^+$ in the extremal case and logarithmic in the subextremal case.
\end{rem}

\begin{proof}
	The proof will proceed along the same lines as the proof of the corresponding Lemma~\ref{tstar_lem_schw} for the Schwarzschild background. We use $(t^*,r)$-coordinates as defined in Section~\ref{rngeometry}. In the entire argument all momenta and radii are considered along the geodesic $\gamma$ and we will frequently omit the dependence on the affine parameter $s$. Since for $s \in [s_1,s_2]$ we have $p^r \neq 0$ we may conclude in an identical fashion to the Schwarzschild case that
	\begin{equation}
		t^*(s_2) - t^*(s_1) = \int_{r_1}^{r_2} \frac{p^{t^*}}{p^r} \, d r.
	\end{equation}
	where we carefully note that if $p^r < 0$, the integration boundaries will be such that $r_2 < r_1$ so that the orientation of the integral ensures the correct sign. Note also that we are slightly abusing notation and using the radius $r$ as an integration variable. We will now use Lemma~\ref{prptstar_ern} to bound the integrand in different regions of spacetime. 
	
	It follows immediately from Lemma~\ref{prptstar_ern} that away from the event horizon the quotient $\frac{p^{t^*}}{p^r}$ is comparable to the quotient in the Schwarzschild case apart from the apparent difference in location of the event horizon and photon sphere. Therefore if $r_1,r_2 \geq (1+\delta)m$ for $0 < \delta < 1$, the computations take on an almost identical form to those in the proof of Lemma~\ref{tstar_lem_schw}. We therefore focus our attention on the region close to the event horizon and assume that $m \leq r_1,r_2 \leq (1+\delta) m$ for some fixed $0 < \delta = \delta_1 < 1$. We will omit the index of $\delta_1$ for simplicity of notation. 
	
	Let us begin by considering the case that $p^r \leq 0$ so that $r_1 \geq r_2$. We further distinguish the two cases that $\trapern < 0$ and $\trapern \geq 0$. If $\trapschw \geq 0$ let us apply Lemma~\ref{prptstar_ern} and note that $(r-2m)^2 + \trapschw \mathfrak{a} \sim_\delta m^2$ and $\frac{1}{(2- \trapschw)} \sim 1$, so that $\frac{p^{t^*}}{|p^r|} \sim_\delta 1$. Hence
	\begin{equation}
		\int_{r_1}^{r_2} \frac{p^{t^*}}{p^r} \, d r = \int_{r_2}^{r_1} \frac{p^{t^*}}{\left| p^r \right|} \, d r \sim_\delta r_1 - r_2 \leq r_1 - m.
	\end{equation}
	In particular it is evident that if $r_2 = m$ and $\trapern \geq 0$ then the bound~\eqref{tstar_lowerbound_ern} holds. If $\trapschw < 0$, let us again apply Lemma~\ref{prptstar_ern} and write
	\begin{equation} \label{eqn::yexplained}
		(r-2m)^2 + \trapschw \mathfrak{a} = (r-2m)^2 \left( 1 + \trapschw (r-m)^2 \mathfrak{b} \right) \sim_\delta m^2 \left( 1 + \trapschw (r-m)^2 \mathfrak{b} \right) = m^2 \left( 1 + \trapschw x^2 \mathfrak{b} \right),
	\end{equation}
	where we have introduced the change of coordinates $x=r-m$ and defined the coefficient $\mathfrak{b}$ such that $\mathfrak{a} = (r-2m)^2 (r-m)^2 \mathfrak{b}$. Explicitly in $x$-coordinates, $\mathfrak{b}(x) = 16m^2 (x-m)^{-2} (x^2+6mx+m^2)^{-1}$, which may be seen by a simple computation. Note that $m^2 \mathfrak{b} \sim_\delta 1$ in the region of integration. We find
	\begin{equation*}
		\int_{r_2}^{r_1} \frac{p^{t^*}}{|p^r|} \, dr \sim_\delta \int_{r_2}^{r_1} \frac{2-\trapschw}{\sqrt{1+\trapschw (r-m)^2 \mathfrak{b}}} \, dr = \int_{r_2-m}^{r_1-m} \frac{2-\trapschw}{\sqrt{1+\trapschw x^2 \mathfrak{b}}} \, dx \sim_\delta m \int_{y_2}^{y_1} \frac{2-\trapschw}{\sqrt{1+\trapschw y^2}} \, dy,
	\end{equation*}
	where we have made the two changes of coordinate $x = r-m$ and $y = x \sqrt{\mathfrak{b}(x)}$ and have furthermore correspondingly set $y_i =  (r_i-m) \sqrt{\mathfrak{b}(r_i-m)}$ for $i=1,2$. In the last step we made use of the fact that the Jacobian
	\begin{equation}
		\left| \sqrt{\mathfrak{b}} \left( 1 + \frac{x}{2} \frac{\mathfrak{b}'}{\mathfrak{b}} \right) \right|^{-1} = \mathfrak{b}^{-\frac{1}{2}} \frac{(x-1) (x^2+6 x+1)}{(x+1)^3} \sim_{\delta} m.
	\end{equation}
	Equation~\eqref{eqn::yexplained} implies $1+ \trapschw y^2 = 1 - \left| \trapschw \right| y^2 = (r-2m)^{-2}((r-2m)^2 + \trapern \mathfrak{a}) \geq 0$, where in the last step we applied Lemma~\ref{prptstar_ern}. Therefore $0 \leq \sqrt{\left| \trapschw \right|} y_i \leq 1$ for $i=1,2$. We then compute
	\begin{align}
		&\int_{y_2}^{y_1} \frac{2-\trapern}{\sqrt{1+\trapern y^2}} \, dy = \frac{2+ \left| \trapern \right|}{\sqrt{\left| \trapern \right|}} \int_{\sqrt{\left| \trapern \right|} y_2}^{\sqrt{\left| \trapern \right|} y_1} \frac{1}{\sqrt{1-z^2}} \, dz \sim \frac{2+ \left| \trapern \right|}{\sqrt{\left| \trapern \right|}} \int_{\sqrt{\left| \trapern \right|} y_2}^{\sqrt{\left| \trapern \right|} y_1} \frac{1}{\sqrt{1-z}} \, dz \\
		&= 2 \frac{2+ \left| \trapern \right|}{\sqrt{\left| \trapern \right|}} \left( \sqrt{1-\sqrt{\left| \trapern \right|} y_2}-\sqrt{1-\sqrt{\left| \trapern \right|} y_1} \right) \leq 2 \frac{2+ \left| \trapern \right|}{\sqrt{\left| \trapern \right|}} \left( 1-\sqrt{1-\sqrt{\left| \trapern \right|} y_1} \right) \label{inequ_proof_tstart_ern} \\
		&\sim \frac{2+ \left| \trapern \right|}{\sqrt{\left| \trapern \right|}} \sqrt{\left| \trapern \right|} y_1 = \left( 2+ \left| \trapern \right| \right) y_1 \sim_\delta \left( 1 + \left| \trapern \right| \right) \sqrt{\Osqern(r_1)}.
	\end{align}
	Note carefully that in line~\eqref{inequ_proof_tstart_ern} we are being somewhat wasteful. In fact, whenever we assume that $(\sqrt{\left| \trapern \right|} y_1,\sqrt{\left| \trapern \right|} y_2) \notin [1-\kappa,1]^2$ for some fixed $0 < \kappa < 1$ we have the estimate
	\begin{equation}
		\frac{2+ \left| \trapern \right|}{\sqrt{\left| \trapern \right|}} \left( \sqrt{1-\sqrt{\left| \trapern \right|} y_2}-\sqrt{1-\sqrt{\left| \trapern \right|} y_1} \right) \sim_\kappa \left( 1+ \left| \trapern \right| \right) \left( y_1 - y_2 \right),
	\end{equation}
	so that it follows in particular when we consider the special case that the geodesic arrives at the event horizon at affine parameter time $s_2$ (in which case $y_2 = 0$ or equivalently $r_2 = m$) we have the estimate
	\begin{equation}
		\int_{m}^{r_1} \frac{p^{t^*}}{|p^r|} \, dr \sim_\delta \left( 1+ \left| \trapern \right| \right) (r_1-m),
	\end{equation}
	which readily allows us to conclude inequality~\eqref{tstar_lowerbound_ern}. Next notice that if we assume $r_1 \geq r_2 > m$, we may further bound
	\begin{equation}
		\left( 1 + \left| \trapern \right| \right) \sqrt{\Osqern(r_1)} \lesssim_\delta \sqrt{1 + \left| \trapern \right|} \lesssim_\delta \frac{1}{\sqrt{\Osqern(r_2)}},
	\end{equation}
	as claimed in equation~\eqref{tstar_upperbound_triv}. Alternatively, one may also readily deduce the bound claimed in equation~\eqref{tstar_upperbound_triv} from the computation above. 
	
	Let us now consider the case that $p^r \geq 0$ and let $0 < \delta = \delta_2 < 1$ now. We will again omit the index of $\delta_2$ for simplicity of notation. Then depending on the sign of $\trapschw$, the geodesic will eventually either cross the photon sphere or scatter off it, however we are only interested in the region close to the horizon here. Note that since $p^r \geq 0$ we have
	\begin{equation}
		\int_{r_1}^{r_2} \frac{p^{t^*}}{p^r} \, d r = \int_{r_1}^{r_2} \frac{p^{t^*}}{\left| p^r \right|} \, d r.
	\end{equation}
	If $\trapschw \geq 0$ we begin by noting that Lemma~\ref{prptstar_ern} implies that $\frac{p^{t^*}}{|p^r|} \sim_\delta \frac{1}{\Osqern}$ in a similar fashion to the case that $p^r < 0$. We therefore find
	\begin{equation}
		\int_{r_1}^{r_2} \frac{p^{t^*}}{|p^r|} \, dr \sim_\delta \int_{r_1}^{r_2} \frac{1}{\Osqern} \, dr \sim \left[ - \frac{m^2}{r-m} \right]_{r_1}^{r_2} = \frac{m^2}{r_1-m} - \frac{m^2}{r_2-m} \leq \frac{m}{\sqrt{\Osqern(r_1)}}.
	\end{equation}
	When $\trapschw < 0$ the geodesic must scatter off the photon sphere. We apply Lemma~\ref{prptstar_ern} as above and again rewrite $(r-2m)^2 + \trapschw \mathfrak{a} = (r-2m)^2 \left( 1 + \trapschw (r-m)^2 \mathfrak{b} \right) \sim_\delta m^2 \left( 1 + \trapschw (r-m)^2 \mathfrak{b} \right) = m^2 \left( 1 + \trapschw x^2 \mathfrak{b} \right)$ where we have introduced the change of coordinates $x=r-m$ and defined $\mathfrak{b}$ as above. We compute
	\begin{equation}
		\int_{r_1}^{r_2} \frac{p^{t^*}}{|p^r|} \, dr \sim_\delta \int_{r_1-m}^{r_2-m} \frac{m^2}{x^2 \sqrt{1+\trapschw x^2 \mathfrak{b}}} \, dx \sim_\delta m \int_{y_1}^{y_2} \frac{1}{y^2 \sqrt{1+\trapschw y^2}} \, dy,
	\end{equation}
	where we have made the change of variables $y = x \sqrt{\mathfrak{b}(x)}$ and defined $y_i = (r_i-m) \sqrt{\mathfrak{b}(r_i-m)}$ for $i=1,2$. In the last step we have also made use of the fact that the Jacobian
	\begin{equation}
		\left| \sqrt{\mathfrak{b}} \left( 1 + \frac{x}{2} \frac{\mathfrak{b}'}{\mathfrak{b}} \right) \right|^{-1} \sim_{\delta} m,
	\end{equation}
	as well as $m^2 \mathfrak{b} \sim_{\delta} 1$ as already shown above. We further compute
	\begin{equation}
		\begin{aligned}
			\int_{y_1}^{y_2} \frac{1}{y^2 \sqrt{1+\trapschw y^2}} \, dy &= \sqrt{\left| \trapschw \right|} \int_{\sqrt{\left| \trapschw \right|} y_1}^{\sqrt{\left| \trapschw \right|} y_2} \frac{1}{z^2 \sqrt{1-z^2}} \, dz = \sqrt{\left| \trapschw \right|} \left[ - \frac{\sqrt{1-z^2}}{z} \right]_{\sqrt{\left| \trapschw \right|}y_1}^{\sqrt{\left| \trapschw \right|}y_2} \\
			&\leq \sqrt{\left| \trapschw \right|} \frac{\sqrt{1-\left| \trapschw \right| y_1^2}}{\sqrt{\left| \trapschw \right|}y_1} \leq \frac{\sqrt{1-\left| \trapschw \right| y_1^2}}{y_1} \lesssim_{\delta} \frac{1}{\sqrt{\Osqern(r_1)}}.
		\end{aligned}
	\end{equation}
	Next note that since $\left| \trapern \right| \Osqern(r_{\text{min}}^-(\trapern)) \sim_\delta 1$ we find that if $\trapschw < 0$ is large enough in absolute value, we have $r_{\text{min}}^-(\trapern) \leq \min((1+\delta) m, \frac{3}{2}m)$. By definition, if we set $r_2 = r_{\text{min}}^-(\trapern)$ then $1 - \left| \trapern \right| y_2^2 = 0$. Therefore the integral in this case evaluates to
	\begin{equation}
		\int_{y_1}^{y_2} \frac{1}{y^2 \sqrt{1+\trapschw y^2}} \, dy = \frac{\sqrt{1-\left| \trapschw \right| y_1^2}}{y_1}.
	\end{equation}
	Recall that $m^2 \mathfrak{b} \sim_\delta 1$. In fact one may easily see that the explicit bound $11 \leq m^2 \mathfrak{b} \leq 16$ holds for $m \leq r \leq \frac{3}{2} m$. If we now assume in addition that the bound $r_1 \leq \frac{1}{2}(r_2+m)$ holds, then it follows directly that $r_1 -m \leq \frac{1}{2}(r_2-m)$ and we furthermore find that
	\begin{equation}
		y_1 = (r_1-m) \sqrt{\mathfrak{b}(r_1-m)} \leq \frac{1}{2} \frac{\sqrt{16}}{\sqrt{11}} (r_2-m) \sqrt{\mathfrak{b}(r_2-m)} < \frac{7}{10} y_2 = \frac{7}{10} \frac{1}{\sqrt{\left| \trapern \right|}}.
	\end{equation}
	Therefore we find that $\sqrt{1-\left| \trapschw \right| y_1^2} \geq \frac{1}{2}$ and conclude that
	\begin{equation}
		\int_{y_1}^{y_2} \frac{1}{y^2 \sqrt{1+\trapschw y^2}} \, dy \gtrsim \frac{1}{y_1}.
	\end{equation}
	This concludes the proof.
\end{proof}

\subsection{Upper bounds for the momentum support} \label{psupport_section_ERN}
In this section we give the proof of Proposition~\ref{psupport_propERN}, the extremal analogue of Proposition~\ref{psupport_prop}, which establishes an upper bound on the size of the momentum support of a solution $f$ to the massless Vlasov equation as $\tau \rightarrow \infty$. We remind the reader that the definitions of the sets $\trappedsupportset, \smallsupportset \subset \mathcal{P}$ involve a choice of constants $\bigc, \decayrate$ and $\tauzero$, while the definition of $\badsetaplarge$ involves the choice of a constant $\constbsl$, see Section~\ref{sec_subsets}.

We remark that Lemma~\ref{E_eps_bound_schw} and~\ref{psupport_bounded}, which concern the boundedness of the momentum support of a solution to the massless Vlasov equation with initially compact support, hold verbatim in the extremal case and we will refer to their extremal analogues here and in later sections.

\begin{proof}[Proof of Proposition~\ref{psupport_propERN}]
	The proof follows the same outline as the proof of Proposition~\ref{psupport_prop} in the subextremal case and we will omit certain details which are virtually identical to the subextremal case. Let $\gamma: [0,s] \rightarrow \Mern$ be an affinely parametrised future-directed null geodesic segment expressed in $(t^*,r)$-coordinates. Assume $(\gamma(0),\dot{\gamma}(0)) \in \supp(f_0)$ and let us for simplicity of notation suppress the dependence on $s$, so we write $(\gamma(s),\dot{\gamma}(s)) = (t^*,r,\omega,p^{t^*},p^r,\pslash)$ and $\tau(x(s)) = \tau$. Let us choose $\tauzero$ like in Lemma~\ref{tauestimate_ERN} and let $0 < \delta < 1$ be fixed. For the sake of clarity, the reader may make an explicit choice, such as $\delta = \frac{1}{2}$. We distinguish between the two cases that $p^r \leq 0$ and $p^r > 0$ at time $\tau$.
	
	%In each of these cases, we will begin by showing a bound for the $p^r$-component. Although the way in which we show the bound for $p^r$-component itself does not depend on its sign, the distinction based on the sign of $p^r$ becomes necessary when we next show a bound for the $p^{t^*}$-component. Finally we will show bounds for the angular momentum components, whose proof also does not depend on the sign of $p^r$.
	
	\paragraph{Case 1: $p^r \leq 0$ at time $\tau$.}
	In this case Lemma~\ref{tauestimate_ERN} allows us to conclude the bounds
	\begin{equation} \label{Case1ERN}
		\frac{1}{m} \left( \tau - \tauzero \right) \lesssim \begin{cases}
			\left( 1 + \left| \trapschw \right| \right) \sqrt{\Osqern(r(0))} + \mathfrak{s} \frac{1}{\sqrt{\Osqern(r(0))}} + \left( \log \left| \trapschw \right| \right)_-  & m \leq r \leq (1+\delta)m \\
			\mathfrak{s} \frac{1}{\sqrt{\Osqern(r(0))}} + \left( \log \left| \trapschw \right| \right)_-  & r \geq (1+\delta)m
		\end{cases},
	\end{equation}
	where we have defined $\mathfrak{s} = \chi_{(0,\infty)}\left( p^r(0) \right)$ as in the statement of the lemma. Similarly to the proof of Proposition~\ref{psupport_prop} above, this bound implies that we must be in one of the following three cases.
	
	\paragraph{Case 1.1: $\gamma$ is almost trapped at the photon sphere.}
	In this case we have $\left| \trapern \right| < 1$ and
	\begin{equation}
		\frac{1}{m} \left( \tau - \tauzero \right) \lesssim \log \left| \trapern \right|,
	\end{equation}
    which implies the bound
    \begin{equation}
        \left| \trapern \right| \leq e^{- \frac{\decayrateERN}{m} (\tau - \tauzero)}
    \end{equation}
	for an appropriate constant $\decayrateERN > 0$ arising from the constant implicit in inequality~\eqref{Case1ERN}. The argument in this case proceeds virtually identically to the corresponding Case~1.1 in the proof of Proposition~\ref{psupport_prop} and we conclude that $p \in \ERNtrappedsupportsetx$ for a suitable choice of constants $\bigc, \decayrateERN$ and $\tauzero$.
	
	\paragraph{Case 1.2: $\gamma$ is initially outgoing and starts close to $\mathcal{H}^+$.}
	In this case $p^r(0) > 0$ and
	\begin{equation} \label{eqn_case12_ern}
		\frac{1}{m} (\tau - \tauzero) \lesssim \frac{1}{\sqrt{\Osqern(r(0))}},
	\end{equation}
    which implies the bound
    \begin{equation}
        \Osqern(r(0)) \lesssim \frac{m^2}{(\tau-\tauzero)^2}.
    \end{equation}
	If we assume that $\tau \geq \tauzero + C' m$ for a constant $C' > 0$ then $\Osqern(r(0)) \leq \frac{1}{(C')^2}$, in particular we may assume $C' > 2$ so that $r(0) < 2m$. From the definition of energy we find
	\begin{equation}
		E = \Osqern(r(0)) p^{t^*}(0) - \left(1 - \Osqern(r(0)) \right) p^r(0) \leq \Osqern(r(0)) \psuppconst \lesssim \frac{m^2}{(\tau-\tauzero)^2} \psuppconst,
	\end{equation}
	where we have used the fact that $p^r(0) > 0$. Conservation of energy~\eqref{Schw::energy_conservation} immediately implies
	\begin{equation}
		\left| p^r \right| \leq E  \lesssim \frac{m^2}{(\tau-\tauzero)^2} \psuppconst.
	\end{equation}
	To obtain a bound on the $p^{t^*}$ component apply Lemma~\ref{ptstar_bounds_ern} noting that $p^r \leq 0$ to find
	\begin{equation}
		p^{t^*} \lesssim \left(1 + \frac{m^2}{r^2} \left| \trapern \right| \right) E \leq  \left(1 + \left| \trapern \right| \right) \Osqern(r(0)) \psuppconst \leq \psuppconst.
	\end{equation}
	Note carefully that the bound we have shown for the energy $E$ only depends on our choice of $\delta$, which is fixed, say $\delta = \frac{1}{2}$. By making use of Lemma~\ref{ptstar_bounds_ern} again, we conclude the bound
	\begin{equation}
		p^{t^*} \lesssim \frac{E}{\Osqern}  \lesssim  \frac{1}{\Osqern} \frac{m^2}{(\tau-\tauzero)^2} \psuppconst.
	\end{equation}
    \vskip -0.2cm  
    \noindent
	We see that one can obtain a decay estimate (with a constant that degenerates as $r \rightarrow m$) at an arbitrarily small distance to the event horizon. We claim however that the bound $p^{t^*} \lesssim \psuppconst$ is generically sharp up to a constant multiple along the event horizon, see Proposition~\ref{sharpness_lemma} below for a discussion. Finally to obtain a bound on the angular momentum components, we find from conservation of energy that
    \vskip -0.2cm
	\begin{equation}
		\frac{\Osqern(r(0))}{r(0)^2} L^2 \leq E^2 \leq \Osqern(r(0))^2 \psuppconst^2,
	\end{equation}
	so that we find
	\begin{equation} \label{eqn_case12_L_ern}
		\left| \pslash \right|_{\gslash} = \frac{L}{r} \leq \sqrt{\Osqern(r(0))} \frac{r(0)}{r} \psuppconst \lesssim \frac{m}{\tau - \tauzero} \frac{\rsuppconst}{r} \psuppconst.
	\end{equation}
	Now assume that $\tau(x) > \bar{\tau} > \tauzero$. We show that $(x(0),p(0)) \in \badsetaplarge_{\constbsl,\bar{\delta}}$ for $\bar{\delta} = \bar{\tau}^{-1}$ and $\constbsl = 1$, first note that from~\eqref{eqn_case12_ern} it follows that $r(0) \leq (1+\bar{\delta})m$ if we assume $\bar{\tau}$ is large enough. Next note that since $p^r(0) > 0$, it directly follows that
    \vskip -0.3cm
	\begin{equation}
		p^r(0) \leq E \leq \Osqern(r(0)) \psuppconst.
	\end{equation}
    %\vskip -0.2cm
    %\noindent
	To bound the angular component, we argue precisely as in~\eqref{eqn_case12_L_ern} to find that
	\begin{equation}
		\left| \pslash(0) \right|_{\gslash} \leq  \sqrt{\Osqern(r(0))} \psuppconst.
	\end{equation}
    Finally, the inequality $p^{t^*}(0) \leq \psuppconst$ follows immediately since $f_0$ satisfies Assumption~\ref{assumption_support}.
 
	\paragraph{Case 1.3: $\gamma$ is slowly in-falling.}
	Note carefully that in the subextremal case discussed above, there were only the analogues of Cases~1.1 and 1.2 just discussed. This third case is unique to extremal black holes, although the proof proceeds in an almost identical manner to Case~1.2 above. Here we may assume that $m \leq r \leq (1+\delta)m$ and we have the bound
	\begin{equation} \label{eqn_case13_timebound}
		\frac{1}{m}(\tau - \tauzero) \lesssim \left( 1 + \left| \trapern \right| \right) \sqrt{\Osqern(r(0))}.
	\end{equation}
	Using the bound $\left( 1 + \left| \trapern \right| \right) \Osqern(r(0)) \lesssim 1$ we conclude $m^{-1}(\tau - \tauzero) \lesssim \sqrt{ 1 + \left| \trapern \right|}$ and therefore
	\begin{equation} \label{eqn_case13_ern}
		\Osqern(r(0)) \lesssim \frac{1}{1 + \left| \trapern \right|} \lesssim \frac{m^2}{(\tau - \tauzero)^2}.
	\end{equation}
	Using the extremal analogue of Lemma~\ref{E_eps_bound_schw}, we find for the energy
	\begin{equation} \label{eqn::Eboundextremal}
		E \lesssim \frac{\psuppconst}{1 + \left| \trapern \right|} \lesssim \psuppconst \frac{m^2}{(\tau - \tauzero)^2}.
	\end{equation}
	Using conservation of energy~\eqref{Schw::energy_conservation} again, we readily conclude
	\begin{equation}
		\left| p^r \right| \leq E \lesssim \frac{m^2}{(\tau - \tauzero)^2} \psuppconst.
	\end{equation}
	To bound the $p^{t^*}$-component, we proceed as in Case~1.2 above and apply Lemma~\ref{ptstar_bounds_ern} together with the first inequality in equation~\eqref{eqn::Eboundextremal} to find that the bound
	\begin{equation}
		p^{t^*} \lesssim \left(1 + \frac{m^2}{r^2} \left| \trapern \right| \right) E \lesssim \psuppconst
	\end{equation}
	holds for all $r \geq m$. For $r > m$ we may alternatively combine Lemma~\ref{ptstar_bounds_ern} with the second inequality in equation~\eqref{eqn::Eboundextremal} to conclude
	\begin{equation}
		p^{t^*} \lesssim \frac{E}{\Osqern} \lesssim \frac{1}{\Osqern} \frac{m^2}{(\tau - \tauzero)^2} \psuppconst.
	\end{equation}
	Like in Case~1.2 we claim that the bound $p^{t^*} \lesssim \psuppconst$ is sharp up to a constant for generic initial data along the event horizon and refer to Section~\ref{psupporteventhorizon} for a discussion. To obtain a bound on the angular momentum components, we apply the extremal analogue of Lemma~\ref{E_eps_bound_schw} together with the definition of the trapping parameter in Definition~\ref{def_eps_def} to find
	\begin{equation} \label{eqn_case13_L_ern}
		\frac{L^2}{m^2} \lesssim \left( 1 + \left| \trapern \right| \right) E^2 \lesssim E \psuppconst \lesssim \frac{m^2}{(\tau - \tauzero)^2} \psuppconst^2.
	\end{equation}
	From the definition of angular momentum we therefore find
	\begin{equation}
		\left| \pslash \right|_{\gslash} = \frac{L}{r} \lesssim \frac{m}{\tau-\tauzero} \frac{m}{r} \psuppconst.
	\end{equation}
	Now assume that $\tau(x) > \bar{\tau} > \tauzero$. In order to prove that $(x(0),p(0)) \in \badsetaplarge_{\constbsl,\bar{\delta}}$ with $\bar{\delta} = \bar{\tau}^{-1}$ and $\constbsl=1$, note again that from~\eqref{eqn_case13_ern} it follows $r(0) \leq (1+\bar{\delta})m$ if $\bar{\tau}$ is large enough. In order to bound the $p^r(0)$-component, apply the extremal analogue of Lemma~\ref{E_eps_bound_schw} combined with~\eqref{eqn_case13_timebound} to find
	\begin{equation}
		\left| p^r(0) \right| \leq E \lesssim \frac{1}{1+\left| \trapschw \right|} \psuppconst \lesssim \sqrt{\Osqern(r(0))} \frac{m}{\tau-\tauzero} \psuppconst.
	\end{equation}
	If we assume that $\bar{\tau}$ is large enough, we may conclude
	\begin{equation}
		\left| p^r(0) \right| \leq  \sqrt{\Osqern(r(0))} \psuppconst.
	\end{equation}
	Note that this bound suffices according to the definition of $\badsetaplarge_{\constbsl,\bar{\delta}}$ in Definition~\ref{def_badsetapprox} since we may assume $p^r(0) \leq 0$ in the current case. To bound the angular component over $\Sigma_0$, we make use of~\eqref{eqn_case13_L_ern} to find
	\begin{equation}
		\left| \pslash(0) \right|_{\gslash}^2 \lesssim E \psuppconst \lesssim \sqrt{\Osqern(r(0))} \frac{m}{\tau-\tauzero} \psuppconst^2.
	\end{equation}
	If we again assume $\bar{\tau}$ to be large enough, we can conclude
	\begin{equation}
		\left| \pslash(0) \right|_{\gslash} \leq (\Osqern(r(0)))^{\frac{1}{4}} \psuppconst.
	\end{equation}
	
	\paragraph{Case 2: $p^r > 0$ at time $\tau$.}
	In this case we may assume that $p^r > 0$ at time $\tau$. Lemma~\ref{tauestimate_ERN} allows us to conclude the bound
	\begin{equation} \label{Case2ERN}
		\frac{1}{m}(\tau-\tauzero) \lesssim \begin{cases}
			\mathfrak{s} \frac{1}{\sqrt{\Osqern(r(0))}} & m \leq r \leq (1+\delta)m \\
			\mathfrak{s} \frac{1}{\sqrt{\Osqern(r(0))}} + \left( \log \left| \trapern \right| \right)_- & r \geq (1+\delta)m
		\end{cases},
	\end{equation}
	where $\mathfrak{s} = \chi_{(0,\infty)}(p^r(0))$ and $\chi_{(0,\infty)}$ denotes the characteristic function of the set $(0,\infty) \subset \R$. Like in the Schwarzschild case, we therefore distinguish two cases.
	
	\paragraph{Case 2.1: $\gamma$ is almost trapped at the photon sphere.}
	In this case we may assume that the lower bound $r \geq (1+\delta)m$ holds, that $\left| \trapern \right| < 1$ and in addition
	\begin{equation}
		\frac{1}{m} \left( \tau - \tauzero \right) \lesssim \log \left| \trapern \right| .
	\end{equation}
    This immediately implies that
    \begin{equation}
        \left| \trapern \right| \leq e^{-\frac{\decayrateERN}{m} (\tau - \tauzero)}
    \end{equation}
	for an appropriate constant $\decayrateERN > 0$ arising from the constant implicit in inequality~\eqref{Case2ERN}. We again proceed in a manner identical to Case~2.1 in the proof of Proposition~\ref{psupport_prop} to conclude $p \in \ERNsmallsupportsetx$.
 
	\paragraph{Case 2.2: $\gamma$ is initially outgoing and starts close to $\mathcal{H}^+$.}
	In this case we may assume that $p^r(0) > 0$ and we have
	\begin{equation}
		\frac{1}{m}(\tau-\tauzero) \lesssim \frac{1}{\sqrt{\Osqern(r(0))}} ,
	\end{equation}
    which may be rewritten to
    \begin{equation}
        \Osqern(r(0)) \lesssim \frac{m^2}{(\tau-\tauzero)^2}.
    \end{equation}
	We initially proceed in a manner identical to Case~1.2 above and note that if we assume $\tau \geq C'( \tauzero + m )$ with a large enough constant $C' > 0$, we may assume $r(0) < 2m$. As above we conclude
	\begin{equation}
		\left| p^r \right| \leq E \leq \Osqern(r(0)) p^{t^*}(0) \lesssim \frac{m^2}{(\tau-\tauzero)^2} \psuppconst.
	\end{equation}
	To obtain a bound on the $p^{t^*}$-component let us first assume $r \geq m$ and apply Lemma~\ref{ptstar_bounds_ern}, using the fact that $p^r > 0$ at time $\tau$, to find
	\begin{equation}
		p^{t^*} \lesssim \frac{E}{\Osqern(r)} \leq \frac{\Osqern(r(0))}{\Osqern(r)} \psuppconst \leq \psuppconst,
	\end{equation}
	where in the last step we have used that $r \geq r(0)$, since $p^r = p^r(s) > 0$ implies that $p^r(s') > 0$ for all parameter times $s' \in [0, s]$. Of course we have already obtained this bound in the extremal analogue of Lemma~\ref{psupport_bounded} above. If we assume $r > m$, then from Lemma~\ref{ptstar_bounds_ern} we may conclude the better bound
	\begin{equation}
		p^{t^*} \lesssim \frac{E}{\Osqern(r)} \lesssim  \frac{1}{\Osqern(r)} \frac{m^2}{(\tau-\tauzero)^2} \psuppconst.
	\end{equation}
	As in Case~1.2 above we claim that the bound $p^{t^*} \lesssim \psuppconst$ is sharp up to a constant along the event horizon and defer to Section~\ref{psupporteventhorizon} for a discussion. We obtain a bound on the angular momentum components in an identical manner to Case~1.2. Finally, the proof that $(x(0),p(0)) \in \badsetaplarge_{\constbsl,\bar{\delta}}$ with $\bar{\delta} = \bar{\tau}^{-1}$ and $\constbsl=1$ if $\tau(x) > \bar{\tau} > \tauzero$ proceeds in an identical fashion to Case~1.2 above.
\end{proof}

\subsection{Lower bounds for the momentum support} \label{psupporteventhorizon}
In this section we will study the momentum support of a solution along the event horizon and provide the proof of Proposition~\ref{sharpness_prop_rough}. In Proposition~\ref{psupport_propERN} above we have shown that for any $(x,p) \in \ERNsmallsupportset$ the $p^{t^*}$-component satisfies the bound
\begin{equation}
	p^{t^*} \lesssim \psuppconst \min \left( 1, \frac{1}{\Osqern} \frac{m^2}{(\tau(x)-\tauzero)^2} \right).
\end{equation}
Therefore, we have shown that at any positive distance from the event horizon, i.e. $r \geq (1+\delta)m$ for some $\delta > 0$, the $p^{t^*}$-component decays as $\tau(x) \rightarrow \infty$, with a constant that degenerates like $\delta^{-2}$ as $\delta \rightarrow 0$. We will now establish the existence of a family of geodesics with initial data in the sets $\badsetapprox_\delta$ which eventually cross the event horizon at arbitrarily late times $\tau$, while their momentum satisfies $p^{t^*} \sim \psuppconst$ on the event horizon. 

We break up Proposition~\ref{sharpness_prop_rough} into three separate results: Propositions~\ref{properties_of_badsetapprox}-\ref{ptstar_size_slowsupport}. Proposition~\ref{properties_of_badsetapprox} establishes some basic properties of the family $\badsetapprox_{c_1,c_2,\delta}$. Then we prove in Proposition~\ref{sharpness_lemma} that the sets $\badsetapprox_\delta$ and $\badsettau$ are included in one another if we choose $\tauslow \sim m \delta^{-1}$ and make an appropriate choice for all other constants involved in the definition. Roughly speaking, this implies that geodesics with initial data in $\badsetapprox_\delta$ populate the set $\slowsupportset$ along the event horizon. Proposition~\ref{sharpness_lemma} also shows that all defining constants may be chosen in such a way that $\badsetapprox_\delta \subset \badsetaplarge_\delta$ and in addition every point $(x,p) \in \badsetaplarge_\delta$ satisfies Assumption~\ref{assumption_support}. Finally, we estimate the volume of the set $\slowsupportsetx$ along the event horizon in Proposition~\ref{ptstar_size_slowsupport}. Taken together, these propositions constitute the proof of Proposition~\ref{sharpness_prop_rough}.

\subsubsection{Properties of $\badsetapprox$}
In this subsection we prove a quantitative estimate for the volume of the set $\badsetapprox_{\delta}$ and establish that every $p = (p^{t^*},p^r,\pslash) \in \badsetapprox_{\delta}$ satisfies $p^{t^*} \sim \psuppconst$.

\begin{prop}[Properties of $\badsetapprox$] \label{properties_of_badsetapprox}
	Let $0 < c_1<c_2$ and $0< \delta < \frac{1}{2}$. Then every $(x,p) \in \badsetapprox_{c_1,c_2,\delta}$ expressed in $(t^*,r)$-coordinates as $(x,p) = (t^*,r,\omega,p^{t^*},p^r,\pslash)$ satisfies $p^{t^*} \sim_{c_1,c_2} \psuppconst$. Furthermore, we have $\vol \badsetapprox_{c_1,c_2,\delta} = \int_{\badsetapprox_{c_1,c_2,\delta}} 1 \, d \mu_x \sim_{c_1,c_2} \delta^3 m^3 \psuppconst^2$.
\end{prop}
\begin{proof}
	Let $(x,p) = (t^*,r,\omega,p^{t^*},p^r,\pslash) \in \badsetapprox$. We first obtain a bound on energy and angular momentum by using conservation of energy to find
	\begin{gather}
		c_1 \Osqern \psuppconst \leq E = \sqrt{(p^r)^2 + \Osqern \left| \pslash \right|_{\gslash}^2} \leq \sqrt{2} c_2 \Osqern \psuppconst, \label{proof_badsetapprox_1} \\
		c_1 \sqrt{\Osqern} \psuppconst \leq \left| \pslash \right|_{\gslash} \leq \frac{L}{m} \leq 2 \left| \pslash \right|_{\gslash} \leq 2 c_2 \sqrt{\Osqern} \psuppconst. \label{proof_badsetapprox_2}
	\end{gather}
	In order to obtain a bound on $p^{t^*}$ we need to distinguish the two cases that $p^r > 0$ and $p^r \leq 0$. Let us turn to the case that $p^r \leq 0$ first. We apply Lemma~\ref{express_ptstar} to find
	\begin{equation} \label{ptstar_bound_equn_proof_badset}
		\frac{1}{4} \frac{c_1^2}{c_2} \psuppconst \leq p^{t^*} = \left| p^r \right| + \frac{\left| \pslash \right|_{\gslash}^2}{E + \left| p^r \right|} \leq \left( c_2 \delta^2 + \frac{c_2^2}{c_1} \right) \psuppconst,
	\end{equation}
	where we made use of the fact that $\Osqern \leq \delta^2$. Let us now turn to the remaining case that $p^r > 0$. In this case Lemma~\ref{express_ptstar} allows us to conclude
	\begin{equation}
		c_1 \psuppconst \leq \frac{E}{\Osqern} \leq p^{t^*} = \frac{E + \left[ 1-\Osqern \right] p^r}{\Osqern} \leq \frac{2 E}{\Osqern} \leq 4 c_2 \psuppconst.
	\end{equation}
	Next fix a point $x \in \Sigma_0$ with $m \leq r \leq (1+\delta)m$ and let us compute $\vol \badsetapprox_x = \vol (\badsetapprox \cap \mathcal{P}_x)$. We parameterise the null-cone by using the coordinates induced by $(t^*,r)$-coordinates and eliminating the $p^{t^*}$-component of the momentum, as discussed in Section~\ref{sec_parametrising_massshell}. We further make the change of variables $(t^*, r, \omega, p^r, \pslash) \mapsto (t^*, r, \omega, p^r, L, \pslashangle)$ introduced in Section~\ref{sec_angular_coords_massshell} and we recall the explicit expression of the volume form $\dmux$ derived in equation~\eqref{eqn_volume_form_pslashangle}. We then readily find
	\begin{equation}
		\vol \badsetapprox_x = \int_{\badsetapprox_x} 1 \, \dmux \sim \int_{c_1 m \psuppconst \sqrt{\Osqern(r)}}^{c_2 m \psuppconst \sqrt{\Osqern(r)}} \int_{0}^{c_2 \psuppconst \Osqern(r)} \frac{1}{m^2} \frac{L}{E} \, d p^r d L.
	\end{equation}
	According to~\eqref{proof_badsetapprox_1} and~\eqref{proof_badsetapprox_2} above we have that in the set $\badsetapprox_x$ the bound
	\begin{equation}
		\frac{1}{m} \frac{L}{E} \sim_{c_1,c_2} \frac{1}{\sqrt{\Osqern}}
	\end{equation}
	holds, so that we may readily compute
	\begin{equation}
		\int_{\badsetapprox_x} 1 \, \dmux \sim_{c_1,c_2} \Osqern \psuppconst^2.
	\end{equation}
	Recalling that the hypersurface $\Sigma_0$ coincides with the set $\{ t^* = 0 \}$ when $r \leq (1+\delta)m$, we find that
	\begin{equation}
		\vol \badsetapprox = \int_{m}^{(1+\delta)m} \int_{\sphere} \int_{\badsetapprox_x} r^2 \, \dmux d \omega dr \sim_{c_1,c_2} m^2 \psuppconst^2 \int_{m}^{(1+\delta)m} \Osqern \, dr \sim_{c_1,c_2} \delta^3 m^3 \psuppconst^2.
	\end{equation}
	This concludes the proof.
\end{proof}

\subsubsection{Geodesics with initial data in $\badsetaplarge_\delta$ have lower-bounded momenta when crossing the event horizon}

We next show that $\badsetapprox_\delta \subset \badsetaplarge_\delta$ for all $0 < \delta < \frac{1}{2}$ and that $\badsettau \approx \badsetapprox_{\delta}$, if  $\tauslow \sim \delta^{-1}$ and all remaining constants in the definition are chosen appropriately. Heuristically, the second approximate equality means that the family of geodesics with initial data in the set $\badsetapprox_\delta$, when intersected with the event horizon, spans a subset of momenta which satisfy $p^{t^*} \sim 1$. This is the crucial part to establishing a lower bound for the momenta in the support of a solution to the massless Vlasov equation, which in turn is crucial to showing a lower bound for the transversal derivative of the energy-momentum tensor. The inclusion $\badsetapprox_\delta \subset \badsetaplarge_\delta$ on the other hand heuristically means that the initial data required to generate these slowly in-falling geodesics is in fact contained within the set $\badsetaplarge_\delta$ and therefore does not violate our assumption of compact initial support.

\begin{prop}[$\badsettau$ and $\badsetapprox_\delta$ are comparable] \label{sharpness_lemma}
	There exist dimensionless constants $C_1,C_2, \ubar{c}_1, \ubar{c}_2, \bar{c}_1, \bar{c}_2, \constbsl > 0$ independent of $\psuppconst,\rsuppconst$ such that for any $0 < \bar{\delta} < \frac{1}{2}$, there exist constants $\tauslow > m$ and $0< \ubar{\delta} < \bar{\delta}$ such that $\tauslow \sim m \bar{\delta}^{-1} \sim m \ubar{\delta}^{-1}$ and the following holds
	\begin{equation}
		\badsetapprox_{\ubar{c}_1, \ubar{c}_2, \ubar{\delta}} \subset \badsetallconst \subset \badsetapprox_{\bar{c}_1, \bar{c}_2, \bar{\delta}} \subset \badsetaplarge_{\constbsl,\bar{\delta}}.
	\end{equation}
	Furthermore, each point $(x,p) \in \badsetaplarge_{\constbsl,\bar{\delta}}$ satisfies Assumption~\ref{assumption_support} and the choice of constant $\constbsl$ is compatible with the choice made in Proposition~\ref{psupport_propERN}.
\end{prop}
\begin{proof}
	We first show the existence of constants such that $\badsetallconst \subset \badsetapprox_{c_1,c_2,\delta}$, then we show that $\badsetapprox_{c_1,c_2,\delta} \subset \badsetaplarge_{\constbsl,\delta}$ and finally we show that the reverse inclusion $\badsetapprox_{c_1,c_2,\delta} \subset \badsetallconst$ holds for an appropriate and compatible choice of constants. For simplicity we will often write $\badsettau = \badsetallconst$, $\badsetapprox_{\delta} = \badsetapprox_{c_1,c_2,\delta}$ and $\badsetaplarge_{\constbsl,\delta} = \badsetaplarge_{\delta}$ when no confusion can arise.
	
	\paragraph{Step 1: The inclusion $\badset \subset \badsetapprox$.}
	We begin by showing the existence of constants such that $\badsetallconst \subset \badsetapprox_{c_1,c_2,\delta}$. Let $0 < \delta < \frac{1}{2}$ and $\gamma: [0,s] \rightarrow \Mern$ be an affinely parametrised future-directed null geodesic segment expressed in $(t^*,r)$-coordinates. We will denote $\gamma(s') = x(s')$ and $\dot{\gamma}(s') = p(s')$ for $s' \in [0,s]$. Assume $(x(0),p(0)) \in \badsettau$ and that we have chosen the parameter time $s$ such that $x(s) \in \mathcal{H^+}$. We now aim to derive bounds on $r(0)$ and the components of $p(0)$. Let us begin by deriving estimates on the energy, angular momentum and trapping parameter of the geodesic. As usual we abbreviate $E = E(\gamma,\dot{\gamma})$ and $L = L(\gamma,\dot{\gamma})$.

	Recall that the relation $E = - p^r(s) = \left| p^r(s) \right|$ is implied by conservation of energy along the event horizon $\mathcal{H}^+$. Using the definition of $\slowsupportset$ and this relation we therefore immediately obtain the bounds
	\begin{equation}
		\begin{gathered}
			C_1 \frac{\psuppconst}{\tau(x(s))^2} \leq E \leq C_2 \frac{\psuppconst}{\tau(x(s))^2}, \\
			C_1 \frac{\psuppconst}{\tau(x(s))} \leq \frac{L}{m} \leq C_2 \frac{\psuppconst}{\tau(x(s))},
		\end{gathered}
	\end{equation}
	We conclude further that
	\begin{equation} \label{proof_slow_epsestimate}
		\left( \frac{C_1}{C_2} \right)^2 \tau(x(s))^2 \leq \frac{1}{m^2} \frac{L^2}{E^2} \leq \left( \frac{C_2}{C_1} \right)^2 \tau(x(s))^2.
	\end{equation}
	Note that by definition of $\slowsupportset$, we have $\tau(x(s)) \geq \tauslow$. If we assume $\tauslow$ is large enough, then inequality~\eqref{proof_slow_epsestimate} allows us to conclude $\trapern < - \delta^{-2}$. Let us henceforth assume that $\tauslow$ satisfies this bound.
 
	We next want to obtain a relationship between the time of arrival at the event horizon $\tau(x(s))$ and the values $r(0)$ and $p(0)$. Combined with our bound on the trapping parameter $\trapern$ this will allow us to conclude a bound on $\Osqern(r(0))$. First note that since $\trapern < 0$ the geodesic must certainly remain inside the photon sphere for all times, so $m \leq r(s') < 2m$ for $s' \in [0,s]$. In fact, from inequality~\eqref{epsbound} it immediately follows that for all times
	\begin{equation}
		\Osqern(r(s')) \leq \frac{1}{1 + \left| \trapern \right|} \frac{r(s')^2}{16m^2} \leq  \frac{1}{1 + \left| \trapern \right|} \frac{1}{4} < \delta^{-2},
	\end{equation}
	where we have used that $r(s') < 2m$ for all $s' \in [0,s]$ and our assumption on $\tauslow$. Therefore
	\begin{equation}
		m < r(s') < (1+\delta)m
	\end{equation}
	for all affine parameter times $s' \in [0,s]$. In particular it follows that $m < r(0) < (1+\delta)m < \frac{3}{2}m$. Next note that since $r(s') < \frac{3}{2}m$ for all times, we have $\tau(x(s')) = t^*(s')$ and we may apply the results of Lemma~\ref{tstar_estimate_ern} directly. We need to distinguish between the two cases that $p^r(0) > 0$ and $p^r(0) \leq 0$. Let us first assume that $p^r(0) \leq 0$, so that $p^r(s') < 0$ for all $s' \in [0,s]$. Therefore we may apply Lemma~\ref{tstar_estimate_ern} with $\delta_1 = \frac{1}{2}$ and find
	\begin{equation}
		\tau(x(s)) = \tau(x(s)) - \tau(x(0)) \sim \left( 1 + \left| \trapern \right| \right) \sqrt{\Osqern(r(0))}.
	\end{equation}
	Using inequality~\eqref{proof_slow_epsestimate} combined with the fact that $1 + \left| \trapern \right| = \frac{1}{16m^2} \frac{L^2}{E^2}$ we may now conclude
	\begin{equation}
		\frac{1}{\tau(x(s))^2} \sim \Osqern(r(0)).
	\end{equation}
	Let us now turn to the remaining case that $p^r(0) > 0$. The geodesic must necessarily be initially outgoing, then scatter off the photon sphere and eventually fall into the black hole. Let us denote by $s^* > 0$ the affine time parameter for which $r(s^*) = r_{\text{min}}^-(\trapern)$ or equivalently $p^r(s^*) = 0$. Note that $0 < s^*$ and $r(0) < r_{\text{min}}^-(\trapern)$. In order to obtain a bound $\Osqern(r(0))$ let us now distinguish further between the two cases that $r(0) < \frac{1}{2} (r_{\text{min}}^-(\trapern) + m)$ and $r(0) \geq \frac{1}{2} (r_{\text{min}}^-(\trapern) + m)$. In the latter case, we may immediately conclude
	\begin{equation}
		\Osqern(r(0)) \sim \Osqern(r_{\text{min}}^-(\trapern)) \sim \frac{1}{1+\left| \trapern \right|} \sim \frac{1}{\tau(x(s))^2},
	\end{equation}
	where we have made use of inequality~\eqref{proof_slow_epsestimate}. In the former case that $r(0) < \frac{1}{2} (r_{\text{min}}^-(\trapern) + m)$ an application of Lemma~\ref{tstar_estimate_ern} with $\delta_1 = \frac{1}{2}$ yields
	\begin{equation}
		\begin{gathered}
			\tau(x(s^*)) = \tau(x(s^*)) - \tau(x(0)) \sim \frac{1}{\sqrt{\Osqern(r(0))}}, \\
			\tau(x(s)) - \tau(x(s^*)) \sim \left( 1 + \left| \trapern \right| \right) \sqrt{\Osqern(r(0))}.
		\end{gathered}
	\end{equation}
	Recall that $\left( 1 + \left| \trapern \right| \right) \Osqern(r(0)) \lesssim 1$ so that
	\begin{equation}
		\tau(x(s)) \sim \frac{1}{\sqrt{\Osqern(r(0))}} + \left( 1 + \left| \trapern \right| \right) \sqrt{\Osqern(r(0))} \sim \frac{1}{\sqrt{\Osqern(r(0))}}.
	\end{equation}
	We therefore find that
	\begin{equation}
		\frac{1}{\tau(x(s))^2} \sim \Osqern(r(0))
	\end{equation}
	holds for both signs of $p^r(0)$. Finally we turn to deriving bounds on $p(0)$. We find
	\begin{equation}
		\left| p^r(0) \right| \leq E \leq C_2 \frac{\psuppconst}{\tau(x(s))^2} \lesssim \Osqern(r(0)) \psuppconst.
	\end{equation}
	For the angular momentum component we note $\left| \pslash(0) \right|_{\gslash} = \frac{L}{r(0)}$, so that
	\begin{equation}
		\sqrt{\Osqern(r(0))} \psuppconst \lesssim \frac{C_1}{2} \frac{\psuppconst}{\tau(x(s))} \leq \left| \pslash(0) \right|_{\gslash} \leq C_2 \frac{\psuppconst}{\tau(x(s))} \lesssim \sqrt{\Osqern(r(0))} \psuppconst.
	\end{equation}
	Therefore for any choice of $C_1<C_2$ we may conclude that $(x(0),p(0)) \in \badsetapprox_{c_1,c_2,\delta}$ for an appropriate choice of $\tauslow, c_1, c_2$. This concludes the proof that $\badsetallconst \subset \badsetapprox_{c_1,c_2,\delta}$.

    \paragraph{Step 2: The inclusion $\badsetapprox \subset \badsetaplarge$.}
    Let $C_1 < C_2$ as above. Let us now show that we may choose constants $C_1 < C_2$ (and thereby $c_1,c_2$) as well as $\constbsl > 0$ such that for all $0 < \delta < \frac{1}{2}$, the inclusion $\badsetapprox_{c_1,c_2,\delta} \subset \badsetaplarge_{\constbsl,\delta}$ holds and every $(x,p) \in \badsetaplarge_{\constbsl,\delta}$ satisfies Assumption~\ref{assumption_support}. First note that the second condition will be satisfied if $\sqrt{\delta} \constbsl \leq 1$. Since we assume $\delta < \frac{1}{2}$, it suffices to assume $\constbsl \leq \sqrt{2}$. To see that $\badsetapprox_{c_1,c_2,\delta} \subset \badsetaplarge_{\constbsl,\delta}$, note that for the radial and angular momentum components it suffices to show that we may choose $c_2 \leq \min(1,\sqrt{2} \constbsl)$. Therefore, the bounds for the radial and angular momentum components will be satisfied for any $\constbsl$ satisfying $1 \leq \sqrt{2} \constbsl \leq 2$. For the $p^{t^*}$-component, recall equation~\eqref{ptstar_bound_equn_proof_badset}, wherein we showed the bound
	\begin{equation} \label{eqn:ptstartbound_b_sets}
		p^{t^*} \leq \max \left( 4c_2, c_2 \delta^2 + \frac{c_2^2}{c_1}  \right) \psuppconst .
	\end{equation}
    By choosing $C_2$ sufficiently small, we may assume without loss of generality that 
	\begin{equation}
		c_2 < \frac{1}{4}, \quad \frac{c_2^2}{c_1} < \frac{1}{4}.
	\end{equation}
	From these bounds and inequality~\eqref{eqn:ptstartbound_b_sets} it easily follows that $p^{t^*} \leq \psuppconst$, which concludes the proof.
	
	\paragraph{Step 3: The inclusion $\badsetapprox \subset \badset$.}
	Let us now show that there exists a compatible choice of $C_1,C_2$, as well as $c_1,c_2$ such that for any $\tauslow > m$ suitably large there exists a $\delta > 0$ such that the reverse inclusion $\badsetapprox_{c_1,c_2,\delta} \subset \badsetallconst$ holds. Suppose that $\gamma: [0,s] \rightarrow \Mern$ is a maximally defined geodesic segment as above with $(x(0),p(0)) \in \badsetapprox_\delta$. We begin by bounding the energy and angular momentum of $\gamma$. Conservation of energy implies
	\begin{equation} \label{proof_approx_badset_below_E}
		c_1 \Osqern(r(0)) \psuppconst \leq \sqrt{\Osqern(r(0))} \left| \pslash(0) \right|_{\gslash} \leq E \leq \sqrt{2} c_2 \Osqern(r(0)) \psuppconst.
	\end{equation}
	Similarly we immediately obtain the bound
	\begin{equation} \label{proof_approx_badset_below_L}
		c_1 \sqrt{\Osqern(r(0))} \psuppconst \leq \frac{L}{m} \leq 2 c_2 \sqrt{\Osqern(r(0))} \psuppconst.
	\end{equation}
	Note that for $0 < c_1 < c_2$ chosen arbitrarily, we may always choose $0 < \delta < 1$ so that $\frac{1}{16 m^2} \frac{L^2}{E^2} > 3$. Using the definition of the trapping parameter we may therefore assume that $\trapern = \trapern(\gamma) < -2$ and
	\begin{equation} \label{proof_approx_badset_below_eps}
		\frac{(c_1)^2}{(c_2)^2} \frac{1}{\Osqern(r(0))} \sim \left| \trapern \right|.
	\end{equation}
	Next we want to show that $\gamma$ indeed crosses the event horizon after a finite time and obtain a relationship between the time of crossing and $r(0)$. Let us distinguish the two cases that $p^r(0) > 0$ and $p^r(0) \leq 0$. In the latter case, we note that necessarily $p^r(s') < 0$ for all $s' \in (0,s]$ so that the geodesic must fall into the black hole. Since we have assumed $\gamma$ to be maximally defined, we find $r(s) = m$. We argue in an identical manner to above to find
	\begin{equation} \label{eqn:tau_estimate}
		\tau(x(s)) \sim m (1 + \left| \trapern \right|) \sqrt{\Osqern(r(0))}.
	\end{equation}
	Using equations~\eqref{proof_approx_badset_below_eps} and~\eqref{eqn:tau_estimate} we may argue similarly to Step 1 above and conclude the estimate
	\begin{equation} \label{proof_badset_approx_below_time}
		\frac{m^2}{\tau(x(s))^2} \sim \Osqern(r(0)) .
	\end{equation}
    In order to show that $(x(0),p(0)) \in \badsettau$ we need to consider the values of $\tau(x(s))$ and of $p^r(s), \pslash(s)$. Recall that we consider $\tauslow > m$ to be given and $C_1 < C_2$ to satisfy the assumptions placed on these constants in the argument above. First note that by inequality~\eqref{proof_badset_approx_below_time} we find
	\begin{equation}
		\tau(x(s))^2 \gtrsim \frac{m^2}{\delta^2}.
	\end{equation}
	Assuming we choose $c_1,c_2$ as a function of $C_1,C_2$ only, we may choose $0 < \delta < 1$ small enough as a function of $\tauslow$ so that $\tau(x(s)) > \tauslow$. Next note that since $r(s) = m$ we have $\left| \pslash(s) \right|_{\gslash} = \frac{L}{m}$ and $\left| p^r(s) \right| = E$. Combining inequalities \eqref{proof_approx_badset_below_E} and \eqref{proof_approx_badset_below_L} with inequality~\eqref{proof_badset_approx_below_time} we see that $p^r(s)$ and $\pslash(s)$ satisfy the bounds
	\begin{equation}
		C_1 \psuppconst \frac{m}{\tau(x(s))} \leq \left| \pslash(s) \right|_{\gslash} \leq C_2 \psuppconst \frac{m}{\tau(x(s))}, \quad C_1 \psuppconst \frac{m^2}{\tau(x(s))^2} \leq \left| p^r(s) \right| \leq C_2 \psuppconst \frac{m^2}{\tau(x(s))^2}
	\end{equation}
	if we choose the constants $c_1, c_2$ as appropriate functions of $C_1,C_2$. This proves the existence of $C_1,C_2$ compatible with the assumptions above, as well as $c_1,c_2$ chosen as a function of $C_1,C_2$ and $\delta$ chosen as a function of $\tauslow,c_1,c_2$ such that $\badsetapprox_{c_1,c_2,\delta} \subset \badsetallconst$. Finally note that we only require an upper bound on $C_2$, so that we may choose it sufficiently small to ensure that $\slowsupportset \subset \smallsupportset$.
\end{proof}

\subsubsection{Properties of $\slowsupportset$}
In this subsection, we establish a bound on the phase-space volume of each fibre of the set $\slowsupportset$ and prove in addition that every momentum $p = (p^{t^*},p^r,\pslash) \in \slowsupportsetx$ satisfies $p^{t^*} \sim \psuppconst$.

\begin{prop}[Properties of $\slowsupportset$] \label{ptstar_size_slowsupport}
	In the notation of Proposition~\ref{sharpness_lemma}, the constants $C_1, C_2$ may be chosen such that the conclusions of Proposition~\ref{sharpness_lemma} hold true and in addition for every $(x,p) \in \slowsupportset$ expressed in $(t^*,r)$-coordinates as $(x,p) = (t^*,r,\omega,p^{t^*},p^r,\pslash)$ the following bound holds
	\begin{equation}
		\frac{(C_1)^2}{2 C_2} \psuppconst \leq p^{t^*} \leq \psuppconst.
	\end{equation}
	Furthermore, we have $\vol \slowsupportsetx \sim_{C_1,C_2} \psuppconst^2 m^2 \tau(x)^{-2}$.
\end{prop}
\begin{proof}
	Recall from the proof of Proposition~\ref{sharpness_lemma} that we chose the constants $C_1,C_2$ by first choosing $C_2$ small enough, concretely $C_2 < \bar{C}$ for some constant $\bar{C} < \frac{1}{8}$ defined in the proof and then choosing $C_1$ so that in particular the following bound holds:
	\begin{equation} \label{C1C2bound_slowset_proof}
		2C_1 < C_2, \quad \frac{(C_2)^2}{C_1} \leq \frac{(C_2)^{12}}{(C_2)^{11}} \leq \frac{1}{8},
	\end{equation}
	Recall also that in the notation of Proposition~\ref{sharpness_lemma}, the constants $C_1,C_2$ were chosen independently of $\bar{\delta}$. Let us now in addition assume that $C_2$ also satisfies the lower bound $C_2 > \frac{1}{2} \bar{C}$. Recall from Lemma~\ref{express_ptstar} that along the event horizon $\mathcal{H}^+$ we have $- p^r = E$ and
	\begin{equation}
		p^{t^*} = E + \frac{\left| \pslash \right|_{\gslash}^2}{2E}.
	\end{equation}
	This immediately allows us to conclude the bounds
	\begin{equation}
		\left( \frac{C_1}{\tau(x)^2} + \frac{(C_1)^2}{2 C_2} \right) \psuppconst \leq p^{t^*} \leq \left(\frac{C_2}{\tau(x)^2} + \frac{(C_2)^2}{2 C_1} \right) \psuppconst.
	\end{equation}
	If we assume $\tau(x) > \tauslow > 1$ then inequality~\eqref{C1C2bound_slowset_proof} immediately implies that $p^{t^*} \leq \psuppconst$. Let us now turn to estimating the volume of $\slowsupportsetx$. Note that the bound $C_2 > \frac{1}{2} \bar{C}$ and inequality~\eqref{C1C2bound_slowset_proof} imply
	\begin{equation} \label{proof_ern_slow_volume}
		\bar{C}^2 < 4 (C_2)^2 < 2 C_1 < C_2 < \bar{C}. 
	\end{equation}
	Let us note at this point that for $C_1,C_2$ satisfying inequality~\eqref{proof_ern_slow_volume} the following relations may be verified to hold
	\begin{gather}
		\frac{\bar{C}}{2} \left( \sqrt{C_2} - \sqrt{C_1} \right) < C_2 - C_1 < \sqrt{C_2} - \sqrt{C_1}, \label{proof_slow_CC_1}\\
		\frac{\bar{C}}{4} < \frac{1}{2} C_2 < C_2 - C_1 < C_2 < \bar{C}, \label{proof_slow_CC_2}\\
		\frac{C_2}{C_1} < \frac{1}{2 C_2} < \frac{1}{\bar{C}} .\label{proof_slow_CC_3}
	\end{gather}
	As in the proof of Proposition~\ref{properties_of_badsetapprox} above, we parametrise the null-cone using $(t^*,r)$-coordinates by eliminating the $p^{t^*}$-coordinate using the mass-shell relation. We again introduce the change of coordinates $\pslash \mapsto (L,\pslashangle)$ defined in Section~\ref{sec_angular_coords_massshell}. Therefore an element $p \in \slowsupportsetx$ is now parametrised by $(p^r,L,\pslashangle)$. For any $x \in \mathcal{H}^+$ we may therefore parametrise the set $\slowsupportsetx$ in the above coordinates as the set of points $p$ which satisfy
	\begin{equation}
		\begin{gathered}
			C_1 \frac{\psuppconst}{\tau(x)^2} \leq \left| p^r \right| \leq C_2 \frac{\psuppconst}{\tau(x)^2}, \\
			m C_1 \frac{\psuppconst}{\tau(x)} \leq L \leq m C_2 \frac{\psuppconst}{\tau(x)}, \\
			0 \leq \pslashangle < 2\pi.
		\end{gathered}
	\end{equation}
	Recalling $r=m$ and $\left| p^r \right| = E$ here and using the explicit expression for $\dmux$ derived in equation~\eqref{eqn_volume_form_pslashangle},
	\begin{equation}
		\int_{\slowsupportsetx} 1 \, \dmux = \int_{ \frac{C_1 \psuppconst}{\tau(x)^2}}^{ \frac{C_2 \psuppconst}{\tau(x)^2}} \int_{ \frac{m C_1 \psuppconst}{\tau(x)}}^{ \frac{m C_2 \psuppconst}{\tau(x)}} \int_0^{2 \pi} \frac{1}{m^2} \frac{L}{E} \, d \pslashangle d L d E.
	\end{equation}
	The definition of $\slowsupportset$ and relation~\eqref{proof_slow_CC_3} derived above readily implies the bound
	\begin{equation}
		\psuppconst \lesssim \frac{C_1}{C_2} \psuppconst \leq \frac{1}{m^2} \frac{L^2}{E} \leq \frac{C_2}{C_1} \psuppconst \leq 8 \psuppconst.
	\end{equation}
	Therefore we conclude the relation
	\begin{equation}
		\frac{1}{m} \frac{L}{E} = \sqrt{\frac{1}{m^2} \frac{L^2}{E}} \frac{1}{\sqrt{E}} \sim \frac{\sqrt{\psuppconst}}{\sqrt{E}}.
	\end{equation}
	Finally this allows us to conclude
	\begin{equation}
		\begin{aligned}
			\int_{\slowsupportsetx} 1 \, \dmux &\sim \sqrt{\psuppconst} \left( \int_{ \frac{C_1 \psuppconst}{\tau(x)^2}}^{ \frac{C_2 \psuppconst}{\tau(x)^2}} \frac{1}{\sqrt{E}} \, dE \right) 
			\left( \int_{ \frac{C_1 \psuppconst}{\tau(x)}}^{ \frac{C_2 \psuppconst}{\tau(x)}} 1 \, d L \right) \\
			&\sim \frac{\psuppconst^2}{\tau(x)^2} \left( \sqrt{C_2} - \sqrt{C_1} \right) \left( C_2 - C_1 \right) \sim \frac{\psuppconst^2}{\tau(x)^2},
		\end{aligned}
	\end{equation}
	where in the last step we have made use of the relations~\eqref{proof_slow_CC_1} and~\eqref{proof_slow_CC_2} described above. This concludes the proof.
\end{proof}

\subsection{Polynomial decay of moments} \label{estimating_moments_section_ERN}
We are now well equipped to prove Theorem~\ref{maintheoremERNprecise}, showing decay for moments of a solution $f$ to the massless Vlasov equation. We remark that, in direct analogy to Lemma~\ref{lem_boundedness_moments_schw}, we can show a quantitative bound for the volume of the momentum support of a solution to the massless Vlasov equation with compactly supported initial data. Since the proof is virtually identical, we limit ourselves to noting that the conclusion of Lemma~\ref{lem_boundedness_moments_schw} holds verbatim in the extremal case.

% lem_boundedness_moments_ERN

\begin{proof}[Proof of Theorem~\ref{maintheoremERNprecise}]
	Like in the proof of Theorem~\ref{maintheorem_precise}, we parameterise $\mathcal{P}$ using $(t^*,r)$-coordinates by expressing the $p^{t^*}$-coordinate as a function of the remaining ones. If we introduce spherical coordinates on the sphere $\sphere$ the null-cone is then explicitly parametrised by $(t^*,r,\theta,\phi,p^r,p^\theta,p^\phi)$. According to equation~\eqref{dmu_first_comp_schw} the induced volume form on each fibre of the mass-shell $\mathcal{P}_x$ then takes the form
	\begin{equation}
		\dmu =  \frac{r^2 \sin \theta}{\Osqern p^{t^*} - \left( 1 - \Osqern \right) p^r} \, d p^r d p^\theta d p^\phi =  \frac{r^2 \sin \theta}{E} \, d p^r d p^\theta d p^\phi.
	\end{equation}
	By Proposition~\ref{psupport_propERN} we know that analogously to the Schwarzschild case
	\begin{equation}
		\supp(f) \cap \left\{ (x,p) \in \mathcal{P} \; | \; \tau(x) \gtrsim \tauzero + 1 \right\} \subset \ERNtrappedsupportset \cup \ERNsmallsupportset.
	\end{equation}
	Let us assume in this proof that $w \geq 0$, since otherwise we may replace $w$ by $\left| w \right|$. Recall our assumption that $w$ is bounded in $x$ according to Definition~\ref{boundedness_in_x} and apply the extremal analogue of Lemma~\ref{psupport_bounded} to conclude that any $p \in \supp(f(x,\cdot))$ satisfies $\left| p \right| \lesssim \frac{\rsuppconst \psuppconst}{m}$, so that
	\begin{equation}
		W := \max_{(x,p) \in \supp(f)} \left| w(x,p) \right| < \infty.
	\end{equation}
	We immediately conclude the bound
	\begin{equation} \label{proof_ERN_moments_1}
		\int_{\mathcal{P}_x} w f \, \dmux \leq \| f_0 \|_{L^\infty} W \left( \int_{\ERNtrappedsupportsetx} 1 \, \dmux + \int_{\ERNsmallsupportsetx} 1 \, \dmux \right) .
	\end{equation}
	Let us now estimate the two integrals on the right hand side of equation~\eqref{proof_ERN_moments_1}. For the first summand, we proceed in an identical manner to the proof of Theorem~\ref{maintheorem_precise} and conclude
    \begin{equation}
		\int_{\trappedsupportsetx} 1 \, \dmux \lesssim \rsuppconst^{2} \psuppconst^{2} \frac{1}{r^2} e^{-\frac{c}{2m}(\tau(x) - \tauzero)}.
	\end{equation}
    In order to estimate the integral $\int_{\ERNsmallsupportsetx} 1 \, \dmux = \vol \ERNsmallsupportsetx$, we introduce the change of coordinates $(p^r,\pslash) \mapsto (p^r, L, \pslashangle)$ on each fibre $\mathcal{P}_x$, which we defined in Section~\ref{sec_angular_coords_massshell}. If we assume that $p \in \ERNsmallsupportsetx$, the angular momentum $L$ and radial momentum $p^r$ satisfy the bounds
	\begin{equation}
		L = r \left| \pslash \right|_{\gslash} \leq \bigc \rsuppconst \psuppconst \frac{m}{\tau(x) - \tauzero}, \quad \left| p^r \right| \leq \bigc \psuppconst \frac{m^2}{(\tau(x) - \tauzero)^2}.
	\end{equation}
	Combining these bounds with the explicit expression for the form of the volume form $\dmux$ derived in equation~\eqref{eqn_volume_form_pslashangle}, this allows us to estimate the volume of the set $\ERNsmallsupportsetx$ as follows:
	\begin{equation}
		\int_{\ERNsmallsupportsetx} 1 \, \dmux \lesssim \int_0^{2 \pi} \int_{0}^{\bigc \rsuppconst \psuppconst \frac{m}{\tau(x) - \tauzero}} \int_0^{\bigc \psuppconst \frac{m^2}{(\tau(x) - \tauzero)^2} } \frac{1}{r^2} \frac{L}{E} \, d p^r d L d \pslashangle.
	\end{equation}
	The extremal analogue of Lemma~\ref{E_eps_bound_schw} implies $\left( 1 + \left| \trapschw \right| \right) \left| p^r \right| \leq \left( 1 + \left| \trapschw \right| \right) E \lesssim \frac{\rsuppconst^2}{m^2} \psuppconst$ for $(x,p) \in \supp(f)$. Therefore
	\begin{equation}
		\frac{1}{r^2} \frac{L}{E} \lesssim \frac{m}{r^2} \sqrt{1 + \left| \trapschw \right|} \lesssim \frac{\rsuppconst}{r^2} \sqrt{\psuppconst} \frac{1}{\sqrt{\left| p^r \right|}}.
	\end{equation}
	We conclude the bound
	\begin{equation}
		\begin{aligned}
			\int_{\ERNsmallsupportsetx} 1 \, \dmux &\lesssim \frac{\rsuppconst}{r^2} \sqrt{\psuppconst} \left( \int_{0}^{\bigc \rsuppconst \psuppconst \frac{m}{\tau(x) - \tauzero}} 1 \, d L \right) \left( \int_0^{\bigc \psuppconst \frac{m^2}{(\tau(x) - \tauzero)^2} } \frac{1}{\sqrt{\left| p^r \right|}} \, d p^r \right) \\
			&\sim \frac{\rsuppconst^2 \psuppconst^2}{r^2} \frac{m^2}{(\tau(x) - \tauzero)^2}.
		\end{aligned}
	\end{equation}
	As opposed to the Schwarzschild case, the rate of decay for $\vol \ERNsmallsupportsetx$ is only quadratic and therefore strictly slower than the exponential decay rate of $\vol \ERNtrappedsupportsetx$. We have therefore shown the bound
	\begin{equation}
		\int_{\supp(f(x,\cdot))} 1 \, \dmux \leq C \frac{\rsuppconst^2 \psuppconst^2}{r^2} \frac{m^2}{(\tau(x) - \tauzero)^2},
	\end{equation}
	for an appropriate constant $C > 0$. Let us now assume that $f_0$ is initially supported away from the event horizon. Consider a geodesic $\gamma$ such that $(\gamma(0), \dot{\gamma}(0)) \in \supp(f_0)$ like in the proof of Proposition~\ref{psupport_propERN} and assume that $\tau(x) \geq \bar{C}(1 + \tauzero)$ with a sufficiently large constant $\bar{C} > 0$. It follows readily from the proof of Proposition~\ref{psupport_propERN} that there exists a constant $C'$ such that either $\gamma$ populates the set $\ERNtrappedsupportset$ or its initial data must satisfy $\Osqern(r(0)) \lesssim \frac{m^2}{(\tau(x) - \tauzero)^2}$ for $\tau(x) \geq C' (1 + \tauzero)$. For $\tau(x) > \tauzero + \frac{m}{\delta}$ however, this implies $r(0) < (1+\delta)m$, which contradicts our assumption on the support of $f_0$. Therefore we conclude that
	\begin{equation}
		\supp(f) \cap \left\{ (x,p) \in \mathcal{P} : \tau(x) \gtrsim 1 + \tauzero + \frac{m}{\delta} \right\} \subset \ERNtrappedsupportset.
	\end{equation}
	This immediately implies that for $x \in \Mern$ with $\tau(x) \gtrsim 1 + \tauzero + \frac{m}{\delta}$ the better decay rate
	\begin{equation}
		\int_{\supp(f(x,\cdot))} 1 \, \dmux \leq C \frac{\rsuppconst^2 \psuppconst^2}{r^2} e^{-\frac{c}{2m} \left( \tau(x) - \tauzero \right)}
	\end{equation}
	holds. The same remarks about improve the exponential rate of decay apply as in the Schwarzschild case. For the remaining part of the proof, we will no longer assume that $f$ is supported away from the event horizon. 
	
	We now show that we can improve the rate of decay for weights of the form $w = \left| p^r \right|^a (p^{t^*})^b \left| \pslash \right|_{\gslash}^c$ for $a,b,c \geq 0$. First note that for all $(x,p) \in \supp(f)$, the extremal analogue of Lemma~\ref{lem_boundedness_moments_schw} implies
	\begin{equation}
		\left| p^r \right|^a (p^{t^*})^b \left| \pslash \right|_{\gslash}^c \lesssim_{a,b,c} \frac{\psuppconst^{a+b+c} \rsuppconst^c}{r^c}.
	\end{equation}
	From our above computations we conclude
	\begin{equation}
		\int_{\ERNtrappedsupportsetx} \left| p^r \right|^a (p^{t^*})^b \left| \pslash \right|_{\gslash}^c \, \dmux \lesssim_{a,b,c} \frac{\psuppconst^{2+a+b+c} \rsuppconst^{2+c}}{r^{2+c}} e^{- \frac{c}{2m} (\tau(x)-\tauzero)}.
	\end{equation}
	Therefore the contribution from the almost trapped set still decays at an exponential rate. From the definition of the set $\ERNsmallsupportset$ it follows that
	\begin{equation*}
		\int_{\ERNsmallsupportsetx} \left| p^r \right|^a (p^{t^*})^b \left| \pslash \right|_{\gslash}^c \, \dmux \lesssim_{a,b,c} \frac{\psuppconst^{2+a+b+c} \rsuppconst^{2+c}}{r^{2+c}} \left( \frac{m}{\tau(x) - \tauzero} \right)^{2+2a+c} \left[ \min \left( \frac{1}{\Osqern} \frac{m^2}{(\tau(x)-\tauzero)^2}, 1 \right) \right]^b.
	\end{equation*}
	If we assume that $r \geq (1 + \delta) m $ for some $0 < \delta < 1$, then we may conclude
	\begin{equation} \label{inequ_proof_momentsERN}
		\int_{\supp(f(x,\cdot))} \left| p^r \right|^a (p^{t^*})^b \left| \pslash \right|_{\gslash}^c \, \dmux \lesssim_{a,b,c} \frac{1}{\delta^2} \frac{\psuppconst^{2+a+b+c} \rsuppconst^{2+c}}{r^{2+c}} \left( \frac{m}{\tau(x) - \tauzero} \right)^{2(1+a+b)+c},
	\end{equation}
	whereas if we assume $m \leq r \leq (1+\delta)m$ then the factor of $p^{t^*}$ does not improve the rate of decay and we find
	\begin{equation}
		\int_{\supp(f(x,\cdot))} \left| p^r \right|^a (p^{t^*})^b \left| \pslash \right|_{\gslash}^c \, \dmux \lesssim_{a,b,c} \frac{\psuppconst^{2+a+b+c} \rsuppconst^{2+c}}{r^{2+c}} \left( \frac{m}{\tau(x) - \tauzero} \right)^{2(1+a)+c}.
	\end{equation}
	Next let us turn to weights of the form $w = \left| p^v \right|^a \left| p^u \right|^b \left| \pslash \right|_{\gslash}^c$ for $a,b,c > 0$, where we made use of double null coordinates and the induced coordinates on $\mathcal{P}$ to express $\mathcal{P} \ni (x,p) = (u,v,\omega,p^u,p^v,\pslash)$. Making the change of coordinates we compute $p^v = p^{t^*} + p^r$, so that $p^v \lesssim p^{t^*} \lesssim \psuppconst$ for all $(x,p) \in \supp(f)$ by the extremal analogue of Lemma~\ref{psupport_bounded}. Since the asymptotically flat regions of extremal Reissner--Nordstr\"om and Schwarzschild are comparable, an appropriate analogue of Lemma~\ref{tauestimate_far} holds here. We conclude that for all $(x,p) \in \supp(f)$ with $\tau(x) \geq \tauzero$ and $r \geq R$:
	\begin{equation}
		p^u \lesssim \frac{\rsuppconst^2}{r^2} p^v, \quad p^v \lesssim \frac{\rsuppconst^2}{r^2} p^{t^*}.
	\end{equation}
	This allows us to conclude that
	\begin{equation}
		\int_{\supp(f(x,\cdot))} \left| p^v \right|^a \left| p^u \right|^b \left| \pslash \right|_{\gslash}^c \, \dmux \lesssim_{a,b,c} \frac{\rsuppconst^{2 b}}{r^{2 b}} \int_{\supp(f(x,\cdot))} (p^{t^*})^{a+b} \left| \pslash \right|_{\gslash}^c \, \dmux,
	\end{equation}
	which in combination with inequality~\eqref{inequ_proof_momentsERN} shown above immediately allows us to conclude the desired bound.
\end{proof}

\subsection{Sharpness of polynomial decay along the event horizon} \label{sec_proof_lowerboundERN}
In this section we provide the proof of Theorem~\ref{ERN_slowdecayprop}, in which we show a lower bound on the rate of decay for moments of a solution to the massless Vlasov equation on extremal Reissner--Nordstr\"om.

\begin{proof}[Proof of Theorem~\ref{ERN_slowdecayprop}]
	We begin by recalling that we have shown $\badsetapprox_\delta \subset \badsetaplarge_\delta$ for all $0 < \delta < \frac{1}{2}$ in Proposition~\ref{sharpness_lemma}. Therefore, we may assume that $\inf\nolimits_{\badsetapprox_\delta} f_0 > 0$, for otherwise there is nothing to prove. Recall from Proposition~\ref{sharpness_lemma} that by choosing $\tauslow \sim \delta^{-1}$ and all remaining constants appropriately, we may also assume that $\badsettau \subset \badsetapprox_\delta$. This implies $\badsettau \subset \supp(f_0)$ and therefore by definition that $\slowsupportsettau \subset \supp(f)|_{\mathcal{H}^+}$. Note that by definition of $\slowsupportsettau$ and Proposition~\ref{ptstar_size_slowsupport} we have that for any $(x,p) \in \slowsupportsettau$ the bound
	\begin{equation}
		w_{a,b,c}(x,p) = \left| p^r \right|^a \left| \pslash \right|_{\gslash}^b (p^{t^*})^c \geq \frac{(C_1)^{a+b+2c}}{(2C_2)^c} \psuppconst^{a+b+c} \frac{m^{2a + b}}{\tau(x)^{2a+b}}
	\end{equation}
	holds. Now let $x \in \mathcal{H}^+$ with $\tau(x) > \tauslow$. Then since $f \geq 0$ we find that
	\begin{equation}
		\begin{aligned}
			\int_{\mathcal{P}_x} w_{a,b,c} f \, \dmux &\geq \int_{\slowsupportsettaux} w_{a,b,c} f \, \dmux \\
			&\geq \left( \inf_{\slowsupportsettaux} w_{a,b,c} \right) \left( \inf_{\slowsupportsettaux} f \right) \vol(\slowsupportsettaux) \\
			&\gtrsim_{a+b+c} \psuppconst^{2+a+b+c} \frac{m^{2(a+1) + b}}{\tau(x)^{2(a+1)+b}} \inf\nolimits_{\badsettau} f_0 \\
			&\gtrsim_{a+b+c} \psuppconst^{2+a+b+c} \frac{m^{2(a+1) + b}}{\tau(x)^{2(a+1)+b}} \inf\nolimits_{\badsetaplarge_\delta} f_0,
		\end{aligned}
	\end{equation}
	where we made use of the chain of inclusions $\badsettau \subset \badsetapprox_\delta \subset \badsetaplarge_\delta$ in the last step and in the penultimate estimate we used  Proposition~\ref{ptstar_size_slowsupport} to estimate the volume of $\slowsupportsettaux$ and the size of momentum components in the set $\slowsupportsettaux$.
\end{proof}

\subsection{Non-decay of transversal derivatives along the event horizon} \label{section_derivativesERN}
In this section we provide the proof of Theorem~\ref{ERN_nondecaytransversal}. Note carefully that our proof is not based on the existence of a conservation law. Instead we will make careful use of the precise characterisation of the sets $\badsettau$ obtained above.

\begin{proof}[Proof of Theorem~\ref{ERN_nondecaytransversal}]

The argument proceeds along the lines explained in Section~\ref{subsec_growth_overview}. We will assume without loss of generality that $\partial_{t^*} f |_{\mathcal{P}_0} \geq 0$ on the set $\badsetaplarge_\delta$, otherwise replace $f$ with $-f$. Throughout the proof we employ the convention that Latin letters refer to spherical components.

\paragraph{Step 1: Moving the $\partial_r$-derivative inside the energy-momentum tensor.}
We begin by showing that for all points along the extremal event horizon $x \in \mathcal{H}^+$, the following identity holds:
\begin{equation} \label{proof_nondecay_decomposition}
\left( \partial_{r} \int_{\sphere} T^{t^* t^*}[f] \, d \omega \right) \bigg|_{r=m} = \int_{\sphere} T^{t^* t^*}\left[ \frac{p^{t^*}}{\left| p^r \right|} \partial_{t^*} f \right] \, d \omega \bigg|_{r=m} + \mathcal{E},
\end{equation}
where for all $x \in \mathcal{H}^+$ with $\tau(x) > \tauzero$ the error term $\mathcal{E}$ satisfies
\begin{equation} \label{proof_nondecay_errorbound}
    \left| \mathcal{E} \right| \lesssim \psuppconst^4 \rsuppconst^2 \left( \| f_0 \|_{L^{\infty}} + \| \partial_{t^*} f_0 \|_{L^{\infty}} \right) \frac{1}{(\tau(x) - \tauzero)^2}.
\end{equation}
Let us define $\tilde{f} = \frac{p^{t^*}}{\left| p^r \right|} f$ and note that $\tilde{f}$ is well-defined on $\{ (x,p) \in \mathcal{P} : p^r \neq 0 \}$. We remind the reader of the discussion in Section~\ref{rngeometry} and recall that the coordinate system given by $(p^r,\pslash)$ on a fibre $\mathcal{P}_x$ degenerates at points where $r=m$ and $p^r=0$. We will therefore assume in our computations that $p^r \neq 0$. Note however that $\tilde{f}$ is not locally integrable for general $f$ so that the moments $T^{\mu \nu}[\tilde{f}]$ are in general not finite. Therefore we introduce $\tilde{f}_\epsilon = \frac{p^{t^*}}{\left| p^r \right| + \epsilon} f$ for $\epsilon > 0$, a regularised version of $\tilde{f}$ which is pointwise bounded and compactly supported (and therefore locally integrable). Note immediately that along the event horizon, $p^r \leq 0$ so that $\tilde{f}_\epsilon = \frac{p^{t^*}}{-  p^r + \epsilon} f$. We find
\begin{align}
    X \left( \frac{p^{t^*}}{\epsilon - p^r} \right) = - \left( \Gamma^{t^*}_{\alpha \beta} p^\alpha p^\beta \right) \frac{1}{\epsilon - p^r} - \left( \Gamma^{r}_{\alpha \beta} p^\alpha p^\beta \right) \frac{p^{t^*}}{(p^r-\epsilon)^2},
\end{align}
where $\Gamma^{\mu}_{\alpha \beta}$ denote the Christoffel symbols of the extremal Reissner--Nordstr\"om metric in $(t^*,r)$-coordinates. An explicit computation shows that along the event horizon
\begin{equation}
    X \left( \frac{p^{t^*}}{\epsilon - p^r} \right) \Bigg|_{r=m} = - \frac{2}{m} \frac{\left| p^r \right|}{\epsilon + \left| p^r \right|} \left( p^r + p^{t^*} \right).
\end{equation}
We next recall the standard identity $\nabla_\mu T^{\mu \nu} [\tilde{f}_\epsilon] = \int_{\mathcal{P}_x} p^{\nu} X(\tilde{f}_\epsilon) \, d \mu_x$, see~\cite{martin,riosecosarbach,sarbachzannias} for an in-depth discussion on how to relate derivatives of the energy momentum tensor of a solution to derivatives of the solution itself. Applying this identity, we find that for all $x \in \mathcal{H}^+$
\begin{equation}
    \nabla_\mu T^{t^* \mu} [\tilde{f}_\epsilon] = \int_{\mathcal{P}_x} p^{t^*} X(\tilde{f}_\epsilon) \, d \mu_x = \int_{\mathcal{P}_x} p^{t^*} X \left( \frac{p^{t^*}}{\epsilon - p^r} \right) f \, d \mu_x.
\end{equation}
We compute the limit as $\epsilon \rightarrow 0$ and readily find
\begin{equation}
    \lim_{\epsilon \rightarrow 0} \nabla_\mu T^{t^* \mu} [\tilde{f}_\epsilon] = - \frac{2}{m} \left( T^{t^* t^*}[f] + T^{t^* r}[f] \right).
\end{equation}
Next, we show that the angular derivatives integrate to zero when integrated over the unit sphere. This argument does not rely on the specific form of $\tilde{f}_\epsilon$, so let us for the moment omit the dependence on $\tilde{f}_\epsilon$ and write $T^{\mu \nu} = T^{\mu \nu}[\tilde{f}_\epsilon]$. Then it follows that
\begin{equation}
    \nabla_A \left( T^{A \mu} (\partial_{t^*})_\mu \right) = \left( \nabla_A T^{A \mu} \right) (\partial_{t^*})_\mu + g_{\mu \nu} \Gamma^{\nu}_{A t^*} T^{A \mu} = \nabla_A T^{A t^*},
\end{equation}
where we used that in the whole exterior $\Gamma^{\nu}_{A t^*} = 0$ for all $\nu$. Note carefully that by choosing a time $t^* >0$ and $r=m$ we obtain an embedding of $\iota: \sphere \hookrightarrow \Mern$ such that the Lorentzian metric $\gERN$ restricts to the Riemannian metric $\iota^* \gERN = m^2 d \omega$ where $d \omega$ denotes the standard round metric on $\sphere$. This induces the natural restriction map $\Gamma(T \Mern) \rightarrow \Gamma(T\sphere)$ of vector fields on $\Mern$ to vector fields on $\sphere$, by virtue of the orthogonal decomposition $\iota^* T \Mern = 
T \Mern|_{\sphere} = T \sphere \oplus T^{\perp} \sphere$. Expressed in coordinates, this means that for any choice of coordinates $\omega$ on $\sphere$ and fixed $t^* > 0$ and $r=m$ the restriction of the vector field with components $(T^{t^* \mu})_{\mu = t^*,r,A,B}$ has the components $(T^{\mu t^*})_{\mu = A,B}$, where as usual we denote the spherical indices by capital letters $A,B$. Therefore the spherical part of the divergence may be rewritten as the divergence of a vector field on $\sphere$ so that upon integration
\begin{equation} \label{proof_nondecay_equ}
    \int_{\sphere} \nabla_\mu T^{t^* \mu} [\tilde{f}_\epsilon] \, d \omega \bigg|_{r=m} = \int_{\sphere} \nabla_{t^*} T^{t^* t^*} [\tilde{f}_\epsilon] + \nabla_{r} T^{t^* r} [\tilde{f}_\epsilon] \, d \omega \bigg|_{r=m} .
\end{equation}
Let us now study the two remaining terms and notice that by the definition of the covariant derivative we find that along the event horizon
\begin{equation}
    \nabla_{t^*} T^{t^* t^*} \Big|_{r=m} = \partial_{t^*} T^{t^* t^*} + 2 \Gamma^{t^*}_{\alpha t^*} T^{\alpha t^*} = \partial_{t^*} T^{t^* t^*}.
\end{equation}
Recall that $\partial_{t^*} \in \Gamma(T \Mern)$ is the timelike Killing vector field associated to the stationarity of the Reissner--Nordstr\"om solution. This implies the identity $\partial_{t^*} T^{t^* t^*}[f] = T^{t^* t^*}[ \partial_{t^*}  f]$, where by a slight abuse of notation we use the same symbol $\partial_{t^*}$ to denote the coordinate vector field of the coordinates $(t^*,r,\omega)$ on $\Mern$ on the left hand side, and the coordinate vector field of $(t^*,r,\omega,p^r,\pslash)$ on $\mathcal{P}$ on the right hand side. More formally, we consider the complete lift of the Killing vector field $\partial_{t^*}$ to the mass-shell here, see again~\cite{martin,riosecosarbach,sarbachzannias} for a discussion in greater generality. It also readily follows that $[X,\partial_{t^*}] = 0$ so that $\partial_{t^*} f$ is again a solution to the massless Vlasov equation, where we consider $\partial_{t^*}$ as a vector field on $\mathcal{P}$. Therefore
\begin{equation}
    \int_{\sphere} \nabla_{t^*} T^{t^* t^*}[\tilde{f}_\epsilon] \, d \omega \bigg|_{r=m} = \int_{\sphere} T^{t^* t^*}\left[ \frac{p^{t^*}}{\epsilon - p^r} \partial_{t^*} f \right] \, d \omega \bigg|_{r=m}.
\end{equation}
Taking the limit as $\epsilon \rightarrow 0$ we then find
\begin{equation}
    \lim_{\epsilon \rightarrow 0} \int_{\sphere} \nabla_{t^*} T^{t^* t^*}[\tilde{f}_\epsilon] \, d \omega \bigg|_{r=m} = \int_{\sphere} T^{t^* t^*}\left[ \frac{p^{t^*}}{\left| p^r \right|} \partial_{t^*} f \right] \, d \omega \bigg|_{r=m}.
\end{equation}
For the second term we note that by definition of the covariant derivative
\begin{equation}
    \nabla_{r} T^{t^* r} \Big|_{r=m} = \partial_{r} T^{t^* r} + \Gamma^{t^*}_{\alpha r} T^{\alpha r} + \Gamma^{r}_{\alpha r} T^{\alpha t^*} = \partial_{r} T^{t^* r},
\end{equation}
which allows us to conclude 
\begin{equation}
    \nabla_{r} T^{t^* r}[\tilde{f}_\epsilon] \Big|_{r=m} = \partial_{r} T^{t^* r}[\tilde{f}_\epsilon] \Big|_{r=m} .
\end{equation}
Finally, we note that $T^{t^* r}[\tilde{f}_\epsilon] \rightarrow T^{t^* t^*}[f]$ and $\partial_r T^{t^* r}[\tilde{f}_\epsilon] \rightarrow \partial_r T^{t^* t^*}[f]$ for every $x \in \mathcal{H}^+$ as $\epsilon \rightarrow 0$. Therefore we find
\begin{equation}
    \lim_{\epsilon \rightarrow 0} \int_{\sphere} \nabla_{r} T^{t^* r}[\tilde{f}_\epsilon] \, d \omega \bigg|_{r=m} = \left( \partial_{r} \int_{\sphere} T^{t^* t^*}[f] \, d \omega \right) \bigg|_{r=m}.
\end{equation}
Finally we use equation~\eqref{proof_nondecay_equ} above and take the limit $\epsilon \rightarrow 0$ to see that
\begin{equation}
    \left( \partial_{r} \int_{\sphere} T^{t^* t^*}[f] \, d \omega \right) \bigg|_{r=m} = \int_{\sphere} T^{t^* t^*}\left[ - \frac{p^{t^*}}{p^r} \partial_{t^*} f \right] \, d \omega \bigg|_{r=m} + \mathcal{E},
\end{equation}
where we have abbreviated
\begin{equation}
    \mathcal{E} := - \frac{2}{m} \int_{\sphere} \left( T^{t^* t^*}[f] + T^{t^* r}[f] \right) \, d \omega \bigg|_{r=m}.
\end{equation}
Note that for $(x,p) \in \supp(f)$ with $x \in \mathcal{H}^+$ we have $p^r \leq 0$, so that $- p^r = \left| p^r \right|$. By Theorem~\ref{maintheoremERNprecise} above, for any $x \in \mathcal{H}^+$ with $\tau(x) > \tauzero$ we have the bounds (recall that $T^{t^* t^*}[f] \geq 0$)
\begin{equation}
    \left| T^{t^* r}[f] \right|, T^{t^* t^*}[f] \lesssim \psuppconst^4 \rsuppconst^2 \| f_0 \|_{L^{\infty}} \frac{1}{(\tau(x) - \tauzero)^2},
\end{equation}
so that $\mathcal{E}$ satisfies inequality~\eqref{proof_nondecay_errorbound} as claimed above.

\paragraph{Step 2: Separating off the decaying part.}

We now have an expression for the transversal derivative of $\int_{\sphere} T^{t^* t^*}[f] \, d \omega$ evaluated along the event horizon. The next step is to identify the part of the expression that is decaying. In what follows, we will then show that the remainder is indeed non-decaying. Recall that Proposition~\ref{psupport_propERN} implies that all $(x,p) \in \supp(f)$ with $\tau(x) \geq \tauzero$ must be either almost trapped at the photon sphere or at the event horizon, in other words $(x,p) \in \ERNtrappedsupportset \cup \ERNsmallsupportset$. For $x \in \mathcal{H}^+$ with $\tau(x) > \tauzero$ we therefore conclude
\begin{equation}
    T^{t^* t^*}\left[ \frac{p^{t^*}}{\left|p^r \right|} \partial_{t^*} f \right] = \int_{\ERNtrappedsupportsetx} \frac{(p^{t^*})^3}{\left| p^r \right|} \partial_{t^*} f \, d \mu_x + \int_{ \ERNsmallsupportsetx \setminus \ERNtrappedsupportsetx} \frac{(p^{t^*})^3}{\left| p^r \right|} \partial_{t^*} f \, d \mu_x.
\end{equation}
By definition for all $(x,p) \in \ERNtrappedsupportset$ with $\tau(x) \geq \tauzero$ we have $m \frac{\left| p^r \right|}{L} \geq \frac{1}{8}$ and $m \frac{p^{t^*}}{L} \leq \frac{9}{2}$. Therefore
\begin{equation}
    \frac{p^{t^*}}{\left| p^r \right|} = \frac{m \frac{p^{t^*}}{L}}{m \frac{\left| p^r \right|}{L}} \lesssim 1.
\end{equation}
This implies that
\begin{equation}
    \left| \int_{\ERNtrappedsupportsetx} \frac{(p^{t^*})^3}{\left| p^r \right|} \partial_{t^*} f \, d \mu_x \right| \lesssim \psuppconst^4 \frac{\rsuppconst^2}{m^2} \| (\partial_{t^*}f)_0 \|_{L^\infty} e^{-\decayrate (\tau(x)-\tauzero)},
\end{equation}
so that we may absorb the spherical average of this term into the error term $\mathcal{E}$.

\paragraph{Step 3: Proving the lower bound.}
It now remains to estimate the term
\begin{equation}
	\int_{ \ERNsmallsupportsetx \setminus \ERNtrappedsupportsetx} \frac{(p^{t^*})^3}{\left| p^r \right|} \partial_{t^*} f \, d \mu_x
\end{equation}
from below. To accomplish this, we first show that $\partial_{t^*} f$ has a sign on the set $\ERNsmallsupportsetx \setminus \ERNtrappedsupportsetx$ for all points $x \in \mathcal{H}^+$ with $\tau(x) > \tauslow \sim \delta^{-1}$. Then we show that for such late times, the support of $f$ contains the set $\slowsupportsetx$ as a subset. This is then used by noting that on $\slowsupportsetx$ we can estimate $\frac{p^{t^*}}{\left| p^r \right|}$ from below. \\

To see that $\partial_{t^*} f \geq 0$ on the set $\ERNsmallsupportsetx \setminus \ERNtrappedsupportsetx$ for every $x \in \mathcal{H}^+$ with $\tau(x) > \tauslow \sim \delta^{-1}$, recall that $\partial_{t^*} f$ solves the massless Vlasov equation. Therefore, we may equivalently show that the statement holds on the set
\begin{equation}
	\mathcal{D}_0 := \pi_0 \big( (\smallsupportset \setminus \trappedsupportset) \cap \supp(f) \cap \left\{ \tau \geq \tauslow \right\} \big),
\end{equation}
where we recall that $\pi_0 : \mathcal{P} \rightarrow \mathcal{P}_0$ transports points backwards in time along the geodesic flow until it intersects the initial hypersurface $\Sigma_0$, see Definition~\ref{defn_pi0}. Proposition~\ref{psupport_propERN} implies that $\mathcal{D}_0 \subset \badsetaplarge_{\delta}$, so that the claim follows from our assumption that $\partial_{t^*} f \geq 0$ on $\badsetaplarge_{\delta}$. \\

Recall that we assume $\badsetaplarge_\delta \subset \supp(f_0)$. Proposition~\ref{sharpness_lemma} implies that $\badsetapprox_\delta \subset \badsetaplarge_\delta$, so that by Proposition~\ref{sharpness_lemma} again we find $\slowsupportset_{\tauslow} \subset \supp(f)|_{\mathcal{H}^+}$, where the time parameter $\tauslow$ satisfies $\tauslow \sim \delta^{-1}$. By definition of $\slowsupportset_\tauslow$ and Proposition~\ref{ptstar_size_slowsupport} we see that for $(x,p) \in \slowsupportset_\tauslow$ the bound
\begin{equation}
	\frac{p^{t^*}}{\left| p^r \right|} \geq \frac{(C_1)^2}{2 (C_2)^2} \tau(x)^2
\end{equation}
holds, where $C_1,C_2$ denote the constants used in the definition of $\slowsupportset_\tauslow$. We therefore find
\begin{equation}
    \int_{ \ERNsmallsupportsetx \setminus \ERNtrappedsupportsetx} \frac{(p^{t^*})^3}{\left| p^r \right|} \partial_{t^*} f \, d \mu_x \geq \int_{ \slowsupportset_{\tauslow,x}} \frac{(p^{t^*})^3}{\left| p^r \right|} \partial_{t^*} f \, d \mu_x \gtrsim \psuppconst^2 \tau(x)^2 \int_{ \slowsupportset_{\tauslow,x}} \left| \partial_{t^*} f \right| \, d \mu_x.
\end{equation}
By definition of $\badsettau$ and since $\partial_{t^*} f$ solves the massless Vlasov equation, for all $(x,p) \in \badsettau$
\begin{equation}
	\left| \partial_{t^*} f (x,p) \right| \geq \inf_{\badsettau} \left| \partial_{t^*} f_0 \right|.
\end{equation}
We conclude
\begin{equation}
    \int_{ \ERNsmallsupportsetx} \frac{(p^{t^*})^3}{\left| p^r \right|} \partial_{t^*} f \, d \mu_x \gtrsim \psuppconst^2 \tau(x)^2 \vol \slowsupportsettaux \left( \inf_{\badsettau} \left| \partial_{t^*} f_0 \right| \right) \sim \psuppconst^4 \left( \inf_{\badsettau} \left| \partial_{t^*} f_0 \right| \right),
\end{equation}
where we made use of Proposition~\ref{ptstar_size_slowsupport} to estimate the volume of $\slowsupportsettaux$ in the last step. Using that for any $\delta > 0$, all remaining constants involved in the definition of the sets may be chosen such that $\badset \subset \badsetapprox_\delta \subset \badsetaplarge_{\delta}$ as proven in Proposition~\ref{sharpness_lemma}, we finally conclude
\begin{equation}
    \left( \partial_{r} \int_{\sphere} T^{t^* t^*}[f] \, d \omega \right) \bigg|_{r=m} \gtrsim \psuppconst^4 \left( \inf_{\badsettau} \left| \partial_{t^*} f_0 \right| \right) + \mathcal{E} \geq \psuppconst^4
    \left( \inf_{\badsetaplarge_\delta} \left| \partial_{t^*} f_0 \right| \right) + \mathcal{E}.
\end{equation}
Given the error bound~\eqref{proof_nondecay_errorbound}, this allows us to conclude the desired lower bound if we assume that $\tau(x) - \tauzero \geq C m \left( \inf\nolimits_{\badsetaplarge_\delta} \left| \partial_{t^*} f_0 \right| \right)^{-\frac{1}{2}}$ with a large enough constant $C > 1$.
\end{proof}

\begin{rem}
Having seen the proof of Theorem~\ref{ERN_nondecaytransversal}, let us elaborate on how the argument would differ if it were to be applied to the subextremal case. Step~1 and step~2 of the argument apply in an identical manner, so that the question remains on how to bound the term
\begin{equation} \label{eqn_whatever}
	\int_{ \ERNsmallsupportsetx \setminus \ERNtrappedsupportsetx} \frac{(p^{t^*})^3}{\left| p^r \right|} \partial_{t^*} f \, d \mu_x.
\end{equation}
We first note that in the subextremal case, if $p \in \ERNsmallsupportsetx$ for $x \in \mathcal{H}^+$ then every component $p^\mu, \mu \in \{ t^*, r, \theta, \phi \}$ decays at an exponential rate. In addition, the volume of $\ERNsmallsupportsetx \setminus \ERNtrappedsupportsetx$ decays exponentially in $\tau$. To establish a lower bound on the integral in~\eqref{eqn_whatever} we would therefore need to demonstrate that $p^r$ decays at a \emph{faster} exponential rate than $p^{t^*}$. However we have shown in the proof of Proposition~\ref{psupport_prop} that $p^r$ and $p^{t^*}$ in general decay at the same exponential rate in the subextremal case. This is where the argument fails if it were applied to the subextremal case.
\end{rem}

\begin{rem} \label{rem:difficulty}
We remark on the difficulties with proving the sharpness of the lower bound in Theorem~\ref{ERN_nondecaytransversal}, or in other words of proving an upper bound for the quantity
\begin{equation}
    \left( \partial_{r} \int_{\sphere} T^{t^* t^*}[f] \, d \omega \right) \bigg|_{r=m}.
\end{equation}
Equality~\eqref{proof_nondecay_decomposition} implies that equivalently, we may establish an upper bound of 
\begin{equation}
    \int_{ \supp(f(x, \cdot)} \frac{(p^{t^*})^3}{\left| p^r \right|} \partial_{t^*} f \, d \mu_x.
\end{equation}
The definition of the volume form $d \mu_x$ in equation~\eqref{dmu_first_comp_schw} implies that, along the event horizon, $d \mu_x \sim \frac{1}{\left| p^r \right|} \, d p^r d p^\theta d p^\phi$. Therefore, an application of Lemma~\ref{ptstar_bounds_ern} yields
\begin{equation}
    \int_{ \supp(f(x, \cdot)} \frac{(p^{t^*})^3}{(p^r)^2} \partial_{t^*} f \, d p^r d p^\theta d p^\phi \sim \int_{ \supp(f(x, \cdot)} E (1 + \left| \trapern \right|)^3 \, \partial_{t^*} f \, d E d \pslash,
\end{equation}
along the event horizon. However, an inspection of Assumption~\ref{assumption_support} reveals that the quantity on the right is in general unbounded in the support of $f$. The author suspects that there is an interaction with the $\partial_{t^*}$ derivative, which may balance out the integral so that it may be shown to be bounded, but at present it is not clear to the author how to exploit this.
\end{rem}

\bibliography{masslessvlasov}% common bib file


\begin{thebibliography}{49}
\ifx \bisbn   \undefined \def \bisbn  #1{ISBN #1}\fi
\ifx \binits  \undefined \def \binits#1{#1}\fi
\ifx \bauthor  \undefined \def \bauthor#1{#1}\fi
\ifx \batitle  \undefined \def \batitle#1{#1}\fi
\ifx \bjtitle  \undefined \def \bjtitle#1{#1}\fi
\ifx \bvolume  \undefined \def \bvolume#1{\textbf{#1}}\fi
\ifx \byear  \undefined \def \byear#1{#1}\fi
\ifx \bissue  \undefined \def \bissue#1{#1}\fi
\ifx \bfpage  \undefined \def \bfpage#1{#1}\fi
\ifx \blpage  \undefined \def \blpage #1{#1}\fi
\ifx \burl  \undefined \def \burl#1{\textsf{#1}}\fi
\ifx \doiurl  \undefined \def \doiurl#1{\url{https://doi.org/#1}}\fi
\ifx \betal  \undefined \def \betal{\textit{et al.}}\fi
\ifx \binstitute  \undefined \def \binstitute#1{#1}\fi
\ifx \binstitutionaled  \undefined \def \binstitutionaled#1{#1}\fi
\ifx \bctitle  \undefined \def \bctitle#1{#1}\fi
\ifx \beditor  \undefined \def \beditor#1{#1}\fi
\ifx \bpublisher  \undefined \def \bpublisher#1{#1}\fi
\ifx \bbtitle  \undefined \def \bbtitle#1{#1}\fi
\ifx \bedition  \undefined \def \bedition#1{#1}\fi
\ifx \bseriesno  \undefined \def \bseriesno#1{#1}\fi
\ifx \blocation  \undefined \def \blocation#1{#1}\fi
\ifx \bsertitle  \undefined \def \bsertitle#1{#1}\fi
\ifx \bsnm \undefined \def \bsnm#1{#1}\fi
\ifx \bsuffix \undefined \def \bsuffix#1{#1}\fi
\ifx \bparticle \undefined \def \bparticle#1{#1}\fi
\ifx \barticle \undefined \def \barticle#1{#1}\fi
\bibcommenthead
\ifx \bconfdate \undefined \def \bconfdate #1{#1}\fi
\ifx \botherref \undefined \def \botherref #1{#1}\fi
\ifx \url \undefined \def \url#1{\textsf{#1}}\fi
\ifx \bchapter \undefined \def \bchapter#1{#1}\fi
\ifx \bbook \undefined \def \bbook#1{#1}\fi
\ifx \bcomment \undefined \def \bcomment#1{#1}\fi
\ifx \oauthor \undefined \def \oauthor#1{#1}\fi
\ifx \citeauthoryear \undefined \def \citeauthoryear#1{#1}\fi
\ifx \endbibitem  \undefined \def \endbibitem {}\fi
\ifx \bconflocation  \undefined \def \bconflocation#1{#1}\fi
\ifx \arxivurl  \undefined \def \arxivurl#1{\textsf{#1}}\fi
\csname PreBibitemsHook\endcsname

%%% 1
\bibitem[\protect\citeauthoryear{Hawking and Ellis}{1973}]{general}
\begin{botherref}
\oauthor{\bsnm{Hawking}, \binits{S.W.}},
\oauthor{\bsnm{Ellis}, \binits{G.F.R.}}:
{The large scale structure of space-time}
\textbf{1}
(1973)
\end{botherref}
\endbibitem

%%% 2
\bibitem[\protect\citeauthoryear{Rodr{\'\i}guez~Amado}{2023}]{uniqueness}
\begin{botherref}
\oauthor{\bsnm{Rodr{\'\i}guez~Amado}, \binits{E.}}:
{Birkhoff’s theorem for Einstein gravity and beyond}
(2023)
\end{botherref}
\endbibitem

%%% 3
\bibitem[\protect\citeauthoryear{Poisson and Israel}{1989}]{poissonisreal1}
\begin{barticle}
\bauthor{\bsnm{Poisson}, \binits{E.}},
\bauthor{\bsnm{Israel}, \binits{W.}}:
\batitle{{Inner-horizon instability and mass inflation in black holes}}.
\bjtitle{Physical review letters}
\bvolume{63}(\bissue{16}),
\bfpage{1663}
(\byear{1989})
\end{barticle}
\endbibitem

%%% 4
\bibitem[\protect\citeauthoryear{Poisson and Israel}{1990}]{poissonisreal2}
\begin{barticle}
\bauthor{\bsnm{Poisson}, \binits{E.}},
\bauthor{\bsnm{Israel}, \binits{W.}}:
\batitle{{Internal structure of black holes}}.
\bjtitle{Physical Review D}
\bvolume{41}(\bissue{6}),
\bfpage{1796}
(\byear{1990})
\end{barticle}
\endbibitem

%%% 5
\bibitem[\protect\citeauthoryear{Moschidis}{2022}]{moschidismasslessvlasov}
\begin{botherref}
\oauthor{\bsnm{Moschidis}, \binits{G.}}:
{A proof of the instability of AdS for the Einstein-massless Vlasov system}.
Inventiones mathematicae,
1--206
(2022)
\end{botherref}
\endbibitem

%%% 6
\bibitem[\protect\citeauthoryear{Moschidis}{2020}]{moschidisnulldust}
\begin{barticle}
\bauthor{\bsnm{Moschidis}, \binits{G.}}:
\batitle{{A proof of the instability of AdS for the Einstein-null dust system
  with an inner mirror}}.
\bjtitle{Analysis \& PDE}
\bvolume{13}(\bissue{6}),
\bfpage{1671}--\blpage{1754}
(\byear{2020})
\end{barticle}
\endbibitem

%%% 7
\bibitem[\protect\citeauthoryear{Dafermos and
  Holzegel}{2006}]{adsinstabilityconj}
\begin{botherref}
\oauthor{\bsnm{Dafermos}, \binits{M.}},
\oauthor{\bsnm{Holzegel}, \binits{G.}}:
{Dynamic instability of solitons in $4 + 1$-dimensional gravity with negative
  cosmological constant}
(2006)
\end{botherref}
\endbibitem

%%% 8
\bibitem[\protect\citeauthoryear{O'neill}{1983}]{oneill}
\begin{botherref}
\oauthor{\bsnm{O'neill}, \binits{B.}}:
{Semi-Riemannian geometry with applications to relativity}
(1983)
\end{botherref}
\endbibitem

%%% 9
\bibitem[\protect\citeauthoryear{Wald}{2010}]{waldbook}
\begin{botherref}
\oauthor{\bsnm{Wald}, \binits{R.M.}}:
{General relativity}
(2010)
\end{botherref}
\endbibitem

%%% 10
\bibitem[\protect\citeauthoryear{Bigorgne}{2020}]{leo}
\begin{botherref}
\oauthor{\bsnm{Bigorgne}, \binits{L.}}:
{Decay estimates for the massless Vlasov equation on Schwarzschild spacetimes}.
arXiv preprint arXiv:2006.03579
(2020)
\end{botherref}
\endbibitem

%%% 11
\bibitem[\protect\citeauthoryear{Dafermos and Rodnianski}{2010}]{rp}
\begin{bchapter}
\bauthor{\bsnm{Dafermos}, \binits{M.}},
\bauthor{\bsnm{Rodnianski}, \binits{I.}}:
\bctitle{{A new physical-space approach to decay for the wave equation with
  applications to black hole spacetimes}}.
In: \bbtitle{XVIth International Congress On Mathematical Physics},
pp. \bfpage{421}--\blpage{432}
(\byear{2010}).
\bcomment{World Scientific}
\end{bchapter}
\endbibitem

%%% 12
\bibitem[\protect\citeauthoryear{Velozo~Ruiz}{2022}]{renato}
\begin{botherref}
\oauthor{\bsnm{Velozo~Ruiz}, \binits{R.}}:
{Linear and non-linear collisionless many-particle systems}.
PhD thesis
(August 2022)
\end{botherref}
\endbibitem

%%% 13
\bibitem[\protect\citeauthoryear{Andersson et~al.}{2018}]{anderssonblue}
\begin{barticle}
\bauthor{\bsnm{Andersson}, \binits{L.}},
\bauthor{\bsnm{Blue}, \binits{P.}},
\bauthor{\bsnm{Joudioux}, \binits{J.}}:
\batitle{{Hidden symmetries and decay for the Vlasov equation on the Kerr
  spacetime}}.
\bjtitle{Communications in Partial Differential Equations}
\bvolume{43}(\bissue{1}),
\bfpage{47}--\blpage{65}
(\byear{2018})
\end{barticle}
\endbibitem

%%% 14
\bibitem[\protect\citeauthoryear{Wald}{1979}]{wald}
\begin{barticle}
\bauthor{\bsnm{Wald}, \binits{R.M.}}:
\batitle{{Note on the stability of the Schwarzschild metric}}.
\bjtitle{Journal of Mathematical Physics}
\bvolume{20}(\bissue{6}),
\bfpage{1056}--\blpage{1058}
(\byear{1979})
\end{barticle}
\endbibitem

%%% 15
\bibitem[\protect\citeauthoryear{Kay and Wald}{1987}]{kaywald}
\begin{barticle}
\bauthor{\bsnm{Kay}, \binits{B.S.}},
\bauthor{\bsnm{Wald}, \binits{R.M.}}:
\batitle{{Linear stability of Schwarzschild under perturbations which are
  non-vanishing on the bifurcation 2-sphere}}.
\bjtitle{Classical and Quantum Gravity}
\bvolume{4}(\bissue{4}),
\bfpage{893}
(\byear{1987})
\end{barticle}
\endbibitem

%%% 16
\bibitem[\protect\citeauthoryear{Dafermos and
  Rodnianski}{2012}]{kerrsubextremalfull}
\begin{bchapter}
\bauthor{\bsnm{Dafermos}, \binits{M.}},
\bauthor{\bsnm{Rodnianski}, \binits{I.}}:
\bctitle{{The black hole stability problem for linear scalar perturbations}}.
In: \bbtitle{The Twelfth Marcel Grossmann Meeting: On Recent Developments in
  Theoretical and Experimental General Relativity, Astrophysics and
  Relativistic Field Theories (In 3 Volumes)},
pp. \bfpage{132}--\blpage{189}
(\byear{2012}).
\bcomment{World Scientific}
\end{bchapter}
\endbibitem

%%% 17
\bibitem[\protect\citeauthoryear{Dafermos
  et~al.}{2016}]{dafermosrodnianskishlap}
\begin{botherref}
\oauthor{\bsnm{Dafermos}, \binits{M.}},
\oauthor{\bsnm{Rodnianski}, \binits{I.}},
\oauthor{\bsnm{Shlapentokh--Rothman}, \binits{Y.}}:
{Decay for solutions of the wave equation on Kerr exterior spacetimes III: The
  full subextremal case $|a|< M$}.
Annals of mathematics,
787--913
(2016)
\end{botherref}
\endbibitem

%%% 18
\bibitem[\protect\citeauthoryear{Moschidis}{2016}]{moschidis2016r}
\begin{barticle}
\bauthor{\bsnm{Moschidis}, \binits{G.}}:
\batitle{{The $r^p$-weighted energy method of Dafermos and Rodnianski in
  general asymptotically flat spacetimes and applications}}.
\bjtitle{Annals of PDE}
\bvolume{2},
\bfpage{1}--\blpage{194}
(\byear{2016})
\end{barticle}
\endbibitem

%%% 19
\bibitem[\protect\citeauthoryear{Angelopoulos et~al.}{2021}]{priceslawRN}
\begin{botherref}
\oauthor{\bsnm{Angelopoulos}, \binits{Y.}},
\oauthor{\bsnm{Aretakis}, \binits{S.}},
\oauthor{\bsnm{Gajic}, \binits{D.}}:
{Price's law and precise late-time asymptotics for subextremal
  Reissner--Nordstr\"om black holes}.
arXiv preprint arXiv:2102.11888
(2021)
\end{botherref}
\endbibitem

%%% 20
\bibitem[\protect\citeauthoryear{Angelopoulos et~al.}{2023}]{priceslawKerr}
\begin{botherref}
\oauthor{\bsnm{Angelopoulos}, \binits{Y.}},
\oauthor{\bsnm{Aretakis}, \binits{S.}},
\oauthor{\bsnm{Gajic}, \binits{D.}}:
{Late-time tails and mode coupling of linear waves on Kerr spacetimes}.
Advances in Mathematics
\textbf{417}
(2023)
\end{botherref}
\endbibitem

%%% 21
\bibitem[\protect\citeauthoryear{Angelopoulos et~al.}{2018}]{priceslaw2018}
\begin{barticle}
\bauthor{\bsnm{Angelopoulos}, \binits{Y.}},
\bauthor{\bsnm{Aretakis}, \binits{S.}},
\bauthor{\bsnm{Gajic}, \binits{D.}}:
\batitle{{Late-time asymptotics for the wave equation on spherically symmetric,
  stationary spacetimes}}.
\bjtitle{Advances in Mathematics}
\bvolume{323},
\bfpage{529}--\blpage{621}
(\byear{2018})
\end{barticle}
\endbibitem

%%% 22
\bibitem[\protect\citeauthoryear{Hintz}{2022}]{hintz2022sharp}
\begin{barticle}
\bauthor{\bsnm{Hintz}, \binits{P.}}:
\batitle{{A sharp version of Price’s law for wave decay on asymptotically
  flat spacetimes}}.
\bjtitle{Communications in Mathematical Physics}
\bvolume{389}(\bissue{1}),
\bfpage{491}--\blpage{542}
(\byear{2022})
\end{barticle}
\endbibitem

%%% 23
\bibitem[\protect\citeauthoryear{Dafermos
  et~al.}{2021}]{schwarzschildnonlinearstable}
\begin{botherref}
\oauthor{\bsnm{Dafermos}, \binits{M.}},
\oauthor{\bsnm{Holzegel}, \binits{G.}},
\oauthor{\bsnm{Rodnianski}, \binits{I.}},
\oauthor{\bsnm{Taylor}, \binits{M.}}:
{The non-linear stability of the Schwarzschild family of black holes}.
arXiv preprint arXiv:2104.08222
(2021)
\end{botherref}
\endbibitem

%%% 24
\bibitem[\protect\citeauthoryear{Giorgi et~al.}{2022}]{klainerman3}
\begin{botherref}
\oauthor{\bsnm{Giorgi}, \binits{E.}},
\oauthor{\bsnm{Klainerman}, \binits{S.}},
\oauthor{\bsnm{Szeftel}, \binits{J.}}:
{Wave equations estimates and the nonlinear stability of slowly rotating Kerr
  black holes}.
arXiv preprint arXiv:2205.14808
(2022)
\end{botherref}
\endbibitem

%%% 25
\bibitem[\protect\citeauthoryear{Aretakis}{2011a}]{aretakis1}
\begin{barticle}
\bauthor{\bsnm{Aretakis}, \binits{S.}}:
\batitle{{Stability and instability of extreme Reissner--Nordstr{\"o}m black
  hole spacetimes for linear scalar perturbations I}}.
\bjtitle{Communications in mathematical physics}
\bvolume{307}(\bissue{1}),
\bfpage{17}
(\byear{2011})
\end{barticle}
\endbibitem

%%% 26
\bibitem[\protect\citeauthoryear{Aretakis}{2011b}]{aretakis2}
\begin{bchapter}
\bauthor{\bsnm{Aretakis}, \binits{S.}}:
\bctitle{{Stability and instability of extreme Reissner--Nordstr{\"o}m black
  hole spacetimes for linear scalar perturbations II}}.
In: \bbtitle{Annales Henri Poincar{\'e}},
vol. \bseriesno{12},
pp. \bfpage{1491}--\blpage{1538}
(\byear{2011}).
\bcomment{Springer}
\end{bchapter}
\endbibitem

%%% 27
\bibitem[\protect\citeauthoryear{Angelopoulos et~al.}{2017}]{aretakis3}
\begin{bchapter}
\bauthor{\bsnm{Angelopoulos}, \binits{Y.}},
\bauthor{\bsnm{Aretakis}, \binits{S.}},
\bauthor{\bsnm{Gajic}, \binits{D.}}:
\bctitle{{The trapping effect on degenerate horizons}}.
In: \bbtitle{Annales Henri Poincar{\'e}},
vol. \bseriesno{18},
pp. \bfpage{1593}--\blpage{1633}
(\byear{2017}).
\bcomment{Springer}
\end{bchapter}
\endbibitem

%%% 28
\bibitem[\protect\citeauthoryear{Aretakis}{2012}]{subextkerr}
\begin{barticle}
\bauthor{\bsnm{Aretakis}, \binits{S.}}:
\batitle{{Decay of axisymmetric solutions of the wave equation on extreme Kerr
  backgrounds}}.
\bjtitle{Journal of Functional Analysis}
\bvolume{263}(\bissue{9}),
\bfpage{2770}--\blpage{2831}
(\byear{2012})
\end{barticle}
\endbibitem

%%% 29
\bibitem[\protect\citeauthoryear{Angelopoulos
  et~al.}{2020}]{angelopoulos2020late}
\begin{botherref}
\oauthor{\bsnm{Angelopoulos}, \binits{Y.}},
\oauthor{\bsnm{Aretakis}, \binits{S.}},
\oauthor{\bsnm{Gajic}, \binits{D.}}:
{Late-time asymptotics for the wave equation on extremal
  Reissner--Nordstr{\"o}m backgrounds}.
Advances in Mathematics
\textbf{375}
(2020)
\end{botherref}
\endbibitem

%%% 30
\bibitem[\protect\citeauthoryear{Taylor}{2017}]{martin}
\begin{barticle}
\bauthor{\bsnm{Taylor}, \binits{M.}}:
\batitle{{The global nonlinear stability of Minkowski space for the massless
  Einstein--Vlasov system}}.
\bjtitle{Annals of PDE}
\bvolume{3}(\bissue{1}),
\bfpage{9}
(\byear{2017})
\end{barticle}
\endbibitem

%%% 31
\bibitem[\protect\citeauthoryear{Joudioux et~al.}{2021}]{minkstab2}
\begin{bchapter}
\bauthor{\bsnm{Joudioux}, \binits{J.}},
\bauthor{\bsnm{Thaller}, \binits{M.}},
\bauthor{\bsnm{Valiente~Kroon}, \binits{J.A.}}:
\bctitle{{The conformal Einstein field equations with massless Vlasov matter}}.
In: \bbtitle{Annales de l'Institut Fourier},
vol. \bseriesno{71},
pp. \bfpage{799}--\blpage{842}
(\byear{2021})
\end{bchapter}
\endbibitem

%%% 32
\bibitem[\protect\citeauthoryear{Bigorgne et~al.}{2021}]{fajman}
\begin{barticle}
\bauthor{\bsnm{Bigorgne}, \binits{L.}},
\bauthor{\bsnm{Fajman}, \binits{D.}},
\bauthor{\bsnm{Joudioux}, \binits{J.}},
\bauthor{\bsnm{Smulevici}, \binits{J.}},
\bauthor{\bsnm{Thaller}, \binits{M.}}:
\batitle{{Asymptotic stability of Minkowski space-time with non-compactly
  supported massless Vlasov matter}}.
\bjtitle{Archive for Rational Mechanics and Analysis}
\bvolume{242}(\bissue{1}),
\bfpage{1}--\blpage{147}
(\byear{2021})
\end{barticle}
\endbibitem

%%% 33
\bibitem[\protect\citeauthoryear{Dafermos}{2006}]{mihalisSpherSymmVlasov}
\begin{barticle}
\bauthor{\bsnm{Dafermos}, \binits{M.}}:
\batitle{{A note on the collapse of small data self-gravitating massless
  collisionless matter}}.
\bjtitle{Journal of Hyperbolic Differential Equations}
\bvolume{3}(\bissue{04}),
\bfpage{589}--\blpage{598}
(\byear{2006})
\end{barticle}
\endbibitem

%%% 34
\bibitem[\protect\citeauthoryear{Rein and Rendall}{1992}]{reinrendall}
\begin{barticle}
\bauthor{\bsnm{Rein}, \binits{G.}},
\bauthor{\bsnm{Rendall}, \binits{A.D.}}:
\batitle{{Global existence of solutions of the spherically symmetric
  Vlasov--Einstein system with small initial data}}.
\bjtitle{Communications in mathematical physics}
\bvolume{150}(\bissue{3}),
\bfpage{561}--\blpage{583}
(\byear{1992})
\end{barticle}
\endbibitem

%%% 35
\bibitem[\protect\citeauthoryear{Lindblad and Taylor}{2020}]{lindbladtaylor}
\begin{barticle}
\bauthor{\bsnm{Lindblad}, \binits{H.}},
\bauthor{\bsnm{Taylor}, \binits{M.}}:
\batitle{{Global stability of Minkowski space for the Einstein--Vlasov system
  in the harmonic gauge}}.
\bjtitle{Archive for rational mechanics and analysis}
\bvolume{235}(\bissue{1}),
\bfpage{517}--\blpage{633}
(\byear{2020})
\end{barticle}
\endbibitem

%%% 36
\bibitem[\protect\citeauthoryear{Fajman et~al.}{2021}]{fajman2}
\begin{barticle}
\bauthor{\bsnm{Fajman}, \binits{D.}},
\bauthor{\bsnm{Joudioux}, \binits{J.}},
\bauthor{\bsnm{Smulevici}, \binits{J.}}:
\batitle{{The stability of the Minkowski space for the Einstein--Vlasov
  system}}.
\bjtitle{Analysis \& PDE}
\bvolume{14}(\bissue{2}),
\bfpage{425}--\blpage{531}
(\byear{2021})
\end{barticle}
\endbibitem

%%% 37
\bibitem[\protect\citeauthoryear{Wang}{2022}]{wang2022global}
\begin{botherref}
\oauthor{\bsnm{Wang}, \binits{X.}}:
{Global stability of the Minkowski spacetime for the Einstein-Vlasov system}.
arXiv preprint arXiv:2210.00512
(2022)
\end{botherref}
\endbibitem

%%% 38
\bibitem[\protect\citeauthoryear{Bieri and Garfinkle}{2015}]{bierigarfinkle}
\begin{bchapter}
\bauthor{\bsnm{Bieri}, \binits{L.}},
\bauthor{\bsnm{Garfinkle}, \binits{D.}}:
\bctitle{{Neutrino radiation showing a Christodoulou memory effect in general
  relativity}}.
In: \bbtitle{Annales Henri Poincar{\'e}},
vol. \bseriesno{16},
pp. \bfpage{801}--\blpage{839}
(\byear{2015}).
\bcomment{Springer}
\end{bchapter}
\endbibitem

%%% 39
\bibitem[\protect\citeauthoryear{Christodoulou}{1991}]{christodoulougravwaves}
\begin{barticle}
\bauthor{\bsnm{Christodoulou}, \binits{D.}}:
\batitle{{Nonlinear nature of gravitation and gravitational-wave experiments}}.
\bjtitle{Physical review letters}
\bvolume{67}(\bissue{12}),
\bfpage{1486}
(\byear{1991})
\end{barticle}
\endbibitem

%%% 40
\bibitem[\protect\citeauthoryear{Angelopoulos et~al.}{2020}]{nonlinearwavesERN}
\begin{barticle}
\bauthor{\bsnm{Angelopoulos}, \binits{Y.}},
\bauthor{\bsnm{Aretakis}, \binits{S.}},
\bauthor{\bsnm{Gajic}, \binits{D.}}:
\batitle{{Nonlinear scalar perturbations of extremal Reissner--Nordstr{\"o}m
  spacetimes}}.
\bjtitle{Annals of PDE}
\bvolume{6}(\bissue{2}),
\bfpage{1}--\blpage{124}
(\byear{2020})
\end{barticle}
\endbibitem

%%% 41
\bibitem[\protect\citeauthoryear{Apetroaie}{2022}]{apetroaie}
\begin{botherref}
\oauthor{\bsnm{Apetroaie}, \binits{M.A.}}:
{Instability of gravitational and electromagnetic perturbations of extremal
  Reissner--Nordstr\"om spacetime}.
arXiv preprint arXiv:2211.09182
(2022)
\end{botherref}
\endbibitem

%%% 42
\bibitem[\protect\citeauthoryear{Aretakis}{2012}]{aretakisgeneralhorizons}
\begin{botherref}
\oauthor{\bsnm{Aretakis}, \binits{S.}}:
{Horizon instability of extremal black holes}.
arXiv preprint arXiv:1206.6598
(2012)
\end{botherref}
\endbibitem

%%% 43
\bibitem[\protect\citeauthoryear{{Teixeira~da~Costa}}{2020}]{rita}
\begin{barticle}
\bauthor{\bsnm{{Teixeira~da~Costa}}, \binits{R.}}:
\batitle{{Mode stability for the Teukolsky equation on extremal and subextremal
  Kerr spacetimes}}.
\bjtitle{Communications in Mathematical Physics}
\bvolume{378}(\bissue{1}),
\bfpage{705}--\blpage{781}
(\byear{2020})
\end{barticle}
\endbibitem

%%% 44
\bibitem[\protect\citeauthoryear{Lucietti and Reall}{2012}]{reall}
\begin{botherref}
\oauthor{\bsnm{Lucietti}, \binits{J.}},
\oauthor{\bsnm{Reall}, \binits{H.S.}}:
{Gravitational instability of an extreme Kerr black hole}.
Phys. Rev. D
\textbf{86}
(2012)
\end{botherref}
\endbibitem

%%% 45
\bibitem[\protect\citeauthoryear{Gajic}{2023}]{dejanforthcoming}
\begin{botherref}
\oauthor{\bsnm{Gajic}, \binits{D.}}:
{Azimuthal instabilities on extremal Kerr}.
arXiv preprint arXiv:2302.06636
(2023)
\end{botherref}
\endbibitem

%%% 46
\bibitem[\protect\citeauthoryear{Dafermos and
  Rodnianski}{2013}]{dafermosETHnotes}
\begin{barticle}
\bauthor{\bsnm{Dafermos}, \binits{M.}},
\bauthor{\bsnm{Rodnianski}, \binits{I.}}:
\batitle{{Lectures on black holes and linear waves}}.
\bjtitle{Clay Math. Proc}
\bvolume{17},
\bfpage{97}--\blpage{205}
(\byear{2013})
\end{barticle}
\endbibitem

%%% 47
\bibitem[\protect\citeauthoryear{Aretakis}{2018}]{aretakisbook}
\begin{botherref}
\oauthor{\bsnm{Aretakis}, \binits{S.}}:
{Dynamics of extremal black holes}
\textbf{33}
(2018)
\end{botherref}
\endbibitem

%%% 48
\bibitem[\protect\citeauthoryear{Rioseco and Sarbach}{2017}]{riosecosarbach}
\begin{botherref}
\oauthor{\bsnm{Rioseco}, \binits{P.}},
\oauthor{\bsnm{Sarbach}, \binits{O.}}:
{Accretion of a relativistic, collisionless kinetic gas into a Schwarzschild
  black hole}.
Classical and Quantum Gravity
\textbf{34}(9)
(2017)
\end{botherref}
\endbibitem

%%% 49
\bibitem[\protect\citeauthoryear{Sarbach and Zannias}{2014}]{sarbachzannias}
\begin{botherref}
\oauthor{\bsnm{Sarbach}, \binits{O.}},
\oauthor{\bsnm{Zannias}, \binits{T.}}:
{The geometry of the tangent bundle and the relativistic kinetic theory of
  gases}.
Classical and Quantum Gravity
\textbf{31}(8)
(2014)
\end{botherref}
\endbibitem

\end{thebibliography}

\end{document}